\documentclass[aps,prx,twocolumn,groupedaddress,showpacs,preprintnumbers,amsmath]{revtex4-1}
\usepackage{epsfig,amsmath,graphics,color,calc,hyperref}
\usepackage{amssymb}
\usepackage{array}
\usepackage{mathtools}
\usepackage{graphicx}
\usepackage{hyperref}
\usepackage{xcolor}
\usepackage{pifont}
\usepackage[bitstream-charter, greekfamily=default]{mathdesign}
\usepackage{lipsum}

\newcommand{\eg}{{\it e.g.}}
\newcommand{\ie}{{\it i.e.}}

\newcommand{\avg}[1]{\left\langle #1 \right\rangle}

\begin{document}

\title{Large-deviation analysis of rare resonances for the Many-Body localization transition}

\author{Giulio Biroli}
\affiliation{Laboratoire de Physique Statistique, Ecole Normale Sup\'erieure,
PSL Research University, 24 rue Lhomond, 75005 Paris, France}

\author{Alexander K. Hartmann}
\affiliation{Institut f\"ur Physik, Universit\"at Oldenburg, 26111 Oldenburg,
Germany}

\author{Marco Tarzia} 
\affiliation{LPTMC, CNRS-UMR 7600, Sorbonne Universit\'e, 4 Pl. Jussieu, F-75005 Paris, France}
\affiliation{Institut  Universitaire  de  France,  1  rue  Descartes,  75005  Paris,  France}

\begin{abstract}
    A central theoretical issue at the core of the current research on many-body localization (MBL) consists in characterizing the statistics of rare long-range resonances in many-body eigenstates. This is of paramount importance to understand: (i) the critical properties of the MBL transition and the mechanism for its destabilization through quantum avalanches; (ii) the unusual transport and anomalously slow out-of-equilibrium relaxation when the transition is approached from the metallic side. In order to study and characterize such long-range rare resonances, we develop a large-deviations approach based on an analogy with the physics of directed polymers in random media, and in particular with their freezing glass transition on infinite-dimensional graphs. The basic idea is to enlarge the parameter space by adding an auxiliary parameter (which plays the role of the inverse temperature in the directed polymer formulation) which allows us to fine-tune the effect of anomalously large outliers in the far-tails of the probability distributions of the transmission amplitudes between far-away many-body configurations in the Hilbert space. We first benchmark our approach onto two non-interacting paradigmatic toy models, 
    namely the single-particle Anderson model on the (loop-less) Cayley tree and the Rosenzweig-Porter random matrix ensemble, and then apply it to the study of a class of disordered quantum spin chains in a transverse field. This analysis shows the existence of a broad disorder range in which rare, long-distance resonances, that may form only for a few specific realizations of the disorder and a few specific choice of the random initial state, destabilize the MBL phase, while the genuine MBL transition is shifted to much larger values of the disorder than originally thought.
\end{abstract}


\maketitle

\tableofcontents

\section{Introduction \label{sec:introduction}}
Transport properties of quantum systems are known to be extremely sensitive to inhomogeneities and spatial disorder. A well known example is provided by the phenomenon of Anderson localization~\cite{anderson1958absence}, where diffusion is suppressed by quantum interference above a critical disorder strength (which turns out to be vanishingly small in low dimensions~\cite{abrahams1979scaling}). The fate of quantum localization in presence of many-body interactions and in regimes far from low temperature equilibrium, has been at the center of theoretical and experimental research  (see \eg~Refs.~\cite{altman2015universal,nandkishore2015many,abanin2017recent,alet2018many,abanin2019colloquium} for recent reviews), both for its fundamental interest for our basic understanding of quantum statistical mechanics, and for its practical implications in the search for mechanisms to protect quantum information: The long-time properties of a many-body localized system cannot be described by the conventional ensembles of quantum statistical mechanics, as they can remember, forever and locally, information about their initial conditions. 

Although the possibility of many-body localization (MBL) was already suggested by Anderson in his pioneering work~\cite{anderson1958absence}, the  stability of the Anderson insulator under the effect of small interactions was first thoroughly investigated only 15 years ago by the breakthrough of Basko, Aleiner, and Altshuler~\cite{basko2006metal} (see also Ref.~\cite{gornyi2005interacting}). Since then this area of research has attracted considerable interest, 
as MBL represents a completely new mechanism for ergodicity breaking: Differently from integrable systems, the MBL phase is stable to perturbations and, although an effective integrability emerges in the insulating regime~\cite{imbrie2016diagonalization,imbrie2017local,serbyn2014quantum,serbyn2013local,ros2015integrals}, the conserved quantities turn out to be all linear combinations of completely local operators. Differently from standard phase transitions, MBL is not associated with any spontaneous symmetry breaking, and occurs without any signature in the static observables (and in isolated systems only). The MBL state is also different from (classical or quantum) glasses. In fact ergodicity breaking in glassy systems is due to the presence of an underlying complex free-energy landscape, which contains a huge number of metastable states separated by large barriers~\cite{berthier2011theoretical}, whereas MBL manifests in highly excited eigenstates. 
In fact the glass transition occurs when the collective modes involved in the jumps between metastable states freeze, while long-range vibrations around the minima of the landscape are still possible. Hence, differently from the MBL case, glasses have a finite thermal conductivity  and the coupling to environment has no effect on classical glassiness (beyond a rescaling on the microscopic timescale).

Achieving a complete and satisfactory comprehension of MBL has been incredibly challenging so far. Although significant and exciting progress has been made in our comprehension of MBL, both in theory and experiments, many questions remain open. Beside the perturbative analysis of Refs.~\cite{basko2006metal,gornyi2005interacting,gornyi2017spectral}, analytical approaches can only be employed on phenomenological models~\cite{vosk2015theory,potter2015universal,dumitrescu2017scaling,thiery2017microscopically,goremykina2019analytically,dumitrescu2019kosterlitz,morningstar2019renormalization,morningstar2020many}, and numerical simulations are limited to relatively small system sizes that might lack the asymptotic physics displayed by large systems~\cite{vznidarivc2008many,pal2010many,kjall2014many,luitz2015many}. 
In this context, some recent works have even questioned the existence of a truly MBL phase in the thermodynamic limit~\cite{vsuntajs2020quantum,vsuntajs2020ergodicity,sierant2022challenges,sierant2020polynomially,sierant2022slow,sels2023thermalization,sels2021markovian,sierant2020thouless,sels2021dynamical,sels2022bath,kiefer2021slow,kiefer2020evidence}, due to the possibility of the existence of a runaway avalanche instability seeded by rare thermal inclusions~\cite{de2017stability,thiery2018many,luitz2017small,goihl2019exploration,crowley2020avalanche,leonard2023probing,peacock2023many,morningstar2022avalanches,sels2022bath,Ha2023many} that may destroy the localized phase in the infinite-time, infinite-size limit. In fact, rare regions where the disorder is particularly small necessarily exist in large systems. These locally thermal region, often called ``thermal bubbles'', may thermalize the neighboring degrees of freedom, which are typically within a MBL portion, provided that the relaxation rates of these spins are smaller than the level spacing of the thermal inclusion,  such that the spins can not resolve the discreteness of the spectrum~\cite{de2017stability} (see also Refs.~\cite{sels2022bath,Ha2023many}). In this case the thermal bubble becomes bigger and can more effectively thermalize the adjacent regions. This leads to an avalanche that may spread through the whole system, destroying the MBL phase. Recent results indicates that such avalanche instability persists at much stronger randomness than had been previously thought~\cite{morningstar2022avalanches,sels2022bath,Ha2023many,sierant2022slow,leonard2023probing}. These observations and strong numerical finite size effects raise the question whether  MBL is only a finite-size crossover or a bonafide phase transition takes place at some large value of disorder~\cite{sierant2020thouless,sels2021dynamical,morningstar2022avalanches,long2022phenomenology,doggen2021many}. 

Actually, as in the field of the glass transition, the existence of an underlying phase transition is an interesting question, but it's even more interesting to characterize the microscopic mechanisms responsible for the different physical regimes which one observes before the putative phase transition. Although these may be just cross-overs---which  complicate the theoretical analysis---they seems to be associated to unusual and novel physical behaviors. In fact, another set of very important open questions that have attracted particular interest in the last years concern the anomalous transport and thermalization properties on the metallic side of the MBL transition (see Refs.~\cite{luitz2017ergodic,agarwal2017rare} for recent reviews). This line of research has been triggered by the evidence of sub-diffusive transport and anomalously slow out-of-equilibrium relaxation toward thermal equilibrium, which is described by power laws with exponents that gradually approach zero upon increasing the disorder, in a broad range of parameter before the putative MBL transition. These phenomena have been observed, both numerically~\cite{agarwal2015anomalous,vznidarivc2016diffusive,lev2014dynamics,luitz2016anomalous,doggen2018many,bera2017density} and experimentally~\cite{schreiber2015observation,bordia2017probing,luschen2017observation,smith2016many,xu2018emulating}, and appear to be remarkably robust and much less affected by finite-size effects. An appealing phenomenological interpretation of these anomalies has been proposed in terms of the existence around the MBL transition of a quantum Griffiths phase~\cite{agarwal2015anomalous,luitz2017ergodic,agarwal2017rare}. This is characterized by rare inclusions of the insulating phase with an anomalously small localization length. Several phenomenological proposals have been put forward to describe this physics, from purely classical resistor-capacitor models with power-law distributed resistances~\cite{agarwal2015anomalous,schulz2020phenomenology,schiro2020toy}, to stochastic models for merging thermal and insulating regions which are motivated by strong-disorder renormalization group arguments~\cite{vosk2015theory,potter2015universal,dumitrescu2017scaling,thiery2017microscopically,goremykina2019analytically,dumitrescu2019kosterlitz,morningstar2019renormalization,morningstar2020many}. 

Interestingly enough, both the ascertainment of the instability of the MBL phase with respect to quantum avalanches seeded by rare thermal bubbles, as well as  the comprehension of the physical mechanisms at the origin of the anomalous relaxation in the bad metal regime require to understand the nature of rare long-range resonances in many-body eigenstates and to characterize their statistics, although in different disorder regimes. There is by now strong numerical evidence of the fact that rare many-body resonances play a crucial role in determining the physical properties of the MBL transitions and the associated crossover regimes~\cite{long2022phenomenology,morningstar2022avalanches,crowley2020avalanche,garratt2021local,villalonga2020eigenstates,khemani2017critical,Ha2023many,de2021rare}. Yet, a precise characterization of their statistics is still missing.
The aim of this paper goes precisely in this direction. In particular, we develop a novel set of tools that allows one to take into account the effect of rare resonances in terms of the rarefaction of the paths in configuration (or Fock) space that may lead to decorrelation from a random initial state. 
We follow a perspective that is different and complementary to the one more commonly explored in the current literature: Instead of focusing on real space, we study the problem directly on configuration space and develop a large-deviation approach 
to study paths in configuration space for many body quantum disordered systems. This approach is based on techniques to study the physics of directed polymers in random media (DPRM), and in particular their freezing glass transition on infinite-dimensional graphs~\cite{derrida1988polymers}. The mapping of Anderson localization onto DPRM has already been discussed and exploited in the past~\cite{monthus2008anderson,monthus2011anderson,biroli2020anomalous,kravtsov2018non,lemarie2019glassy}. Here we generalize it to the $2^n$-dimensional Hilbert space of an interacting spin chain of $n$ spins. The central objects that we focus on are the rate functions associated with the statistics of ``{\it generalized fractional conductivities}'' (or ``{\it generalized fractional Landauer transmissions}''), defined as $\sum_{t} |G_{0,t} |^\beta$, where $G_{0,t}$ is the propagator between a randomly chosen initial spin configuration $|0 \rangle = | \! \uparrow \downarrow \downarrow \cdots \rangle$ and all the spin configurations $| t \rangle$ at large distance from $|0 \rangle$ (\eg, in which half of the spins have flipped). 
The configuration $|0 \rangle$ is taken from the middle of the many-body spectrum and hence correspond to a typical infinite temperature wavefunction. 
These objects have been already introduced in the mathematical literature on Anderson localization on hierarchical lattices~\cite{aizenman1993localization,warzel2013resonant} within the context of the so-called ``fractional moment method''. In our approach we use them as an efficient computational tool to individuate an characterize rare system-wide resonances between $|0 \rangle$ and faraway configurations. In particular, by analogy with DPRM, the auxiliary parameter $\beta$ (that plays the role of the inverse temperature in the context of directed polymers) allows us to fine tune the relative weight of strong resonances responsible for the rare outliers in the far tails of the probability distributions of the generalized conductivities, and hence to characterize their statistics.


In order to illustrate and benchmark our method, in the first part of the paper we study two benchmark and paradigmatic non-interacting cases, namely single-particle Anderson localization on loop-less Cayley trees~\cite{monthus2008anderson,monthus2011anderson,tikhonov2016fractality,sonner2017multifractality,biroli2020anomalous}, and the Rosenzweig-Porter (RP) random-matrix ensemble~\cite{Kravtsov_2015,vonSoosten_2019,Facoetti_2016,Truong_2016,Bogomolny_2018,DeTomasi_2019,amini2017spread,pino2019ergodic,berkovits2020super,venturelli2022replica}. The purposes of this analysis are listed below: 
\begin{itemize}
\item[(i)] Both the Anderson model on hierarchical lattices and the RP model have been argued to provide a pictorial description of MBL (see below and Refs.~\cite{altshuler1997quasiparticle,tikhonov2021anderson,de2013ergodicity,biroli2017delocalized,logan2019many,biroli2021out,tarzia2020many,faoro2019non}) and can give valuable insights to grasp an intuitive understanding of the many-body problem; 
\item[(ii)] The phase diagram of these models in the infinite-size limit is well known,  
thus providing useful guidelines to rationalize the numerical results of finite-size samples and to determine the range of validity of the approximations that may be used to treat the interacting case (such as the forward-scattering approximation (FSA)~\cite{anderson1958absence,pietracaprina2016forward,abou1973selfconsistent});
\item[(iii)] The mapping to directed polymers in a random potential can be carried out exactly for the Anderson model on a loop-less tree~\cite{monthus2008anderson,monthus2011anderson,biroli2020anomalous,kravtsov2018non}, without resorting to any approximation. Thus all the methodological aspects of our approach can be constructed and discussed in a transparent way for this model; 
\item[(iv)] The analysis of the benchmark cases shows that the large-deviation approach based on the analogy with DPRM is able to take into account the effect of rare system-wide resonances in a compact and efficient way, providing a helpful set of novel observables to distinguish between the different phases. 
\end{itemize}
In the second part of the paper we build on these results and apply the large-deviation analysis 
to a many-body disordered (isolated) quantum spin chain. In particular we consider the disordered spin chain for which the existence
of the MBL transition has been proven rigorously in Ref.~\cite{imbrie2016many} under the minimal assumption of absence of level attraction. We perform exact diagonalizations  varying the length of the chain between $n=10$ and $n=17$, and complement the exact diagonalizations results with the FSA, which allows us to study larger systems in the strong disorder limit. Our approach boils down to the computation of two rate functions 
which are essentially related to the typical and average values of the generalized fractional Landauer transmissions $\sum_{t} |G_{0,t} |^\beta$. A thorough analysis of their behaviors and shapes shows the existence of a broad disorder range in which rare, long-distance resonances, that may form only for a few specific realizations of the disorder and a few specific choice of the random initial state, destabilize the MBL phase. 

The paper is organized as follows: We start by presenting below a brief summary of the physical picture emerging from our analysis; In Sec.~\ref{sec:DPRM} we review a few basic properties of DPRM on the tree~\cite{derrida1988polymers}, with a special emphasis on their freezing-glass transition. In Sections~\ref{sec:CT} and~\ref{sec:RP} we focus on two non-interacting benchmark models, namely the Anderson tight-binding model on the loop-less Cayley tree~\cite{monthus2008anderson,monthus2011anderson,tikhonov2016fractality,sonner2017multifractality,biroli2020anomalous}, and the Rosenzweig-Porter random-matrix ensemble~\cite{Kravtsov_2015,vonSoosten_2019,Facoetti_2016,Truong_2016,Bogomolny_2018,DeTomasi_2019,amini2017spread,pino2019ergodic,berkovits2020super,venturelli2022replica}; In Sec.~\ref{sec:MBL} we present the analysis of  a disordered quantum interacting spin chain using the large-deviation approach; Finally, in Sec.~\ref{sec:conclusions} we provide a few concluding remarks and some perspectives for future investigations. In the Appendix sections~\ref{app:DPRM}--\ref{app:PG} we  present  some  technical details and supplementary  information  that  complement the results discussed in the main text.

\subsection{Summary of the main results} \label{sec:summary}

\begin{figure}
\includegraphics[width=0.49\textwidth]{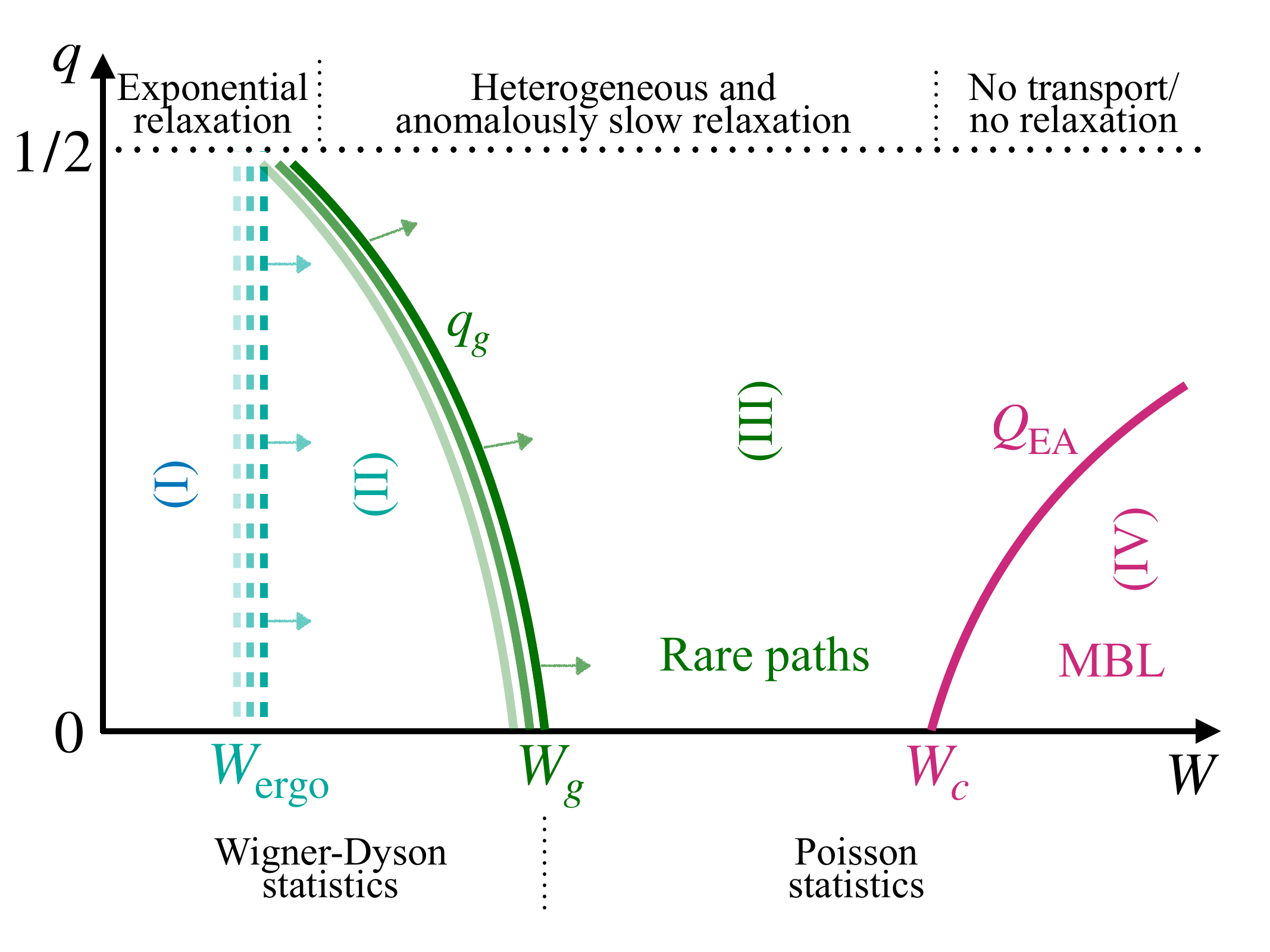} 
\vspace{-0.5cm}
\caption{Pictorial sketch of the (finite-size) out-of-equilibrium phase diagram of the one-dimensional disordered quantum spin chain described by Eq.~\eqref{eq:HMBL}. $W$ is the disorder strength and $q$ is the overlap of the target spin configuration with a random initial configuration (see Eq.~\eqref{eq:def:overlap} and Fig.~\ref{fig:hilbert}). The phase diagram shows the regimes (I)-(IV) described in the text as well as the transition/crossover lines between them. The arrows indicate pictorially how the lines drift when the length of the chain $n$ is increased (darker lines correspond to larger sizes). For the largest accessible sizes (\ie~$n=17$ spins) we find $W_{\rm ergo} \approx 2.5$, $W_g \approx 4.3$, and $W_c \approx 10.5$.
\label{fig:Lbetastar}}
\end{figure}

The main results of our analysis of the disorder quantum spin chain~\eqref{eq:HMBL} by means of the large-deviation approach described above are summarized in a pictorial way in the phase diagram of Fig.~\ref{fig:Lbetastar}. $W$ is the amplitude of the disorder in the random fields, while $q$ is the overlap between the initial and the final spin configurations $\vert 0 \rangle$ and $\vert t \rangle$ in the propagator $G_{0,t}$ (\ie, $n(1-q)/2$ is the Hamming distance on the $n$ dimensional hypercube which corresponds to the Fock space of the problem, see Fig.~\ref{fig:hilbert}). Thus, smaller $q$ means that we are studying rare resonances between more distant configurations. Varying the disorder strength modifies the statistics of these rare resonances, which allows us to distinguish {\it four} different regimes in the range of the numerically accessible system sizes (see Sec.~\ref{sec:MBL} and in particular Secs.\ref{sec:phasediagram} and~\ref{sec:speculations} for more details): 
\begin{itemize}
    \item[{\bf I:}]  At small enough disorder, $W<W_{\rm ergo}(n)$, 
    all paths connecting the initial and final configurations contribute to transport and decorrelation in the Hilbert space. The number of paths is exponential in $n$ and increases with the distance between the two configurations.
    Although, we do not study directly level spacing, this region should correspond to a fully ergodic regime in which the many-body wave-functions are described by random-matrix theory and satisfy the eigenstate thermalization hypothesis (ETH)~\cite{srednicki1994chaos,rigol2008thermalization}. 
    \item[{\bf II:}] At moderately small disorder, $W_{\rm ergo}(n)<W<W_g(n)$, 
    the number of paths that contribute to transport and decorrelation is still exponentially large in $n$ but with a smaller prefactor in front of $n$ compared to the total number of paths available. The total entropy of relevant paths is therefore reduced.  
     This regime is similar to the intermediate phase of the RP model~\cite{Kravtsov_2015} (see Sec.~\ref{sec:RP}), and corresponds to a delocalized but only partially ergodic regime since not all typical infinite temperature configurations are connected via resonances. 
    Locally the level statistics is expected to be described by the universal GOE distributions (likely with strong finite-size corrections at finite $n$, see Fig.~\ref{fig:rstat}). 
    \item[{\bf III:}] At moderately strong disorder, $W_g(n)<W<W_c$,
     the number of paths that contribute to transport and decorrelation becomes of order one. The system is still delocalized, in the sense that the probability of reaching a configuration far away from a random initial condition stays finite as the system size is increased, but quantum many-body states only hybridize with a few rare resonances far away in the Hilbert space. Since this is the analogous of the frozen phase in the DPRM analogy (see Sec.~\ref{sec:DPRM}), we will hereafter often refer to this regime as ``glassy''. Although the quantum dynamics should be able to decorrelate from the initial configuration in the infinite size, infinite time limit due to such rare resonances, 
     out of equilibrium relaxation and dynamics are presumed to be extremely heterogeneous and slow. The many-body eigenstates are expected to exhibit strong violations of ETH, 
     and the level statistics to be closer to the Poisson one than to that of the GOE ensemble, since two wave-functions close in energy have in general very different support sets. 
Furthermore, in this regime {\it finite-size} (not large enough) systems typically lack of the rare outliers of $G_{0,t}$ that dominate the transport at large $n$. In this case, these rare outliers are only found in rare samples and/or for a few specific choice of the initial state.
    Hence, above a certain value of the disorder most of the samples of not large enough size (and in particular of the sizes that can be simulated with the current computational resources) 
    appear as many-body localized to all practical purposes. Yet, upon increasing the system size the emergence of these rare resonances becomes progressively more prevalent even in typical samples, ultimately manifesting as delocalization in the limit of large $n$.
    \item[{\bf IV:}] Finally, at strong enough disorder, $W>W_c$, we find a genuine many-body localized phase, which is characterized by complete absence of transport and dissipation in the thermodynamic limit. Many-body eigenstates are exponentially localized around specific site orbitals in the Hilbert space and the level statistics is of Poisson type.
\end{itemize}
With increasing system sizes we find that both the crossover from the fully ergodic regime (I) to the partially ergodic one (II), $W_{\rm ergo} (n)$, and the crossover to the regime dominated by rare paths (III), $W_g(n)$, exhibit a pronounced tendency to shift towards higher disorder values. Conversely, the position of the MBL transition $W_c$ does not appear to be markedly influenced by variations in $n$. Yet, the MBL transition takes place at much larger values of the disorder compared to the one estimated in previous studies~\cite{abanin2021distinguishing,roy2021fock,creed2022probability,tarzia2020many}, 
almost by a factor $3$, in agreement with the idea that rare resonances push the genuine MBL transition to much larger values of the disorder, while the apparent localisation found for finite systems is indicative of a pre-thermal regime~\cite{sierant2020thouless,sels2021dynamical,morningstar2022avalanches,long2022phenomenology,herre2023ergodicity}. In fact, we argue that the standard diagnostics used in the literature to locate the MBL transition, such as the average spectral statistics, the eigenstates' participation entropies, and the entanglement entropy ~\cite{luitz2015many,mace2019multifractal}, are not suitable to capture the effect of these long-range rare resonances~\cite{morningstar2022avalanches}, at least for the range of systems sizes currently accessible. The reason for this are twofold: First, the change of behavior of the spectral statistics and of the scaling of the participation entropies is likely to occur at the onset of the regime (III), and not at the MBL transition. Second, even if the ``glassy'' regime turned out to be a crossover region which eventually becomes fully thermal, the samples that can be currently simulated lack the relevant long-range resonances that allow the system to decorrelate from a random initial state and appear as localized for most practical purposes. In contrast, by introducing a bias on the weights of large transition amplitudes via the auxiliary parameter $\beta$, our large-deviation approach inspired by the analogy with directed polymers is suitably designed to unveil the presence and quantify the effect of these rare resonances already from the analysis of the behavior of moderately small samples for which they are typically absent---as clearly shown by the analysis of the benchmark models (see Secs~\ref{sec:CT} and \ref{sec:RP}). 

The scenario presented above, characterized by these four distinct regimes (in finite-size systems), is very similar, at least qualitatively, to the one recently discussed in Ref.~\cite{morningstar2022avalanches}, although the analysis of the statistics of the rare long-range resonances and the instability of the MBL phase towards quantum avalanches are performed in different ways (and for different models). 
It is important to stress that, in absence of an exact solution of the problem in the infinite-size limit, we cannot foresee whether the intermediate regimes (II) and (III) observed at finite sizes will persist in the thermodynamic limit, or if in the limit of very large system sizes they eventually crossover towards a fully ergodic phase (a similar crossover is for instance observed in the tight-binding Anderson model on the random-regular graph, see, \eg, Refs.~\cite{biroli2012difference,altshuler2016nonergodic,kravtsov2018non,tikhonov2016anderson,Tikhonov_2019,garcia2017scaling,biroli2018delocalization,herre2023ergodicity,garcia2022critical,vanoni2023renormalization}). 
However, even if this is the case, the crossover is likely to occur on length and time scales which are far beyond the ones that are currently accessible, both in numerical simulations and in experiments. 


\section{A brief review of DPRM on the Cayley tree} \label{sec:DPRM}

Since our methodical approach is strongly based on an analogy with DPRM in large dimensions, we start by briefly revisit this problem, focusing in particular on the Cayley tree~\cite{derrida1988polymers}. The aim of this short section is not to provide an exhaustive review, but just to introduce the basic tools and ideas that we will use in the rest of the paper, when mapping the transport properties of quantum (interacting and non-interacting) disordered system onto DPRM.

Let us consider a Cayley tree, \ie~a loop-less hierarchical lattice with fixed branching ratio $k$. Each sites on the lattice has $k+1$ neighbors, apart from the leaves at the boundary which have only one neighbor belonging to the previous generation of the tree. A finite fraction of the nodes of a Cayley tree is located at the boundary: for a tree of branching ratio $k$ and $n$ generations the number of leaves is $(k+1)k^{n - 1}$, while the total number of nodes is $N(n) = [k^{n} (k + 1) - 2]/(k-1)$. On each edge $(i_m \leftrightarrow i_{m+1})$ of the lattice, there is a random potential $\omega_{i_m \leftrightarrow i_{m+1}}^m$,
independently and identically distributed (iid) according to a certain probability density $\rho(\omega)$. A directed polymer on the tree is a self-avoiding walk of $n$ steps starting from the root at $0$. The total number of paths is equal to the number of leaves, \ie, $(k+1) k^{n-1}$. By definition, the total energy of a walk is the sum of the random potentials on the edges crossed by the path ${\cal P}$:
\[
E_{\cal P} = \omega_{0 \leftrightarrow i_1}^0 + \omega_{i_1 \leftrightarrow i_2}^1 + \ldots + \omega_{i_{n-1} \leftrightarrow i_n}^{n-1} \, , 
\]
where the indices $i_m = 1, \ldots, (k+1) k^{m-1}$ denotes the nodes of the $m$-th generation. To describe the statistical properties of the directed polymer we thus need to compute the partition function:
\begin{equation} \label{eq:DPRM}
Z = \sum_{{\cal P}} e^{- \beta E_{\cal P}} =  \sum_{{\cal P}} e^{- \beta \sum_{(i_m \leftrightarrow i_{m+1} ) \in {\cal P}}  \omega^m_{i_m \leftrightarrow i_{m+1}}}\, .
\end{equation}
Assuming that the random potentials have zero mean and a finite variance $\sigma_\omega^2$, in the limit of large trees the distribution of the $E_{\cal P}$ converges to a Gaussian with zero mean and variance $n \sigma_\omega^2$. In this respect the partition function~\eqref{eq:DPRM} is very similar to the one of the celebrated Random Energy Model (REM)~\cite{derrida1980random}, which is the simplest and most paradigmatic model of mean-field structural glasses, featuring a low-temperature glassy phase described by a one-step replica symmetry breaking~\cite{mezard1987spin}. 

The partition function~\eqref{eq:DPRM} is a sum over $\sim k^n$ terms. There is therefore a competition between an entropic factor (large number of terms) and an energetic factor (the Boltzmann weight). At high temperature the polymer visits an exponential number of configurations from which $Z$ receives a significant contribution. At low temperature, instead, in the glassy phase, the polymer can only visit a few $O(1)$ configurations having very large weights (\ie~very low energies), and the sum over the paths is dominated by these few outliers. As we will see in the following, within the analogy between DPRM and (many-body) localization such a glassy phase corresponds to a regime in which quantum transport can only occur through rare, specific, disorder-dependent paths. 

It is clear that such a glass transition is driven by the condensation of the Gibbs' measure on the extreme values of the (correlated) random variable $e^{-\beta E_{\cal P}}$. Neglecting, at first, the correlations of the $E_{\cal P}$'s on different paths, one obtains a REM-like problem. 
The total number of configurations of the polymer at  energy (per unit length) $\epsilon_{\cal P} = E_{\cal P}/n$ in the large $n$ limit is:
\[
\Omega(\epsilon_{\cal P}) = (k+1) k^{n-1} \sqrt{\frac{n}{2 \pi \sigma_\omega^2}} e^{-n \epsilon_{\cal P}^2/(2 \sigma_\omega^2)} \, ,
\]
corresponding to a configurational entropy (per unit length) equal to
\[
\Sigma (\epsilon_{\cal P}) = \frac{1}{n} \ln \Omega(\epsilon_{\cal P}) \simeq \ln k - \frac{\epsilon_{\cal P}^2}{2 \sigma_\omega^2} \, 
\]
(where we have set $k_B=1$). This expression is only valid for intensive energies such that $\Sigma > 0$, \ie~for $\epsilon_{\cal P}$ within the interval $|\epsilon_{\cal P}| \le \epsilon_\star = \sqrt{2 \ln k} \, \sigma_\omega$. For $|\epsilon_{\cal P}| > \sqrt{2 \ln k} \, \sigma_\omega$, instead, the entropy $\Sigma$ is negative, implying that the number of walks at those values of the energy is exponentially small in $n$.

\begin{figure*}
\includegraphics[width=0.332\textwidth]{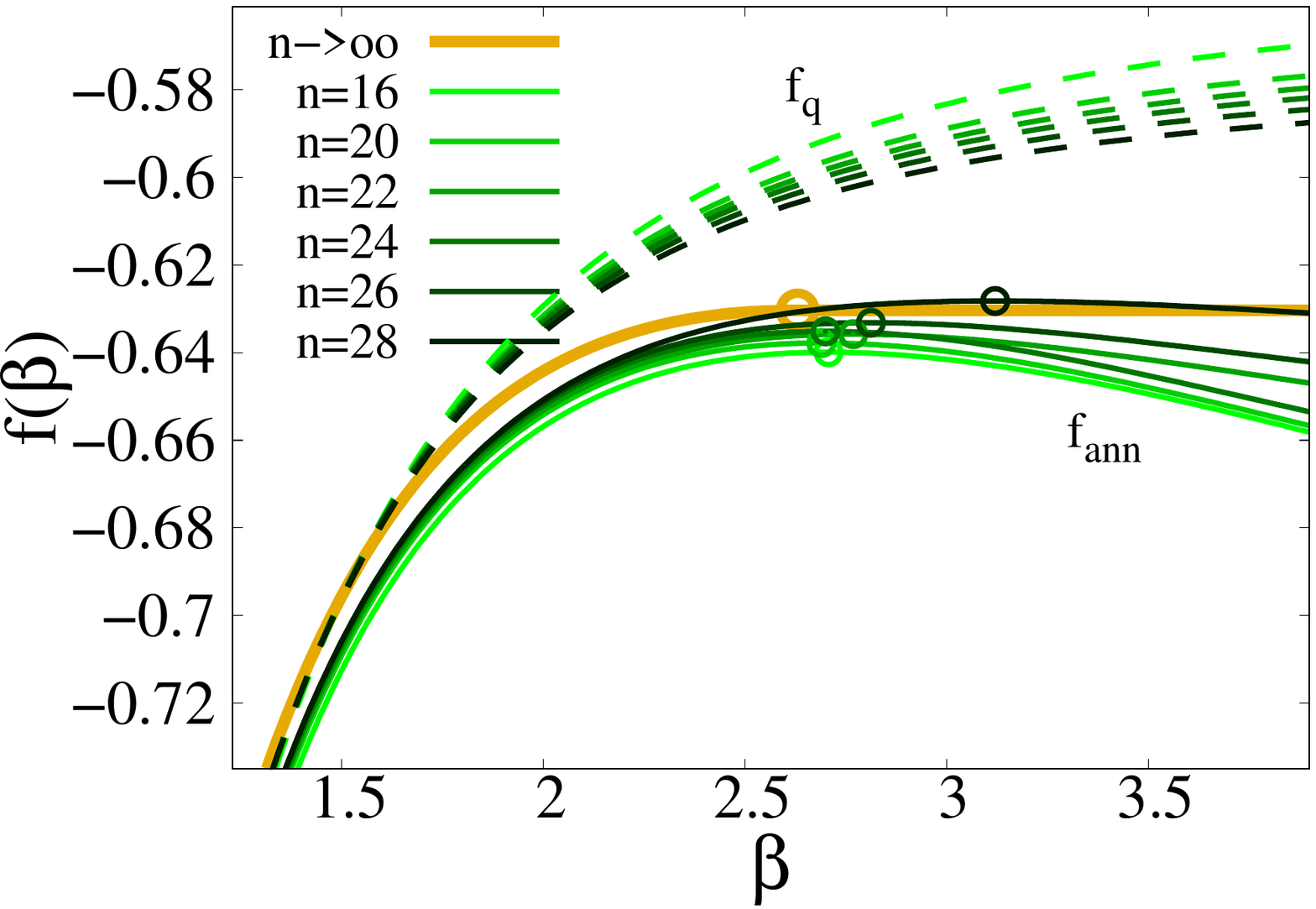} \hspace{-0.03cm} \includegraphics[width=0.332\textwidth]{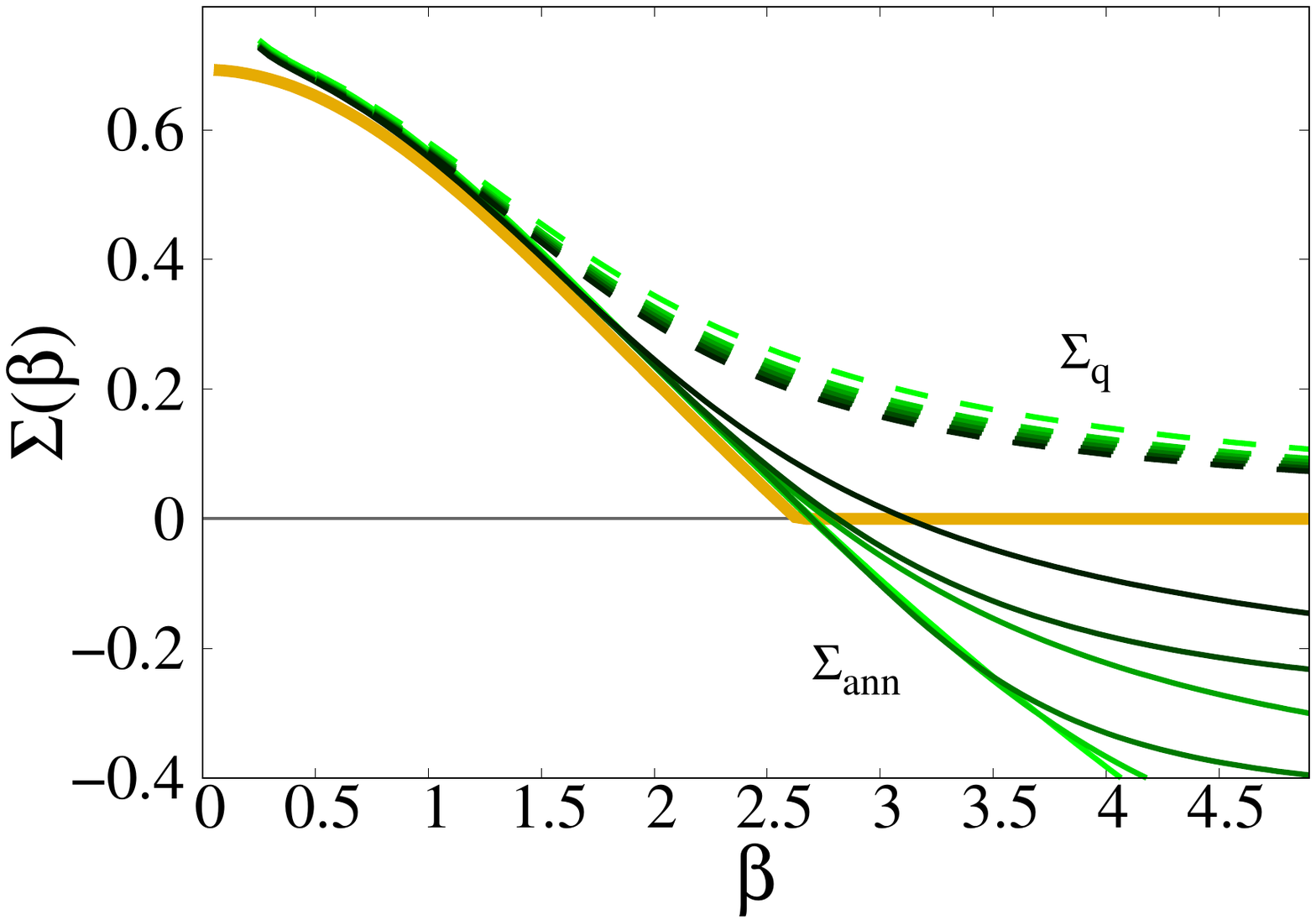} \hspace{-0.07cm} \includegraphics[width=0.32\textwidth]{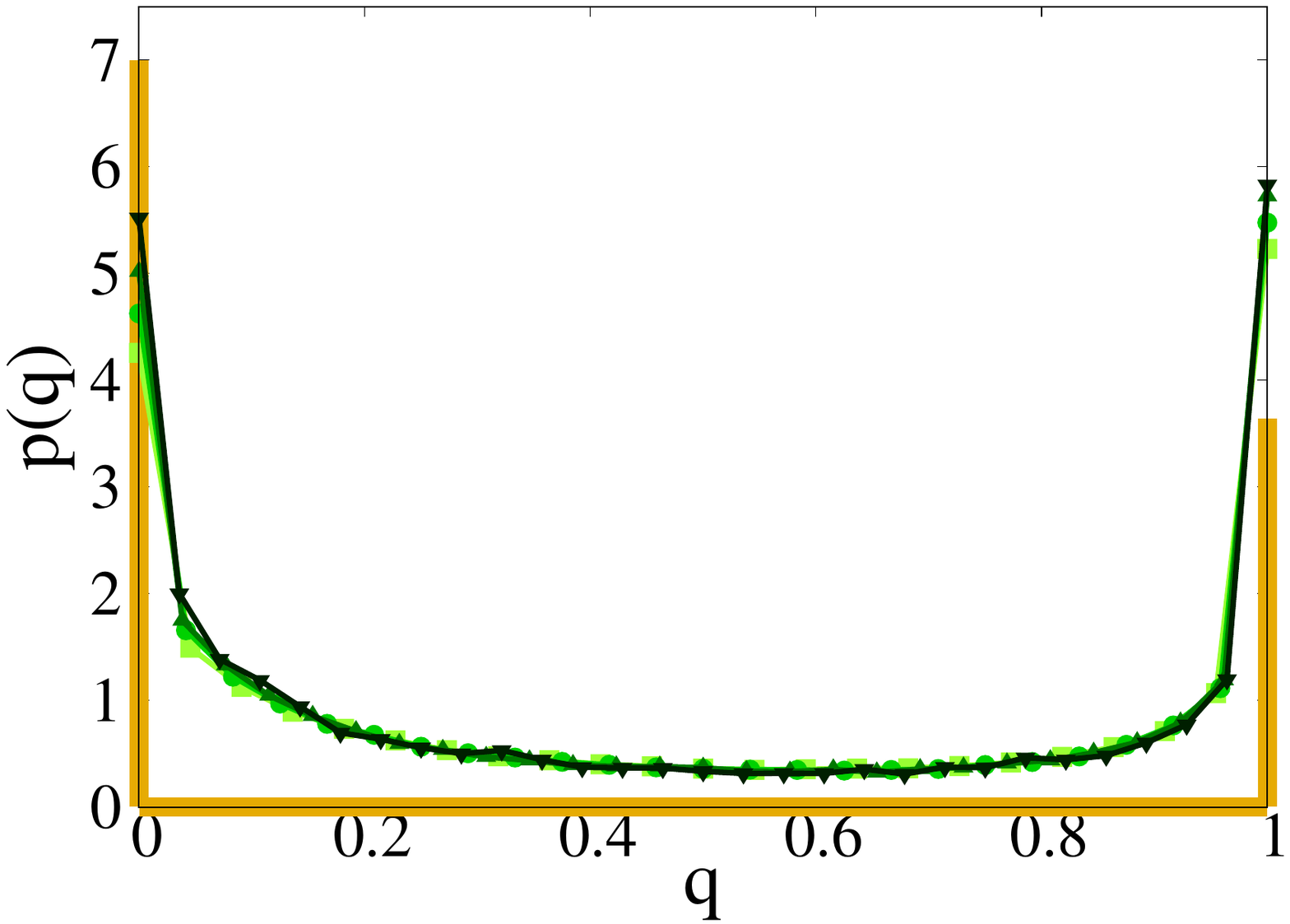} 
\caption{Left: Quenched ($f_{\rm q} = - \avg{\ln Z}/(\beta n)$, dashed curves) and annealed ($f_{\rm ann} = - \ln \avg{Z}/(\beta n)$, continuous curves) free-energies (per unit length) as a function of the inverse temperature $\beta$ of a direct polymer on Cayley trees of $n$ generations in presence of random bond disorder taken from a uniform distribution of width $2$. The green circles spot the position of the minima of the annealed free-energy at finite $n$.  The thick yellow curve corresponds to the exact solution in the thermodynamic limit, given by Eq.~\eqref{eq:fqDPRM}, and the circle gives the position of the critical temperature $\beta_\star \approx 2.631$ above which the free-energy becomes flat. Middle: Quenched (dashed curves) and annealed (continuous curves) estimation of the configurational entropy at finite $n$ as a function of $\beta$, $\Sigma_{\rm q,ann}(\beta) = \beta^2 {\rm d} f_{\rm q,ann}/{\rm d}\beta$. The thick yellow curve corresponds to the analytic expression of the configurational entropy in the thermodynamic limit. Right: Overlap probability distribution $p(q)$, Eq.~\eqref{eq:pqDPRM}, for $\beta=4>\beta_\star$. The thick yellow lines represent the limiting distribution in the $n \to \infty$ limit, which converges to two $\delta$-peaks in $q=0$ and $q=1$ with coefficients given in Eq.~\eqref{eq:pq1rsb}. These data are obtained sampling over $2 \cdot 10^6$ realizations for $n=16$, $4 \cdot 10^5$ realizations for $n=20$, $n=22$, and $n=24$, $10^5$ realizations for $n=26$, and $3 \cdot 10^4$  realizations for $n=28$.
\label{fig:DPRM}}
\end{figure*}

The partition function~\eqref{eq:DPRM} can be then rewritten in the large $n$ limit as an integral over all paths giving a contribution characterized by a value of the energy between $\epsilon_{\cal P}$ and $\epsilon_{\cal P} + {\rm d} \epsilon_{\cal P}$ times the number of such paths:
\begin{equation} \label{eq:Zint}
Z = \int \!\! {\rm d} \epsilon_{\cal P} \, e^{-n \beta \epsilon_{\cal P} + n \Sigma(\epsilon_{\cal P})} \, .
\end{equation}
In the limit $n \to \infty$ the value of the energy $\overline{\epsilon}_{\cal P}$ that dominates the integral is given by the saddle-point condition:
\[
\left . \frac{{\rm d} \Sigma}{{\rm d} \epsilon_{\cal P}} \right \vert_{\epsilon_{\cal P}= \overline{\epsilon}_{\cal P}} \!\!\!\!\!\!\!\! = \beta \, \Longrightarrow \, \overline{\epsilon}_{\cal P} = - \beta \sigma_\omega^2  \, .
\]
Yet, this solution is only correct provided that $\overline{\epsilon}_{\cal P} \ge - \epsilon_\star$, where the entropy vanishes. This condition yields the critical temperature $\beta_\star = \sqrt{2 \ln k} / \sigma_\omega$ below which the energy of the polymer freezes at the minimal value $-\epsilon_\star$. The corresponding (quenched) free-energy (per unit length) is thus:
\begin{equation} \label{eq:FDPRM}
f (\beta) = - \lim_{n \to \infty} \frac{\langle \ln Z \rangle}{\beta n}  = 
\left \{
\begin{array}{ll}
- \beta \sigma_\omega^2/ 2 - \ln k/ \beta  \,\, & \textrm{for~} \beta< \beta_\star \, , \\
-\epsilon_\star & \textrm{for~} \beta \ge \beta_\star \, .
\end{array}
\right .
\end{equation}
The transition taking place at $\beta_\star$ corresponds therefore to a transition to a frozen phase. The free-energy varies with temperature until it reaches its maximum, and sticks there at lower temperatures (see Fig.~\ref{fig:DPRM}). Also the shape of the whole free-energy distribution does not change with temperature in the frozen phase~\cite{derrida1988polymers}. 

The configurational entropy, which encodes the exponential number of paths typically visited by the polymer, is related to the free-energy through a Legendre transform:
\begin{equation} \label{eq:ScDPRM}
\Sigma (\beta) = \beta^2 \frac{{\rm d} f}{{\rm d} \beta} = \left \{
\begin{array}{ll}
\ln k - \beta^2 \sigma_\omega^2/ 2   \,\, & \textrm{for~} \beta< \beta_\star \, , \\
0 & \textrm{for~} \beta \ge \beta_\star \, ,
\end{array}
\right .
\end{equation}
corresponding to the fact that in the high-temperature phase, $\beta<\beta_\star$, the partition function receives contribution from an exponential number of paths, while in the low-temperature phase, $\beta>\beta_\star$, only a few, specific, disorder-dependent paths, corresponding to the extreme values of $E_{\cal P}$ out of the $k^n$ realizations, dominate the sum. Note that for $\beta=0$ the configurational entropy attains its maximum value $\Sigma = \ln k$, signaling that all $O(k^n)$ configurations contribute equally to $Z$.

The problem of DPRM is slightly more complicated than the REM since, although the $\omega_{i_m}$'s are iid random variables, the energies $E_{\cal P}$ are correlated due to the fact that different walks have in general some nodes in common. Nonetheless, DPRM undergo a freezing glass transition at low temperature of the very same type the one of the REM described above. As shown in App~\ref{app:DPRM}, this can be shown analytically following Ref.~\cite{derrida1988polymers}, by mapping DPRM onto the study of a certain nonlinear equation of reaction-diffusion type, the so-called Kolmogorov-Petrovsky-Piscounov (KPP) equation~\cite{kpp1937}. The exact result for the (quenched) free-energy per unit length reads (yellow curve in the left panel of Fig.~\ref{fig:DPRM}):
\begin{equation} \label{eq:fqDPRM}
f (\beta) = - \lim_{n \to \infty} \frac{\langle \ln Z \rangle}{\beta n}  = 
\left \{
\begin{array}{ll}
- \ln \left ( k \avg{e^{-\beta \omega}} \right) / \beta  \,\, & \textrm{for~} \beta< \beta_\star \, , \\
- \ln \left ( k  \avg{e^{-\beta_\star \omega}} \right ) / \beta_\star & \textrm{for~} \beta \ge \beta_\star \, .
\end{array}
\right .
\end{equation}
It is easy to check that for the particular case in which $\rho(\omega)$ is a Gaussian distribution with zero mean and variance $\sigma_\omega^2$ one recovers the results~\eqref{eq:FDPRM} and~\eqref{eq:ScDPRM} --- with $\beta_\star = \sqrt{2 \ln k}/\sigma_\omega$ --- corresponding to the approximation in which one neglects the correlations between the energies of the different paths. For a generic probability density the results are quantitatively slightly different, although the physical picture is still exactly the same.

The low-temperature phase of DPRM is very similar to the one of spin glasses, where the phase space breaks up into many disconnected pieces separated by free-energy barriers~\cite{mezard1987spin}. The order parameter function associated with such many-valley structure is the distribution of the overlap $q({\cal P},{\cal P}^\prime)$ between two polymer's configurations ${\cal P}$ and ${\cal P}^\prime$, defined as the fraction of edges that the two walks have in common for a given disorder realization. For a finite tree of $n$ generations $q$ is an integer multiple of $1/n$ varying between $0$ and $1$. The probability of finding a certain value of the overlap at inverse temperature $\beta$ reads:
\begin{equation} \label{eq:pqDPRM}
p(q) = \left \langle \sum_{{\cal P},{\cal P}^\prime}  \frac{ e^{- \beta (E_{\cal P} + E_{{\cal P}^\prime})}}{Z^2(\beta)} \delta \left[ q-q ({\cal P},{\cal P}^\prime)\right] \right \rangle \, ,
\end{equation}
where the average is performed over independent realizations of the quenched disorder. As shown in Ref.~\cite{derrida1988polymers}, for $n \to \infty$ $p(q)$ behaves identically as in the REM~\cite{derrida1980random}: In the high-temperature phase the overlap is identically equal to zero, while in the low-temperature phase the overlap is either zero or one. In particular in the large $n$ limit one finds:
\begin{equation} \label{eq:pq1rsb}
p(q) = \left \{
\begin{array}{ll}
\delta(q) & \textrm{for~} \beta< \beta_\star \, , \\
\frac{\beta_\star}{\beta} \delta(q) + \left( 1 - \frac{\beta_\star}{\beta} \right) \delta (q - 1) \,\, & \textrm{for~} \beta \ge \beta_\star \, ,
\end{array}
\right .
\end{equation}
corresponding to a one-step replica symmetry breaking ansatz~\cite{mezard1987spin} (right panel of Fig.~\ref{fig:DPRM}).

The freezing transition of the paths is associated with the condensation of the Boltzmann weights on the configurations that correspond to the extreme values  of the probability distributions of the energies $E_{\cal P}$. Such anomalously large values of the Boltzmann weights have strong fluctuations from one sample to another, thereby producing a broad distribution $P(Z)$ of the partition function, which exhibits power-law tails (see Ref.~\cite{derrida1988polymers} and App.~\ref{app:DPRM}). This property is related to a very efficient practical way to determine numerically the glass transition temperature in many similar problems, in particular the ones related to quantum transport that we will consider below, in which $\beta_\star$ is not known analytically. This method consists in computing the so-called {\it annealed} free-energy (per unit length), defined as:
\begin{equation} \label{eq:faDPRM}
f_{\rm ann} (\beta,n) = - \frac{1}{\beta n} \ln \avg{Z} \, ,
\end{equation}
and comparing it to the quenched one
\begin{equation} \label{eq:fqqDPRM}
f_{\rm q} (\beta,n) = - \frac{1}{\beta n} \avg{\ln Z} \, .
\end{equation}
In the low-temperature phase $P(Z)$ decays with an exponent smaller than $2$ and $\avg{Z}$ is dominated by the extreme values in the tails of the probability distribution. For $\beta < \beta_\star$, instead, the tails of $P(Z)$ decay fast enough and $\avg{Z}$ is dominated by the typical value. As a consequence, in the high-temperature replica-symmetric phase the annealed free-energy converges to the quenched one upon increasing the system size and displays a maximum in the vicinity $\beta_\star$. This is in fact a generic features of other mean-field glassy systems, such as the REM, characterized by a low-temperature phase in which the Gibbs measure breaks in many disconnected components~\cite{mezard1987spin}. 

Following the position of such minimum in trees of large but finite size allows thus one to obtain a  reliable numerical estimation the critical temperature.  Moreover by replacing the variation of $f_{\rm ann}$ beyond $\beta_\star$ by its value at $\beta_\star$,
\begin{equation} \label{eq:replacement}
\tilde{f}_{\rm ann} (\beta,n) = \left \{
\begin{array}{ll}
f_{\rm ann} (\beta,n) \,\, & \textrm{for~} \beta< \beta_\star(n) \, , \\
f_{\rm ann} (\beta_\star(n),n) & \textrm{for~} \beta \ge \beta_\star(n) \, ,
\end{array}
\right .
\end{equation}
one obtains a function that in the large $n$ limit converge to its quenched counterpart:
\[
\lim_{n \to \infty} f_{\rm q} (\beta,n) = \lim_{n \to \infty} \tilde{f}_{\rm ann} (\beta,n) = f(\beta) \, ,
\]
where $f(\beta)$ is given in Eq.~\eqref{eq:fqDPRM}. In consequence one can use either $f_{\rm q}$ or $f_{\rm ann}$ (but with the replacement~\eqref{eq:replacement}), to study the system. They are complementary, and as we shall see, they allow to assess the role of rare resonances and finite size effects. In the following we will denote as ``annealed'' all the results obtained using the latter procedure; beware that this is a different terminology from the one commonly used for DPRM, where annealed is associated to the entire $f_{\rm ann}$ curve without replacement~\eqref{eq:replacement}.  

This procedure is illustrated in Fig.~\ref{fig:DPRM}. In the left panel we plot the annealed and quenched free-energies computed numerically on Cayley trees of $n$ generations (for random potentials uniformly distributed in the interval $[-1,1]$), together with the exact quenched free-energy~\eqref{eq:fqDPRM} in the $n \to \infty$ limit. This example  clearly shows that the diminution of the number of realizations with the system size limits the influence of rare outliers which are important for the annealed free-energy: Indeed, for the largest available size $n=28$ (for which the number of sampling is reduced due to the larger computational cost) the agreement of $\beta_\star$ with the analytic prediction is less good than for $n=26$. In the middle panel we plot the (quenched and annealed) estimation of the configurational entropy at finite $n$ as a function of the inverse temperature, $\Sigma_{\rm q,ann}(\beta,n) = \beta^2 {\rm d} f_{\rm q,ann}(\beta,n)/{\rm d}\beta$, together with the exact result in the limit of infinite trees. $\Sigma(\beta)$ tends to $\ln k$ at infinite temperature and vanishes at $\beta_\star$ where the freezing of the paths take place. Finally in the right panel we show the overlap distribution at $\beta=4>\beta_\star$, which is consistent with the emergence of two $\delta$-peaks as described by the one-step replica symmetry breaking ansatz.

\section{Benchmark case I: The Anderson model on the Cayley tree} \label{sec:CT}

Our first benchmark consists in the Anderson model on the Cayley tree. 
Since preliminary works on the subject, MBL was related to a form of localization in the Fock space of Slater determinants~\cite{altshuler1997quasiparticle} (see also Refs.~\cite{basko2006metal,gornyi2005interacting,herre2023ergodicity}), which plays the role of lattice sites in a disordered Anderson tight-binding model. A paradigmatic representation which gives a very intuitive picture of MBL is therefore single-particle Anderson localization on an expander graph/hierarchical lattice. Within this mapping many-body configurations are seen as site orbitals on the graph (with strongly correlated diagonal disorder) and the interactions play the role of an effective hopping connecting them. 
The local structure and topology of the graph depend on the specific form of the many-body Hamiltonian and on the choice of the basis. Yet, the Hilbert space of an interacting model (in a basis in which the system is localized in absence of interactions) is generically a very high dimensional disorder lattice, in which the number of sites  grows exponentially with the distance. Although the analogy between MBL and Anderson localization on hierarchical lattices involves several drastic simplifications (\eg, the correlation between random energies are neglected, as well as the specific structure of the loops, and the scaling of the connectivity of the graph with the number of interacting degrees of freedom in the original many-body problem), it is very useful to obtain a qualitative understanding of the problem~\cite{tikhonov2021anderson,de2013ergodicity,biroli2017delocalized,logan2019many,garcia2022critical,herre2023ergodicity}. 

In this section we thus focus on the study of the non-interacting tight-binding Anderson model on the loop-less Cayley tree, 
defined by the following Hamiltonian:
\begin{equation} \label{eq:Handerson}
H = - t \sum_{\langle i, j \rangle } \left( | i \rangle \langle j | +  | j \rangle \langle i | \right) - \sum_i \epsilon_i  | i \rangle \langle i | \, ,
\end{equation}
where $\epsilon_i$ are iid random energies taken from a uniform distribution in the interval $[-W/2,W/2]$. 
Since a finite fraction of the nodes of a Cayley tree are located at the boundary 
one can expect that that the presence of the leaves crucially affects the wave-functions' statistics which turns out to be strongly generation-dependent even in the delocalized phase. Indications of the peculiar character of the eigenstates of the Anderson model on a Cayley tree are in fact given in Refs.~\cite{monthus2008anderson,monthus2011anderson,tikhonov2016fractality,sonner2017multifractality,biroli2020anomalous}, which have shown that the statistics the eigenfunctions' amplitudes at the root of a Cayley tree are distributed fractally in the most of the extended phase (at contrast with the ergodicity of the wave-functions in the extended phase of the Anderson model on the random-regular graph~\cite{tikhonov2016anderson,Tikhonov_2019,garcia2017scaling,biroli2018delocalization}, which essentially represents a finite Cayley tree wrapped onto itself, with no boundary and loops of size $\ln N$). Throughout we will consider a Cayley tree of fixed branching ratio $k=2$ and, without loss of generality, we will only focus on the middle of the single-particle energy band, at $E=0$.

As anticipated in the introduction, the main focus of this paper is the statistics of the propagators between distant nodes. In particular, below we will consider the propagator between the root of the tree, labeled as $0$, and the $(k+1) k^{n-1}$ boundary nodes, labeled $i_n$, of the last generation. Due to the hierarchical structure of the lattice, the propagator  can be expressed as a product of the $n$ ``cavity'' Green's functions on the sites belonging to the {\it unique} path ${\cal P}$ that connects the root with $i_n$
\begin{equation} \label{eq:propagatorCT}
G_{0,i_n} = t^n G_{i_n \to i_{n-1}} G_{i_{n-1} \to i_{n-2}} \cdots  G_{i_{1} \to i_{0}} G_{0,0} \, .
\end{equation}
As first shown in Ref.~\cite{abou1973selfconsistent}, the cavity Green's functions satisfies the following exact recursion relations in terms of the cavity Green's function defined on the neighboring sites in the next generation:
\begin{equation} \label{eq:cavity}
G_{i_m \to i_{m-1}}^{-1} = -\epsilon_{i_m} - t^2 \!\!\! \sum_{j_{m+1} \in \partial i_m}  \!\! G_{j_{m+1} \to i_{m}} \, ,
\end{equation}
with the initial conditions on the leaves $G_{i_n \to i_{n-1}} = -1/\epsilon_{i_n}$.
For completeness, the Green's function on the root is given by:
\[
G_{0,0}^{-1} = -\epsilon_{0} - t^2 \sum_{i_{1}=1}^{k+1}  G_{i_{1} \to 0} \, .
\]
To simplify the notations we will set $t=1$ hereafter.

Physically $|G_{0,i_n}|^2$ represents the probability that a particle starting on the root at time $0$ is found in $i_n$ after infinite time. Therefore, the probability that a particle created at the root at $t=0$ reaches one of the nodes of the boundary is proportional to 
\begin{equation} \label{eq:T2}
{\cal T} = \sum_{{\cal P}} |G_{0,i_n}|^2 = |G_{0,0}|^2 \sum_{{\cal P}} \prod_{(i_m \leftrightarrow i_{m-1}) \in {\cal P}} |G_{i_m \to i_{m-1}}|^2 \, ,
\end{equation}
where the sum runs over all $(k+1)k^{n-1}$ paths ${\cal P}$. 
Note that here we have set the imaginary regulator to zero from the start and work only with the real parts of the Green's functions. In fact ${\cal T}$ also describes the dissipation propagation between the root and the boundary in the linearized regime (see Ref.~\cite{parisi2019anderson} for a detailed discussion). In particular, it is easy to show that introducing a very small imaginary part $\eta \ll 1$ of the Green's function on the boundary of the tree, and linearizing the recursion relations~\eqref{eq:cavity} with respect to the imaginary part, the imaginary part of the Green's function at the root is given by ${\rm Im} G_{0,0} = \eta {\cal T}$. In the localized phase the imaginary part of the Green's function decrease exponentially when the recursion relations~\eqref{eq:cavity} are iterated from one generation of the tree to the next. Conversely, in the delocalized regime the imaginary part of the Green's function grows exponentially under iteration (and then the linearization of the recursion relations is no longer justified). One can thus introduce the Lyapunov exponent $\gamma \equiv \ln {\cal T}/n$ as:
\begin{equation} \label{eq:ImG}
{\rm Im} G_{0,0} = \eta {\cal T} = \eta e^{\gamma n} \, ,
\end{equation}
in such a way that $\gamma>0$ in the metallic phase and $\gamma<0$ in the insulating one. 
Equivalently, ${\cal T}$ can also be interpreted as the Fisher-Lee conductivity~\cite{fisher1981relation} (or the total Landauer transmission) of a Cayley tree of $n$ generations in a scattering geometry when a semi-infinite lead is attached to the root and $(k+1)k^{n-1}$ semi-infinite leads are attached to the nodes of the boundary, as discussed in Refs.~\cite{monthus2008anderson,monthus2011anderson}. 

Interestingly, as already pointed out in Refs.~\cite{biroli2020anomalous,monthus2008anderson,monthus2011anderson,kravtsov2018non,biroli2012difference,miller1994weak} (see also Refs.~\cite{feigel2010superconductor,ioffe2010disorder,chakrabarti2022traveling,ros2021fluctuation} for a similar mapping in the context of quantum interacting disordered models at equilibrium), Eq.\eqref{eq:T2} is formally equivalent to the partition function of a directed polymer on the Cayley tree, Eq.~\eqref{eq:DPRM}, with (correlated) quenched random energies $e^{-  \omega_{i_m \to i_{m-1}}} \equiv |G_{i_m \to i_{m-1}}|$ and $\beta=2$. The only difference with the classical DPRM problem lies in the fact that the random energies are strongly correlated (since the random ``energies'' $\omega_{i_m \to i_{m-1}} - \ln |G_{i_m \to i_{m-1}}|$ need to satisfy the recursion relation~\eqref{eq:cavity}). 
Yet, as discussed in the previous section, this problem can undergo a freezing transition from a 
phase in which the ``polymer'' visits an exponential number of configurations from which ${\cal T}$ receives a significant contribution, to a ``glassy'' phase 
in which the ``polymer'' can only visit a few $O(1)$ configurations having very large  weights, and the sum over the paths is dominated by these few outliers. Our general strategy to address this problem, which will be extended to the interacting case, is detailed in the section below.

\subsection{General strategy and key observables} \label{sec:strategy}

As discussed above, the total Landauer transmission~\eqref{eq:T2} is formally equivalent to the partition function of a directed polymer on the Cayley tree.
In order to make the connection with DPRM even stronger, we formally introduce an extra parameter $\beta$ which plays exactly the same role of the inverse temperature 
(from now on we omit the multiplicative factor $\vert G_{0,0} \vert^2$ appearing in Eq.~\eqref{eq:T2} which does not contribute at large $n$):
\begin{equation} \label{eq:T}
{\cal T} (\beta) = \sum_{{\cal P}} \prod_{(i_m \leftrightarrow i_{m-1}) \in {\cal P}} |G_{i_m \to i_{m-1}}|^\beta \, .
\end{equation}
Enlarging the parameter space in this way boils down to studying ``generalized fractional conductivities'' (or ``fractional Landauer transmissions'') associated with different moments of the propagators, and allows us to fine tune the relative weight of large resonances in the tails of the probability distribution of the $|G_{0,i_n}|^2$. The physical conductivity, Eq.~\eqref{eq:T2}, is recovered for $\beta=2$. There are three main questions that we would like to answer:
\begin{itemize}
\item[1)] For a given disorder strength $W$, is the system in the conducting/delocalized phase or in the insulating/localized one? In other words, does the typical value of the physical conductivity (which is recovered for $\beta=2$) goes to zero with the system size or not?
    \item[2)] For a given disorder strength $W$, the value of the inverse temperature $\beta=2$ associated with the physical transport 
    falls within the high-temperature phase (in the DPRM jargon), in which the sum in~\eqref{eq:T} is dominated by an exponential number of paths, or within the glassy phase, implying that transport can only occur through a few specific, disorder-dependent paths?
    \item[3)] In the case in which ${\cal T} (\beta=2)$ is in the high temperature phase, what is the asymptotic value of the number of paths contributing to the Landauer tranmission as a function of the length of the path?
\end{itemize}
As illustrated by the DPRM problem presented in Sec.~\ref{sec:DPRM}, the answers to all these questions are encoded in the properties of the free-energy in the large $n$ limit, $\phi(\beta) = \lim_{n \to \infty} \avg{\ln {\cal T} (\beta)}/n$. Unfortunately in general $\phi(\beta)$ cannot be computed exactly (as we will see below, for the Anderson model on the Cayley tree an exact computation of $\phi(\beta)$ can be performed in the framework of the FSA, see Eq.~\eqref{eq:phiFSA}). Yet, as discussed above for the DPRM problem, $\phi(\beta)$ can be estimated numerically in two complementary ways, namely from the finite-size behavior of the
quenched and annealed free-energies~\cite{biroli2020anomalous,kravtsov2018non,biroli2012difference,feigel2010superconductor,ioffe2010disorder,ros2021fluctuation}:
\begin{equation} \label{eq:phiCT}
\begin{aligned}
\phi_{\rm ann} (\beta,n) & = \frac{1}{ \beta n} \, \ln \avg{{\cal T} (\beta)} \, , \\
\phi_{\rm q} (\beta,n) & = \frac{1}{ \beta n} \, \avg{\ln {\cal T} (\beta)} \, . \\
\end{aligned}
\end{equation}
The averages here are performed over several independent realizations of the random energies $\epsilon_i$ of the non-interacting tight-binding Hamiltonian~\eqref{eq:Handerson}, which yield the effective (correlated) energy landscape explored by the polymers via Eqs.~\eqref{eq:cavity}. Note that $\phi_{\rm ann}$ and $\phi_{\rm q}$ are in fact defined as {\it minus} the annealed and quenched free-energies of the directed polymer, Eqs.~\eqref{eq:fqDPRM} and~\eqref{eq:faDPRM}, in such a way that they are directly defined as the rate functions of the generalized conductivities, and positive (resp. negative) free-energies correspond to exponentially large (resp. exponentially small) transmissions. 

To answer the second question we just need to follow the position of the minimum of $\phi_{\rm ann}$: If $\beta_\star (n)$ stays smaller than $2$ upon increasing the system size, then the transport of particles from the root to the leaves only occurs through a few specific disorder-dependent paths. Conversely, if $\beta_\star (n)$ becomes larger than $2$ at large $n$, then the transport occurs over an exponential number of paths.

The answer to the first question resides instead in the sign of the free-energy at $\beta=2$: a negative value of $\phi(\beta=2)$ implies that the typical value of the total conductivity from the root to the leaves (in the large $n$ limit) decreases exponentially with the radius of the tree, corresponding to the localized regime. Conversely, a positive value of $\phi(\beta=2)$ implies the probability that a particle that starts on the root at $t=0$ is found on any node of the leaves after infinite time stays finite in the thermodynamic limit, corresponding to the extended regime. In fact, as explained above, the free-energy at $\beta=2$ is proportional to the Lyapunov exponent $\gamma$ describing the exponential growth or the exponential decay of the typical value of the imaginary part of the Green's functions under iteration. In particular, from  Eq.~\eqref{eq:ImG} one has that:
\begin{equation} \label{eq:lyap_def}
\gamma = \lim_{n \to \infty} \frac{\avg{\ln {\rm Im} G_{0,0}}}{n} = \lim_{n \to \infty} \frac{\avg{\ln {\cal T} (2)}}{n} = 2 \phi(2) \, .
\end{equation}
In fact the rate functions~\eqref{eq:phiCT} 
first appeared in the mathematical literature in the context of the so-called ``fractional moment method'' of Refs.~\cite{aizenman1993localization,warzel2013resonant}, where it was rigorously proven that the annealed free-energy $\phi_{\rm ann} (\beta)$ converges to a Lyapunov function that bounds the asymptotic behavior of the expectation value of the $\beta$-th moment of the imaginary part of the Green's function in the linearized regime. 

In order to 
estimate the value of the free-energy at $\beta=2$ in the large $n$ limit 
one has---in principle---to distinguish between two different situations:
\begin{itemize}
\item[-] If $\beta_\star >2$, then $\beta=2$ is in the high-temperature regime in which the annealed and quenched free-energies~\eqref{eq:phiCT} both converge to $\phi(\beta)$. We thus have that:
\[
\lim_{n \to \infty} 2 \phi_{\rm q} (2,n) = \lim_{n \to \infty} 2 \phi_{\rm ann} (2,n) = \gamma \, ,
\]
with $\phi_{\rm q} (2,n) \le \phi_{\rm ann} (2,n)$.
\item[-] If $\beta_\star <2$ the localization transition occurs in a regime in which transport is dominated by few paths. As explained in Sec.~\ref{sec:DPRM} in this case in order to obtain a function that in the large $n$ limit converges to quenched the free-energy one has to replace the variation of $\phi_{\rm ann}$ beyond $\beta_\star$ by a constant given by its value in $\beta_\star$, as in Eq.~\eqref{eq:replacement}:
\begin{equation} \label{eq:replacementCT} 
\tilde{\phi}_{\rm ann} (\beta,n) = \left \{
\begin{array}{ll}
\phi_{\rm ann} (\beta,n) \,\, & \textrm{for~} \beta< \beta_\star(n) \, , \\
\phi_{\rm ann} (\beta_\star(n),n) & \textrm{for~} \beta \ge \beta_\star(n) \, ,
\end{array}
\right .
\end{equation}
We thus obtain:
\[
\lim_{n \to \infty} 2 \phi_{\rm q} (2,n) = \lim_{n \to \infty} 2 \phi_{\rm ann} (\beta_\star,n)  = \gamma \, ,
\]
with $\phi_{\rm q} (2,n) \le \phi_{\rm ann} (\beta_\star,n)$.
\end{itemize}
Hence the numerical estimations of the Lyapunov exponent obtained from the quenched and annealed free-energies of finite-size samples can be represented in a compact form as:
\begin{equation} \label{eq:lyap}
\begin{aligned}
\gamma_{ \rm q} (n) & \equiv  2 \phi_{\rm q} (2) \, ,\\
\gamma_{\rm ann} (n) & \equiv  {\rm min} \{ 2 \phi_{\rm ann} (\beta_\star,n), 2 \phi_{\rm ann} (2,n) \} \, , \\
\lim_{n \to \infty} \gamma_{\rm q} (n) & = \lim_{n \to \infty} \gamma_{\rm ann} (n)  = \gamma\, .
\end{aligned}
\end{equation}
At finite $n$, by construction, one has that $\gamma_{\rm q} (n) \le \gamma_{\rm ann} (n)$.

So far we have seen that the annealed and quenched free-energies encode two crucial information: The position of the breaking point $\beta_\star$ indicates whether physical transport is dominated by an exponential number of paths --- if $\beta_\star >2$ --- or by few $O(1)$ paths --- if $\beta_\star < 2$. The sign of the free-energy at $\beta=2$ indicates whether the system is in the localized phase --- if $\gamma < 0$ --- or in the delocalized phase --- if $\gamma > 0$. As explained in Sec.~\ref{sec:DPRM}, the Legendre transform of the free-energy yields also the answer to the third question, namely the asymptotic value of the number of paths contributing to physical transport, \ie, the number of terms contributing to the sum~\eqref{eq:T} at a given ``temperature'' $\beta$. This can be shown again using  the analogy with DPRM (see Sec.~\ref{sec:DPRM}): Denoting $e^{-n \beta \epsilon_{\cal P}}$ the contribution of a given path of length $n$, one can rewrite Eq.~\eqref{eq:T} as an integral over all paths giving a contribution characterized by a value of $\epsilon_{\cal P}$ between $\epsilon_{\cal P}$ and $\epsilon_{\cal P} + {\rm d} \epsilon_{\cal P}$ times the number of such paths, defined as $e^{n \Sigma(\epsilon_{\cal P})}$. 
The value of $\epsilon_{\cal P}$ that dominates the integral for $n \to \infty$ is given by the condition ${\rm d} \Sigma / {\rm d} \epsilon_{\cal P} |_{\overline{\epsilon}_{\cal P}} = \beta$, and depends on $\beta$. For small enough $\beta$, one finds that the saddle point value of $\overline{\epsilon}_{\cal P} (\beta)$ is such that $\Sigma( \overline{\epsilon}_{\cal P} ) > 0$. In this regime an exponential number of paths contribute to the sum. By increasing $\beta$, $\Sigma(\overline{\epsilon}_{\cal P})$ decreases until the value $\beta_\star$ is reached. At this point the configurational entropy $\Sigma(\overline{\epsilon}_{\cal P}(\beta_\star))$ vanishes. The generalized free-energy of the directed polymer is related to the Legendre transform of $\Sigma$: ${\rm d} \phi / {\rm d} \beta  = - \Sigma(\overline{\epsilon}_{\cal P}(\beta_\star))/\beta^2$. 
Therefore, the number of paths contributing to the sum~\eqref{eq:T} for a given temperature $\beta$ is given by $e^{- n \beta^2 \phi^\prime (\beta)}$ for $\beta < \beta_\star$, and is zero (in the sense that it is sub-exponential in $n$) for $\beta>\beta_\star$. In particular, the number of paths contributing to physical transport for $\beta=2$ is $ e^{- 4 n \phi^\prime (\beta = 2)}$. A finite-$n$ estimation of the configurational entropy (\ie, the logarithm of the number of paths contributing to the total Landauer transmission ${\cal T}(2)$ per unit length) can thus be obtained from the derivatives of the quenched and annealed configurational entropies at $\beta=2$  as:
\begin{equation} \label{eq:sigma}
\Sigma_{\rm ann} (n) = -4 \, \frac{{\rm d} \phi_{\rm ann}}{{\rm d} \beta} \bigg |_{\beta = 2} \, , \qquad
\Sigma_{\rm q} (n) = -4 \, \frac{{\rm d} \phi_{\rm q}}{{\rm d} \beta} \bigg |_{\beta = 2} \, .
\end{equation}
If $W$ and $n$ are such that $\beta_\star < 2$, then $\Sigma_{\rm ann}<0$ by definition, which should be replaced by $\tilde{\Sigma}_{\rm ann} (n) = -4 \, {\rm d} \tilde{\phi}_{\rm ann}/{\rm d} \beta |_{\beta = 2} = 0$. Instead $\Sigma_{\rm q} (n) \ge 0$ for all $W$ and $n$ by construction.

To sum up, our approach, which we will straightforwardly generalise to the interacting case, consists in taking the following steps:
\begin{itemize}
    \item Define the generalized Landauer transmissions~\eqref{eq:T} associated to the auxiliary parameter $\beta$;
    \item Compute numerically the quenched and annealed free-energies~\eqref{eq:phiCT} averaging over many samples of finite size;
    \item Focusing on the behavior (upon varying $W$ and $n$) of the following key observables:
    \begin{itemize}
        \item[-] The position of the minimum of $\phi_{\rm ann}$, $\beta_\star$;
        \item[-] The value of the free-energies at $\beta=2$, $\phi_{\rm q}(2)$ and $\tilde{\phi}_{\rm ann} (2)$, (where $\tilde{\phi}_{\rm ann}$ is defined in Eq.~\eqref{eq:replacementCT}), which provide two estimations of the Lyapunov exponent via Eq.~\eqref{eq:lyap};
        \item[-] The derivatives of $\phi_{\rm q}$ and $\phi_{\rm ann}$ at $\beta=2$ which provides the asymptotic scaling of the number of paths contributing to physical transport via Eq.~\eqref{eq:sigma}.
    \end{itemize}
\end{itemize}
We now proceed to apply this approach to the tight-binding Anderson Hamiltonian~\eqref{eq:Handerson} on the Cayley tree.

One key aspect of our work, that will be detailed below in the benchmark cases, is using annealed results to capture the crucial rare events, which appear only in rare samples for not large enough sizes, but which for large $n$'s are instead present in typical samples. This procedure is rooted in the theory developed for DPRM and the REM transition. As already stressed, we cannot exclude (or prove) that such theory and a bona-fide DPRM-like transition hold for the quantum many body problem. This should be considered as a working hypothesis. The studies in the benchmark cases presented in the following and their comparison with the analysis of the many-body cases bring positive evidence in favor of it. 

\subsection{The quenched and annealed free-energies}

\begin{figure*}
\includegraphics[width=0.338\textwidth]{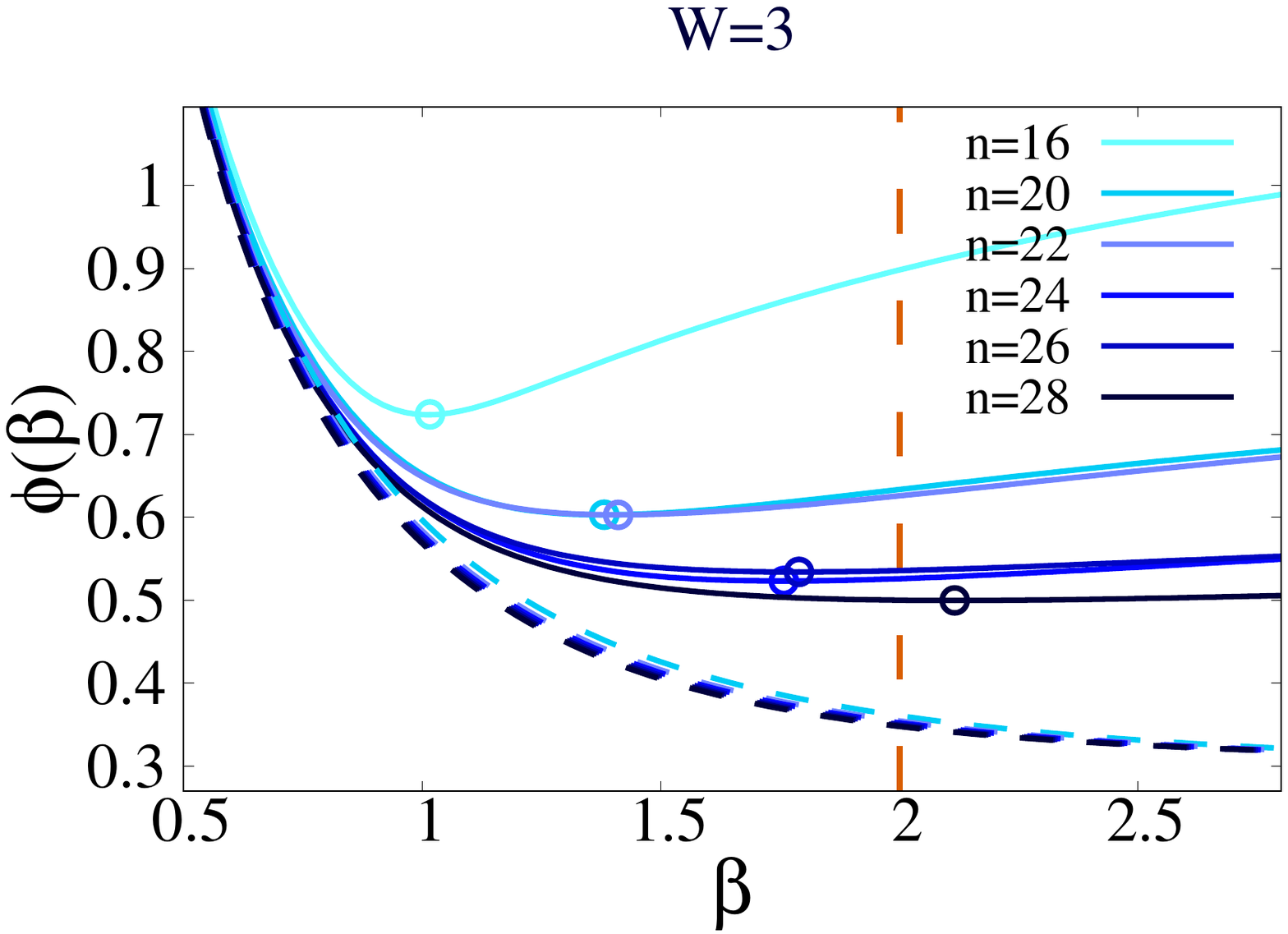} \hspace{-0.36cm} \includegraphics[width=0.338\textwidth]{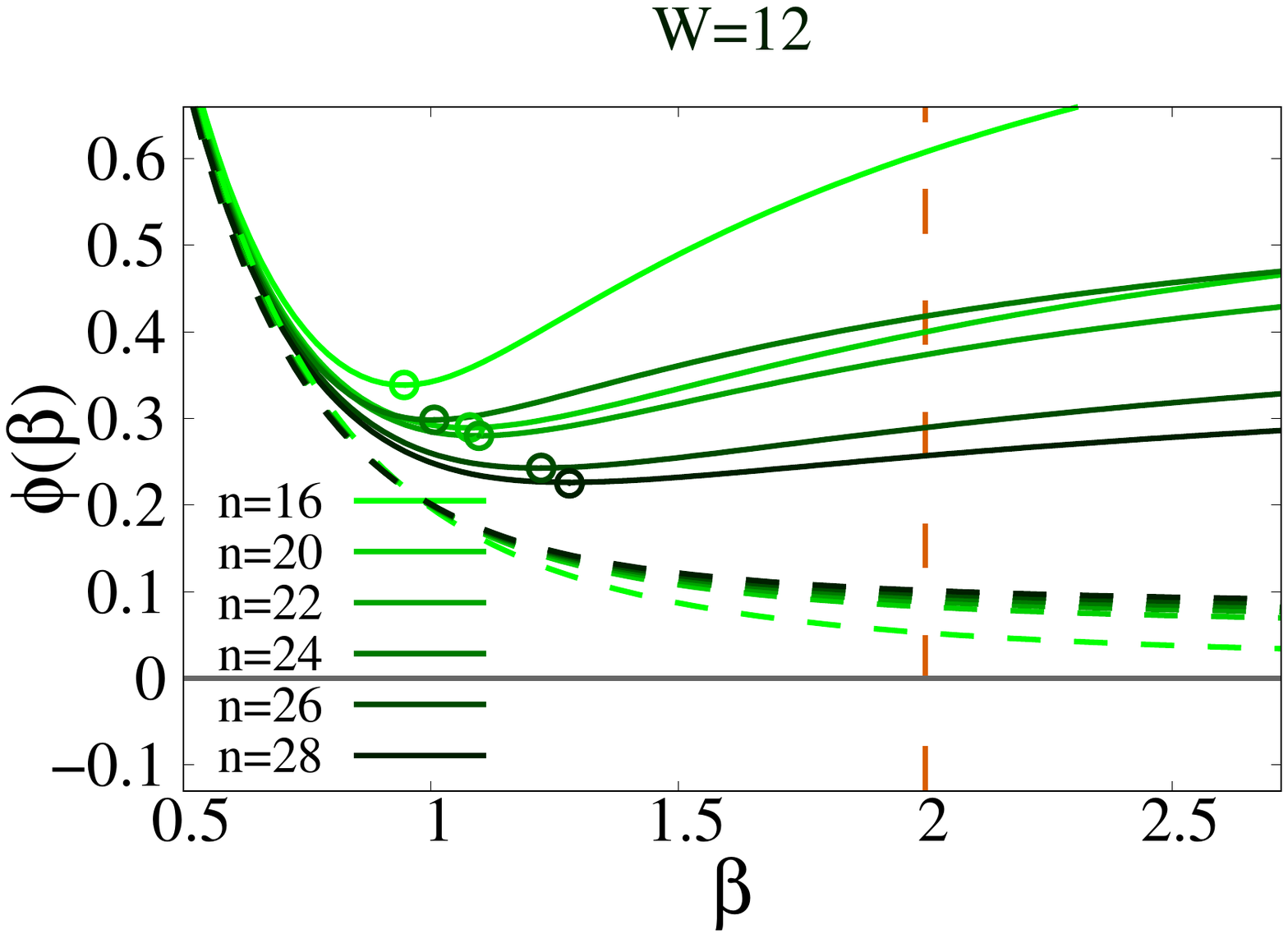} \hspace{-0.36cm} \includegraphics[width=0.338\textwidth]{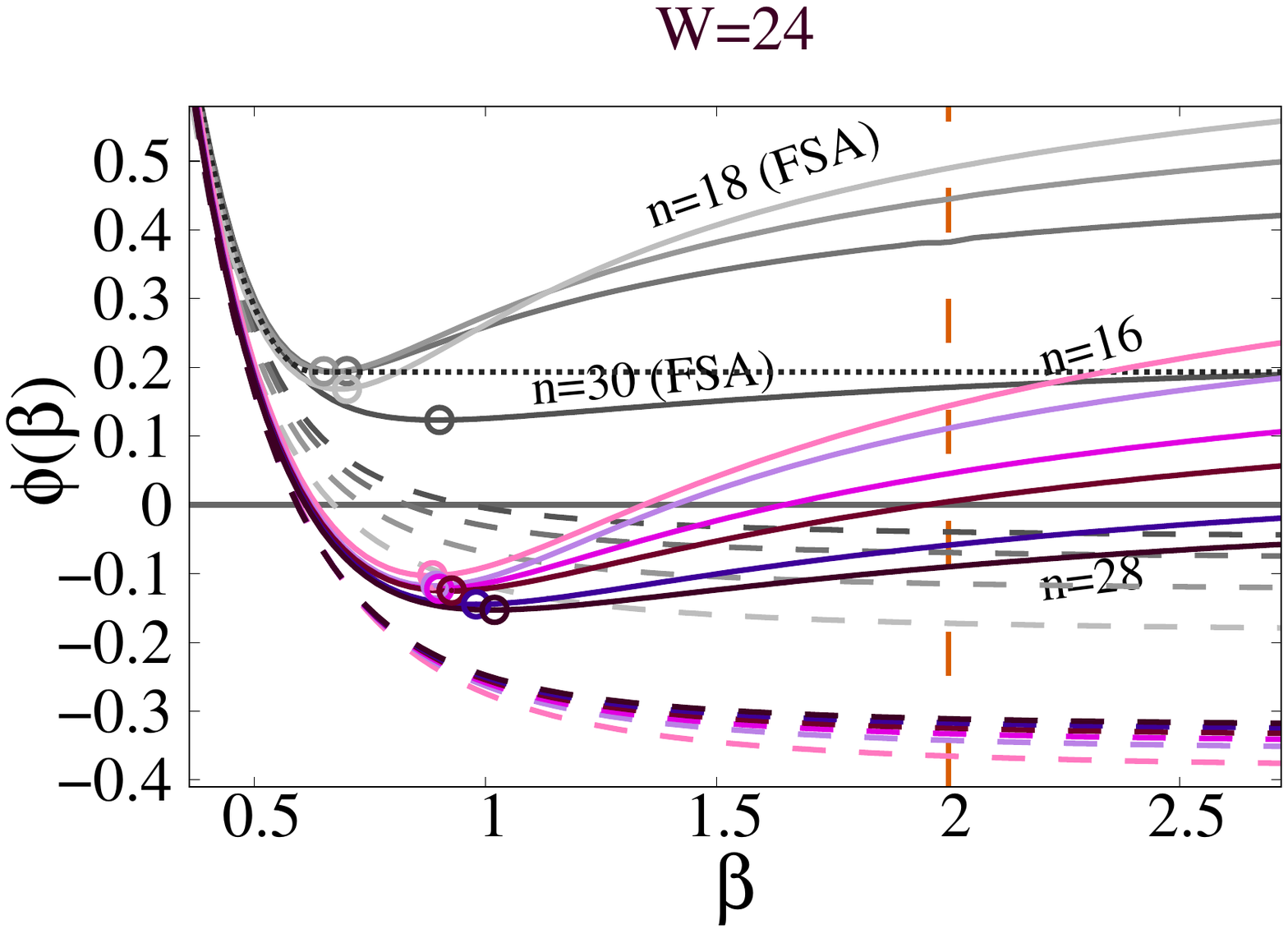} 
\caption{Annealed (continuous curves) and quenched (dashed curves) free-energies, Eq.~\eqref{eq:phiCT}, as a function of the inverse ``temperature'' $\beta$ for several values of the number of generations $n$ and for $W=3$ (left), $W=16$ (middle), and $W=24$ (right). The circles spot the positions of the minima of the annealed free-energy. 
The gray continuous and dashed curves in the right panel show the results obtained for the annealed and quenched free-energy within the FSA for $18 \le n \le 30$. The dotted gray curve shows the FSA free-energy in the $n\to\infty$ limit, Eq.~\eqref{eq:phiFSA}. The vertical dashed line at $\beta=2$ indicate the value of the auxiliary parameter $\beta$ which corresponds to physical transport.
\label{fig:CT}}
\end{figure*}

 The computation of the free-energies~\eqref{eq:phiCT} has been already discussed in several previous studies~\cite{biroli2020anomalous,monthus2008anderson,monthus2011anderson,tikhonov2016fractality,sonner2017multifractality,kravtsov2018non}, which focused directly on the thermodynamic limit (see App.~\ref{app:PD}). These studies have shown the existence of three distinct phases: A weak disorder delocalized phase in which an exponential number of paths contribute to the transport of particles from the root of the tree to the leaves ($\beta_\star > 2$). An intermediate delocalized ``glassy'' regime, in which a particle starting from the root does reach the leaves ($\gamma>0$) but only through a few, rare, disorder-dependent paths ($\beta_\star <2$). A localized phase at strong disorder in which, even on these rare paths, the probability that a particle reaches the boundaries decays exponentially with the number of generations ($\gamma<0$).  Here, with the aim of constructing a guideline for the interacting problem, instead of considering the $n \to \infty$ limit, we employ a different strategy which consists in generating many instances of the random energies on Cayley trees of $n$ generations, solving the self-consistent equations for the Green's functions~\eqref{eq:cavity} for each specific realization, computing the generalized conductivities for many values of $\beta$ according to Eqs.~\eqref{eq:T}, and obtaining the quenched and annealed free-energies~\eqref{eq:phiCT} averaging over many independent instances. In practice this procedure can be carried out for Cayley tress up to $n \le 28$ generations.

As explained above, in the ``glassy'' phase --- at strong disorder and/or large $\beta$ --- 
${\cal T}(\beta)$ is dominated by exponentially rare outliers, \ie~a few rare paths such that $|G_{0,i_n}|$ is very large. 
Crucially, such rare outliers are precisely the long-range rare resonances (that at finite $n$ may exist only in few realization of the disorder) that allow a particle starting on the root of the tree at $t=0$ to reach the boundary after an infinite time. In the context of MBL such rare resonances are expected to play a crucial role  both in the destabilization of the insulating phase via the avalanche mechanism~\cite{morningstar2022avalanches,Ha2023many}, and in the unusual transport and relaxation observed in the vicinity of the transition~\cite{agarwal2015anomalous,luitz2017ergodic,agarwal2017rare}. The analysis of the properties of the quenched and annealed free-energies~\eqref{eq:phiCT} provides therefore a compact and efficient set of observables to characterize the statistics of these rare resonances in a setting in which the physics in the infinite-size limit in known.

\begin{figure}
\includegraphics[width=0.44\textwidth]{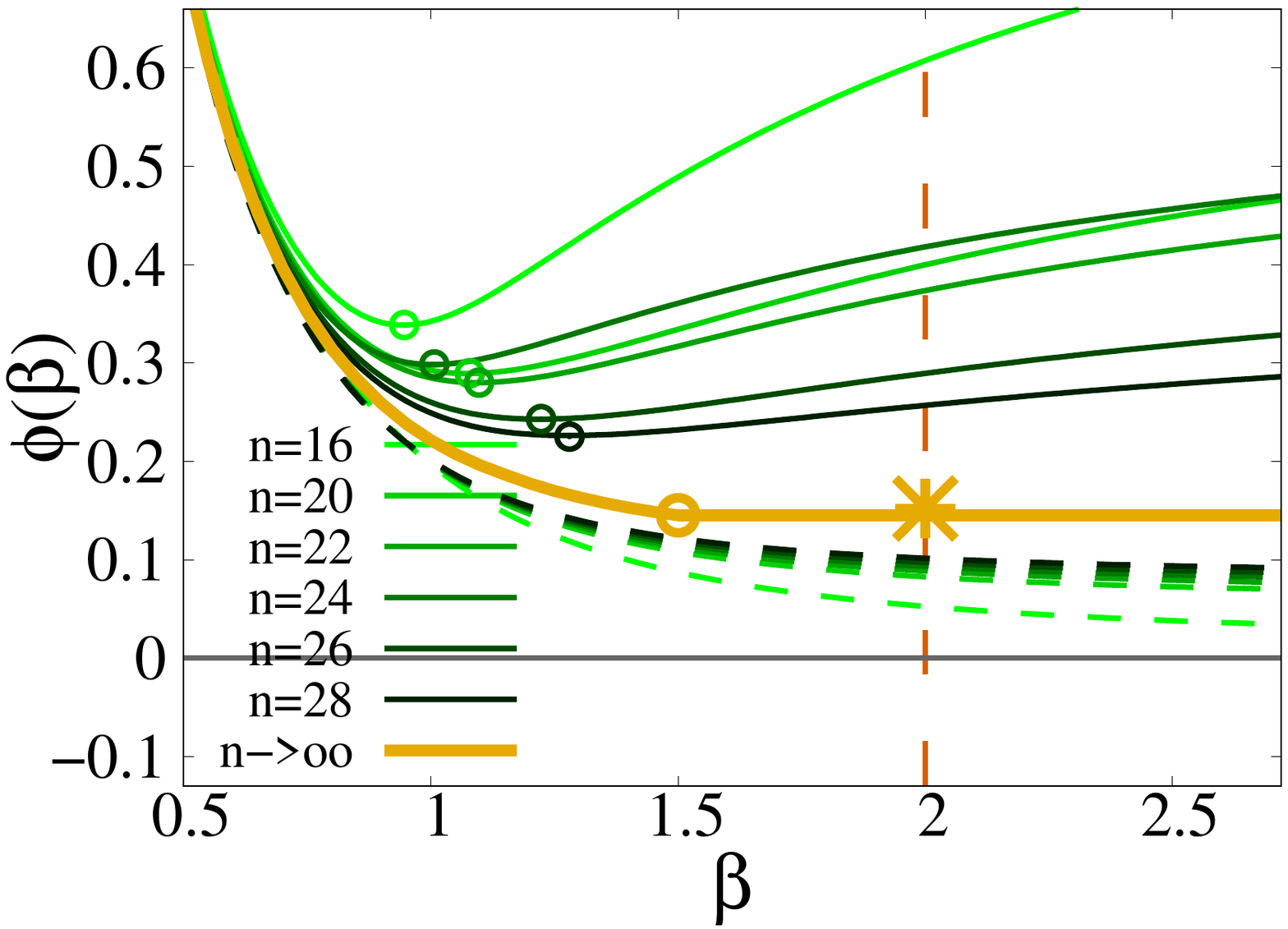} 
\caption{Annealed (continuous curves) and quenched (dashed curves) free-energies as a function of $\beta$ for $W=12$ at finite $n$ (as in the middle panel of Fig.~\ref{fig:CT}). The yellow thick curve represents the extrapolated behavior of the free-energy in the $n \to \infty$ limit obtained using the population dynamics algorithm (see App.~\ref{app:PD} and Ref.~\cite{biroli2020anomalous,kravtsov2018non}). The finite-$n$ quenched free-energy converges to the yellow curve from below. The annealed free-energy, with the replacement of its dependence on $\beta$ by a constant as in Eq.~\eqref{eq:replacementCT}, converges to the yellow curve from above. The star shows the values of the free-energy at $\beta=2$ in the thermodynamic limit, which yields the Lyapunov exponent $\gamma$.  
\label{fig:phiCTPD}}
\end{figure}

In Fig.~\ref{fig:CT} we show the numerical results for the annealed and quenched free-energies for $W=3$ (left), $W=12$ (middle), and $W=24$ (right), and for $n$ between $16$ and $28$.  The shape of $\phi_{\rm ann}$ and $\phi_{\rm q}$ is very similar to that of the classical DPRM problem shown in Fig.~\ref{fig:DPRM} (apart from the minus sign): At finite $n$ the quenched free-energy is a smooth decreasing function of $\beta$ for any $W$. The annealed free-energy, instead, exhibits a minimum at a disorder-dependent value of the inverse temperature $\beta_\star$ (circles), which depends on $W$ and on the number of generations. For $\beta<\beta_\star$ the quenched and annealed free-energies tend to converge to the same function upon increasing $n$. The presence of a minimum of the annealed free-energy in $\beta_\star$ signals the onset of a regime in which the rare outliers of $|G_{0,i_n}|^\beta$ start to dominate the sum~\eqref{eq:T} for $\beta>\beta_\star$. These outliers depend on the specific realization of the random on-site energies. Therefore the value of ${\cal T} (\beta)$ in typical samples differs from its average, which is given by the extreme value statistics of rare samples with particularly large values of ${\cal T}(\beta)$. 

It is important to keep in mind that the computational time is exponential in $n$ and therefore the number of realizations that we could consider numerically to compute the averages in~\eqref{eq:phiCT} is limited by the system size. More precisely, we performed $2 \cdot 10^6$ realizations for $n=16$, $2 \cdot 10^5$ realizations for $n=20$, $2 \cdot 10^5$ realizations for $n=22$, $10^5$ realizations for $n=24$, $3 \cdot 10^4$ realizations for $n=26$, and $8 \cdot 10^3$ realizations for $n=28$. The diminution of the number of realizations limits the influence of rare but important outliers for the annealed result.

At weak disorder --- $W=3$, left panel --- $\beta_\star$ moves to larger values as $n$ is increased and becomes larger than $2$ for large enough $n$. This implies that for large enough trees the physical conductivity, \ie~the total Landauer transmission ${\cal T} (\beta=2)$, receives significant contribution from an exponential number of paths. Conversely, at intermediate disorder --- $W=12$, middle panel --- $\beta_\star$ seems to approach a value between $1$ and $2$ at large $n$. This latter behavior corresponds to a freezing transition of the paths that contribute to transport at $\beta=2$. In Fig.~\ref{fig:phiCTPD} we compare the finite-$n$ results with the ones obtained in the large $n$ limit using the population dynamics algorithm (see App.~\ref{app:PD} and Ref.~\cite{biroli2020anomalous,kravtsov2018non}). This plot shows that in the thermodynamic limit $\beta_\star$ corresponds to a genuine phase transition, associated with a non-analytic behavior of the free-energy. In fact, as in the classical DPRM problem described in Sec.~\ref{sec:DPRM}, the glass transition is driven by the condensation of the Gibbs's measure onto the extreme values of the probability distribution of $|G_{0,i_n}|^\beta$,  
and the free-energy  becomes flat for $\beta \ge \beta_\star$ for $n \to \infty$ (see Fig.~\ref{fig:DPRM}). The finite-$n$ quenched free-energy converges to the large-$n$ result from below. The annealed free-energy, with the replacement of its dependence on $\beta$ by a constant as in Eq.~\eqref{eq:replacementCT}, converges to the large-$n$ limit from above.

At strong disorder --- $W=24$, right panel --- the quenched and annealed free-energies look similar to the intermediate disorder case, with a minimum around $\beta_\star \approx 1$, that moves only very little when $n$ is increased (the fact that $\beta_\star=1$ at the Anderson localization transition is in fact a rigorous result~\cite{warzel2013resonant,abou1973selfconsistent,parisi2019anderson,tikhonov2019critical}). The important difference between the middle panel and the right panel consists in the value of the free-energy at $\beta=2$, which is positive for $W=12$ and negative for $W=24$. The latter, as explained above, is an indicator of localization, since the free-energy at $\beta=2$ is proportional to the Lyapunov exponent describing the exponential growth or the exponential decay of the imaginary part of the Green's functions under iteration, Eqs.~\eqref{eq:lyap_def} and~\eqref{eq:lyap}, or, equivalently, the exponential growth or the exponential decay of the total Landauer transmission of the Cayley tree from the root to the boundary. 
For this specific problem, as shown in the plots of Fig.~\ref{fig:CT}, we observe that at strong enough disorder $\gamma_{\rm q}$ increases with the size of the tree, while $\gamma_{\rm ann}$ decreases with $n$. Hence, the conditions $\gamma_{\rm ann} = 0$ and $\gamma_{\rm q} = 0$ provide respectively an upper and lower bound for Anderson localization on the Cayley tree at finite $n$.

\begin{figure*}
\includegraphics[width=0.338\textwidth]{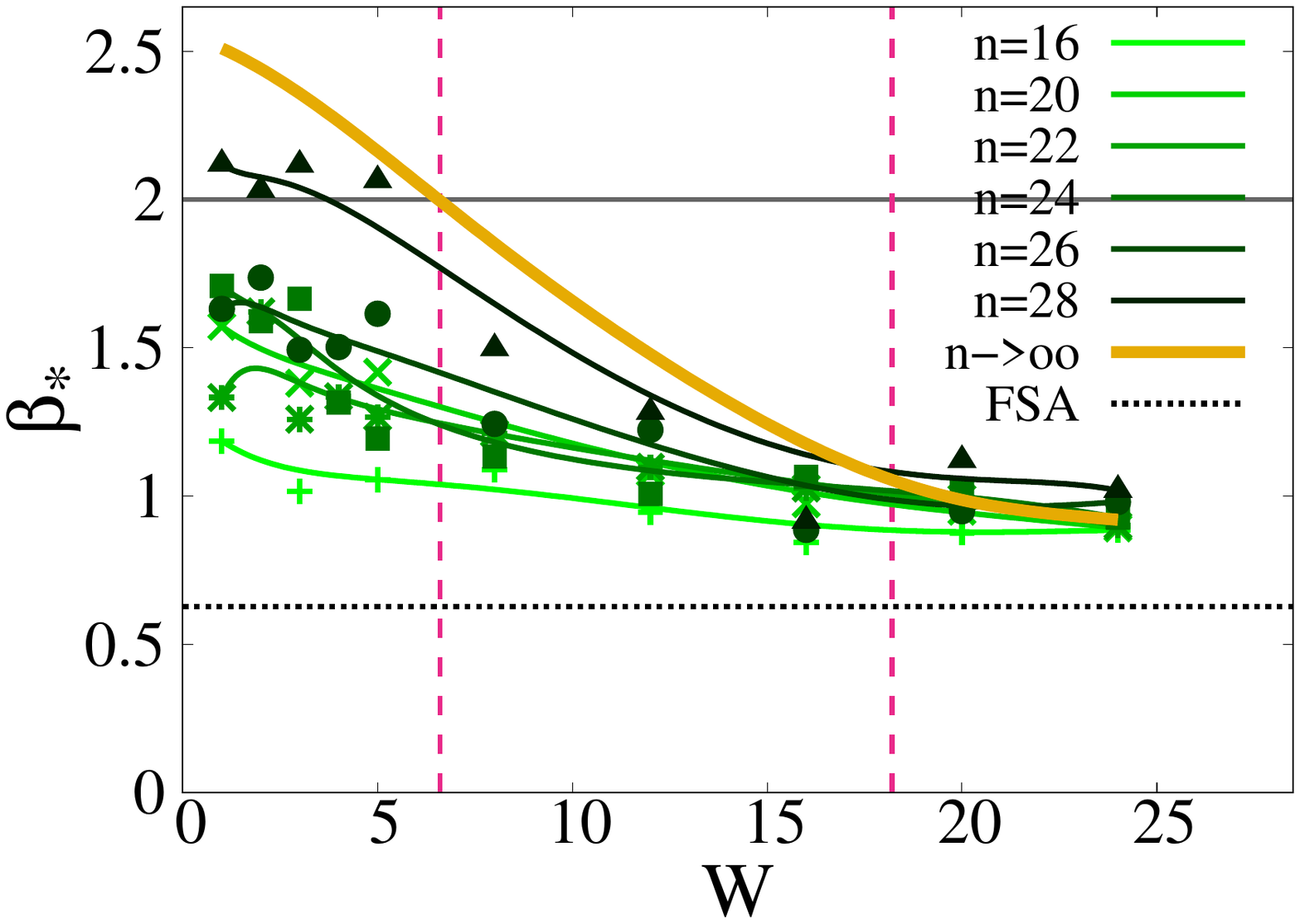} \hspace{-0.36cm} \includegraphics[width=0.338\textwidth]{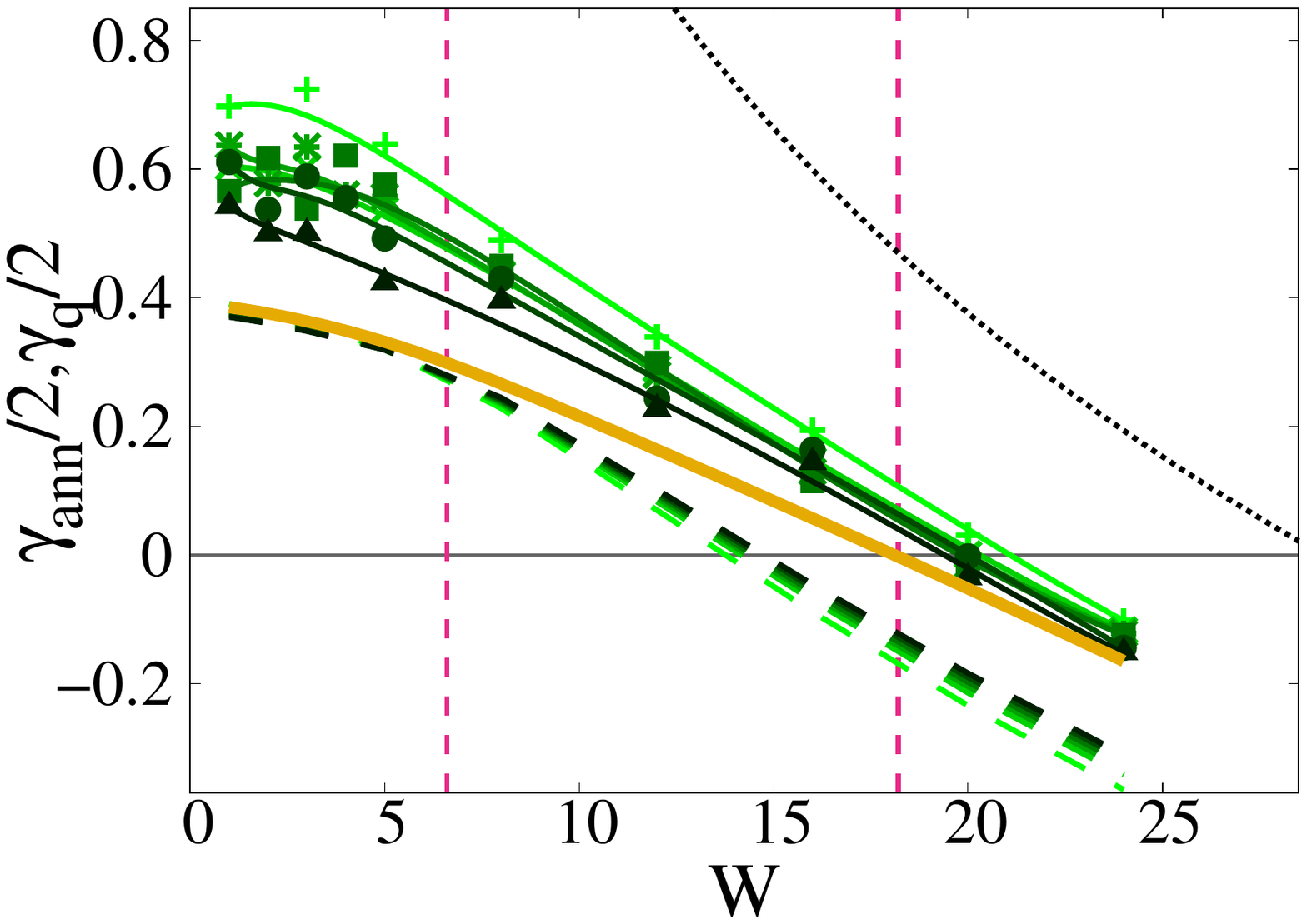} \hspace{-0.36cm}
\includegraphics[width=0.338\textwidth]{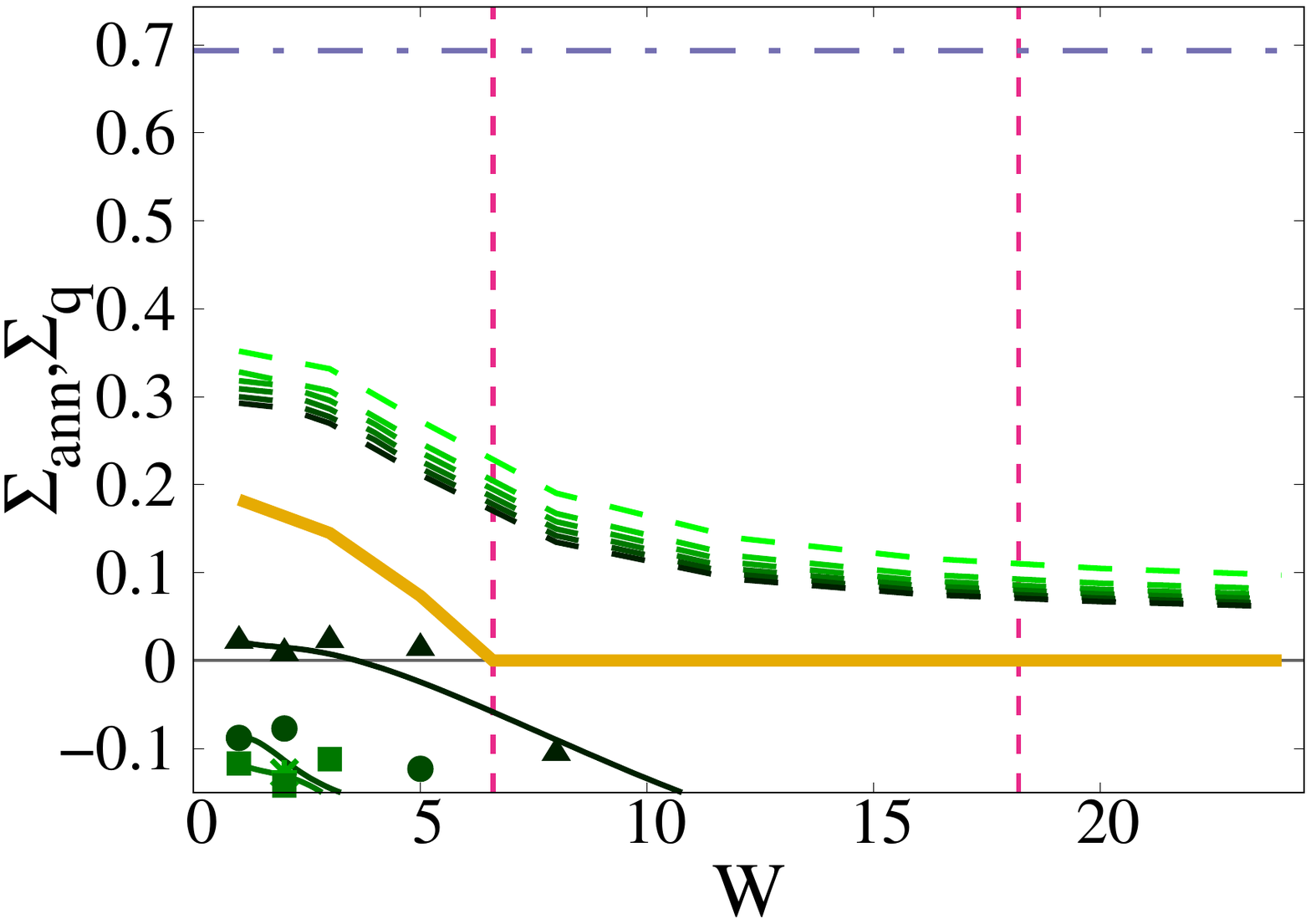} 
\caption{Left: Minimum of the annealed free-energy $\beta_\star$ as a function of $W$ for several sizes $n$ of the tree. 
The freezing 
of the paths contributing to ${\cal T}(2)$ takes place when $\beta_\star$ crosses $2$ at large $n$.  The gray dotted line shows the FSA value of $\beta_\star^{\rm fsa} \approx 0.6266$ in the $n \to \infty$ limit, which does not depend on $W$  (see Eq.~\eqref{eq:phiFSA}). Middle: Value of the free-energies at $\beta=2$ as a function of $W$ for several values of $n$, providing the quenched and annealed estimations of the Lyapunov exponent according to Eq.~\eqref{eq:lyap}. $\gamma_{\rm q}(n)$ (dashed lines) is obtained directly from $\phi_{\rm q} (2,n)$, while $\gamma_{\rm ann}(n)$ (symbols) is obtained from $\tilde{\phi}_{\rm ann}(2,n) = {\rm min} \{ \phi_{\rm ann} (\beta_\star(n),n), \phi_{\rm ann} (2,n) \}$ after replacing the $\beta$-dependence of $\phi_{\rm ann}$ beyond $\beta_\star$ by a constant equal to its value at $\beta_{\rm star}$, Eq.~\eqref{eq:replacementCT}.
In the $n \to \infty$ limit $\gamma_{\rm q}$ and $\gamma_{\rm ann}$ converge to the same value $\gamma$. AL occurs when $\gamma = 0$, $W_c \approx 18.17$~\cite{tikhonov2019critical,parisi2019anderson}. The gray dotted curve show the value of $\phi^{\rm fsa} (2) = \phi^{\rm fsa} (\beta_\star^{\rm fsa})$ obtained within the FSA for $n \to \infty$ (see Eq.~\eqref{eq:phiFSA}), which provide an upper bound for the critical disorder~\cite{abou1973selfconsistent}, $W_c^{\rm fsa} \simeq 29.122$. Right: Annealed (symbols) and quenched (dashed curves) configurational entropy at $\beta=2$ as a function of $W$ for several values of $n$. 
The horizontal dashed-dotted line corresponds to the maximum value $\Sigma_{\rm max} = \ln k$ if all the $k^n$ paths contribute uniformly to ${\cal T}(2)$. 
The continuous green lines in the three panels are guides for the eye obtained performing a Bézier cubic interpolation of the annealed data. The yellow curves show the results found in Ref.~\cite{biroli2020anomalous} in the large $n$ limit using the population dynamics algorithm, see App.~\ref{app:PD}. 
\label{fig:mstarCT}}
\end{figure*}

\subsection{The forward-scattering approximation} \label{sec:CTFSA}

In the strong disorder limit, deep in the localized phase, the self-energy contribution $t^2 \sum_{j_{m+1}} \!\! G_{j_{m+1} \to j_m}$ to the denominators of the cavity equations~\eqref{eq:cavity} are subdominant with respect to the random energies $\epsilon_{i_m}$ and can be neglected. In this limit the propagator between the root and one of the leaves simply becomes:
\begin{equation} \label{eq:FSA}
G_{0,i_n}^{\rm fsa} = \frac{-t}{\epsilon_{i_n}} \, \frac{-t}{\epsilon_{i_{n-1}}} \times \cdots \times \frac{-t}{\epsilon_{i_1}} \, \frac{-t}{\epsilon_{0}}  \, .
\end{equation}
This correspond to the so-called forward-scattering approximation~\cite{anderson1958absence,pietracaprina2016forward,abou1973selfconsistent}, which consists in retaining only the leading term of the perturbative expansion for the propagator starting from the insulator. $G_{0,i_n}^{\rm fsa}$ is obviously broadly distributed and its average over the random energies 
is infinite. However $\log|G_{0,i_n}^{\rm fsa} |$ is a sum of independent and identically distributed random variables with a finite variance and its probability distribution can be computed analytically (see App.~\ref{app:FSA} and Refs.~\cite{pietracaprina2016forward,abou1973selfconsistent}). 

Although for the Anderson model on the Cayley tree the propagators can be computed exactly using Eqs.~\eqref{eq:propagatorCT} (after solving Eqs.~\eqref{eq:cavity}), studying the problem within the FSA is instructive to understand the limitations and the virtues of this approximation, which will be used below in the many-body case to study larger systems compared to exact diagonalizations. In particular, since the random energies on different nodes are uncorrelated, the computation of ${\cal T} (\beta)$ within the FSA becomes formally equivalent to the classical DPRM problem described in Sec.~\ref{sec:DPRM} with the identification $e^{-\beta \omega_{i_m \leftrightarrow i_{m+1}}} = (t/|\epsilon_{i_m}|)^\beta$. As a consequence, the FSA quenched free-energy in the $n \to \infty$ limit can be computed from Eq.~\eqref{eq:fqDPRM} yielding:
\begin{equation} \label{eq:phiFSA}
\phi^{\rm fsa} (\beta)  = \left \{
\begin{array}{ll}
     \ln (2/W)  + \frac{1}{\beta} \ln \left( \frac{k}{1-\beta} \right) \,\, & \textrm{for~} \beta<\beta_\star^{\rm fsa}  \, ,\\
      \ln (2/W)  + \frac{1}{\beta_\star} \ln \left( \frac{k}{1-\beta_\star} \right) \,\, & \textrm{for~} \beta\ge \beta_\star^{\rm fsa} \, ,
\end{array}
\right .
\end{equation}
with $\beta_\star^{\rm fsa} \approx 0.6266$ (independently of $W$) given by the condition ${\rm d} \phi / {\rm d} \beta|_{\beta_\star^{\rm fsa}} = 0$.

For completeness, in the right panel of Fig.~\ref{fig:CT} we also show the quenched and annealed free-energies computed numerically within the FSA for 
Cayley trees of finite size. As explained in App.~\ref{app:FSA} and as indicated in Eq.~\eqref{eq:phiFSA}, 
changing $W$ leads only to a trivial vertical shift of order $\sim \log W$ of the FSA free-energies. 
Thus, we only plot the results for $W=24$ in the right panel of Fig.~\ref{fig:CT}, showing that deep in the localized phase the predictions of the FSA approximation look  qualitatively similar to the exact ones. 
The heights of the FSA free-energies, and in particular $\tilde{\phi}_{\rm ann}^{\rm fsa} (2) = \phi_{\rm ann}^{\rm fsa} (\beta_\star^{\rm fsa}) = \gamma_{\rm ann}^{\rm fsa} / 2$ and $\phi_{\rm q}^{\rm fsa} (2) = \gamma_{\rm q}^{\rm fsa} / 2$, decrease with $W$, and the conditions $\gamma_{\rm ann}^{\rm fsa} = 0$ and $\gamma_{\rm q}^{\rm fsa} = 0$ yield the FSA estimations of the critical disorder at which Anderson localization takes place. $W_c^{\rm fsa}$ can be computed analytically in the $n \to \infty$ limit from Eq.~\eqref{eq:phiFSA} imposing that $\phi^{\rm fsa} (W_c^{\rm fsa}) = 0$. This condition yields $W_c^{\rm fsa} = 2 (k/(1-\beta_\star^{\rm fsa}))^{1/\beta_\star^{\rm fsa}}$, and verifies the relation $W_c^{\rm fsa} = 2 e k \ln (W_c^{\rm fsa}/2)$~\cite{abou1973selfconsistent}, which is known to provide an upper bound for the exact value (\eg, $W_c^{\rm fsa} \simeq 29.122 > W_c \simeq 18.17$ for $k=2$). 

\subsection{Summary of the results}

\begin{figure*}
\includegraphics[width=0.338\textwidth]{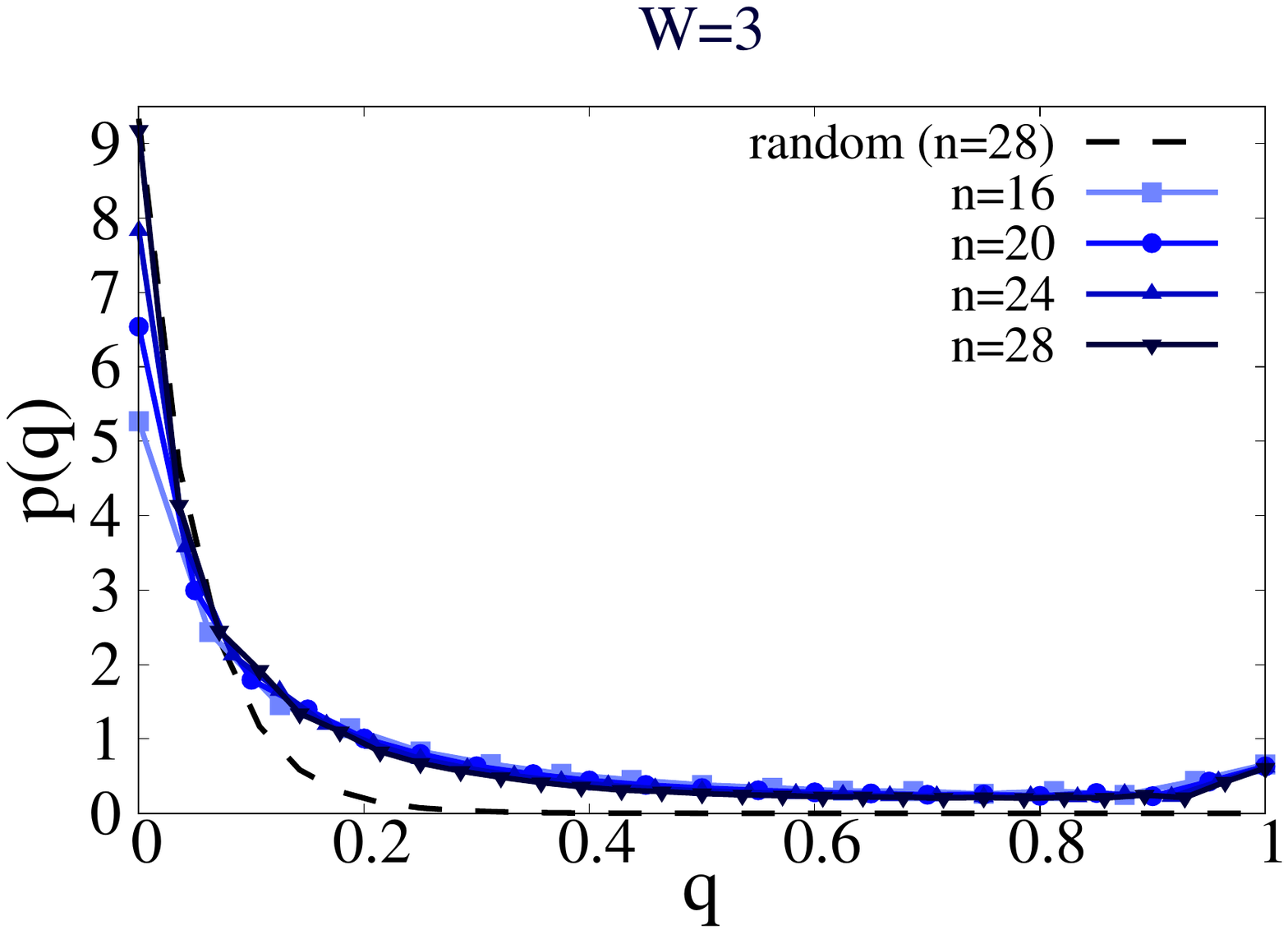} \hspace{-0.36cm} \includegraphics[width=0.338\textwidth]{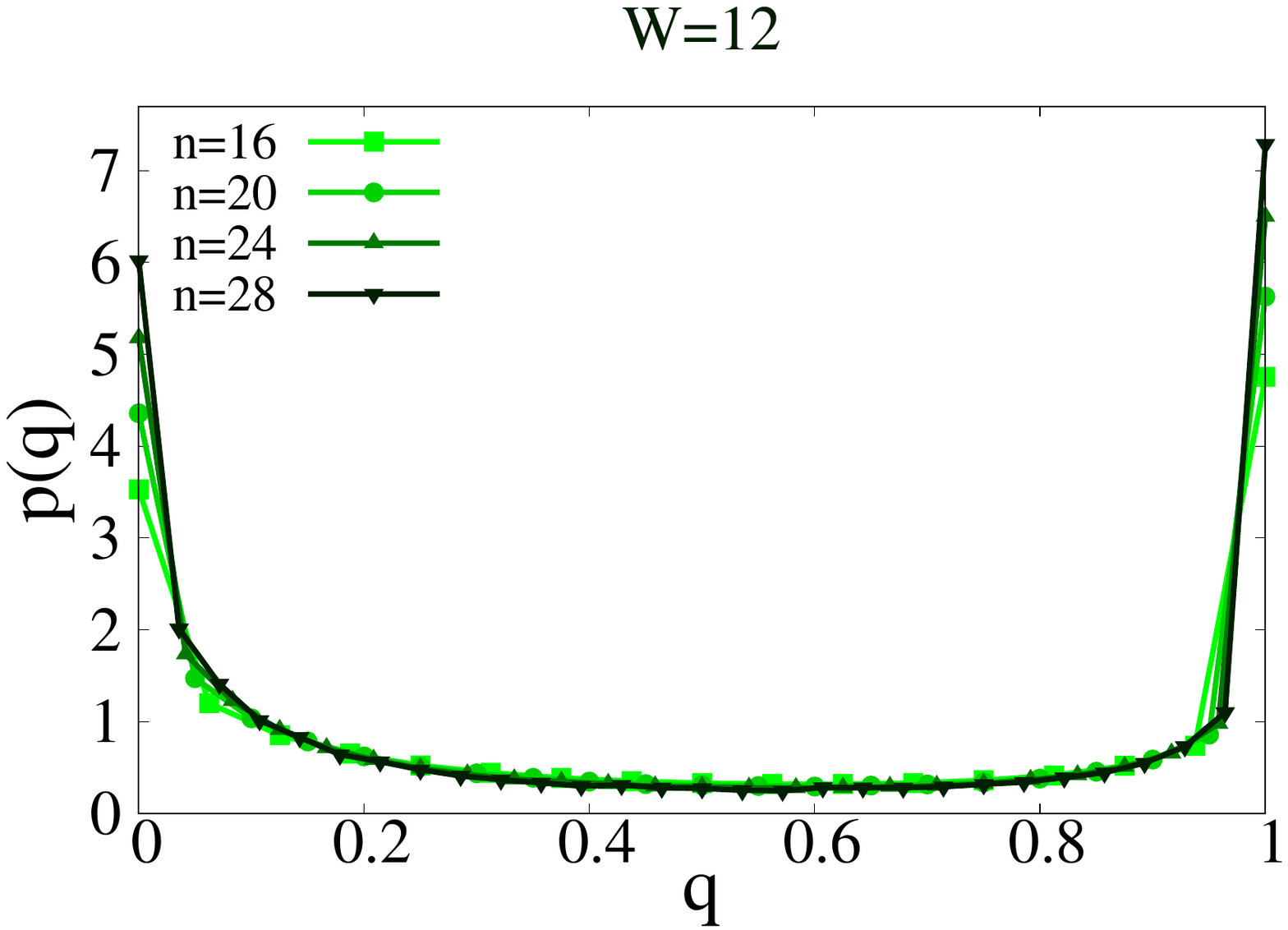} \hspace{-0.36cm} \includegraphics[width=0.338\textwidth]{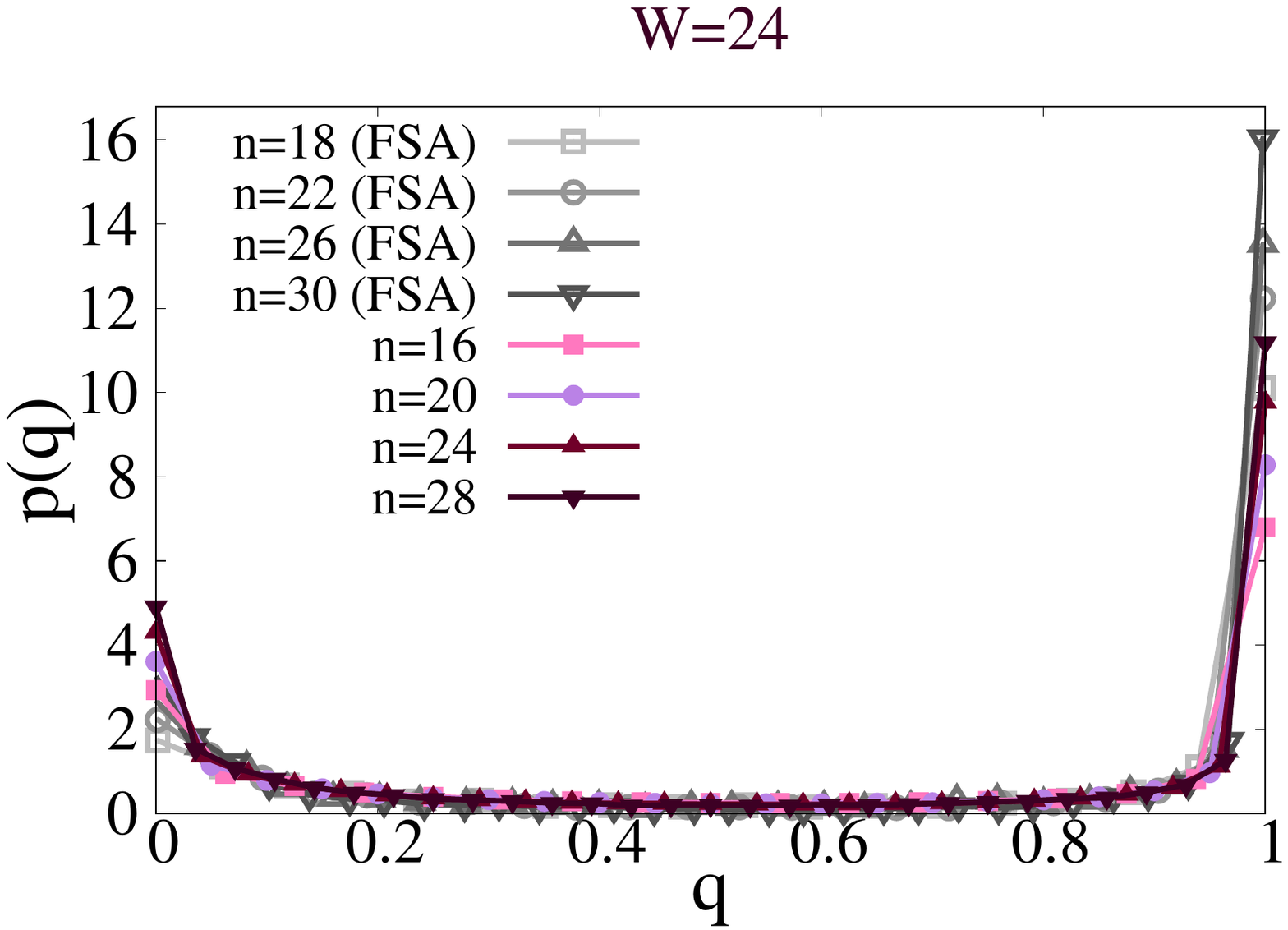} 
\caption{Overlap probability distribution $p(q)$, Eq.~\eqref{eq:pqCT}, for $W=3$ (left), $W=12$ (middle), and $W=24$ (right) for several values of the number of generations $n$. The dashed curve in the left panel represent the limiting exponential distribution in the totally random case, $p(q) \propto k^{-n q}$, for the largest size $n=28$. Note that in order to represent overlap distributions obtained for different values of $n$ on the same plot, we have rescaled the distribution by a factor $n$ in such a way that $\int \!\! p(q) {\rm d} q = 1$ (and not $\sum_q p(q) =1$).  The gray curves in the right panel show the results of the FSA (which are independent of $W$).
\label{fig:pqCT}}
\end{figure*}


In Fig.~\ref{fig:mstarCT} we plot the position of the minimum of the annealed free-energy $\beta_\star$ (left panel), the annealed and quenched estimations of the Lyapunov exponent at finite $n$ defined in Eq.~\eqref{eq:lyap}, and the annealed and quenched configurational entropies at $\beta=2$ defined in Eq.~\eqref{eq:sigma} (right panel), as a function of the disorder strength $W$ for several values of the number of generations $n$. We also show on the same plots the results obtained in Ref.~\cite{biroli2020anomalous} in the large $n$ limit using the population dynamics algorithm (see App.~\ref{app:PD}). These plots allows one to estimate the position of the localization transition $W_c$ (where $\gamma=0$), as well as the position of the freezing transition at which the transport of particles from the root to the leaves is dominated by few specific disorder-dependent paths $W_g$ (where $\beta_\star=2$). Note that for this problem one can establish with high accuracy the value of the critical disorder, $W_c \approx 18.17$ (for $k=2$), by diagonalizing explicitly the integral operator which governs the evolution of the imaginary part of the Green's function under iteration~\cite{tikhonov2019critical,parisi2019anderson}. It is important to notice that while the finite-size estimation of $W_c$ obtained from the values of the disorder at which $\gamma_{\rm q}$ vanishes is still far below the exact value for the accessible sizes (and slowly drifts to larger disorder as $n$ is increased), the estimation obtained from the condition $\gamma_{\rm ann}=0$ is remarkably close to it, even for moderately small trees (and slowly drifts to slightly smaller values of the disorder upon increasing $n$). As a consequence, in Cayley trees of large but finite sizes there is a broad disorder regime  ($15 \lesssim W \lesssim W_c$ for the accessible sizes) in which $\gamma_{\rm ann}>0$ but $\gamma_{\rm q}<0$.  This implies that at small enough $n$ the rare resonances that allow the particle to delocalize only exist for rare realizations of the disorder for which the average conductivity is much larger than the typical one. As one considers increasingly larger systems, the emergence of these rare resonances becomes more prevalent even in typical samples, ultimately manifesting as delocalization in the limit of large $n$.


Also $W_g$ has been previously computed in the literature at large $n$ in Refs.~\cite{biroli2020anomalous,kravtsov2018non}, yielding $W_g \simeq 6.6$, which is compatible with the behavior of $\beta_\star$ and $\Sigma$ at finite $n$. Comparing the middle panel of Fig.~\ref{fig:DPRM} with the right panel of Fig.~\ref{fig:mstarCT} we note that in the small disordered phase, $W < W_g$, the number of paths contributing to ${\cal T} (2)$ is much smaller than the total number of paths $k^n$ and, contrary to the DPRM case, does not tend to $k^n$ even for $W \to 0$. Since $e^{n \Sigma}$ is the number of nodes at the boundary of the tree reached after infinite time by a particle starting at the root at time $0$, the configurational entropy is a lower bound for the fractal dimension of the wave-functions' support set. Thus its behavior partly explains why the eigenstates of the Anderson model on the Cayley tree are always fractal, even in the delocalized phase~\cite{ostilli2022spectrum}, and reflects the fact that the level statistics is non-universal and is not described by the GOE even at small disorder~\cite{sade2003localization}.

As explained above, the freezing transition of the paths is associated with the condensation of the measure $|G_{0,i_n}|^\beta$ onto the extreme value statistics of the probability distribution of the propagators. It is therefore instructive to inspect the behavior of such probability distributions. This analysis is carried out in  App.~\ref{app:PG}, where we show that $P(|G_{0,i_n}|)$ develops power-law tails at large arguments which depend on $W$ and $n$.

\subsection{The overlap distribution}

In analogy with DPRM (see Sec.~\ref{sec:DPRM}) the order parameter function associated with the frozen phase is
the overlap distributions between two directed polymers thrown on a Cayley tree accordingly to their Boltzmann weights for a given  realization of the disorder~\cite{derrida1988polymers,mezard1987spin}. We only consider here the physical value of the inverse ``temperature'' $\beta=2$. The statistical weight of a given configuration of the polymer starting from the root $0$ and terminating on one of the leaves $i_n$ is $|G_{0,i_n}|^2$. The overlap $q({\cal P}_{0 \to i_n},{\cal P}_{0 \to j_n})$ between two configurations of the polymer terminating in $i_n$ and $j_n$ respectively is defined as the fraction of edges that the paths have in common (and is thus an integer multiple of $1/n$ varying between $0$ and $1$). In analogy with Eq.~\eqref{eq:pqDPRM}, the probability distribution of the overlap is  defined as:
\begin{equation} \label{eq:pqCT}
p(q) = \left \langle 
\sum_{{\cal P}, {\cal P}^\prime} \frac{|G_{0,i_n}|^2 \, |G_{0,j_n}|^2  }{{\cal T}^2(2)} \, \delta \left [q-q({\cal P}_{0 \to i_n},{\cal P}^\prime_{0 \to j_n}) \right] \right \rangle \, ,
\end{equation}
where the average is performed over several realizations of the disorder. In Fig.~\ref{fig:pqCT} we plot the overlap distribution for three values of $W$  in the three regimes described above. (We have rescaled the distributions in such a way that $\int \!\! p (q) {\rm d} q = 1$.) 

These plots show that the overlap distributions are very similar to the ones found in the classical DPRM problem, Eq.~\eqref{eq:pq1rsb}. At small disorder, $W=3<W_g$, $p(q)$ is a decreasing function of $q$. The weight of the peak in $q=0$ increases upon increasing the number of generations, although $p(q)$ does not converge to the limiting exponential distribution of the completely random case (\ie, when all the paths have the same statistical weight, $p(q) \propto k^{-n q}$, dashed curve). Conversely, for $W>W_g$ the overlap distribution develops an extra peak in $q=1$. Both in the intermediate regime ($W=12$, middle panel) and in the localized phase ($W=24$, right panel), both peaks in $q=0$ and $q=1$ grow as $n$ is increased, compatibly with the existence of a glass transition in the directed polymer  problem, associated with a one-step replica symmetry breaking~\cite{derrida1988polymers,mezard1987spin,derrida1980random}, well before the localization transition.

In the right panel of Fig.~\ref{fig:pqCT} we also show the distribution of overlaps obtained within the FSA. As mentioned, $p(q)$ is independent of the disorder within this approximation, which yields the strong-disorder limit of the exact distributions.

\begin{figure}
\includegraphics[width=0.44\textwidth]{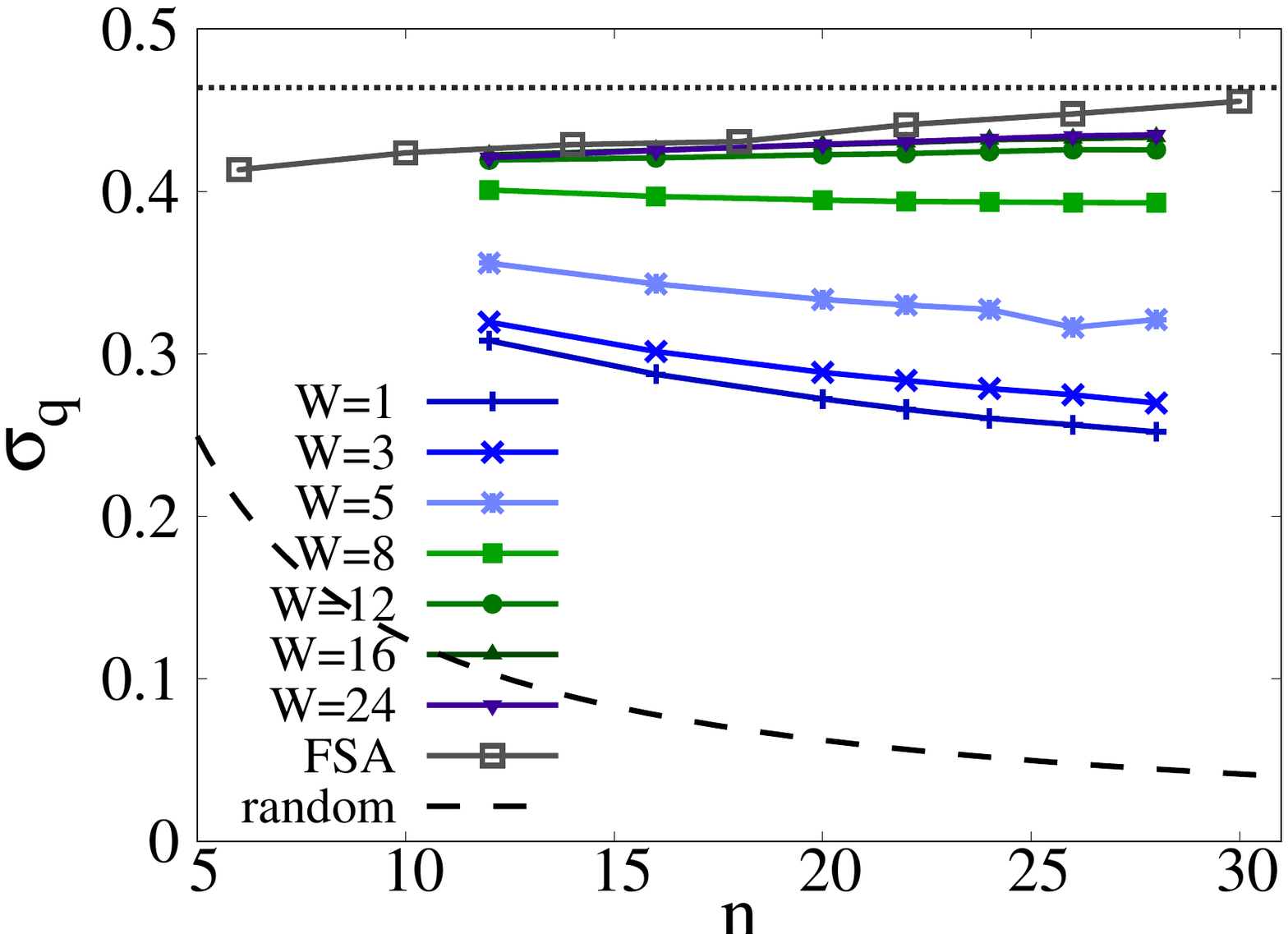} 
\caption{Square root of the variance of the overlap distribution as a function of the size of the Cayley tree for several values of the disorder. The dashed line gives the value of $\sigma_q$ in the completely random case, $\sigma_q \propto 1/n$.  The gray curve with empty squares show the results obtained within the FSA (which are independent of $W$). The gray dotted line corresponds to $\sigma_q^{\rm fsa} = \sqrt{(1 - \beta_\star^{\rm fsa}/\beta) \beta_\star^{\rm fsa} / \beta}$ obtained from Eq.~\eqref{eq:pq1rsb} in the $n \to \infty$ limit for $\beta_\star^{\rm fsa} \approx 0.6266$ and $\beta=2$. 
\label{fig:variance}}
\end{figure}

In order to go beyond the visual impression and obtain a quantitative estimation of the width of the overlap distribution, in Fig.~\ref{fig:variance} we plot the square root of the variance $\sigma_q = \sqrt{\avg{q^2} - \avg{q}^2}$ as a function of the size of the tree for several values of the disorder. The plot shows that, while in the low-disorder phase, $W<W_g$ (blue curves), $\sigma_q$ decreases with $n$ (albeit being still much above the completely random case $\sigma_q \propto 1/n$), for $W > W_g$ (both below and above the localization transition) $\sigma_q$ slightly grows with $n$. A fit to a function $\sigma_q (n) \approx  \sigma_q^\infty - an^{-\zeta}$ yields a result 
clearly distinct from zero for $\sigma_q^\infty$, confirming the existence of a broad distribution also in the limit $n \to \infty$. 

\begin{figure*}
\includegraphics[width=0.338\textwidth]{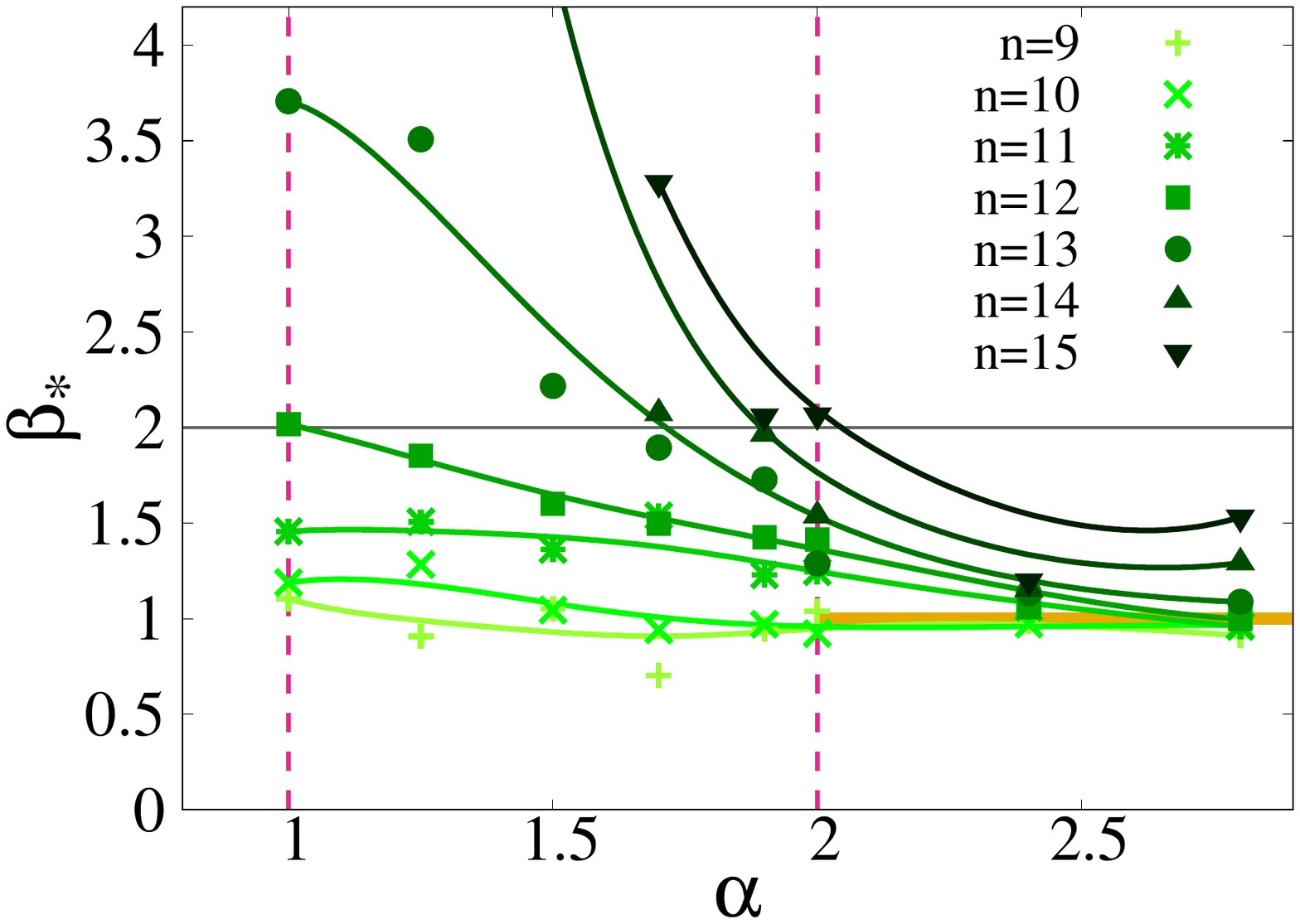} \hspace{-0.36cm} \includegraphics[width=0.338\textwidth]{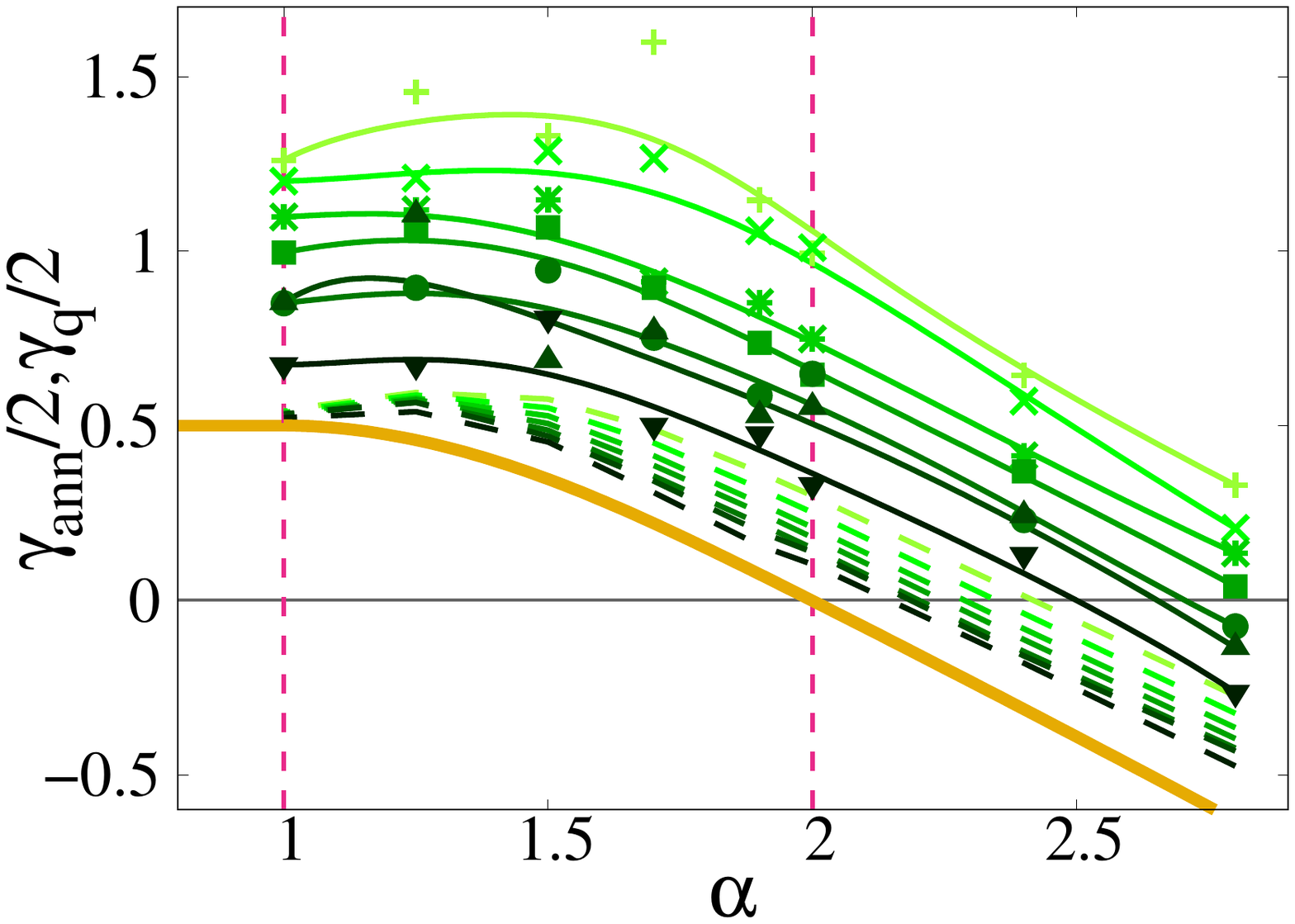} \hspace{-0.36cm}
\includegraphics[width=0.338\textwidth]{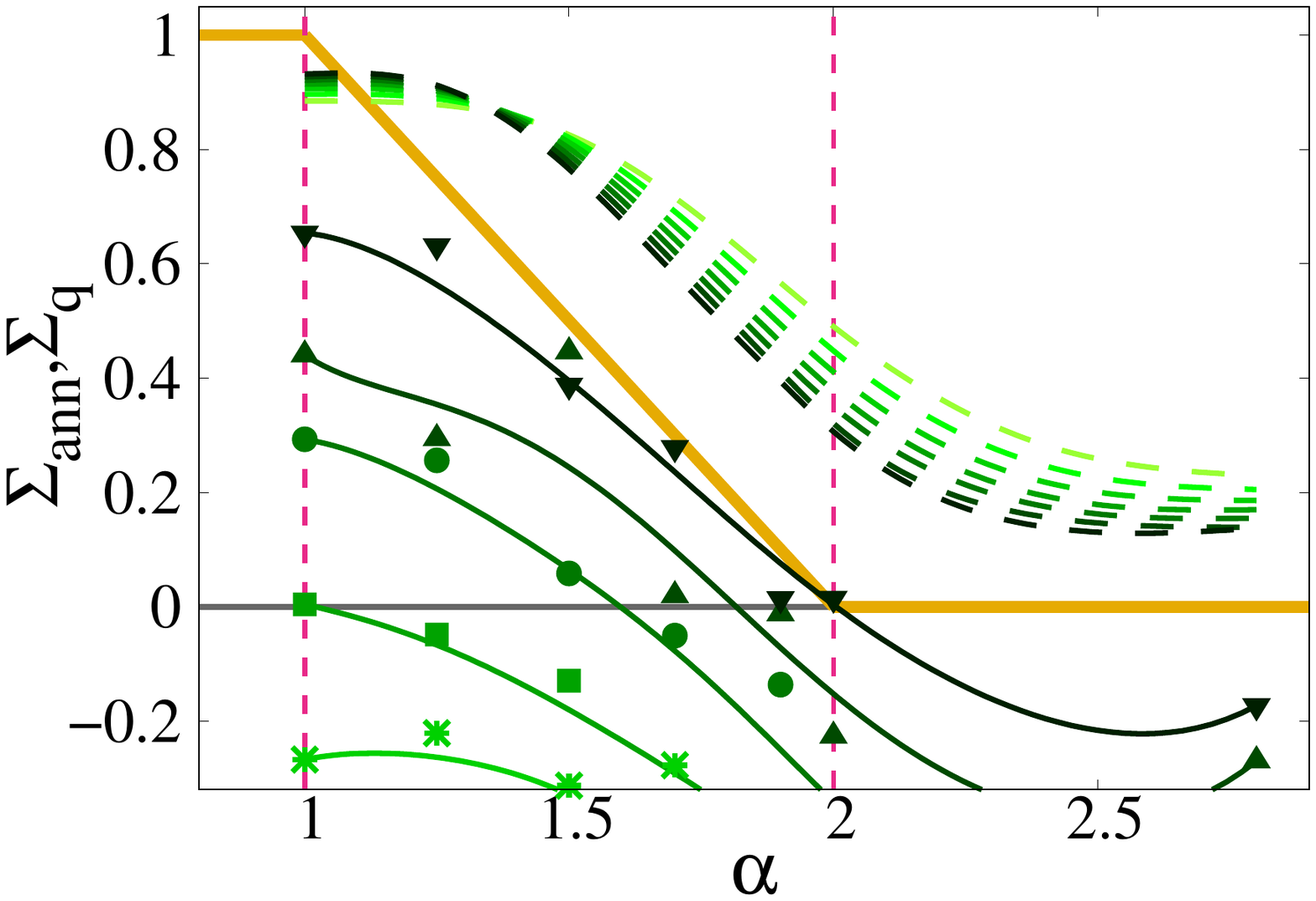} 
\caption{Left: Minimum of the annealed free-energy $\beta_\star$ as a function of $\alpha$ for several sizes $N=2^n$, with $n=10, \ldots, 15$. Middle: Annealed (symbols) and quenched (dashed lines) free-energy computed at $\beta=2$, $\tilde{\phi}_{\rm ann}(2,N) = {\rm min} \{ \phi_{\rm ann} (\beta_\star,N), \phi_{\rm ann} (2,N) \} = \gamma_{\rm ann}(N)/2$ and $\phi_{\rm q} (2,N) = \gamma_{\rm q} (N)/ 2$, as a function of $\alpha$ for several values of $n$. AL occurs when $\gamma_{\rm ann} = \gamma_{\rm q} = 0$ in the $n \to \infty$ limit, \ie~$\alpha=2$~\cite{Kravtsov_2015}. Right: Annealed (symbols) and quenched (dashed curves) configurational entropy at $\beta=2$ as a function of $\alpha$ for several values of $n$. Upon increasing $n$ the configurational entropy converges to the fractal dimension of the support set of the eigenstates (orange curve), \ie~$\Sigma = D =2 - \alpha$ in the intermediate phase, $\Sigma=1$ in the fully ergodic phase, and $\Sigma=0$ in the localized phase. The continuous green lines in the three panels are guides for the eye obtained performing a Bézier cubic interpolation of the annealed data. 
\label{fig:mstarRP}}
\end{figure*}

\section{Benchmark case II: The Rosenzweig-Porter random matrix ensemble} \label{sec:RP}

The paradigmatic Rosenzweig-Porter (RP) model~\cite{Kravtsov_2015} is the simplest random matrix ensemble displaying a phase with fractal eigenstates, sandwiched between a fully delocalized phase and a fully localized one. Due to its connections to MBL, this model has been intensively investigated over the past few years~\cite{Kravtsov_2015,vonSoosten_2019,Facoetti_2016,Truong_2016,Bogomolny_2018,DeTomasi_2019,amini2017spread,pino2019ergodic,berkovits2020super,venturelli2022replica}. Following the analogy with disordered quantum many-body systems (see \eg~Refs.~\cite{biroli2021out,tarzia2020many,faoro2019non}), every site of the reference space is represented by a matrix index $i=1 , \ldots, N$, and can be thought of as a site orbital of the reference Hilbert space. On each site an uncorrelated random potential $\epsilon_i$ is introduced. The difference with the Anderson model on the Cayley tree considered in the previous section is that in the RP case every site is connected to every other site with transition amplitudes distributed according to a Gaussian probability law. More precisely, the Hamiltonian of the RP model is defined as follows:
\[
H = A + \frac{c}{N^\alpha} B \, ,
\]
where $A$ is a diagonal $N \times N$ matrix, $A_{i,j} = \epsilon_i \delta_{i,j}$, with $\epsilon_i$ iid uniformly in the interval $\epsilon_i \in [-W/2,W/2]$, and  $B$ is a $N \times N$ matrix belonging to the GOE ensemble: its elements are Gaussian random variables with zero mean and unit variance (\ie, $\langle B_{i,i}^2 \rangle = 1/2$ and $\langle B_{i,j}^2 \rangle = 1$). ($c$ is a constant of $O(1)$ that we set to  $1$ throughout.)

In the thermodynamic limit the RP model exhibits three distinct phases varying the parameter $\alpha$ (and two transition points between them)~\cite{Kravtsov_2015,vonSoosten_2019,Facoetti_2016,Truong_2016,Bogomolny_2018,DeTomasi_2019,amini2017spread,pino2019ergodic,berkovits2020super,venturelli2022replica}: a fully delocalized phase for $\alpha<1$, a fully Anderson localized phase for $\alpha>2$, and an intermediate fractal phase for $1 < \alpha < 2$. Such intermediate phase is particularly interesting: The eigenvectors are  partially delocalized over a large number of sites $N^D$, which represent, however, a vanishing fraction of the total number of sites $N$ in the thermodynamic limit, their
fractal dimension being $D = 2 - \alpha$.

Albeit this model has undergone extensive investigation in the past and its properties are well known in many details\cite{Kravtsov_2015,vonSoosten_2019,Facoetti_2016,Truong_2016,Bogomolny_2018,DeTomasi_2019,amini2017spread,pino2019ergodic,berkovits2020super,venturelli2022replica}, it is instructive to study it through the approach outlined  in Sec.~\ref{sec:strategy}. Since there is no notion of space in this model, the natural generalization of Eq.~\ref{eq:T} is simply given by:
\begin{equation} \label{eq:TRP}
{\cal T}_i (\beta) = \sum_{j \neq i}^N |G_{i,j}|^\beta \, ,
\end{equation}
where $i$ is a randomly chosen site, and $G_{i,j} = \langle i | H^{-1} | j \rangle$ are the matrix elements of the resolvent. Similarly to the case of the Anderson model on the Cayley tree, for $\beta=2$ this object is related to the probability that a particle that is created in $i$ at $t=0$ reaches any other site after infinite time. Equivalently ${\cal T}_i (2)$ can be interpreted as the Fisher-Lee conductivity (or the total Landauer transmission) of the model in a scattering geometry in which a semi-infinite lead is attached to a randomly chosen site $i$, and $N-1$ semi-infinite leads are attached to {\it all} other $N-1$ sites~\cite{fisher1981relation}. The quenched and annealed free-energies are defined similarly to Eq.~\eqref{eq:phiCT} as:
\begin{equation} \label{eq:phiRP}
\begin{aligned}
\phi_{\rm ann} (\beta, N) & = \frac{1}{ \beta \ln N} \, \ln \avg{{\cal T}_i (\beta)} \, , \\
\phi_{\rm q} (\beta, N) & = \frac{1}{ \beta \ln N} \, \avg{\ln {\cal T}_i (\beta)} \, , \\
\end{aligned}
\end{equation}
where the factor $\ln N$ in the denominator of the previous expressions plays the role of the ``radius'' of the system.

We have computed the quenched and annealed free-energies~\eqref{eq:phiRP} performing exact diagonalizations of matrices of size $N=2^n$, with $10 \le n \le 15$, varying the parameter $\alpha$ (and averaging over sever realizations of the disorder and of the ``initial'' site $i$). The RP free-energies have the same qualitative shape and properties of the quenched and annealed free-energies defined for the Anderson model on the Cayley tree plotted in Fig.~\ref{fig:CT}. The numerical results are summarized in Fig.~\ref{fig:mstarRP} where we plot the same set of observables shown in Fig~\ref{fig:mstarCT}. The total number of samples here is $2 \cdot 10^6$ for $n=10$, $2 \cdot 10^5$ for $n=11$, $2 \cdot 10^5$ for $n=12$, $10^5$ for $n=13$, $3 \cdot 10^4$ for $n=14$, and $8 \cdot 10^3$ for $n=15$. In the left panel we plot the position of the minimum of the annealed free-energy $\beta_\star$. The middle panel shows the quenched and annealed estimations of the Lyapunov exponent, Eq.~\ref{eq:lyap}. The quenched one is simply given by $\gamma_{\rm q} (N) = 2 \phi_{\rm q} (2,N)$, while the annealed estimation is obtained by replacing the $\beta$-dependence of $\phi_{\rm ann}$ beyond $\beta_\star$ by a constant equal to is value at $\beta_\star$, Eq.~\ref{eq:replacementCT}, \ie~ $\gamma_{\rm ann} (N) = 2 \tilde{\phi}_{\rm ann}(2,N) = {\rm min} \{ 2 \phi_{\rm ann} (\beta_\star,N), 2 \phi_{\rm ann} (2,N) \}$. As discussed above, $\gamma_{\rm q}$ and $\gamma_{\rm ann}$ provide two finite-$n$ estimates of the localization transition. In the right panel we plot the annealed and quenched configurational entropies at $\beta=2$. The configurational entropy is defined in analogy with the directed polymer problem presented in Sec.~\ref{sec:DPRM}: Defining the number of terms contributing to ${\cal T}(\beta)$ as $N^{\Sigma (\beta)}$, 
from Eqs.~\eqref{eq:TRP} and~\eqref{eq:phiRP} one immediately finds that the configurational entropy at $\beta=2$ is given by $\Sigma = - 4 \phi^\prime (2)$.

These plots are fully compatible with the well known physical properties of the RP model in its three different phases~\cite{Kravtsov_2015,vonSoosten_2019,Facoetti_2016,Truong_2016,Bogomolny_2018,DeTomasi_2019,amini2017spread,pino2019ergodic,berkovits2020super,venturelli2022replica}, and can be rationalized as follows: In the fully delocalized phase, $\alpha<1$, the wave-functions spread uniformly over the whole volume. Hence a particle created in $i$ at $t=0$ can reach any other site of the reference space, implying that $O(N)$ terms contribute to the sum~\eqref{eq:TRP} for $\beta=2$, and that ${\cal T}_i (2)$ grows with $N$. Thus one finds that $\beta_\star > 2$ and $\gamma_{{\rm q,ann}} >0$. In fact the Lyapunov exponent converges to its maximal value $\gamma=1$ corresponding to a maximally chaotic system. Conversely, in the fully localized phase, $\alpha>2$, a particle created in $i$ can only diffuse to a few sites with on-site energies very close to $\epsilon_i$. As a consequence for $\beta=2$ the sum~\eqref{eq:TRP} only receives contribution from $O(1)$ terms (\ie~$\beta_\star < 2$), and ${\cal T}_i(2)$ vanishes with $N$. We indeed observe that for $N$ large enough $\beta_\star \simeq 1$ and $\gamma_{{\rm q,ann}} <0$. Finally, in the intermediate partially delocalized fractal phase, $1<\alpha<2$, a particle create in $i$ reaches a large but sub-extensive number $N^D \ll N$ of sites $j$ whose on-site random energies are such that $|\epsilon_i - \epsilon_j| \lesssim |H_{ij}|$. 
Thus for $\beta=2$ the sum~\eqref{eq:TRP} receives contribution by a large (but sub-extensive) number of terms and the conductivity is finite. Hence, similarly to the fully delocalized phase, one finds that  for $N$ large enough $\beta_\star>2$ and $\gamma_{{\rm q,ann}}>1$. The difference between the fully delocalized phase and the partially delocalized fractal one resides in the behavior of the configurational entropy which, in the thermodynamic limit, is identically equal to $1$ for $\alpha<1$, while it is equal to $D=2-\alpha$ for $1 < \alpha < 2$. ($\Sigma$ is identically equal to zero in the localized phase, $\alpha>2$, for $n\to \infty$.)

To sum up, the large-deviation analysis of the phase diagram of the RP model using the rate functions~\eqref{eq:phiRP} inspired by the analogy with DPRM has two main virtues. First, this approach allows one to recover the physical properties of the model in a very transparent and compact way. Second, it highlights the differences with the phase diagram of the Anderson model on the Cayley tree presented in the previous section. In fact, while the RP model exhibits a direct transition from a delocalized but non-ergodic phase (with $\Sigma<1$, $\gamma>0$, and $\beta_\star>2$) to a Anderson localized phase (with $\Sigma=0$, $\gamma<0$, and $\beta_\star \simeq 1$), for the Anderson model on the Cayley tree the localization transition is preceded by a frozen phase in which transport is possible but only over a few specific, disorder-dependent paths ($\Sigma=0$, $\gamma>0$, and $1<\beta_\star<2$). Such frozen phase is absent in the RP model. Conversely, the Anderson model on the Cayley tree lacks of a genuinely fully ergodic phase, even at very small disorder. 

It is also interesting to mention that the partially ergodic phases characterized by $\Sigma < \Sigma_{\rm max}$ found in the Anderson model on the Cayley tree for $0 < W <W_g$ and in the RP model for $1 < \alpha < 2$ have completely different spectral properties. In the RP case the statistics of the gaps between neighboring eigenvalues is in fact described by the Wigner-Dyson statistics, due to the formation of compact mini-bands in the local spectrum~\cite{Kravtsov_2015,vonSoosten_2019,Facoetti_2016,Truong_2016,Bogomolny_2018,venturelli2022replica}. Neighboring eigenvectors of the Anderson model on the Cayley tree have instead a very different support sets even at weak disorder, due to the pathological structure of the lattice, yielding a non-universal statistics~\cite{sade2003localization}. 

\section{Analysis of the many-body problem} \label{sec:MBL}

\begin{figure}[t!]
\includegraphics[width=.66\textwidth]{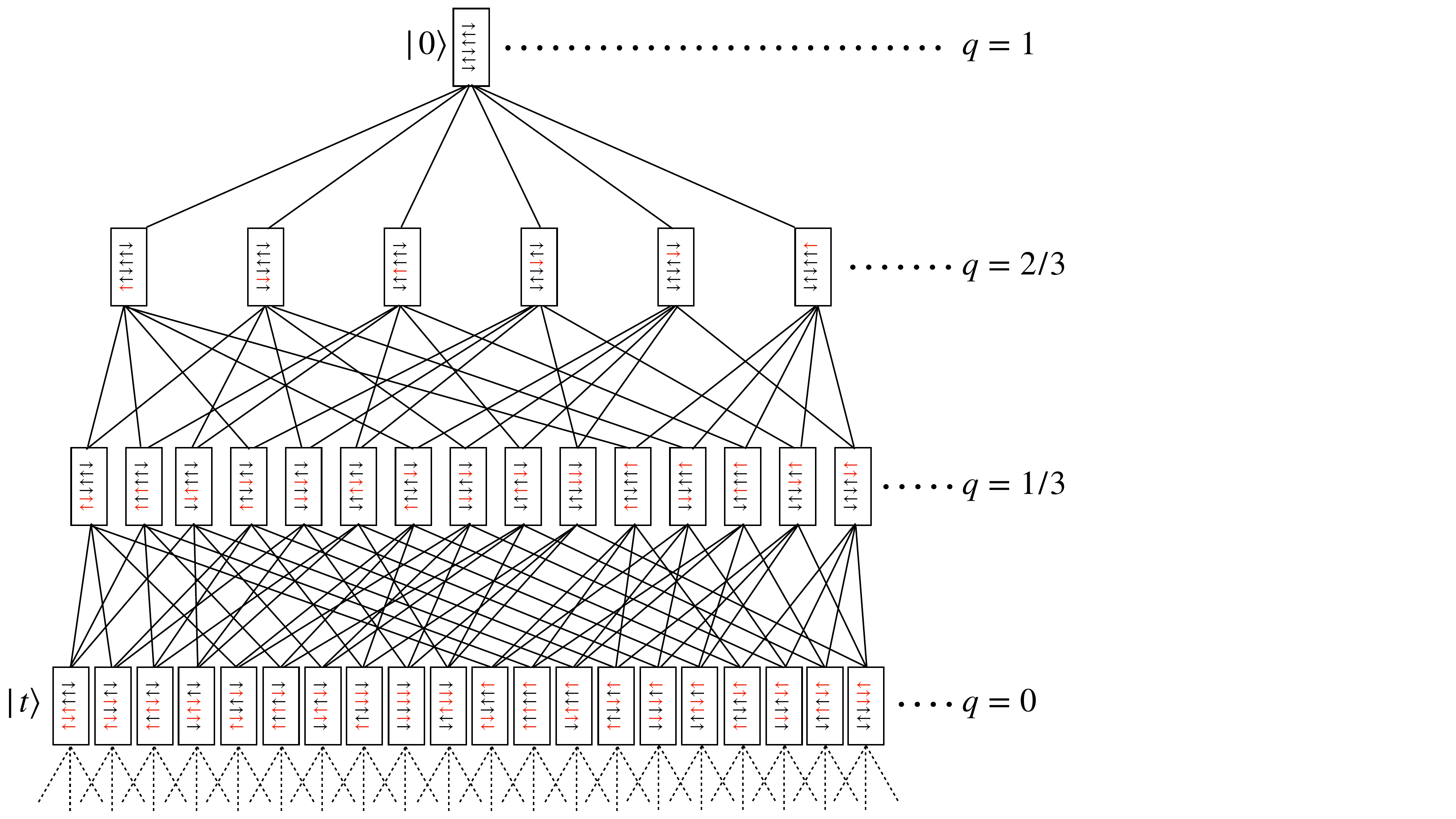} 
\caption{Sketch of half of the Hilbert space of the model~\eqref{eq:HMBL} in the basis of the $\{ \sigma_i^z \}$ operators for $n=6$ spins. The Hilbert space is unfolded starting from a random spin configuration $\vert 0 \rangle$ up to a distance $\ell = n/2=3$, where all the target sites $\vert t \rangle$ having zero overlap with $\vert 0 \rangle$ lie. The red spins are the ones that have been flipped.
\label{fig:hilbert}}
\end{figure}

After completing the analysis of the two benchmark cases, we now switch to the main focus of this work which is study of many-body problem. In the following we will apply the same set of tools and techniques described in the previous sections (and in particular in Sec.~\ref{sec:strategy}) to study a class of quantum disordered spin-chains for which previous studies have indicated the presence of a MBL transition at strong enough disorder~\cite{imbrie2016many,abanin2021distinguishing,roy2021fock,creed2022probability,tarzia2020many}. The model is defined by the following Hamiltonian:
\begin{equation} \label{eq:HMBL}
H = \sum_{i=1}^{n-1} \Delta_i \sigma_i^z \sigma_{i+1}^z + \sum_{i=1}^n ( h_i \sigma_i^z + \Gamma \sigma_i^x ) \, ,
\end{equation}
where $\Gamma=1$, and $\Delta_i$ and $h_i$ are iid random variables taken from the uniform distributions $\Delta_i \in [0.8,1.2]$ and $h_i \in [-W,W]$. This model is sometimes called the ``Imbrie model'', since for this specific Hamiltonian in Ref.~\cite{imbrie2016many} Imbrie has rigorously proven the existence of the MBL transition at strong disorder under the minimal assumption of absence of level attraction. Previous numerical studies of this model based on the analysis of the average spectral statistics of finite-size samples ($n \le 16$) seem to indicate that the MBL transition should occur in the interval $W_c \in [3.5,4]$~\cite{abanin2021distinguishing,roy2021fock,creed2022probability,tarzia2020many}.

We choose as a basis of the Hilbert space the simultaneous eigenstates of the $\sigma_i^z$ operators (\ie, a basis in which the system is fully localized in absence of the transverse field). The resulting Hilbert space is a $n$-dimensional hypercube of $N=2^n$ nodes. Each
configuration $\vert m \rangle = \vert \uparrow \downarrow \uparrow \cdots \rangle$ of the $n$ spins corresponds to a corner of the hypercube by considering $\sigma_i^z = \pm 1$ as the top/bottom face of the cube's $n$-th dimension. The random part of the Hamiltonian is by definition diagonal in
this basis, and gives correlated random energies on each site orbital of the hypercube $E_m = \langle m \vert [ \sum_{i=1}^{n-1} \Delta_i \sigma_i^z \sigma_{i+1}^z + \sum_{i=1}^n  h_i \sigma_i^z ] \vert m \rangle$. $E_m$ are the extensive energies of a many-body (classical) $1d$ Ising model with random antiferromagnetic interactions $\Delta_i$ and in presence of a random field $h_i$: They are Gaussian distributed with zero mean and variance proportional to $n$. 
We then pick a random configuration of the $n$ spins, \eg~$|0\rangle = | \uparrow \downarrow \downarrow \cdots \rangle$ (\ie, a random initial state, which is in the middle of the many-body spectrum, with $E_0/n \sim 1/\sqrt{n}$), and compute the propagator $G_{0,t}$ between such initial configuration and another (distant) spin configuration $|t\rangle$, 
$G_{0,t} = \langle 0 \vert H^{-1} \vert t \rangle$.

Differently from the Cayley tree discussed in Sec.~\ref{sec:CT}, the Hilbert space of the many-body problem in the $\{ \sigma_i^z \}$ basis 
has many loops of all (even) size and no boundary. This implies that there is a huge number of paths (which grows factorially with the distance) going from $|0\rangle$ to $|t\rangle$. For these reasons $G_{0,t}$ cannot be computed analytically. Formally the propagator (at $E=0$) can be still written in a form that is reminiscent of Eq.~\eqref{eq:propagatorCT}~\cite{anderson1958absence}:
\begin{equation} \label{eq:SAW}
G_{0,t} = \!\sum_{{\cal P}_{0 \to t}^\star} \, \prod_{m \in {\cal P}_{0 \to t}^\star} \frac{\Gamma}{E_m+\Sigma_m ({\cal P}_{0 \to t}^\star)} \, ,
\end{equation}
where the sum is over all self-avoiding paths ${\cal P}_{0 \to t}^\star$ from $0$ to $t$, and $\Sigma_m ({\cal P}_{0 \to t}^\star)$ is the self-energy (which depend both on the path ${\cal P}_{0 \to t}^\star$ and on the node $m$). *Yet, the computation of the self-energies is a formidable task. 
Below we proceed in two complementary ways. On the one hand we perform numerical matrix inversion of the Hamiltonian~\eqref{eq:HMBL} (in the basis of the eigenvectors of $\sigma_i^z$) and compute $G_{0,t}$ exactly, but for system of moderate size only (with $10 \le n \le 17)$. Alternatively, in order to access larger sizes, we estimate the propagators using the \emph{forward-scattering approximation}~\cite{anderson1958absence,pietracaprina2016forward,imbrie2017local,ros2015integrals,tarzia2020many} (see Sec.~\ref{sec:CTFSA} and  App.~\ref{app:FSA}). This amounts in  setting $\Sigma_m=0$ in the expression above and considering only the shortest paths between $\vert 0 \rangle$ and $\vert t \rangle$ in the sum. Within the FSA the expression of $G_{0,t}$ becomes again analogous to the partition function of a directed polymer lying on the $n$-dimensional hypercube. However, since the ratios $\Gamma/E_m$ can be positive or negative 
the Boltzmann weights of the polymer are generically complex~\cite{derrida1993mean,lemarie2019glassy}. Since the FSA consists in retaining only the leading term of the perturbative expansion in $\Gamma$ for the propagator starting from the insulator, this approximation is asymptotically exact in the infinite disorder limit and is only justified deep in the MBL phase. 

For the case of FSA, the propagators $G_{0,t}$ can be obtained by a transfer matrix approach, called dynamic programming in the computer science literature. The basic idea is to store the values of $G$ as a function of the $z$ value of the current spin configuration $\{ \sigma^z \}$ along the path $\vert 0 \rangle \to \vert t \rangle$ (to simplify the notation we omit the $z$ in $\sigma^z$ in the description of the algorithm below). Thus one starts with the initial configuration $\vert 0 \rangle \equiv \{ \sigma^{(0)} \} \equiv (\sigma^{(0)}_1, \ldots, \sigma^{(0)}_n)$. Then one considers spin configurations with increasing distance $\ell'=1,\ldots,\ell$. We denote the values of $G$ for spin configurations with distance $\ell'$ as $G_{\ell'}$. We assume that $\Omega_{\ell'}$ is a set of configurations which contains those of distance $\ell'$. Technically we realized this by a C++ \emph{map} data structure. This means, whenever a value $G_{\ell'}(\sigma)$ is assigned for a spin configuration which is yet not member of $\Omega_{\ell'}$, the spin configuration will be automatically added to $\Omega_{\ell'}$. Note that by using the \emph{map} data type, only memory was consumed for actually existing configurations, in contrast to the range of $2^n$ different possible configurations. This allowed us to go to larger sizes in contrast to a straight-forward array-based implementation.

The algorithm shown below calculates $G_l$ for all target spin configurations $\vert t \rangle$ which have distance $\ell$, for any given disorder configuration $\{\Delta_i,h_i\}$. It works by iterating over all spin configurations at a given distance $\ell$, flipping iteratively one yet not flipped spin, respectively, and updating $G_{\ell'+1}$ for the such obtained spin configurations:
\begin{tabbing}
xx \= xx \= xx \= xx \= xx \= xx \= xx \= xxxxx \kill
{\bf algorithm} $G(\sigma^{(0)},\ell)$     \\
\> $G_0( \sigma^{(0)})=\Gamma/{H(\sigma^{(0)})}$ \\
\> {\bf for} $\ell'=1,\ldots \ell $\\
\> \> {\bf for} all $\sigma \in \Omega_{\ell'}$\\
\> \> \> {\bf for} $i=1,\ldots n$ \\
\> \> \> \> $\sigma'=\sigma$ \\
\> \> \> \> {\bf if} $\sigma_i = \sigma^{(0)}_i$ {\bf then}\\
\> \> \> \> \> $\sigma'_i = - \sigma_i$ \\
\> \> \> \> \> $G_{\ell'+1}(\sigma') = G_{\ell'+1}(\sigma')$
$+ G_{\ell'}(\sigma) \Gamma / H(\sigma')$\\
\> \> \> \> {\bf end if}\\
\> \> \> {\bf end for}\\
\> \> {\bf end for}\\
\> \>  Clear $\Omega_{\ell '}$\\
\> {\bf end for}\\
\end{tabbing}
Note that $H(\sigma)$ corresponds to the energies $E_m$ in (\ref{eq:SAW}).  
The above {\bf if} statement takes care that each spin is flipped only once. If one wishes to reach just a single given target configuration $\vert t \rangle \equiv \{ \sigma^{(t)} \}$, then the condition in the {\bf if} statement has to be replaced by  $\sigma_i = \sigma^{(0)}_i$ AND $\sigma_i \neq \sigma^{(t)}_i$.
Clearing the sets $\Omega_{\ell'}$ after they have bee used also saves memory. 

\begin{figure*}
\includegraphics[width=0.338\textwidth]{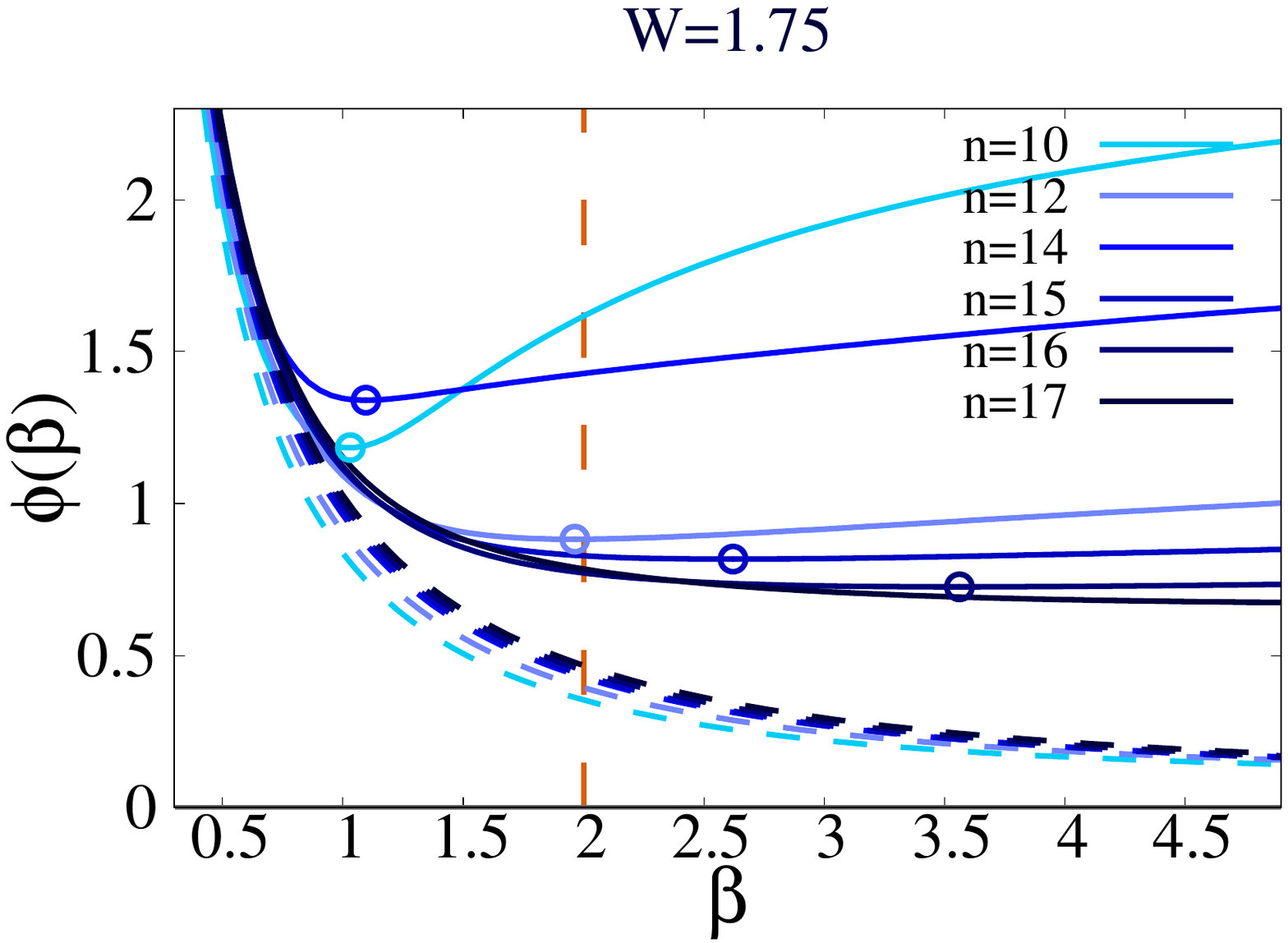} \hspace{-0.36cm} \includegraphics[width=0.338\textwidth]{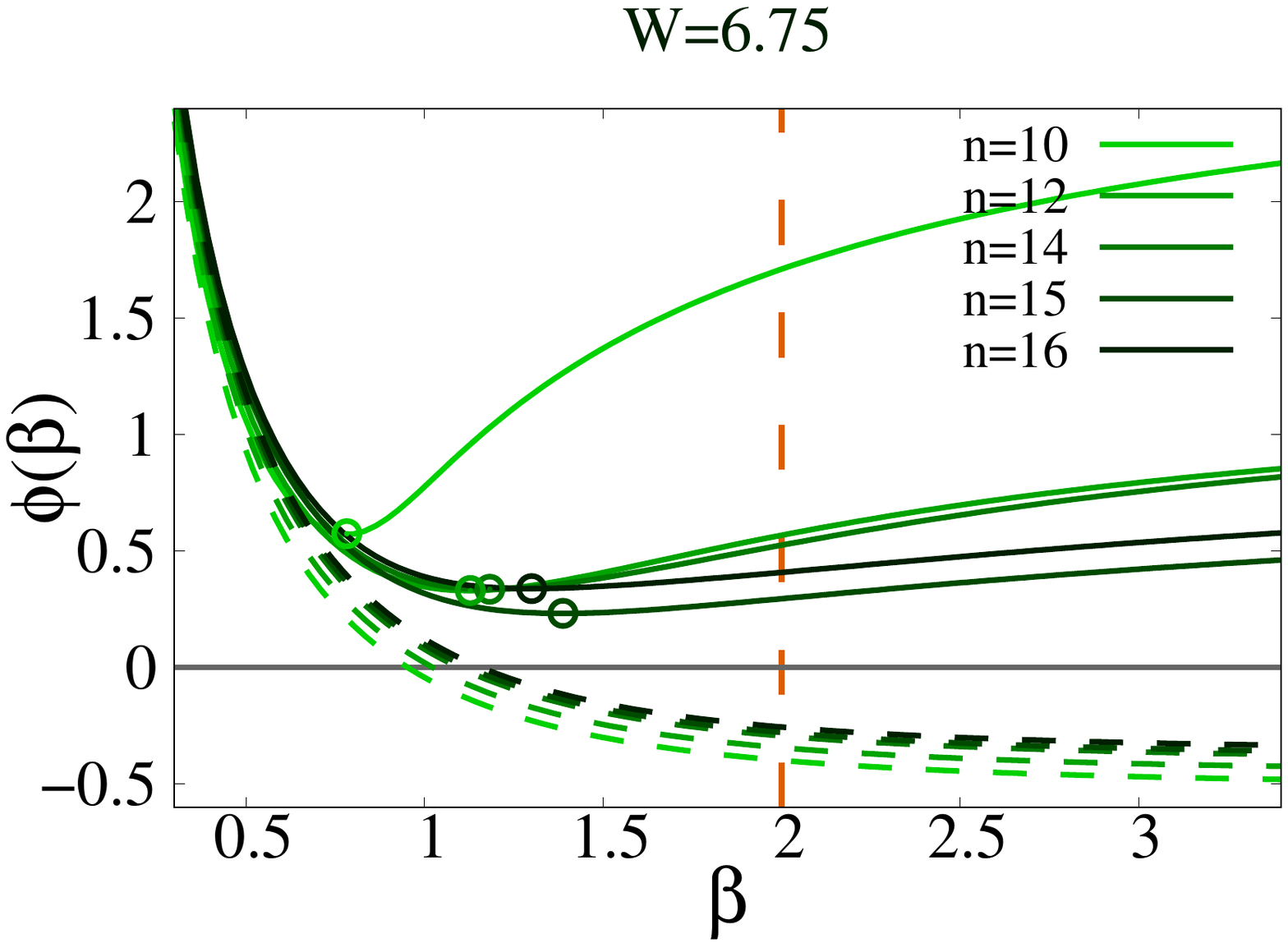} \hspace{-0.36cm} \includegraphics[width=0.338\textwidth]{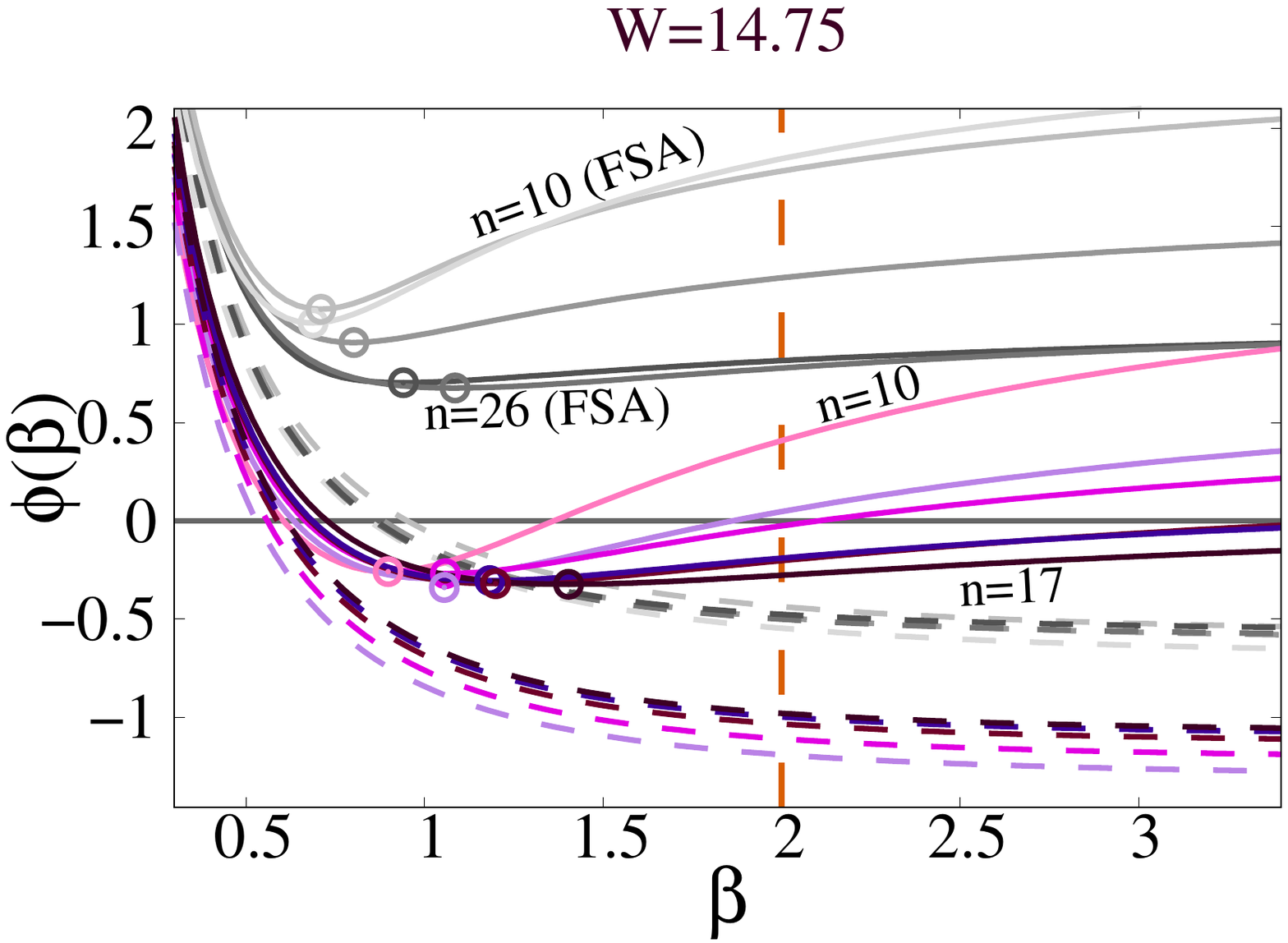}
\caption{Annealed (continuous lines) and quenched (dashed lines) free-energies, Eq.~\eqref{eq:free_energy}, as a function of the inverse temperature $\beta$ for $n=10,\ldots,17$  and for $W=1.75$ (left), $W=6.75$ (middle), and $W=14.75$ (right). The filled circles spot the positions of the minima of the annealed free-energy. Note that the plots of the quenched and annealed free-energies for $n$ odd (\ie~$n=15$ and $n=17$) are obtained as linear interpolations of the quenched and annealed free-energies with target sites at distance $(n-1)/2$ and $(n+1)/2$. The gray continuous and dashed curves in the right panel show the  annealed and quenched free-energy obtained within the FSA for $10 \le n \le 26$. The horizontal dashed orange line shows the value of the auxiliary parameter $\beta=2$ corresponding to physical transport and decorrelation.
\label{fig:phi}}
\end{figure*}

As sketched in Fig.~\ref{fig:hilbert}, a natural choice is to classify the target states $\vert t \rangle$ in terms of their Hamming distance $\ell$ from the initial state, with $0 \le \ell \le n$. The Hamming distance is defined as the minimum number spin flips required to go from one state to the other (\ie, the length of the shortest paths on the hypercube that join them). The number of states at distance $\ell$ from $\vert 0 \rangle$ is  ${{n}\choose{\ell}}$. Equivalently, one can introduce the overlap of the $z$-component of the spins between the states $\vert 0 \rangle$ and $\vert t \rangle$ as:
\begin{equation} \label{eq:overlap}
    q_{0,t} = \frac{1}{n} \sum_{i=1}^n \sigma_i^z (0) \, \sigma_i^z (t) = 1 - \frac{2 \ell}{n} \, .
\end{equation}
Here we start by focusing on the spin configurations $\vert t \rangle$ that are at the ``equator'' from the initial configuration $\vert 0 \rangle$, \ie, at distance $\ell = n/2$ from it 
(the analysis of the dependence on the distance $\ell$ is carried out in Sec.~\ref{sec:length}). Those configurations are such that half of the spins have flipped (see Fig.~\ref{fig:hilbert}), and thus have by definition zero overlap with $|0\rangle$. The number of those configurations for the model~\eqref{eq:HMBL} is ${{n}\choose{n/2}}$. Following the approach outlined in Sec.~\ref{sec:strategy} and applied to the benchmark cases discussed above, we introduce the generalized ``partition function'':
\begin{equation} \label{eq:TMBL}
{\cal T}_0 (\beta) = \sum_{t=1}^{{n}\choose{n/2}} |G_{0,t} |^\beta \, .
\end{equation}
${\cal T}_0 (2)$ is proportional to the probability that the system is initialized in the state $|0\rangle$ at time $0$ and is found in a state at zero overlap from it after infinite time, and is thus a measure of delocalization in the Hilbert space. 
${\cal T}_0 (2)$ has also a physical interpretation in terms of the single-particle tight-binding problem defined on the $n$-dimensional hypercube: one can imagine to attach a semi-infinite lead to $|0\rangle$ where ``particles'' are injected and ${n}\choose{n/2}$ semi-infinite leads to the ``target'' nodes $|t\rangle$ where ``particles'' are extracted. Then ${\cal T}_0 (2)$ is proportional to the Fisher-Lee conductivity (or the total Landauer transmission) between these leads~\cite{lemarie2019glassy,fisher1981relation}.

We again introduce the quenched and annealed free-energies (divided by the length $\ell=n/2$ of the shortest paths in the Hilbert space that connects $|0\rangle$ to $|t\rangle)$, defined as:
\begin{equation} 
\begin{aligned}
\phi_{\rm ann} (\beta,n) & =  \frac{2}{\beta n} \ln \avg{{\cal T}_0 (\beta)} \, , \\
\phi_{\rm q} (\beta,n) & =  \frac{2}{\beta n} \avg{\ln {\cal T}_0 (\beta)} \, .
\end{aligned} 
\label{eq:free_energy}
\end{equation}
The averages here are performed over several independent realizations of the $h_i$'s and $\Delta_i$'s in the Hamiltonian~\eqref{eq:HMBL}, as well as over many independent choices of the random initial state $\vert 0 \rangle$. Since we aim at considering initial configurations close to the middle of the many-body spectrum (corresponding to infinite temperature), we have chosen the node $\vert 0 \rangle$ uniformly at random among $1/32$ of the total nodes whose on-site energies $E_m$ are the closest to $0$. The total number of realizations is $10^6$ for $n=10$, $3 \cdot 10^5$ for $n=12$, $3 \cdot 10^5$  for $n=14$, $6 \cdot 10^4$  for $n=15$, $2 \cdot 10^4$  for $n=16$, and $5 \cdot 10^3$  for $n=17$. Since within the FSA we can study larger sizes compared to exact diagonalizations, in order to avoid sorting the energies of all exponentially numerous configurations, the initial state is instead selected as that one exhibiting minimum absolute value of the energy among 2500$\sqrt{n}$ randomly generated spin configuration, which takes into account the growth of the width of the energy distribution and therefore should exhibit an initial energy sufficiently close to zero.

\begin{figure*}
\includegraphics[width=0.338\textwidth]{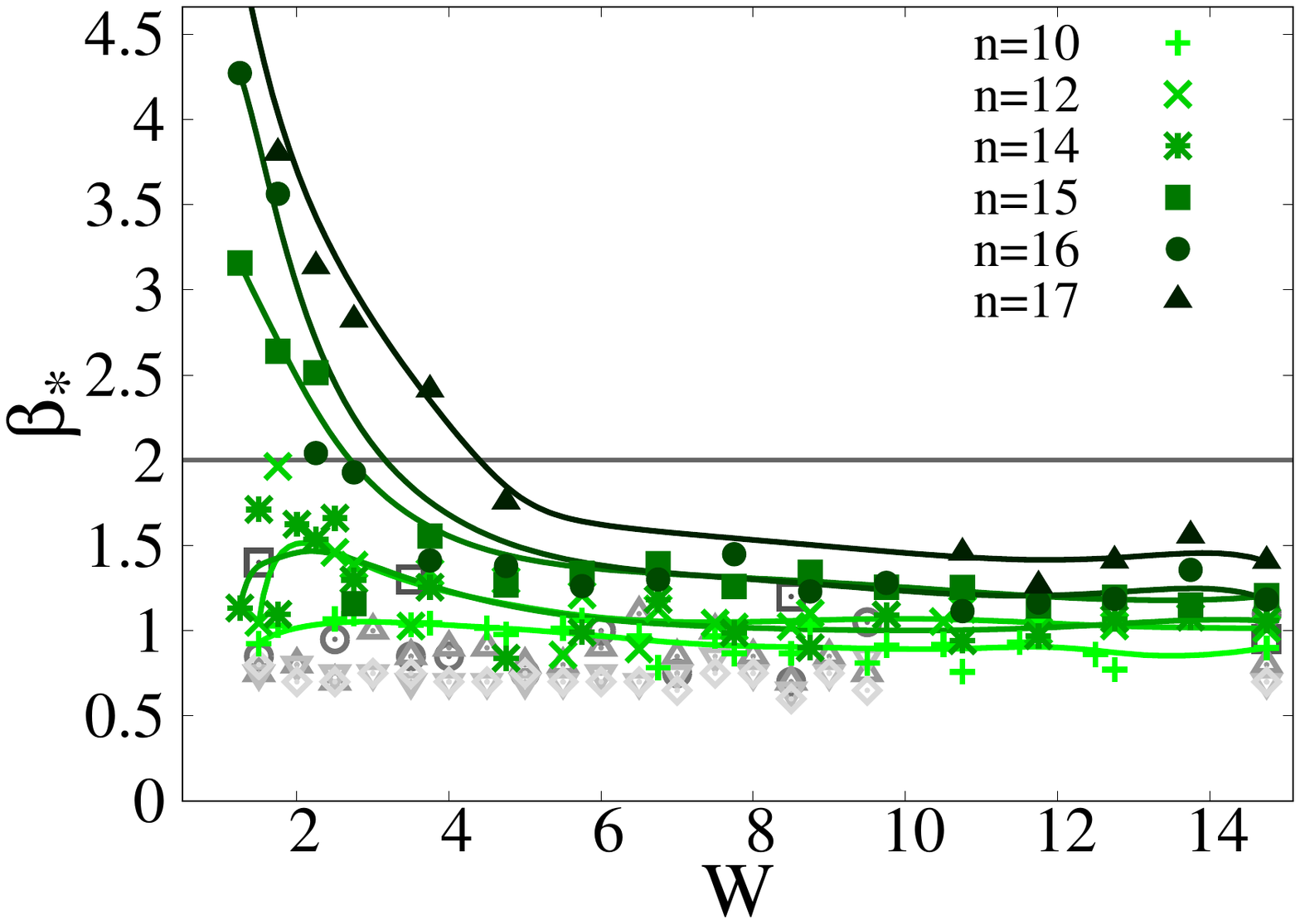} \hspace{-0.36cm} \includegraphics[width=0.338\textwidth]{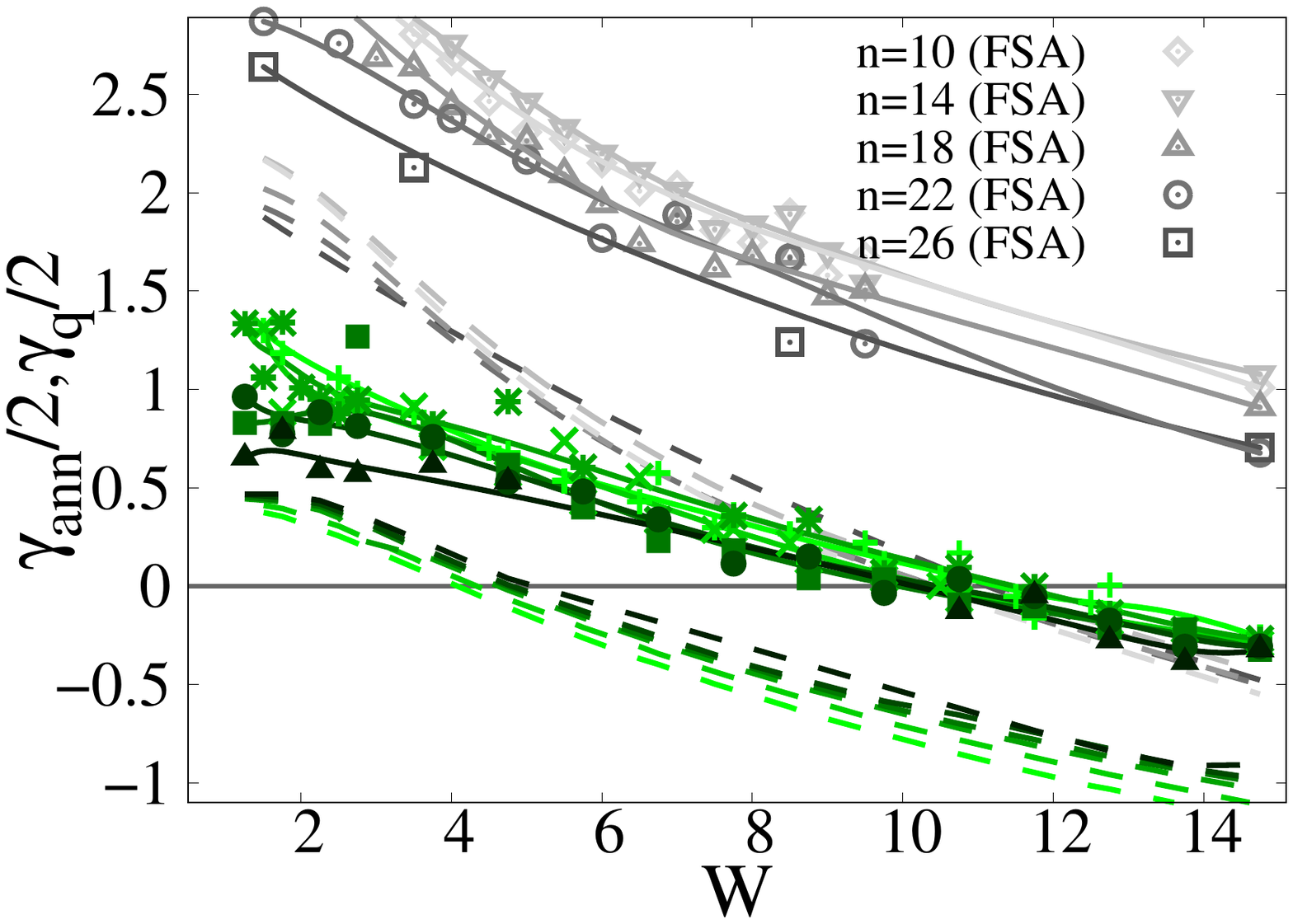} \hspace{-0.36cm}
\includegraphics[width=0.338\textwidth]{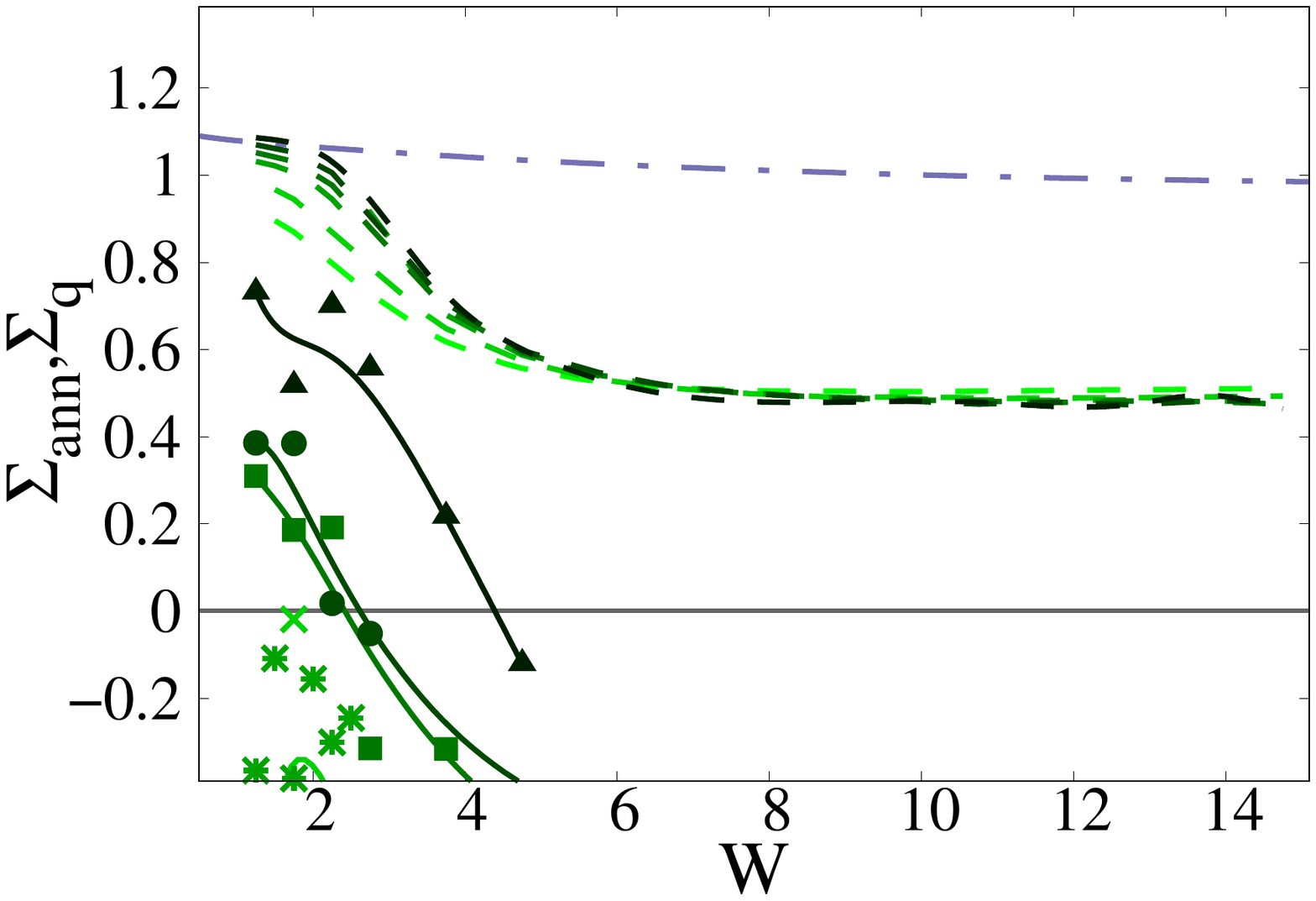} 
\caption{Left: Minimum of the annealed free-energy $\beta_\star$ as a function of $W$ for $n=10, \ldots , 17$ spins. The gray symbols show the FSA value of $\beta_\star^{\rm fsa}$ for $n$ between $10$ and $26$, which, similarly to the non-interacting case, does not depend on $W$.  Middle: Value of the free-energies at $\beta=2$ as a function of $W$ for several values of $n$, providing the quenched and annealed estimations of the Lyapunov exponent according to Eq.~\eqref{eq:lyap}. $\gamma_{\rm q} (n)$ (dashed lines) is given by $2 \phi_{\rm q} (2,n)$, while $\gamma_{\rm ann}(n)$ (symbols) is equal to $2 \tilde{\phi}_{\rm ann}(2, n) = {\rm min} \{ 2 \phi_{\rm ann} (\beta_\star(n),n), 2 \phi_{\rm ann} (2,n) \}$ after replacing the $\beta$-dependence of $\phi_{\rm ann}$ beyond $\beta_\star$ by a constant equal to its value at $\beta_{\rm star}$, Eq.~\eqref{eq:replacementCT}. In the $n \to \infty$ limit $\gamma_{\rm q}(n)$ and $\gamma_{\rm ann}(n)$ converge to the same value $\gamma$. AL occurs when $\gamma = 0$. 
The gray symbols and the gray dashed lines show respectively $\gamma_{\rm ann}^{({\rm fsa})}(n)/2$ and $\gamma_{\rm q}^{({\rm fsa})}(n)/2$ within the FSA for $n$ between $10$ and $26$, which, similarly to the non-interacting case, provide an upper bound for the critical disorder. Right: Annealed (symbols) and quenched (dashed curves) configurational entropy at $\beta=2$ as a function of $W$ for $n=10, \ldots, 17$. The dashed-dotted line gives to the maximum value $\Sigma_{\rm max}$ corresponding to fully ergodic eigenstates, Eq.~\eqref{eq:sigmamax}.  The continuous lines in the three panels are guides for the eye  obtained performing a Bézier cubic interpolation of the annealed data. 
\label{fig:mstar}}
\end{figure*}

The free-energies are shown for three values of the disorder across the putative MBL transition in Fig.~\ref{fig:phi}, where we plot $\phi_{\rm ann}$ (continuous curves) and $\phi_{\rm q}$ (dashed curves) as a function of $\beta$ for $n=10, \ldots, 17$ spins --- the free-energies for target states at distance $\ell=n/2$ for an odd number of spins (\ie, $n=15$ and $n=17$) are obtained by performing a linear interpolation of the results between $\ell = (n+1)/2$ and $\ell = (n-1)/2$. 

Remarkably, these plots are qualitatively very similar to the ones obtained on the loop-less Cayley tree, Fig.~\ref{fig:CT}. 
At weak disorder, $W=1.75$ (left panel), the minimum of the annealed free-energy $\beta_\star$ shifts to larger and larger values of $\beta$ upon increasing $n$ and becomes larger than $2$ already for $n \sim 14$. 

At moderately strong strong disorder, $W=6.75$ (middle panel), $\beta_\star$ does not move much with $n$ and stays far below $2$ even for the largest accessible system size. As discussed above, a measure of localization is obtained from the sign of the free-energy at $\beta=2$ (in the thermodynamic limit). The quenched free-energy at $\beta=2$ is negative for the accessible system sizes, although it increases with $n$. A complementary estimation (see Secs.~\ref{sec:strategy} and~\ref{sec:DPRM}) is obtained by replacing the $\beta$-dependence of $\phi_{\rm ann}$ beyond $\beta_\star$ by a constant equal to its value at $\beta_\star$, Eq.~\eqref{eq:replacementCT}. The figure clearly shows that the value of the annealed free-energy at the minimum stays above $0$ upon increasing the system size. This behavior indicates that the typical value of the total Landauer transmission between a random node of the hypercube and its equator is finite in the thermodynamic limit and that the system does eventually decorrelate from the initial configuration. Yet such decorrelation is due to a few rare resonances found at large distances on the Hilbert space. In other words, if the system is initialized in $\vert 0 \rangle$, it will tunnel to a few states with $n/2$ flipped spins for large enough sizes. Note that $W=6.75$ is far beyond the estimated critical value of the MBL transition obtained in previous studies~\cite{abanin2021distinguishing,roy2021fock,creed2022probability}, which lies in the interval $W_c \in [3.5,4]$. This is due to the fact that such resonances are so rare that they are absent for a typical realization of the quenched disorder and/or for a typical initial state for the accessible sizes.  Upon increasing the system size the emergence of these rare resonances becomes progressively more prevalent even in typical samples, ultimately manifesting as delocalization in the limit of large $n$. This is remarkably similar to the avalanche instability described and analysed in many recent works~\cite{de2017stability,thiery2018many,luitz2017small,goihl2019exploration,crowley2020avalanche,leonard2023probing,peacock2023many,morningstar2022avalanches,sels2022bath,Ha2023many}. The key result here is that enlarging the parameter space by introducing of the auxiliary parameter $\beta$ and studying the functional dependence of the rate functions~\eqref{eq:free_energy} when varying this parameter, enables us to assess the existence and the effect of these rare resonances already from the analysis of the behavior of moderately small samples for which they are typically absent.

At very strong disorder, $W=14.75$ (right panel), one finds that $\beta_\star \simeq 1$ and that both $\phi_{\rm q} (2,n)<0$ and $\tilde{\phi}_{\rm ann} (2,n)<0$, corresponding to a genuinely insulating phase. In this regime we also show the quenched and annealed free-energies obtained within the FSA. (For $n=14,18,22$, a number of 40000 realizations of the disorder could be considered, while for the largest system sizes, $n=26$, up to 9000 realizations for $W=14.75$ are considered,  which limits the influence of rare but important outliers for the annealed result.) As already discussed in the context of single-particle Anderson localization on the Cayley tree, this approximation reproduces the qualitative shape of the free-energies at strong disorder, and overestimates their heights by a global shift (thereby overestimating the critical disorder at which localization occurs).

The information encoded in the behavior of the quenched and annealed free-energies for different values of the disorder are summarized in Fig.~\ref{fig:mstar}, which focuses on the same diagnostics previously used to investigate the benchmark cases of Sec.~\ref{sec:CT} and~\ref{sec:RP} and outlined in Sec.~\ref{sec:strategy}. In the left panel we show the evolution of the position of the minimum of the annealed free-energy $\beta_\star$ with $W$ for several system sizes. In the middle panel we plot the finite-$n$ quenched and annealed estimations of the value of the free-energy at $\beta=2$. The former is directly obtained from the quenched free-energy of finite-size samples, $\phi_{\rm q} (2,n) = \gamma_{\rm q} (n)/2$, while the latter is obtained by replacing the dependence of the annealed free-energy beyond $\beta_\star$ by a constant equal to its value at $\beta_\star$, $\tilde{\phi}_{\rm ann} (2,n) = {\rm min} \{\phi_{\rm ann} (\beta_\star(n),n) , \phi_{\rm ann} (2,n)\}= \gamma_{\rm ann} (n)/2$ (see Eq.~\ref{eq:replacementCT} and Secs.~\ref{sec:strategy} and~\ref{sec:DPRM}). Note that by construction one has that  $\gamma_{\rm ann}$ is strictly larger than $\gamma_{\rm q}$. The conditions $\gamma_{\rm q,ann}=0$ provide two finite-size estimates of localization in the Hilbert space. The right panel shows the annealed and quenched configurational entropies at $\beta=2$. The configurational entropy is defined as before as the logarithm of the number of terms contributing to the sum~\eqref{eq:TMBL} per unit length, as explained in Sec.~\ref{sec:DPRM}. More precisely, denoting $|G_{0,t}| = e^{- n g /2}$, and assuming that the number of terms in the sum~\eqref{eq:TMBL} giving a contribution characterized by a value of $g$ between $g$ and $g + {\rm d} g$ scales as $e^{n \Sigma(g)/2}$, one immediately finds that the number of terms contributing to ${\cal T}_0 (\beta)$ is given by $e^{- n \beta^2 \phi^\prime (\beta) / 2}$. At $\beta=2$ the quenched and annealed configurational entropies are therefore given by $\Sigma_{\rm q} = - 4 \phi^\prime_{\rm q} (2)$ and $\Sigma_{\rm ann} = - 4 \phi^\prime_{\rm ann} (2)$ (provided that $\beta_\star>2$).

In order to check whether the system is fully ergodic or not (at least in the range of the accessible system sizes), one needs to compare the configurational entropy with the logarithm of the maximum number of terms that can possibly contribute to ${\cal T}_0 (2)$. The total number of target sites $\vert t \rangle$ at distance $\ell=n/2$ from $\vert t \rangle$ is ${{n}\choose{n/2}} \simeq 2^n/\sqrt{\pi n /2}$. However, the many-body wave-functions at $E \simeq 0$ that satisfy ETH should spread uniformly over an energy shell with zero intensive energy, while they should have an exponentially small projections outside this energy shells~\cite{srednicki1994chaos,rigol2008thermalization}. Hence, we expect that in the fully ergodic regime a spin chain initiated at $t=0$ in $\vert 0 \rangle$ with energy $E/n \simeq 0$  can only reach the target states $\vert t \rangle$ with the same intensive energy, \ie, $e^{n \Sigma_{\rm max}/2} \approx {{n}\choose{n/2}} \rho(0)$, where $\rho(0)$ is the many-body density of states at zero energy (which grows as $\sqrt{n}$ times a disorder-dependent prefactor of order $1$):
\begin{equation} \label{eq:sigmamax}
\Sigma_{\rm max} \approx 2 \ln 2 + \frac{1}{n} \ln \frac{2}{\pi n} + \frac{2}{n} \ln \rho(0) \, .
\end{equation}
Although in the $n \to \infty$ limit $\Sigma_{\rm max} = 2 \ln 2$, for the limited sizes that we can numerically access the corrections of order $\ln n/n$ represented by the last two terms of the right hand side of the expression above cannot be neglected. $\Sigma_{\rm max}$ for $n=17$ is shown as a dashed-dotted line in the right panel of Fig.~\ref{fig:mstar}. Since $e^{n \Sigma/2}$ is the number of spin configurations at zero overlap from a random initial state reached under the quantum unitary dynamics after infinite time,  $\Sigma/\Sigma_{\rm max}$ provides a lower bound for the fractal dimension of the support set of the many-body wave-functions. As established in Ref.~\cite{de2020multifractality}, this fractal dimension provides, in turn, a lower bound for their entanglement entropy.

All in all the plots of Fig.~\ref{fig:mstar} look qualitatively very similar to the ones found for the Anderson model on the Cayley tree, Fig.~\ref{fig:mstarCT}: Also in the many-body setting  we find  a freezing ``glass'' transition of the paths such that for $W > W_g$ only a few $O(1)$ terms in the sum~\eqref{eq:TMBL} contribute to ${\cal T}_0(2)$ and to the quantum unitary dynamics. $W_g$ drifts significantly to stronger disorder as $n$ is increased and $W_g \approx 4.3$ for the largest accessible system size ($n=17$). A genuine localization transition is instead found at much larger disorder. The criterion obtained from the quenched free-energy, $\gamma_{\rm q} = 0$, yields an estimation for the critical disorder $W_c^{({\rm q})} (n)$ that slowly but progressively moves to larger and larger values of $W$ as $n$ is increased and is close to $\sim 6$ for the largest accessible size ($n=17$). The criterion obtained from the (modified) annealed free-energy, $\gamma_{\rm ann} = 0$, yields a critical disorder around $W_c^{({\rm ann})} \simeq 10.5$, which does not move significantly with the system size (and seems to drift to slightly smaller values upon increasing $n$). Furthermore, in the case of the Anderson model on the Cayley tree $W_c^{({\rm ann})}$ provides an estimation of the localization threshold very close to the exact values in the $n \to \infty$ limit (see the middle panel of Fig.~\ref{fig:mstarCT}). Based on these observations, it is reasonable to assume that $W_c^{({\rm ann})}$ yields the best estimation of the actual value of the critical disorder in the large $n$ limit and in the following we will just denote it by $W_c$. 

There is also a remarkable difference with respect to the Anderson model on the Cayley tree concerning the behavior of the configuration entropy at small disorder (right panel of Fig.~\ref{fig:mstar}). We indeed observe that for $W$ small enough, $W_{\rm ergo} \lesssim 2.5$, the configurational entropy tends to saturate to its maximum value $\Sigma_{\rm max}$ given in Eq.~\eqref{eq:sigmamax}, corresponding to fully ETH eigenstates. Concomitantly the Lyapunov exponent $\gamma_{\rm q,ann}$ seems to converge to its maximally chaotic bound $\gamma=1$, as in the fully ergodic phase of the RP model. These observations indicate that, contrary to the tight-binding model on the Cayley tree, but similarly to the RP model, the quantum disordered spin chain described by the Hamiltonian~\eqref{eq:HMBL} features a fully ergodic phase at small enough disorder described by random matrix theory.

In Fig.~\ref{fig:mstar} we also show the numerical results obtained within the FSA. In the left panel the evolution of $\beta_\star^{\rm fsa}$ where the annealed FSA free-energy attains its minimum, is plotted as a function of the disorder. As in the non-interacting case $\beta_\star^{\rm fsa}$ does not depend on $W$, and yields the strong disorder limit of the value of $\beta_\star$ obtained from exact diagonalizations (which is close to $1$). In the middle panel $\tilde{\phi}^{\rm fsa}_{\rm ann}(2) = \phi^{\rm fsa}_{\rm ann}(\beta_\star^{\rm fsa}) = \gamma_{\rm ann}^{\rm fsa}/2$ and $\phi^{\rm fsa}_{\rm q}(2) = \gamma_{\rm q}^{\rm fsa}/2$ are plotted as function of $W$, showing that, similarly to the non-interacting case, the FSA free-energies provide an upper bound for the localization transition. 

To sum up, these plots leads us to distinguish four different regimes (see Fig.~\ref{fig:Lbetastar}), at least in the range of system sizes numerically accessible~\cite{morningstar2022avalanches}: 
\begin{itemize}
    \item[{\bf I:}] $W \lesssim W_{\rm ergo} (n)$: 
    At small enough disorder $\Sigma \approx \Sigma_{\rm max}$, implying that all paths contribute to transport and decorrelation. This corresponds to a fully ergodic regime in which 
    the many-body eigenfunctions are essentially described by RMT and should satisfy ETH. 
    \item[{\bf II:}] $W_{\rm ergo} (n) \lesssim W \lesssim W_g (n)$: 
    At moderately small disorder  we find that $\beta_\star > 2$ but $\Sigma$ is significantly smaller than $\Sigma_{\rm max}$. This regime is similar to the intermediate phase of the RP model, see Fig.~\ref{fig:mstarRP}. The number of paths contribute to transport and decorrelation from a random initial condition grows exponentially with $n$ yet being a sub-extensive fraction of the total number of paths. 
    \item[{\bf III:}] $W_g (n) \lesssim W \lesssim W_c$: 
    In this disorder window one has that $\beta_\star$ becomes smaller than $2$, but $\gamma_{\rm ann} = 2 \phi_{\rm ann} (\beta_\star)>0$. This regime is the analogous of the frozen glassy phase of directed polymers (see Sec.~\ref{sec:DPRM}): The system is delocalized (the probability of reaching a configuration with zero overlap with the initial condition is finite for $n$ large enough), but only a few, $O(1)$, of these configurations contribute to the relaxation. In other words, quantum many-body configurations only hybridize with a few $O(1)$ resonances far away in the Hilbert space. In a broad portion of this ``glassy'' regime (\ie, for  $W>W_c^{({\rm q})} (n)$ but $W<W_c$) one finds that $\gamma_{\rm ann}>0$ while $\gamma_{\rm q}<0$ for the accessible system sizes. Hence the probability that {\it typical} finite-$n$ samples decorrelate from $\vert 0 \rangle$ 
     is exponentially small in $n$, since they typically lack of the rare system-wide resonances. Hence, the spin chains that can be simulated with the current computational resources appear as many-body localized for all practical purposes, since the conventional observables commonly used in the literature to characterize the MBL transition are insensitive to the rare outliers in the tails of the distributions of $|G_{0,t}|$, in contrast with  our large-deviation approach (an in particular the annealed estimations) that allows us to unveil the presence and quantify the effect of the rare resonances already for systems of moderate size. Such rare long-distance resonances become more and more prominent upon increasing $n$, and eventually yield to delocalization and relaxation in the thermodynamic limit also for typical samples. This is precisely the scenario predicted by the avalanche instability mechanism discussed in many recent works~\cite{de2017stability,thiery2018many,luitz2017small,goihl2019exploration,crowley2020avalanche,leonard2023probing,peacock2023many,morningstar2022avalanches,sels2022bath,Ha2023many}. It would be very interesting to establish a direct connection between this scenario and the one we are putting forward based on rare events in configuration space.
    \item[{\bf IV:}] $W \gtrsim W_c$: 
    A strong disorder the system enters in the genuine MBL phase, characterized by complete absence of transport and dissipation in the thermodynamic limit. Both $\gamma_{\rm q}$ and $\gamma_{\rm ann}$ are negative, corresponding to the fact that the probability to decorrelate from the initial condition goes to zero exponentially with the number of spins, even after taking into account the largest resonances. 
\end{itemize}
In the preceding discussion, we have explicitly denoted the dependence on the system size $n$ in delineating the crossovers between various regimes. This emphasis aims to underscore the observation that, with increasing system sizes, $W_{\rm ergo}$, $W_g$, and $W_c^{({\rm q})}$ exhibit a pronounced and systematic tendency to shift towards higher disorder values. Conversely, the position of the MBL transition $W_c$ does not appear to be markedly influenced by variations in $n$.

As anticipated in the introduction, this phase diagram is very similar to the one recently discussed in Ref.~\cite{morningstar2022avalanches}, although the analysis of the rare system-wide resonances is performed in a different way. Furthermore, the physical properties of the regime (III) characterized by rare resonances seem to be tightly related with the recent results of Ref.~\cite{Ha2023many}, which show that the dominant processes in the spreading of a quantum avalanche in the pre-thermal regime involves only a few pairs of eigenstates with a strong near-resonance. The possible physical implications on transport and spectral statistics of the scenario described above are discussed in more details in Sec.~\ref{sec:phasediagram}.

Similarly to the case of the Anderson model on the Cayley tree, the freezing transition of the paths is connected to the condensation of the measure appearing in~\eqref{eq:TMBL} on the extreme values of the probability distributions of the propagators. In App.~\ref{app:PG} we inspect closely the shape of these distributions varying $W$ and $n$, showing that upon increasing the disorder strength $P(|G_{0,t}|)$ develops power-law tails that do not decrease fast enough to zero at large arguments.

\begin{figure*}
\includegraphics[width=0.338\textwidth]{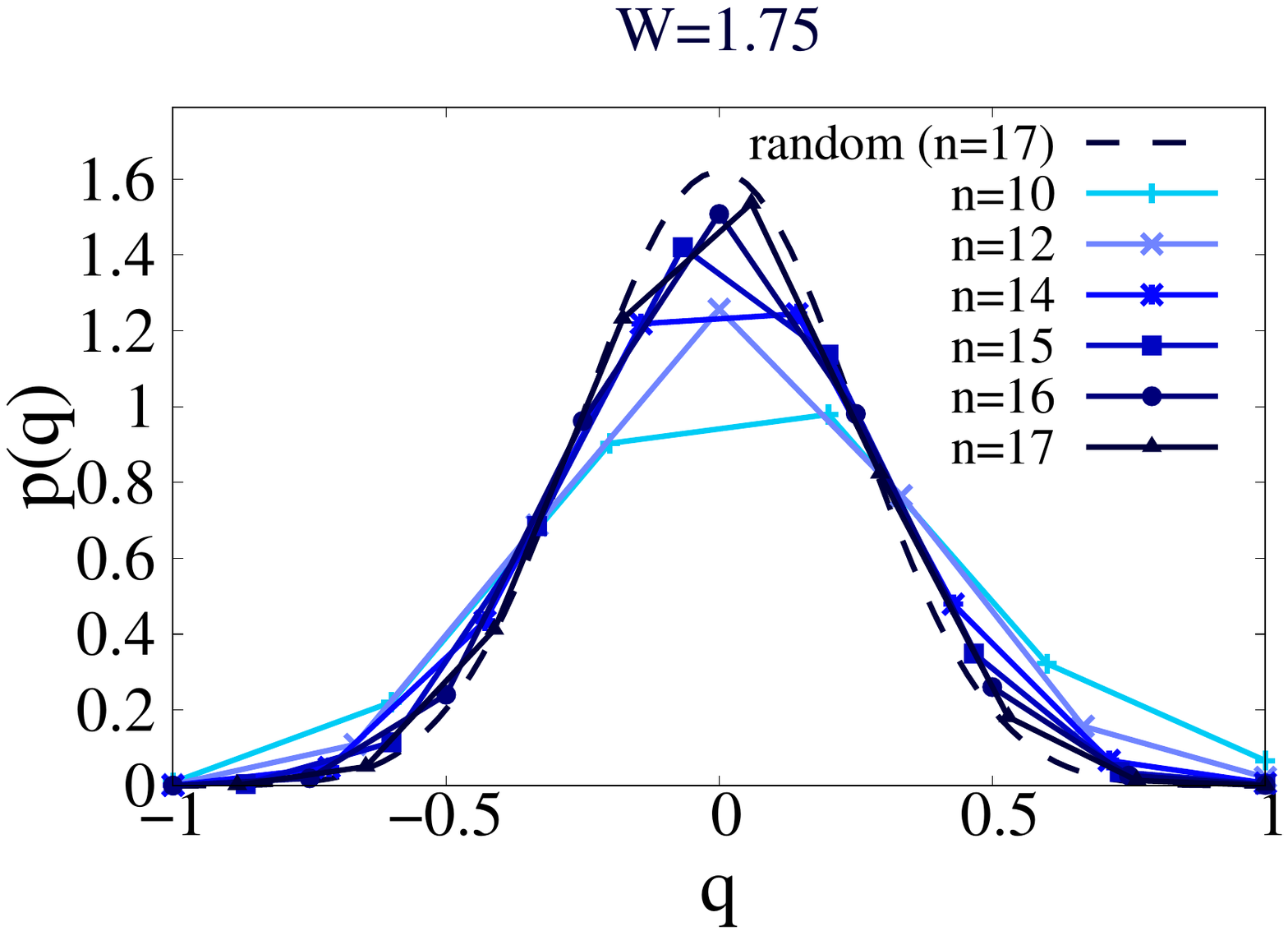} \hspace{-0.36cm} \includegraphics[width=0.338\textwidth]{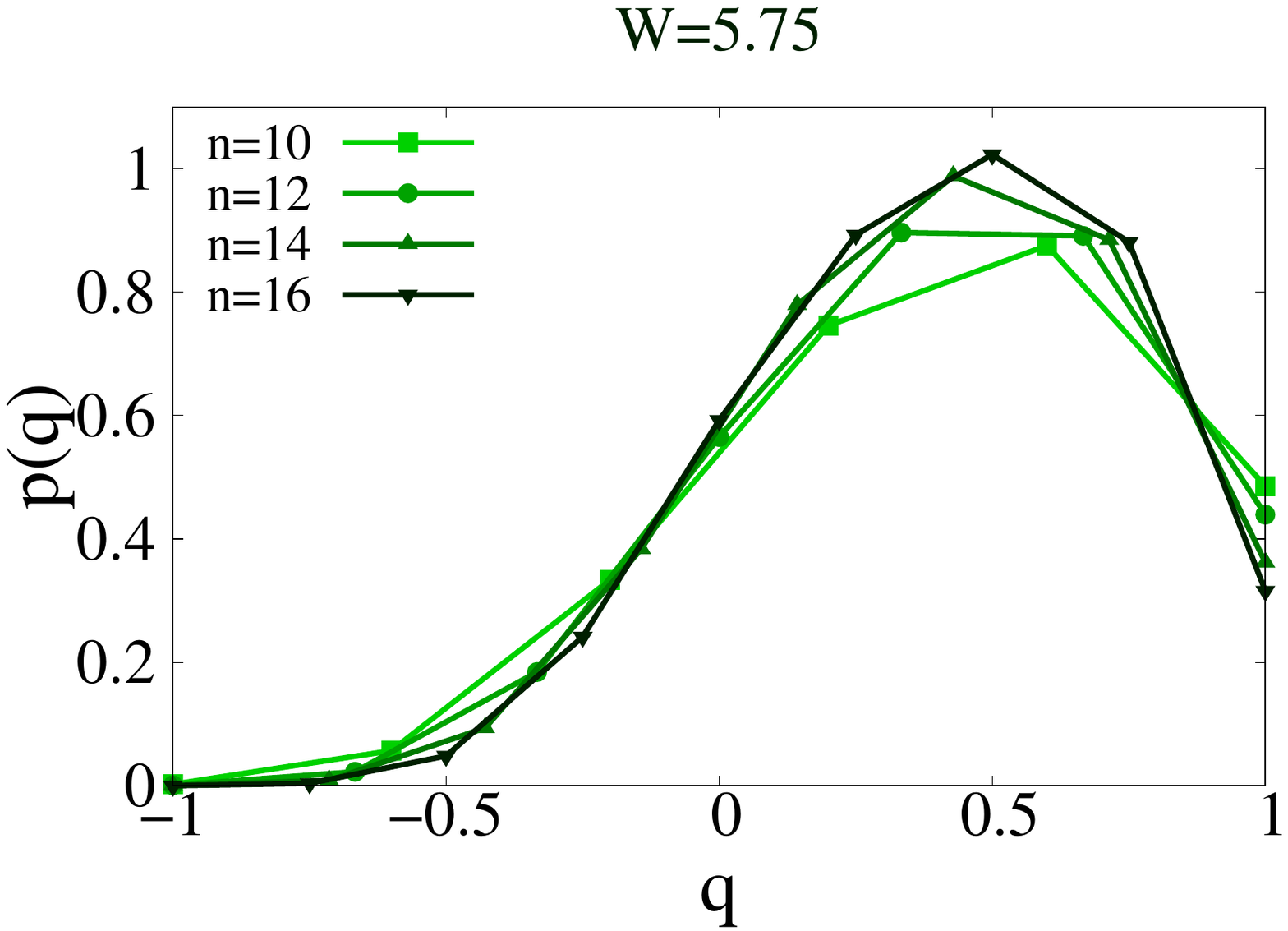} \hspace{-0.36cm} \includegraphics[width=0.338\textwidth]{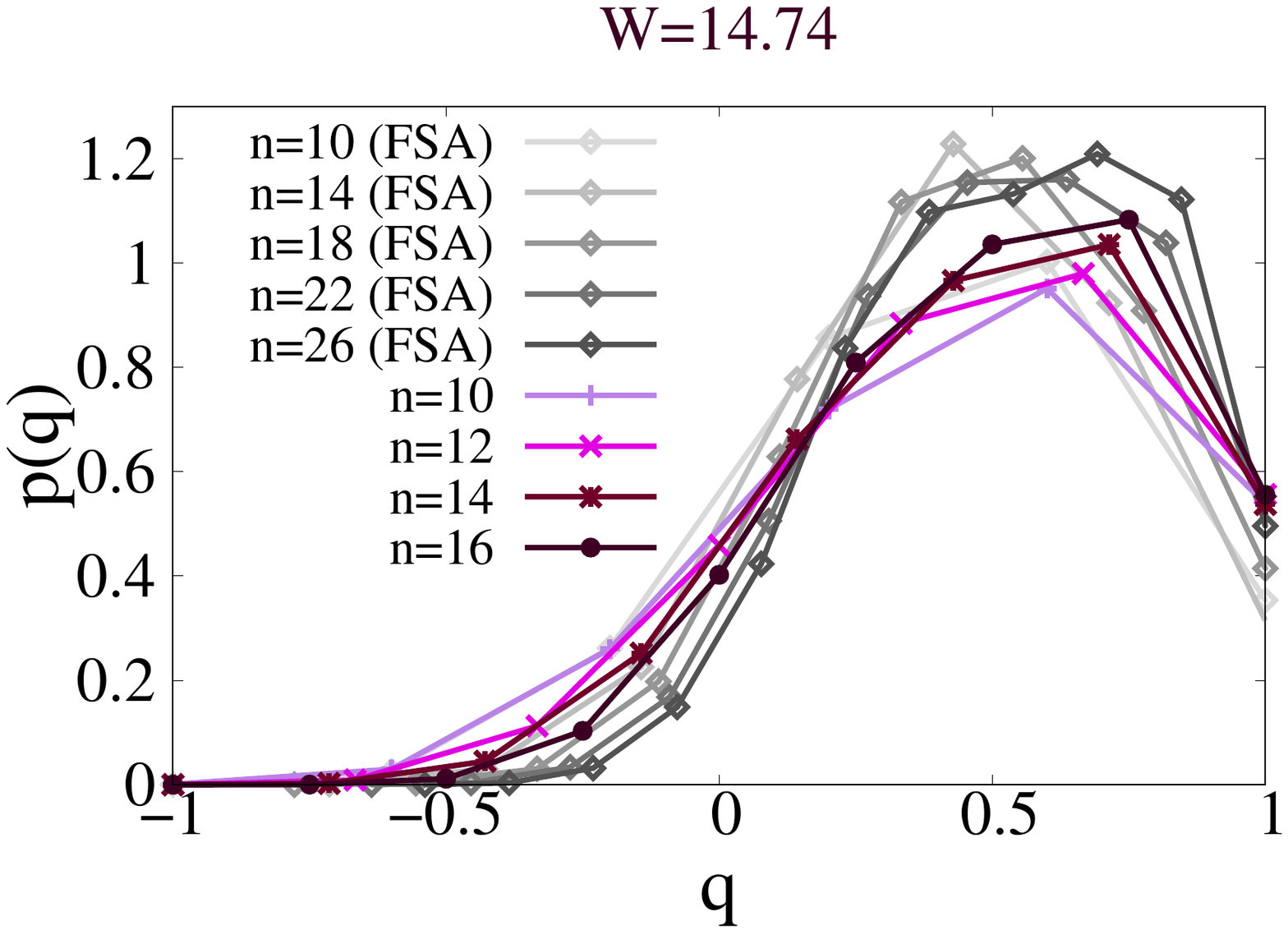} 
\caption{Probability distribution $p(q)$ of the overlap $q$, defined in Eq.~(\ref{eq:def:overlap}), between targets configurations $\vert t \rangle$ at distance $\ell = n/2$ from a randomly chosen initial state $\vert 0 \rangle$. The corresponding disorder strengths are $W=1.75$ (left panel), $W=5.75$ (middle panel), and $W=14.75$ (right panel). The dashed curve in the left panel correspond to the overlap distribution in the completely random case, $p(q) = 2^{-n} {{n(q+1)/2}\choose{n}}$, for $n=17$. The gray curves in the right panel correspond to the overlap distributions obtained within the FSA. Note that in order to represent the overlap distributions obtained at different $n$ on the same plots, we have rescaled $p(q)$ by a factor $n/2$ in such a way that $\int p(q) {\rm d} q = 1$ (and not $\sum_q p(q) = 1$). 
\label{fig:pqMBL}}
\end{figure*}

\subsection{The overlap distribution}

Pursuing the analogy with DPRM a step further, we study the probability distribution of the overlaps between the many-body target configurations $|t\rangle$ that contribute to the generalized partition function ${\cal T}_0$ for the ``physical'' value of the ``inverse temperature'' $\beta=2$. The overlap $q_{t_a,t_b}$ between two target spin configurations is defined according to Eq.~\eqref{eq:overlap} as
\[
q_{t_a,t_b} = \frac{1}{n} \sum_{i =1}^n \sigma_i^z(t_a) \, \sigma_i^z(t_b) \, ,
\]
where $\sigma_i^z(t_{a,b})$ is the value of the $z$-component of the spin on site $i$ in the many-body configuration $|t_{a,b}\rangle$. Note that, differently from the Cayley tree, here the overlap takes $n+1$ discrete values between $-1$ and $1$ with steps of $2/n$ ($q_i =-1+2i/n$, with $i=0, \ldots, n$). All the target configurations $\vert t \rangle$ are constrained to have zero overlap with the initial state, but the ones that contribute the most to transport and dissipation for a given disorder realization and a for given choice of the initial condition may be either close to each other (\ie~have a large mutual overlap) or occupy distant nodes on the hypercube (\ie~have low mutual overlap). The statistical weight in the sum~\eqref{eq:TMBL} corresponding to a given target configuration when the system is initialized in $|0 \rangle$ is by definition  $|G_{0,t}|^2$. In analogy with Eqs.~\eqref{eq:pqDPRM} and~\eqref{eq:pqCT} we then have:
\begin{equation}
p(q) = \left \langle \sum_{a,b}^{{n}\choose{n/2}} \frac{ |G_{0,t_a}|^2 \, |G_{0,t_b}|^2 }{{\cal T}_0^2(2)} \, \delta(q - q_{t_a,t_b}) \right \rangle \, ,
\label{eq:def:overlap}
\end{equation}
where the average is performed over the quenched disorder in the Hamiltonian and over the initial configuration $\vert 0 \rangle$. Note that due to the exponential number of states, which is  particularly severe for large system sizes, 
we have sampled according to the weights $|G_{0,t_0}|^2$ instead of performing the exact sum in Eq.~(\ref{eq:def:overlap}).

In Fig.~\ref{fig:pqMBL} we plot the overlap distribution for three values of the disorder across the phase diagram. 
(Note that in order to compare the distributions obtained for different  $n$ we have chosen to normalize the probability distribution in such a way that $\int \!\! p (q) {\rm d} q = 1$.) At small disorder ($W=1.75 < W_{\rm ergo}$, left panel) $p(q)$ converges to $p(q) = 2^{-n} {{n(q+1)/2}\choose{n}}$ (dashed curve), which tends to a Gaussian of zero mean and variance $1/n$. This is the overlap distribution expected for fully ergodic systems,  for which the amplitudes are all roughly equal and the unitary quantum dynamics spreads the wave-packet uniformly in the Hilbert space. 
Upon increasing the disorder strength ($W=5.75$ and $W=14.75$, middle and right panels), $p(q)$ becomes more skewed and asymmetric, with a bias towards values of the overlap closer and closer to $1$ as the system size is increased, indicating that for $W>W_g$, in the region where only rare paths contribute to decorrelation, the portion of the Hilbert space that the system can explore shrinks.

\begin{figure*}
\includegraphics[width=0.42\textwidth]{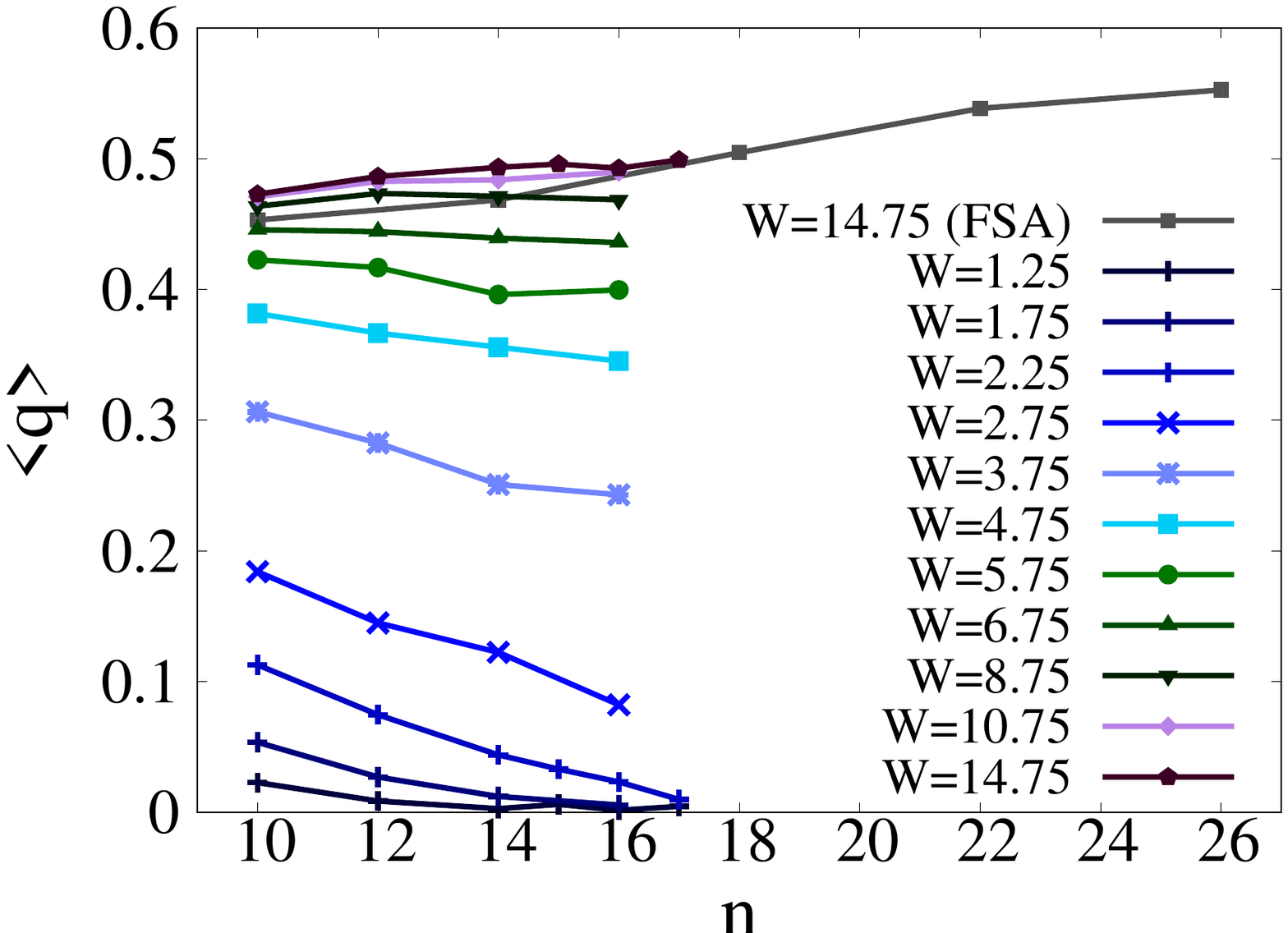} \hspace{0.5cm}
\includegraphics[width=0.42\textwidth]{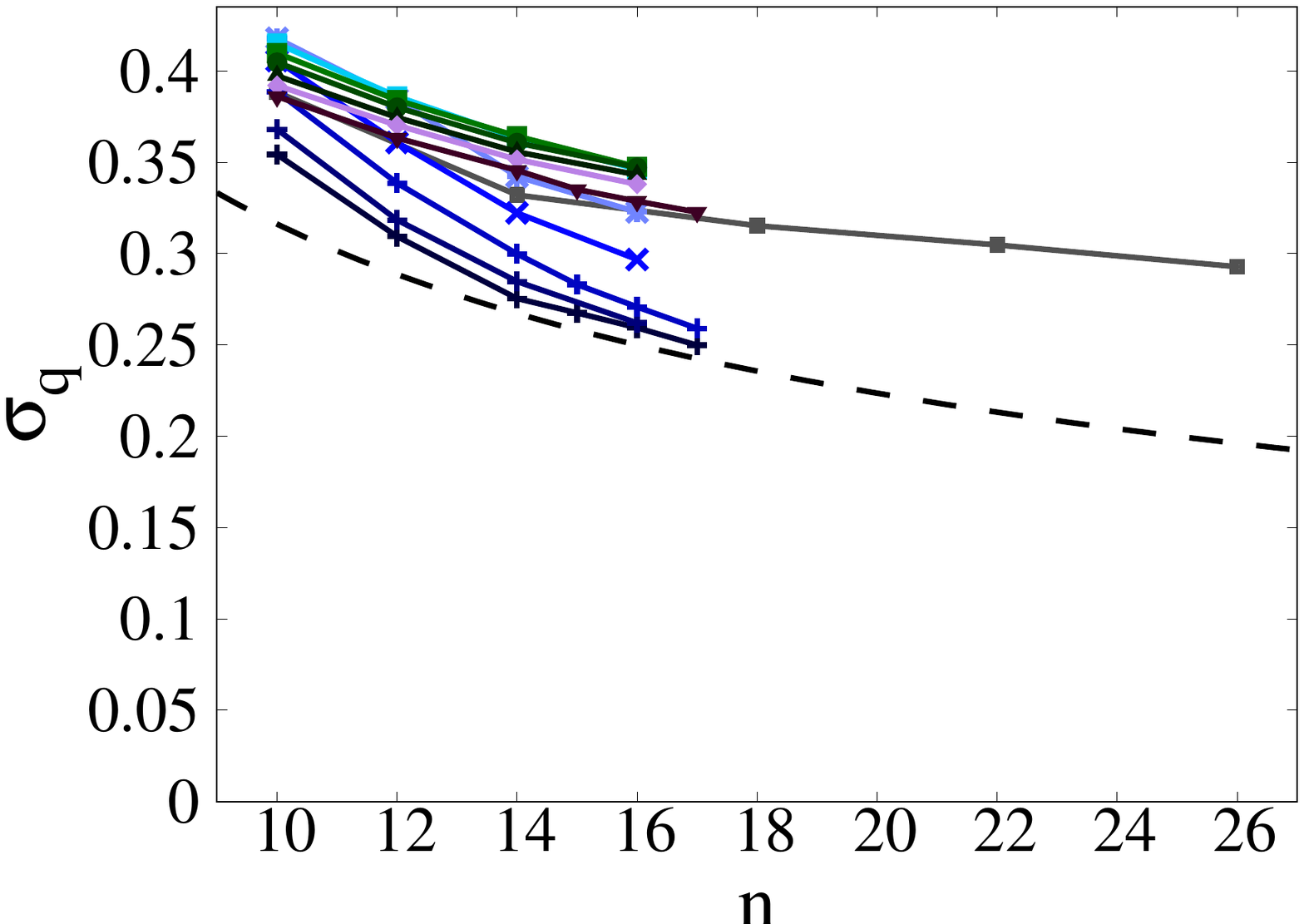} 
\caption{Average overlap $\avg{q}$ (left) and square root of the variance of the overlap distribution $\sigma_q$ (right) as a function of the number of spins $n$ for several values of the disorder. The gray curves show the FSA results for $W=14.75$, which, similarly to the non-interacting case, provides a strong disorder limiting behavior of the data obtained from exact diagonalizations. The dashed curve in the right panel correspond to $\sigma_q = 1/\sqrt{n}$, which is the expected scaling in the completely random case. 
\label{fig:avq}}
\end{figure*}

In the right panel of Fig.~\ref{fig:pqMBL} we also show the overlap distribution computed using the FSA. In contrast to the exact result, within this approximation the shape of $p(q)$ does not change much when varying the disorder width $W$. This is compatible with the nature of the approximation, which is meant to describe the strong disorder limit, deep into the localized phase. 

To evaluate the finite-size behavior of the distributions more quantitatively, we have analyzed the dependence of the average overlap, $\avg{q} = \sum_{q = -1}^{+1} q \, p(q)$ (where $q = -1 + 2i/n$ with $i=0, \ldots, n$), and  of the variance of the distributions, $\sigma_q = \sqrt{\avg{q^2} - \avg{q}^2}$, as function of the system size $n$ for several values of the disorder across the phase diagram. The results are shown in the left and right panels of Fig.~\ref{fig:avq} respectively. At weak disorder, in the  regimes (I) and (II) --- blue curves --- $\avg{q}$ goes to zero with $n$ and the variance approaches $\sigma_q \sim 1/\sqrt{n}$ (dashed line), which is the expected behavior in the completely random case. Conversely, in the regimes (III) and (IV) --- green and violet curves --- $\avg{q}$  
and $\sigma_q$ seem to approach a finite limiting value for large $n$. In fact we have fitted the functions $\sigma_q(n)=\sigma_q^\infty + an^{-\eta}$ to the data, 
obtaining a strictly positive value for $\sigma_q^\infty$ for $W$ large enough.
In Figs.~\ref{fig:avq} we also show the results for $\avg{q}$ and $\sigma_q$ obtained within the FSA which, similarly to the non-interacting case, provide the strong disorder limiting behavior of the data obtained from exact diagonalizations.

\subsection{Dependence on the length of the path} \label{sec:length}

In this section, instead of considering only target configurations at distance $\ell=n/2$ (\ie, with zero overlap from the initial state), we analyze the dependence of the quenched and annealed free-energies as a function of the distance $\ell$ (\ie, the number of flipped spins) between $\vert 0 \rangle$  and $\vert t \rangle$. It is convenient to introduce the variable $x = \ell/n$, defined as the fraction of flipped spins between $\vert 0 \rangle$  and $\vert t \rangle$. The overlap between the initial configurations and a spin configuration at distance $\ell$ from it is thus given by $q_{0,t} = 1 - 2 x$. The number of configurations at distance $\ell$ from $\vert 0 \rangle$ is simply given by the combinatorial factor ${n}\choose{\ell}$, and scales exponentially with the length of the path approximately as $e^{-n[x \ln x + (1-x) \ln (1-x)]}/\sqrt{2 \pi n x (1-x)}$. The ``generalized fractional Landauer transmissions'' of the problem starting from a random initial state $\vert 0 \rangle$ are then defined as
\begin{equation} \label{eq:Tell}
{\cal T}_0 (\ell,\beta) = \sum_{t=1}^{{n}\choose{\ell}} |G_{0,t} |^\beta \, .
\end{equation}
We recall that ${\cal T}_0(\ell,2)$ is proportional to the probability that the system is initialized in a state $|0\rangle$ at time $0$ and is found in any of the states $\vert t \rangle$ at $\ell$ spin flips away from it after infinite time (or, equivalently, to the Fisher-Lee conductivity of the $n$-dimensional hypercube in a scattering geometry in which a semi-infinite lead is attached to $\vert 0 \rangle$ and ${{n}\choose{\ell}}$ semi-infinite leads are attached to the target nodes $\vert t \rangle$~\cite{fisher1981relation}).

\begin{figure*}
\includegraphics[width=0.342\textwidth]{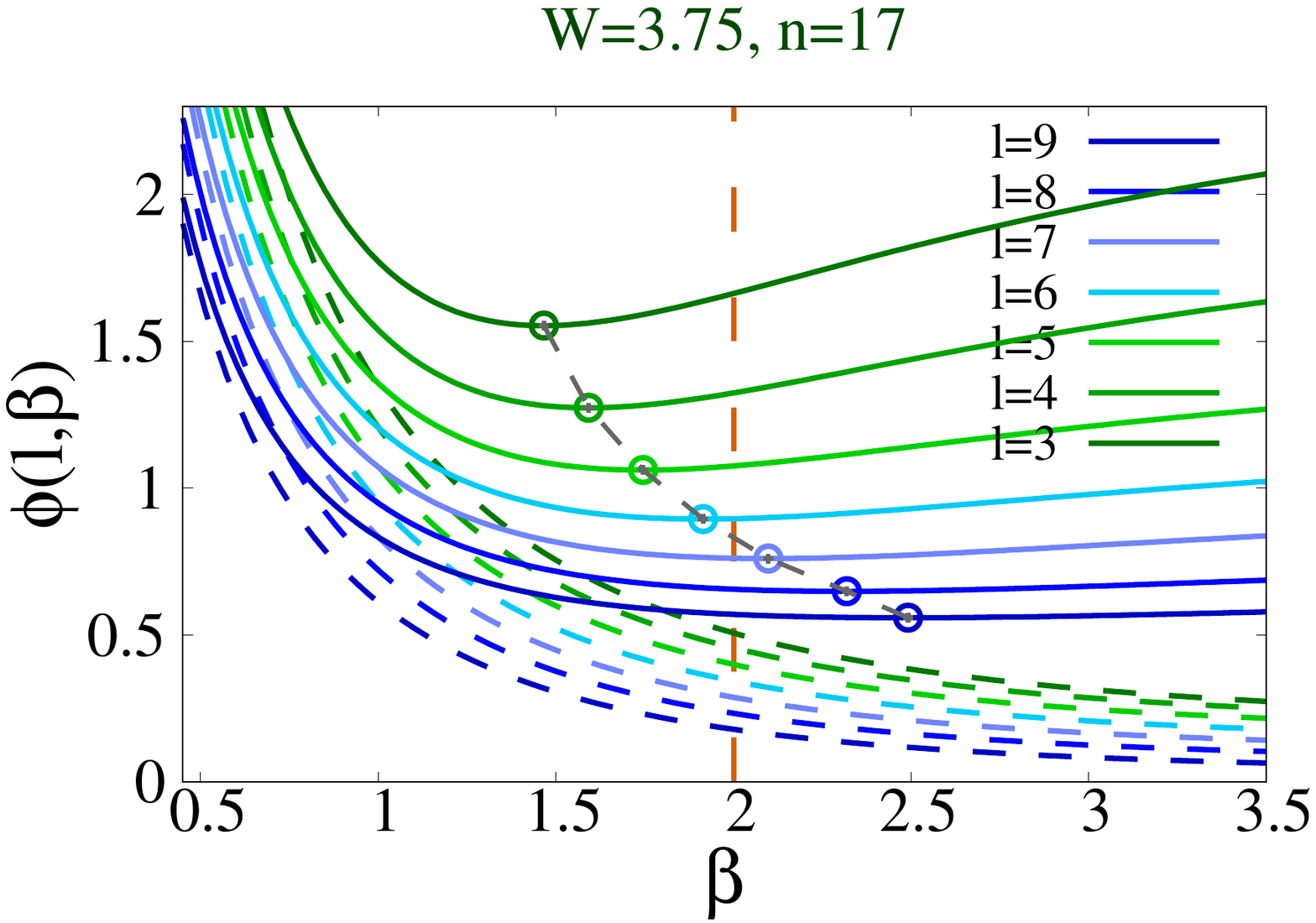} \hspace{-0.32cm} \includegraphics[width=0.342\textwidth]{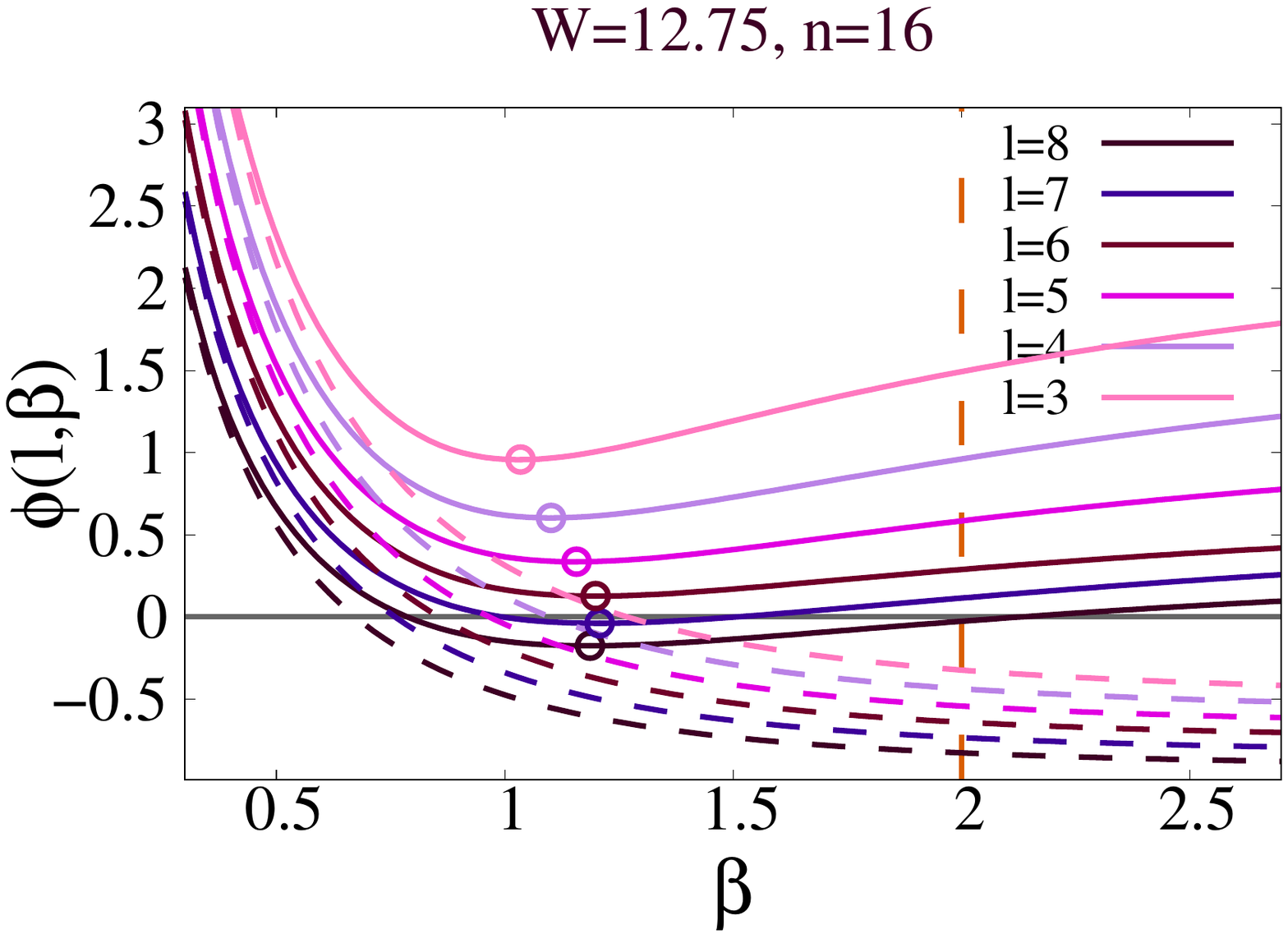} \hspace{-0.28cm}
\includegraphics[width=0.305\textwidth]{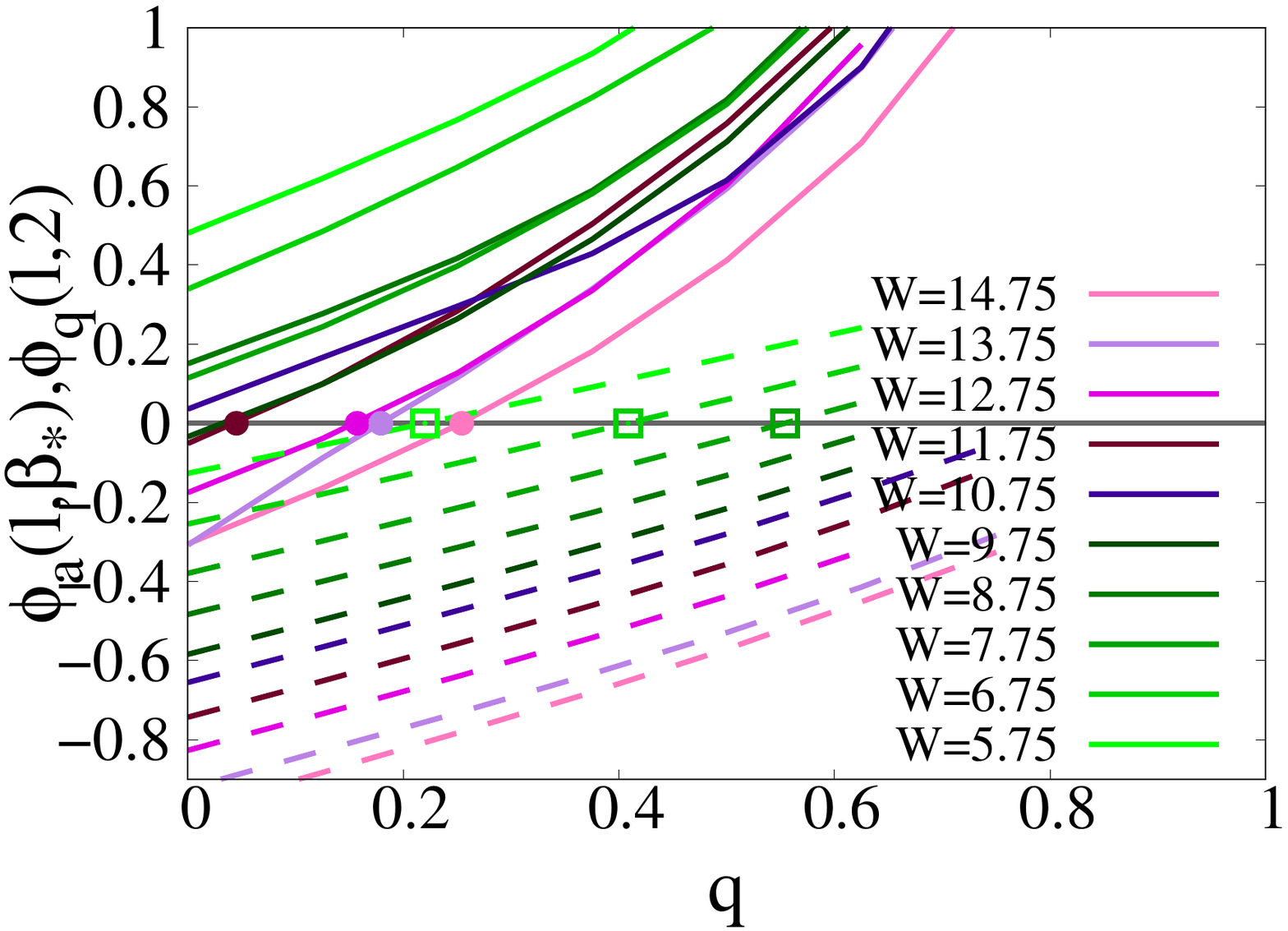} 
\caption{Left: Annealed (continuous curves) and quenched (dashed curves) free-energies, Eq.~(\ref{eq:phil}), as a function of $\beta$ for $W=3.75$ and $n=17$ spins, and for several values of the number of allowed spin flips $\ell=3, \ldots, (n+1)/2$. The circles spot the position of the minima of $\phi_{\rm ann} (\ell,\beta)$ in $\beta = \beta_\star$. $q_g$ is obtained from Eq.~\eqref{eq:qg} as the value of the overlap at which $\beta_\star$ crosses $2$. Middle: Annealed (continuous curves) and quenched (dashed curves) free-energies, Eq.~(\ref{eq:phil}), as a function of $\beta$ for $W=12.75$ and $n=16$ spins, and for several values of the number of allowed spin flips $\ell=3, \ldots, n/2$. The circles spot the position of the minima of $\phi_{\rm ann} (\ell,\beta)$ in $\beta = \beta_\star$. Right: $\phi_{\rm q} (\ell,2)$ and $\tilde{\phi}_{\rm ann} (\ell,2)= \phi_{\rm ann} (\ell,\beta_\star)$ (since $\beta_\star<2$) 
as a function of the overlap $q = 1 - 2 \ell/n$ for several values of $W$ (and for $n=16$ spins). The circles and the squares spot respectively the points where $\phi_{\rm ann} (\beta_\star)$ and $\phi_{\rm q}(2)$ vanish, which yield $Q_{\rm EA}^{({\rm ann})}$ (denoted as $Q_{\rm EA}$ throughout) and  $Q_{\rm EA}^{({\rm q})}$ from the conditions~\eqref{eq:QEA}. 
\label{fig:Lphi}}
\end{figure*}

We have computed numerically the quenched and annealed free-energies (averaging over several realizations of the disorder and over many random initial states close to the middle of the many-body spectrum) which, in analogy with Eq.~\eqref{eq:free_energy}, are defined as:
\begin{equation}
\begin{aligned}
\phi_{\rm ann} (\ell,\beta;n) & =  \frac{1}{\beta \ell} \ln  \avg{{\cal T}_0(\ell,\beta)} \, , \\
\phi_{\rm q} (\ell,\beta;n) & =  \frac{1}{\beta \ell} \avg{\ln {\cal T}_0(\ell,\beta)} \, .
\end{aligned} 
\label{eq:phil}
\end{equation}
(The explicit dependence of the number of spins $n$ will be omitted most of the times to simplify the notation.) In Fig.~\ref{fig:Lphi}(left) we plot $\phi_{\rm ann}$ (continuous curves) and $\phi_{\rm q}$ (dashed curves) as a function of $\beta$ for two values of the disorder. 
At strong disorder, $W = 12.75$ (middle panel), within the MBL phase, the annealed free-energy has a minimum close to $\beta_\star \simeq 1$ whose position does not depend much on $\ell$. Both the quenched and the annealed free-energies are shifted to lower values when the length of the path is increased. At small $\ell$ the height of the minimum of the annealed free-energy (which provides an estimation of the value of the free-energy at $\beta=2$, since $\tilde{\phi}_{\rm ann} (2) = \phi_{\rm ann} (\beta_\star)$ for $\beta_\star < 2$) is positive, implying that the probability that the system reaches a spin configuration at distance $\ell$ from $\vert 0 \rangle$ stays finite. As $\ell$ is increased, however, the value of the annealed free-energy in $\beta_\star$ becomes smaller than zero, signifying that the probability that the system reaches spin configurations at such distances is exponentially small in the distance. The value of $\ell$ at which $\phi_{\rm ann} (\beta_\star,\ell) = 0$ thus yields the maximum Hamming distance in the Hilbert space that the system can explore under the unitary quantum dynamics starting from a random initial configuration (\ie~with intensive energy close to zero). 
Similarly, one finds that the height of the quenched free-energy computed at $\beta=2$ decreases as $\ell$ is increased and crosses $0$ for a given value of $\ell$ which depends on $W$. We repeated this analysis for many values of $W$ across the localized phase and for several system sizes $n$. The values of $\tilde{\phi}_{\rm ann}(\ell,2) = \phi_{\rm ann}(\ell,\beta_\star)$ and $\phi_{\rm q}(\ell,2)$ as a function of the overlap $q = 1 - 2\ell/n$ for several values of $W$ are plotted in the right panel of  Fig.~\ref{fig:Lphi} for $n=16$ spins.  This plots allows us to estimate $Q_{\rm EA}^{({\rm ann})} (W,n)$ and $Q_{\rm EA}^{({\rm q})}(W,n)$, corresponding to the value of the overlap with the initial configuration that the system reaches after infinite time, defined from 
the conditions:
\begin{equation} \label{eq:QEA}
\begin{aligned}
& \phi_{\rm q} \left( \frac{n(1-Q_{\rm EA}^{({\rm q})} )}{2} ,2 \right) = 0 \, , \\
& \phi_{\rm ann} \left( \frac{n(1-Q_{\rm EA}^{({\rm ann})} )}{2}, \beta_\star \right) = 0 \, .
\end{aligned}
\end{equation}
$Q_{\rm EA}$ is essentially proportional to the height of the plateau attained by the spin-spin correlation function at long times and is therefore the equivalent of the Edwards-Anderson order parameter in the context of the spin glass transition~\cite{mezard1987spin}. 
A value of $Q_{\rm EA}$ above zero corresponds to MBL, since it implies that the systems remembers forever its initial state, while $Q_{\rm EA}=0$ signals that the system is able to decorrelate from it (although it might still not be ergodic due to the freezing 
of the paths that lead to such decorrelation, as discussed above). $Q_{\rm EA}^{({\rm q})} (W,n)$ is the value of the overlap typically observed in samples of small enough size, which typically lack of the rare system-wide resonances, and should converge to $Q_{\rm EA}^{({\rm ann})} (W,n)$ upon increasing $n$. Therefore in the following we will simply denote $Q_{\rm EA}^{({\rm ann})} \equiv Q_{\rm EA}$, as the relevant one at large enough $n$.

We have carried out the same analysis also within the FSA. The behavior of $Q_{\rm EA}^{({\rm ann})}$ and $Q_{\rm EA}^{({\rm q})}$ is qualitatively very similar to the one found from exact diagonalizations, with the MBL regime shifted to larger values of the disorder, as expected. 
The the general trend 
is the same, \ie, when increasing $\ell$ an intersection with $\phi^{\rm fsa}=0$ appears. 
In comparison to the exact result, the curves are further above 
$\phi=0$, which reflects in a shift of 
the values of $\phi^{\rm fsa}$  at the minimum $\beta_{\star}$ and at $\beta=2$, respectively,  as function of the overlap $q=1-2\ell/n$, 
and in the intersections of these values with $\phi=0$ as function of $W$. 

At moderately weak disorder, $W=3.75$ (left panel of Fig.~\ref{fig:Lphi}), the dependence of the position of the minimum of the annealed free-energy on the distance is much stronger. At small $\ell$ one has that $\beta_\star<2$, which means the configurational entropy (\ref{eq:ScDPRM}) is zero for the value of the auxiliary parameter corresponding to physical transport and relaxation, $\beta=2$. 
This implies that the number of spin configurations with $\ell$ spin-flips that are in resonance with the initial configuration and contribute significantly to the spreading of the wave-packet under the unitary quantum dynamics does not grow exponentially with $\ell$. Upon increasing the distance, $\beta_\star$ increases and becomes larger than $2$ at large enough $\ell$. This means that the number of resonant spin configurations that contribute to the sum~\eqref{eq:Tell} for $\beta=2$ proliferate and become exponential in $\ell$ (although it might not saturate to their maximum value). Repeating this analysis for several values of $W$ and $n$ yields the Hamming distance $\ell_g (W,n)$ below which the quantum dynamics is dominated by rare resonances and the wave-packet only spreads over a few specific disorder-dependent paths. Using the relation~\eqref{eq:overlap} between the number of spin-flips and the overlap with the initial state one obtains the overlap $q_g(W,n) = 1 - 2 \ell_g(W,n)/n$ above which ${\cal T}_0(\ell,2)$ is in the ``glassy'' phase, corresponding to the condition:
\begin{equation} \label{eq:qg}
\phi^\prime_{\rm ann} \left( \frac{n(1-q_g)}{2}, 2 \right) = 0 \, .
\end{equation}
We find that $q_g$ displays strong finite size effects, and moves systematically to larger values of the disorder as the system size is increased. Also $Q_{\rm EA}^{({\rm q})}$ slightly drifts to larger $W$ upon increasing $n$, due to the fact that considering larger chains enhance the probability of finding long-range resonances which allow the system to decorrelate from the initial state. Conversely the position of $Q_{\rm EA}{({\rm ann})}$ does not exhibit a significant drift. On the contrary, $Q_{\rm EA}{({\rm ann})}$ seems even to move to slightly smaller values of the disorder for larger $n$, supporting the idea that a genuine MBL phase persists in the thermodynamic limit~\cite{imbrie2016many} (at least for the model described by the Hamiltonian~\eqref{eq:HMBL}).


\subsection{The phase diagram} \label{sec:phasediagram}

The results discussed so far on the statistics of the paths in the Hilbert space contributing to transport and relaxation from a random initial configuration are summarized in the (finite-size) schematic phase diagram of Fig.~\ref{fig:Lbetastar} (see also the descriptions of the four regions (I)--(IV) in Sec.~\ref{sec:summary} and at the end of Sec.~\ref{sec:MBL}). The phase diagram  shows $q_g$ and $Q_{\rm EA}$ in the ($W$--$q$) plane, as well as the different regimes described above.

The region of the plane below $Q_{\rm EA}$ corresponds to the genuine MBL phase (IV) 
in the large $n$ limit. For $W > W_c \sim 10.5$ the spin-spin correlation function would decay to a strictly positive plateau value, $Q_{\rm EA}$, corresponding to localized eigenstates in the Hilbert space. In other words, for large $n$ the probability that one finds the system in a spin configuration with overlap smaller than $Q_{\rm EA}$ from a random initial state is exponentially small in $n$.

The left portion of the plane at small disorder and below $q_g$ corresponds to the region in which the eigenfunctions are either fully (I) or partially (II) ergodic:  In particular, for $W \lesssim W_{\rm ergo} (n)$ 
all paths contribute to transport and decorrelation; 
The region (II) is similar to the intermediate phase of the RP model, since an exponential number of resonances with a random initial state are found at large distance in the Hilbert space,  yet they represent a sub-extensive fraction of the accessible volume (at least for the accessible sizes). 

The region (III) above $q_g (n)$ is characterized by rare long-range resonances, and the spreading of the wave-packet only occurs through a few disorder-dependent paths that connect the random initial state with those resonances. This regime is the analogous of the frozen glassy phase of directed polymers (see Sec.~\ref{sec:DPRM}). As discussed above, there is a broad disorder interval within this regime (where $Q_{\rm EA}^{({\rm q})} >0$ but $Q_{\rm EA} = 0$, \ie~ for $W_c^{({\rm q})} < W < W_c$), in which {\it typical} samples of small enough size appear as MBL in most respects, 
as they lack of the rare 
resonances that allows the system to decorrelation from the initial configuration through rare specific paths~\cite{morningstar2022avalanches}. These resonances becomes more and more relevant upon increasing $n$ and eventually lead to the instability of such apparent insulating regime, exactly as discussed in Refs.~\cite{de2017stability,thiery2018many,luitz2017small,goihl2019exploration,crowley2020avalanche,leonard2023probing,peacock2023many,morningstar2022avalanches,sels2022bath,Ha2023many}. We therefore argue that typical properties of the spectral statistics and of the statistics of many-body eigenstates computed by previous work are affected by very strong finite-size effects, as they miss rare long-range resonances which only appear at larger (inaccessible $n$). Our large-deviation approach allows us to circumvent this difficulty, since the the annealed free-energy encodes the effect of such resonances found in rare samples that we are expected to become typical at larger $n$.

It is important to stress once again that this phase diagram is extrapolated from the behavior of system of {\it finite sizes}. We cannot establish whether the intermediate phases (II) and (III) will persist in the thermodynamic limit or if they are just crossover regimes and full ergodicity is eventually restored for all $W<W_c$ upon increasing the length of the chain. For instance, such a crossover from apparent delocalized but non-ergodic eigenstates observed in finite-size systems to fully ergodic eigenstates for systems much larger than a characteristic size that grows exponentially upon approaching Anderson localization, takes places in the Anderson model on the random-regular (\ie, a locally tree-like graph with no boundaries and loops of size $\ln N$)~\cite{biroli2012difference,altshuler2016nonergodic,kravtsov2018non,tikhonov2016anderson,Tikhonov_2019,garcia2017scaling,biroli2018delocalization,garcia2022critical,vanoni2023renormalization}. In fact, as mentioned above, $q_g$ and $W_g$ are affected by strong finite-size effects and drift significantly to larger values of the disorder as $n$ is increased (as schematically represented by the arrows in Fig.~\ref{fig:Lbetastar}). Yet, a partial indication in support of the existence of a phase dominated by rare resonances in the large $n$ limit comes from the close inspection of the behavior of the quenched estimation of the configurational entropy (dashed lines of the right panel of Fig.~\ref{fig:mstar}): The curves for different system sizes cross around $W \sim 6.5$ (\ie~$\Sigma_{\rm q}$ decreases with $n$ for $W \gtrsim 6.5$), 
suggesting that $\Sigma$ does not saturate to its maximum value even for very large $n$ if the disorder is too strong. In addition, the Lyapunov exponent is bounded by its finite-$n$ annealed estimation from above, and $\gamma_{\rm ann}$ seems to decrease upon increasing $n$ (middle panel of Fig.~\ref{fig:mstar}). This suggests that in the large $n$ limit $\gamma$ converges to a value which is strictly smaller than its maximally GOE chaotic bound $\gamma=1$ at strong enough disorder. 
Moreover, since on finite time scales the quantum evolution can only explore finite regions of the hypercube,  it is natural to expect that, even for very large samples, the dynamics at moderately large times is still controlled by the ``glassy'' features of the statistics of the resonances observed at finite distances for all practical purposes~\cite{biroli2017delocalized,biroli2020anomalous}.

The finite-size effects on the MBL transition seem instead to be much weaker: $W_c$ and $Q_{\rm EA}$ do not drift with $n$ in a significant way (at least for the accessible sizes). This, together with the observation that in both benchmark cases studied in Sec.~\ref{sec:CT} and~\ref{sec:RP} the condition $\gamma_{\rm ann} = 0$ provides an upper bound for the localization transition in the thermodynamic limit (see the middle panels of Figs.~\ref{fig:mstarCT} and~\ref{fig:mstarRP}), supports the existence of a genuine MBL phase in the $n \to \infty$ limit for the disordered spin chain described by~\eqref{eq:HMBL}, as predicted by Ref.~\cite{imbrie2016many}, although at a much stronger value of the disorder than previously thought. 

As anticipated above, the scenario discussed here shares many similarities with the one recently put forward in Ref.~\cite{morningstar2022avalanches}, although the analysis of the statistics of the rare long-range resonances and the instability of the MBL phase towards quantum avalanches are performed in different ways (and for different models). Interestingly enough, also in Ref.~\cite{morningstar2022avalanches} the landmarks of the different regimes are identified from the extreme values of some suitable observables. In particular, the landmark of the onset of the avalanche instability destabilizing the MBL phase is estimated by coupling the chain to an infinite bath at one end, and determining at which disorder strength the {\it slowest} decay rate toward the steady state becomes larger than the level spacing of the thermal regions. Rare system-wide resonances are instead identified by individuating the eigenstate with the {\it largest} quantum mutual information between far-away spins. This allows the authors to discriminate between a regime where a vanishing fraction of eigenstates begin to be involved in long-range isolated resonances with atypically large matrix elements in the tails of the distributions---but typical samples have no such resonances---and a regime where the number of such resonances increases exponentially with increasing $n$. Another indication of the presence of long-range resonances introduced in Ref.~\cite{morningstar2022avalanches} is obtained comparing the deviations of the {\it smallest} gap between two subsequent eigenvectors from the prediction of Poisson statistics, which reveals a (system-size-dependent) landmark at which the minimum gap begins to typically undergo significant level repulsion. We argue that, in our terminology, these two landmarks identify the crossover region inside the ``glassy'' phase where $\gamma_{\rm ann}>0$ but $\gamma_{\rm q}<0$, in which the probability to observe decorrelation for {\it typical} samples from a {\it typical} initial state $\vert 0 \rangle$ is exponentially small in $n$ for small enough systems---since small samples typically lack of rare system-wide resonances---while the outliers of the probability distributions which are produced by 
rare long-distance resonances becomes more and more relevant upon increasing the system size, eventually leading to delocalization and relaxation in the large $n$ limit.
The weaker hallmark of MBL discussed in Ref.~\cite{morningstar2022avalanches} is the crossover of the mean spectral gap $\avg{r}$ from GOE to Poisson statistics, which in our scenario signals the onset of the regime in which quantum transport is dominated by rare paths.

\subsection{Speculation on the physical implications on the spectral statistics and on the out-of-equilibrium relaxation} \label{sec:speculations}

Here we combine the results on the statistics of the paths discussed above and pictorially illustrated in the phase diagram of Fig.~\ref{fig:Lbetastar} with the recent numerical studies of the model~\eqref{eq:HMBL} of Refs.~\cite{abanin2021distinguishing,roy2021fock,creed2022probability} to speculate on the possible physical implications of our findings both on the spectral statistics and the out-of-equilibrium dynamics.

\begin{figure*}
\includegraphics[width=0.42\textwidth]{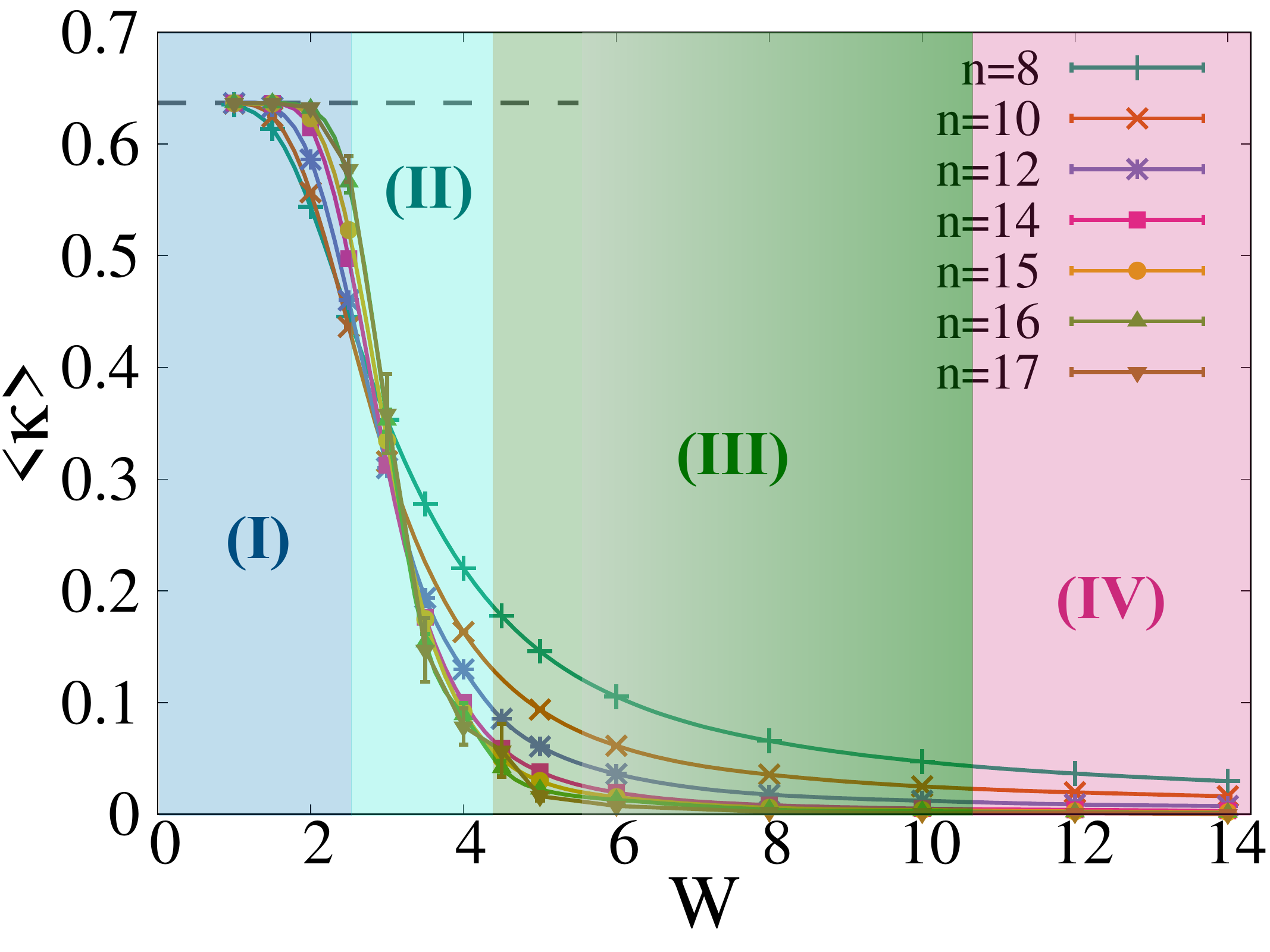} \hspace{0.5cm}
\includegraphics[width=0.42\textwidth]{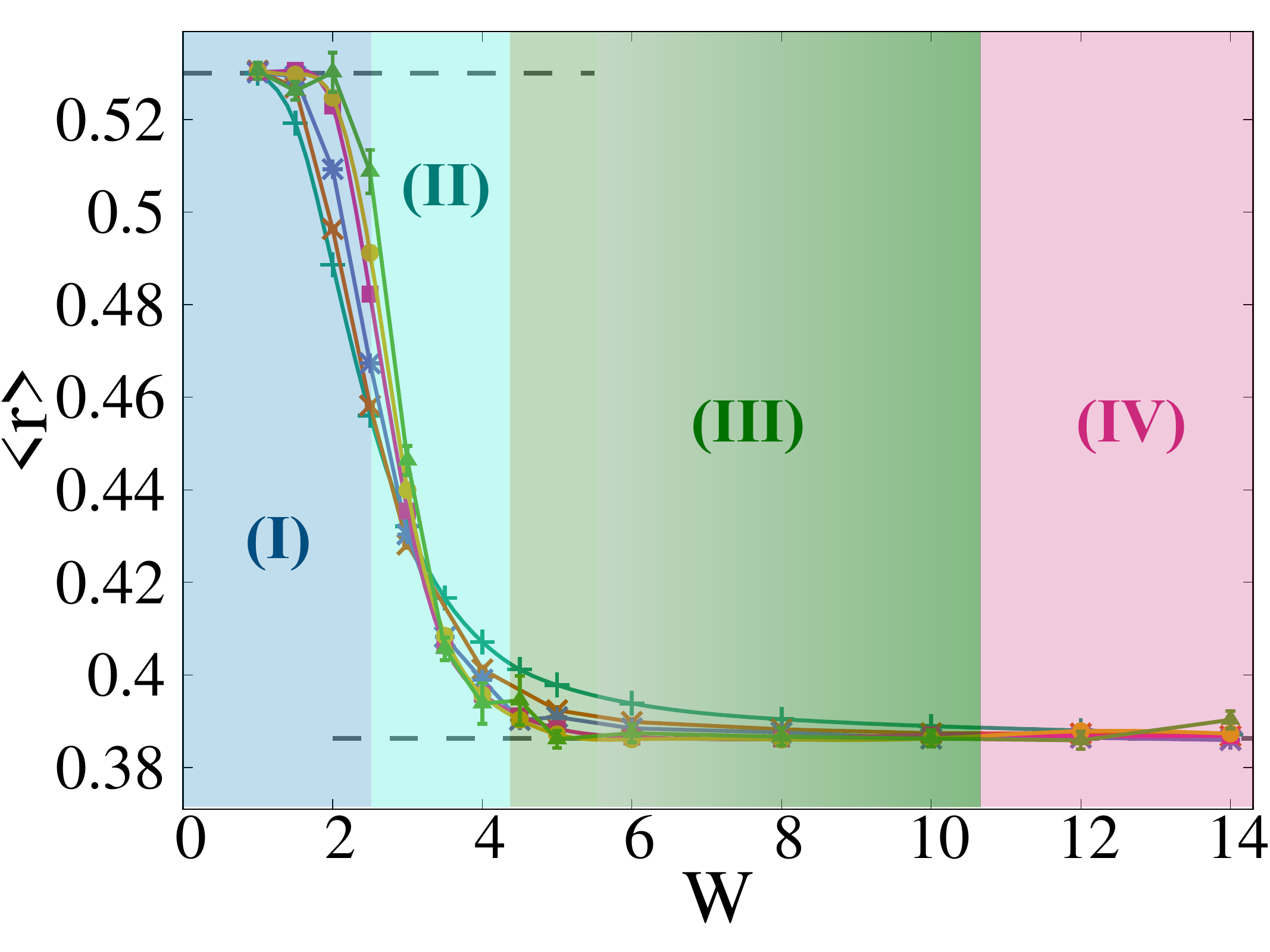} 
\caption{Average overlap between subsequent eigenvectors $\avg{\kappa}$ (left) and average gap ratio $\langle r \rangle$ (right) as a function of $W$ for several system sizes, $n=8, \ldots, 17$ spins. In both panels we highlight with different colors the four different regimes (I)-(IV) discussed in the text.
\label{fig:rstat}}
\end{figure*}

In the fully ergodic region (I) we expect that 
the many-body wave-functions are essentially described by random-matrix theory and satisfy the ETH; In the region (II) 
the eigenfunctions occupy an exponentially large volume which represents an exponentially vanishing fraction of the total accessible Hilbert space (at least for the accessible sizes). Based on the analogy with the RP model, both in the regions (I) and (II) the level statistics should be of Wigner-Dyson type, at least locally.  
Instead in the regime (III) dominated by rare resonances many-body wave-functions are expected to be very ramified on the hypercube, and to exhibit bumps localized in small regions of the Hilbert space (where a resonance is found), separated by large regions where the amplitude is very small~\cite{de2021rare}. 
The support set of two eigenstates close in energy is in general expected to be very different and level repulsion to be weak or absent, implying that the statistics of the gaps should be closer to the Poisson statistics rather than to the GOE one. This is in agreement with the observations of Ref.~\cite{morningstar2022avalanches}, indicating that this regime is characterized by rare strongly repulsive pairs of eigenvectors in an otherwise Poisson-like spectrum (see also Ref.~\cite{garratt2021local}). Therefore, compatibly with the arguments given in Ref.~\cite{morningstar2022avalanches}, it is reasonable to assume that the transition for the level statistics should take place at $W_g$ (which for the largest systems sizes that we have studied numerically is $W_g \sim 4.3$, much before our estimation of the genuine transition to the insulating phase, $W_c \sim 10.5$). This implies that the actual MBL transition is shifted to much larger values of the disorder compared to the one estimated in previous studies~\cite{abanin2021distinguishing,roy2021fock,creed2022probability,tarzia2020many}, due to the fact that the diagnostics based on the statistics of the energy levels 
commonly used in the literature to individuate the onset of MBL do not allow one to distinguish between the ``glassy'' (delocalized) phase from the genuine MBL one, at least for small enough samples~\cite{morningstar2022avalanches}. Similarly, we argue that the diagnostics based on the scaling of the participation entropies and of the entanglement entropy of many-body eigenstates~\cite{luitz2015many,mace2019multifractal} are also ineffective in discriminating between the regimes (III) and (IV), as their support set might be a vanishing small fraction of the Hilbert space. (In fact the configurational entropy $\Sigma/\Sigma_{\rm max}$ provides a lower bound for the fractal dimension of the support set of the many-body wave-functions which, as discussed in Ref.~\cite{de2020multifractality}, yields, in turn, a lower bound for their entanglement entropy.)



To illustrate these ideas, in Fig.~\ref{fig:rstat} we superimpose the four different regimes of the phase diagram discussed above on the plots of the average gap ratio $\langle r \rangle$ and of the average overlap between subsequent eigenvectors $\avg {\kappa}$. Both $\avg{r}$ and $\avg{\kappa}$ are commonly used as diagnostics for the statistics of the energy levels~\cite{abanin2021distinguishing,luitz2015many,oganesyan2007localization}, as they converge to different universal values if the level statistics is described by the GOE ensemble or is of Poisson type. The average gap ratio is defined as~\cite{oganesyan2007localization}:
\[
\avg{r} = \avg{ \frac{\min \{ E_{\alpha + 1} - E_\alpha, E_{\alpha+2} - E_{\alpha + 1} \}}{\max \{ E_{\alpha + 1} - E_\alpha, E_{\alpha+2} - E_{\alpha + 1} \}} }\, ,
\]
where $E_\alpha$ are the eigenvalues of the many-body Hamiltonian~\eqref{eq:HMBL} labeled in ascending order ($E_{\alpha + 1} > E_\alpha$). The average is performed within a small energy window around $E=0$. If the level statistics is described by the GOE ensemble $\langle r \rangle$ converges $\langle r \rangle \simeq 0.53$, while for independent energy levels described by Poisson statistics one has $\langle r \rangle \simeq 2 \ln 2 - 1\simeq 0.386$. Similarly, the overlap between subsequent eigenvectors is defined as:
\[
\avg{\kappa} = \avg{ \sum_{m=1}^{2^n} | \psi_\alpha (m)  | | \psi_{\alpha+1} (m) | } \, ,
\]
where the $\psi_\alpha (i)$'s are the coefficients of the $\alpha$-th eigenvector of the Hamiltonian on the node $m$ of the hypercube (the average is performed over the eigenvectos within a small energy window centered around $E=0$). In the GOE regime the wave-functions amplitudes are i.i.d. Gaussian random variables of zero mean and variance $1/2^n$, hence $\langle \kappa \rangle$ converges to $\langle \kappa \rangle = 2/\pi$. Conversely in the localized phase two successive eigenvector are typically peaked around very distant sites and do not overlap, and therefore $\langle \kappa \rangle \to 0$. 

Fig.~\ref{fig:rstat} shows that $\avg{r}$ and $\avg{\kappa}$ behave in a very similar way and the two panels essentially convey the same information: Upon increasing the strength of the disorder we observe a crossover from the GOE to the Poisson universal values. The crossover becomes sharper as the system size is increased, yet it slowly but systematically drifts to larger values of the disorder as $n$ is increased (see Refs.~\cite{vsuntajs2020quantum,vsuntajs2020ergodicity,abanin2021distinguishing} for a recent discussion about the drift of the crossing point and also Refs.~\cite{tikhonov2016anderson,biroli2018delocalization} for the  analysis of a similar phenomenon on the Anderson model on the RRG). In fact, the crossing point of the curves corresponding to two subsequent sizes moves to the right upon increasing $n$, and roughly coincides with the transition to regime (III), in which transport and decorrelation is dominated by rare paths in the Hilbert space. 
More generally, the figure shows that our approach allows one to rationalize the behavior of $\langle r \rangle$ and $\langle \kappa \rangle$ as a function of $W$ and its dependence on the system size: At small disorder, in the regime (I), the both $\langle r \rangle$ and $\langle \kappa \rangle$ converge to the GOE universal values; 
In the regime (II) strong finite-size corrections are observed, although $\avg{r}$ and $\langle \kappa \rangle$ seem to approach their GOE limiting values upon increasing $n$.  In the phase (III) $\avg{r}$ and $\langle \kappa \rangle$ decrease with $n$ and seem to converge to the Poisson limits, 
albeit our analysis shows that rare system-wide many-body resonances that eventually allows the system to decorrelate from the initial state are present and in the large enough systems. In agreement with Ref.~\cite{morningstar2022avalanches}, we argue that the finite-$n$ crossover in the spectral statistics is insensitive to these rare resonances and signals instead the onset on a pre-thermal regime with apparent localization in small samples~\cite{sierant2020thouless,sels2021dynamical,morningstar2022avalanches,long2022phenomenology}, while the genuine MBL transition is shifted to much larger values of the disorder.

We now turn to the out-of-equilibrium relaxation from a random initial state. In the region (I) the system is fully chaotic and we expect that the relaxation to equilibrium is exponential and fast. Also in the regime (II) we expect the relaxation to equilibrium to be exponential. In fact, a similar phase in which eigenfunctions occupy an exponentially large volume which represents an exponentially vanishing fraction of the total accessible Hilbert space has been reported in the out-of-equilibrium phase diagram of the quantum Random Energy Model~\cite{faoro2019non,baldwin2018quantum,smelyanskiy2019intermittency,parolini2020multifractal,biroli2021out}. In such intermediate regime the decay spin–spin correlation function is found to be a simple exponential~\cite{faoro2019non}, although with a characteristic time that grows in a non-trivial way with the system size.

Conversely, when entering the regime (III) dominated by rare paths we expect that the quantum dynamics and the out-of-equilibrium relaxation become anomalous, highly heterogeneous, and slow~\cite{luitz2017ergodic,agarwal2017rare,biroli2017delocalized,biroli2020anomalous}. Since on finite time scales the dynamical evolution can only explore finite regions of the Hilbert space, with $q$ close to $1$, we expect that the onset of the anomalous transport and relaxation (at least on time scales of order $1$) occurs already at a smaller value of the disorder than $W_g$, namely when $q_g (W)$ starts to deviate from $1$. This is compatible with the recent results of Ref.~\cite{creed2022probability}, which show that the spin-spin correlation function starts to to decay more slowly in time already above $W \sim 3$, with anomalous power-law exponents which reflect the rarefaction of resonances in the Hilbert space.

\section{Summary and outlook} \label{sec:conclusions}

In this paper we put forward a large-deviation approach to investigate the statistics of rare system-wide resonances that may destabilize the MBL phase. The underlying assumption of our work is that the behavior of physical systems exhibiting an MBL phase, including the stability of the latter, is strongly influenced by rare configurations of the disorder~\cite{long2022phenomenology,morningstar2022avalanches,crowley2020avalanche,garratt2021local,villalonga2020eigenstates,khemani2017critical,Ha2023many}. We therefore put forward a sophisticated biased sampling scheme designed to enhance the effect of a special class of configurations that are responsible for the creation of strong long-distance resonances, making accessible otherwise exponentially rare events which are located in the tails of probability distributions of the propagators. More precisely, the basic idea consists in enlarging the parameter space by adding an auxiliary parameter ($\beta$), which  allows us to fine-tune the effect of anomalously large outliers in the far-tails of the probability distribution of the transmission amplitudes. 

On the one hand our results allows one to reconcile the systematic drift of the transition point with the system size reported in many recent works (see \eg~Refs.~\cite{vsuntajs2020quantum,vsuntajs2020ergodicity,sierant2022challenges,sierant2020polynomially,sierant2022slow,leonard2023probing,sels2023thermalization,sels2021markovian,sierant2020thouless,sels2021dynamical,sels2022bath,kiefer2021slow,kiefer2020evidence}) with the existence of a genuine MBL regime at (very) strong disorder in the infinite-size, infinite-time limit~\cite{imbrie2016many}. On the other hand, our analysis suggests that the transition/crossover to the regime dominated by rare resonances should probably be studied as a distinct phenomenon from the MBL phase transition. For example, such a ``glassy'' regime may occur quite generally even in higher dimensions and in systems with longer-range interactions, while the MBL phase transition is suppressed due to the avalanche instability in these situations.

Interestingly enough, the mapping onto directed polymers has also been applied to describe the properties of the strong disorder regime of Anderson localization in two dimensions~\cite{lemarie2019glassy}, revealing some glassy properties such as  pinning, avalanches, and chaos. In this case strong localization confines quantum transport along paths that are pinned by disorder but can change abruptly and suddenly (avalanches) when the energy is varied. The emergence of macroscopic and abrupt jumps of the preferred rare paths when a parameter like the energy is varied has been also discussed for the Anderson model on the Cayley tree~\cite{biroli2020anomalous}. The depinning transition of the polymers through avalanches can be directly related to the singular behavior of the overlap correlation function between eigenstates at different energies which, in turn, is the correlation between the total Landauer transmission at different values of the energy (for a given disorder realization). It would be therefore interesting to investigate whether some signatures of such avalanches and shocks are present also in the many-body problem.

We believe that this work can open the way to other future studies and can be naturally extended and generalized to address several important open problems in the context of the MBL transition. For instance it would be useful to apply our analysis to other models, such as the paradigmatic random-field XXZ Heisenberg chain which has been used as a prototype for the MBL transition (see \eg~Refs.~\cite{vznidarivc2008many,pal2010many,kjall2014many,luitz2015many}), and also varying the (intensive) energy of the initial spin configuration. An important question, already mentioned above, is to understand whether the intermediate regimes (II) and (III) persist in the infinite-size limit or they are just a pre-asymptotic pre-thermal regimes that eventually crossover towards a fully ergodic phase described by random-matrix theory and ETH. Valuable insights might be obtained by benchmarking our approach onto the single-particle Anderson model on the random-regular graph for which such crossover is known to take place and has been investigated in great details~\cite{biroli2012difference,altshuler2016nonergodic,kravtsov2018non,tikhonov2016anderson,Tikhonov_2019,garcia2017scaling,biroli2018delocalization}. In order to make connections with other recent works that focused on the effect of rare long-range resonances~\cite{long2022phenomenology,morningstar2022avalanches,crowley2020avalanche,garratt2021local,villalonga2020eigenstates,khemani2017critical,Ha2023many} in many-body eigenstates, it would also be instructive to establish a precise correspondence between the hallmarks of rare resonances proposed in those works with the the properties and the shape of the rate functions discussed in our paper.

Another set of crucial questions in the current literature concerns the fate of MBL in the presence of a quasiperiodic potential and/or in dimension larger than $1$. In particular, the same techniques and tools described here can be applied to a model of interacting (spinless) fermions in a quasiperiodic potential, similar to the one actually realized in cold-atom experiments~\cite{schreiber2015observation,bordia2017probing,luschen2017observation}. In this case the only source of randomness comes from the choice of the initial configuration. It is therefore useful to compare the properties of the annealed and quenched free-energies found in the quasiperiodic case with the case of uncorrelated random fields, in order to discriminate between the effect of rare resonances created by large segments with anomalously small values of the disorder in real space, and rare resonances due to rare paths with anomalously strong transmission amplitudes in Hilbert space. Along a similar line, it is generally accepted that the MBL phase should be eventually destroyed in $d>1$ in the infinite-size, infinite-time limit by rare thermal regions where the disorder is small~\cite{de2017stability}. It is thus important to check whether this destabilizing mechanism emerges within our approach, \eg~as a strong systematic drift {\it also} of the transition point $W_c$ with the size of the system. 

Another interesting line of research concerns the study of how a localized system reacts when coupled to a thermal region~\cite{morningstar2022avalanches,sels2022bath,Ha2023many,sierant2022slow,leonard2023probing,crowley2020avalanche}. To this aim one could extend the analysis presented above to a $1d$ disordered spin chain coupled to a thermal system at one end (modeled, for instance, by a portion of the system where the disorder is particularly weak), in order to have a clear grasp of the signature of rare ergodic bubbles in real space on the shape and on the properties of the rate functions.

Finally, rare system-wide resonances necessarily exist in large systems. They can be either due to large portions of the system where the disorder is particularly small, or to rare paths in the Hilbert space characterized by anomalously large matrix elements. On general grounds one expects that upon increasing the disorder and approaching the MBL transition, the stronger these resonances are, the rarer they become. Characterizing their statistics and their effect using a standard sampling of the quenched disorder is an extremely difficult task from a computational point of view. A full and complete characterization of rare resonances in the MBL setting possibly can only be achieved through a truly large-deviation approach, in which a biased sampling of the disorder realizations that favors the formation of rare resonances is considered. Complementing our approach with such bias of the sampling is certainly a very promising direction for future investigations.

\section{Acknowledgements}

MT warmly thanks G. Lemarié and I. M. Khaymovich for many stimulating discussions. Some of the simulations were performed at the
  the HPC cluster CARL, located at the University of Oldenburg
  (Germany) and
    funded by the DFG through its Major Research Instrumentation Program
    (INST 184/157-1 FUGG) and the Ministry of
    Science and Culture (MWK) of the
    Lower Saxony State. GB was supported by the Simons Foundation Grant 
No. 454935 (G.B.)
    
\appendix

\section{The Derrida's traveling-wave solution of directed polymers on the Cayley tree} \label{app:DPRM}

As explained in the main text, the problem of DPRM is slightly more complicated than the REM since, although the $\omega_{i_m}$'s are iid random variables, the energies $E_{\cal P}$ are correlated due to the fact that different walks have in general some nodes in common. Nonetheless, an exact solution of the problem in the thermodynamic limit can still be found following the mapping proposed in Ref.~\cite{derrida1988polymers} onto the KPP equation~\cite{kpp1937}.

Thanks to the hierarchical structure of the lattice, the partition function restricted to a branch of the tree originating from a node of the $m$-th generation satisfies the following exact relation in terms of the partition functions restricted to the branches originating from the $k$ neighboring nodes of the next generation:
\[
Z_{i_m} = e^{- \beta \omega^m_{i_m \leftrightarrow i_{m+1}}} \!\!\!\! \sum_{i_{m+1} \in \partial i_m} \! \! Z_{i_{m+1}} \, .
\]
The partition function~\eqref{eq:DPRM} of the DPRM is, by definition, the one computed on the root, $Z_0$. Since $Z_{i_m}$ are random variables, the equations above must be in fact interpreted as a set of complicated integral equations for the probability distributions of the restricted partition functions on each generation:
\begin{equation} \label{eq:PZ}
P_m(Z) = \int \! \! {\rm d} \rho (\omega) \prod_{i=1}^k {\rm d}P_{m+1} (Z_i) \,  \delta \left( Z - e^{-\beta \omega} \sum_i Z_i \right) \, .
\end{equation}
In order to solve the problem Derrida and Spohn introduced an appropriate generating function of $Z_{i_m}$ defined as:
\[
G_m (x) = \int \! \! {\rm d} P_m (Z) \exp \left[ e^{-\beta x} Z \right] \, ,
\]
in terms of which, using Eq.~\eqref{eq:PZ}, one immediately obtains a simpler recursive integral equation:
\[
G_m(x) = \int \! \! {\rm d} \rho (\omega) \left[ G_{m+1} (x + \omega) \right]^k \, ,
\]
which only depends on the temperature via the initial condition on the leaves of the tree: $G_n (x) =  \exp [ e^{-\beta x} ]$. As shown in~\cite{derrida1988polymers}, in the large $n$ limit the solution of the equation above admits a traveling wave solution of the form $G_m(x) = G(x - c m)$, where the ``velocity'' of the front $c$ depends on the initial condition, \ie~on the inverse temperature $\beta$, as $e^{\beta c(\beta)} = k \avg{e^{-\beta \omega}}$. This relation is valid for temperatures larger than a critical value $\beta_\star$ given by the condition that $c(\beta)$ is minimal, \ie~${\rm d} c(\beta) / {\rm d} \beta|_{\beta = \beta_\star} \! = 0$. At lower temperatures, $\beta > \beta_\star$, the front instead moves at the minimal velocity $c (\beta_\star)$. It is possible to show that in fact the velocity of the front yields the free-energy (per unit length) of the DPRM~\cite{derrida1980random}, given in Eq.~\eqref{eq:fqDPRM}. 

The front of the traveling wave $G(x - cm)$ determines also the shape of the full probability distribution of the free-energy and thus of the partition function. In particular Derrida and Spohn have shown that the $q$-th moment $\avg{Z^q}/\avg{Z}^q$ diverges at a $q$-dependent temperature $\beta_q = \beta_\star/\sqrt{q}$~\cite{derrida1988polymers}. This implies that the probability distribution of the partition function has a peak at the typical value $Z_{\rm typ} = e^{- n \beta f}$ (where $f$ is the free-energy given in Eq.~\eqref{eq:fqDPRM}) followed by power-law tails of the form :
\[
P(Z) \simeq \frac{e^{-\beta_\star^2 n f/\beta}}{Z^{1 + \left( \! \frac{\beta_\star}{\beta} \! \right)^2}} \, .
\]
Hence, in the low-temperature phase $P(Z)$ decays with an exponent smaller than $2$ and $\avg{Z}$ is dominated by the extreme values in the tails of the probability distribution. For $\beta < \beta_\star$, instead, the tails of $P(Z)$ decay fast enough and $\avg{Z}$ is dominated by the typical value. As a consequence, in the high-temperature replica-symmetric phase the annealed free-energy, defined in Eq.~\eqref{eq:faDPRM}, converges to the quenched one upon increasing the system size and displays a minimum in the vicinity $\beta_\star$.

\section{Computation of the free-energy for the Anderson model on the Cayley tree with population dynamics} \label{app:PD}

In this appendix we describe the computation of the free-energies~\eqref{eq:phiCT} for the Anderson model on the Cayley tree in the large $n$ limit, already described in Refs.~\cite{biroli2020anomalous,kravtsov2018non}.

Due to the randomness of the on-site potential, the Cavity Green's functions are random variables, and the cavity recursive equations~\eqref{eq:cavity} should be in fact interpreted as an integral recursive equations for their probability distributions on subsequent generations of the tree:
\begin{equation} \label{eq:PG}
P_m (G) = \int \!\! {\rm d}\rho(\epsilon) \prod_{i=1}^k {\rm d}P_{m+1} (G_i) \, \delta \left(  G^{-1} + \epsilon +\sum_{i =1}^k  G_{i}\right) \, .
\end{equation}
This equation can be solved numerically with high precision  using the so-called ``population dynamics'' algorithm~\cite{mezard2001bethe} (also sometimes called the ``pool method''):  The probability distributions are approximated by the empirical distributions of large pools of $\Omega$ elements $G_\alpha^{(m)}$, $P_m(G) \simeq \sum_{\alpha=1}^\Omega \delta (G - G_\alpha^{(m)})$; At each iteration step $k$ instances of $G^{(m+1)}$ are extracted from the sample $P_{m+1}(G)$, and a value of $\epsilon$  is taken from the uniform distribution $\rho(\epsilon)$; A new instance of $G^{(m)}$ is generated using Eq.~(\ref{eq:cavity}) and inserted in a random position of the pool $P_m(G)$ until the whole process converges.

In the computation of the partition function of the associated directed polymer problem, Eq.~\eqref{eq:T}, the cavity Green's functions play the role of the effective random energy landscape seen by the polymer. Once the solutions of Eqs.~\eqref{eq:PG} are found, in order to compute ${\cal T} (\beta)$ we can thus proceed in analogy with Derrida and Spohn, and write the following exact recursion relations between the partition functions restricted to a branch of the tree originating from a node of the $m$-th generation and the partition functions restricted to the branches originating from the $k$ neighboring nodes of the next generation:
\[
{\cal T}_{i_m} (\beta) =  \left| G_{i_m \to i_{m-1}} \right|^\beta \!\! \sum_{i_{m+1} \in \partial i_m} \! \! {\cal T}_{i_{m+1}} (\beta)\, .
\]
Since ${\cal T}_{i_m} (\beta)$ and $G_{i_m \to i_{m-1}}$ are correlated random variables, for each value of the inverse temperature $\beta$ the equation above translates into the following recursive integral equation for their {\it joint} probability distributions on subsequent generations of the tree:
\begin{widetext}
\[
\begin{aligned}
Q_m^{(\beta)} (G,\tau) & = \int \!\! {\rm d}\rho(\epsilon) \prod_{i=1}^k {\rm d} Q_{m+1}^{(\beta)} (G_i,\tau_i) \, \delta \left(  G^{-1} + \epsilon +\sum_{i =1}^k  G_{i}\right) \, \delta \left( \tau - |G|^\beta \sum_{i=1}^k \tau_i \right) \, ,
\end{aligned}
\]
\end{widetext}
with the initial condition:
\[
Q_n^{(\beta)} (G,\tau) = \int \!\! {\rm d}\rho(\epsilon) \, \delta \left(  G^{-1} + \epsilon \right) \, \delta \left (\tau - 1 \right ) \, .
\]
The quenched and annealed free-energies can be finally computed as averages over the probability distributions $Q_{0}^{(\beta)}(G,\tau)$ on the root of the tree:
\[
\begin{aligned}
\phi_{\rm ann} (\beta) &= \frac{1}{\beta n} \ln \left[ \int \!\! {\rm d} G \, {\rm d} \tau \, Q_{0}^{(\beta)} (G,\tau) \, \tau \right] \, ,\\
\phi_{\rm q} (\beta) &= \frac{1}{\beta n} \int \!\! {\rm d} G \, {\rm d}  \tau \, Q_{0}^{(\beta)} (G,\tau) \, \ln \tau \, .
\end{aligned}
\]
These equations can also be solved numerically from the leaves to the root using the pool method described above. Special care should be paid when analyzing the effect of the finiteness of the size pool. In fact, as explained in the main text, the glass transition occurring at $\beta_\star$ is driven by the freezing of the Gibbs' measure on the rare outliers of the probability distributions, and a larger size of the pool allows one to account for these outliers in a more accurate way. We thus need to vary the number of elements of the pool $\Omega$ and carefully extrapolate the results to the $\Omega \to \infty$ limit (see Refs.~\cite{biroli2020anomalous,biroli2012difference}). This approach can be used to study very large trees, up to $n \sim 128$~\cite{biroli2020anomalous}. The outcome of this procedure is shown in Fig.~\ref{fig:phiCTPD} for $W=12$, where the finite-$n$ free-energies are compared to the free-energy computed using the population dynamics algorithm at large $n$. 

\section{Forward scattering approximation on the Cayley tree} \label{app:FSA}

Within the FSA $\log|G_{0,i_n}^{\rm fsa} |$ is a sum of iid random variables with a finite variance, see Eq.~\eqref{eq:FSA}, and its probability distribution can be computed exactly. To this aim it is convenient to introduce the variable $w_i = \log(|\epsilon_i|/t) -\log(W/2 t)$. Its probability distribution is then simply given by $P(w) = e^w \theta(-w)$, with $\langle w \rangle = -1$ and $\sigma_w^2 = 1$. One thus has:
\[
{\cal Y}_n \equiv \frac{\log|G_{0,i_n}^{\rm fsa} |}{n} = -\frac{1}{n} \sum_i w_i -  \log \left ( \frac{W}{2t} \right ) \, .
\]
One immediately finds that $\langle {\cal Y}_n \rangle = - \log (W / 2 t e)$ and thus:
\[
\langle \log|G_{0,i_n}^{\rm fsa} | \rangle = - \frac{n}{\xi_{\rm typ}}, \textrm{~~~~with~~~~} \xi_{\rm typ}^{-1} = 2 \log \left ( \frac{W}{2 t e} \right) \, . 
\]
$\xi_{\rm typ}$ diverges at $W = 2 e t$. However, it is not the localization length, since the latter  must be determined by the decay rate of the maximal amplitude of $|G_{0,i_n}^{\rm fsa}|$ over the full set of $(k+1) k^{n-1}$ sites at distance $n$ from the origin. In order to do this, we compute the probability distribution of the sum ${\cal W}_n = -(1/n) \sum_i w_i$, which can be easily obtained by inverting its characteristic function:
\[
P({\cal W}_n) = \frac{n^n {\cal W}_n^{n-1}}{\Gamma(n)} \, e^{- n {\cal W}_n} \theta({\cal W}_n) \, .
\]
From this analysis we immediately see that the effect of the disorder $W$ on the probability distribution of ${\cal Y}_n$ is trivial and just results in a global shift by a factor $\log(W/2 t)$. Finally, by changing variable to ${\cal W}_n = \log (|G_{0,i_n}| )/n + \log(W/2 t)$, one obtains the probability distribution of the propagators within the FSA approximation:
\begin{widetext}
\[
P(|G_{0,i_n}|) = \frac{1}{\Gamma(n+1)} \left( \frac{2 t n}{W} \right)^n \left( \frac{\log |G_{0,i_n}|}{n} + \log \left( \frac{W}{2 t} \right ) \right)^{n-1} \frac{1}{|G_{0,i_n}|^{3/2}} \, \theta\left( \frac{\log |G_{0,i_n}|}{n} + \log \left( \frac {W}{2 t} \right ) \right) \, .
\]
\end{widetext}

\section{Probability distributions of the propagators} \label{app:PG}

\begin{figure*}
\includegraphics[width=0.43\textwidth]{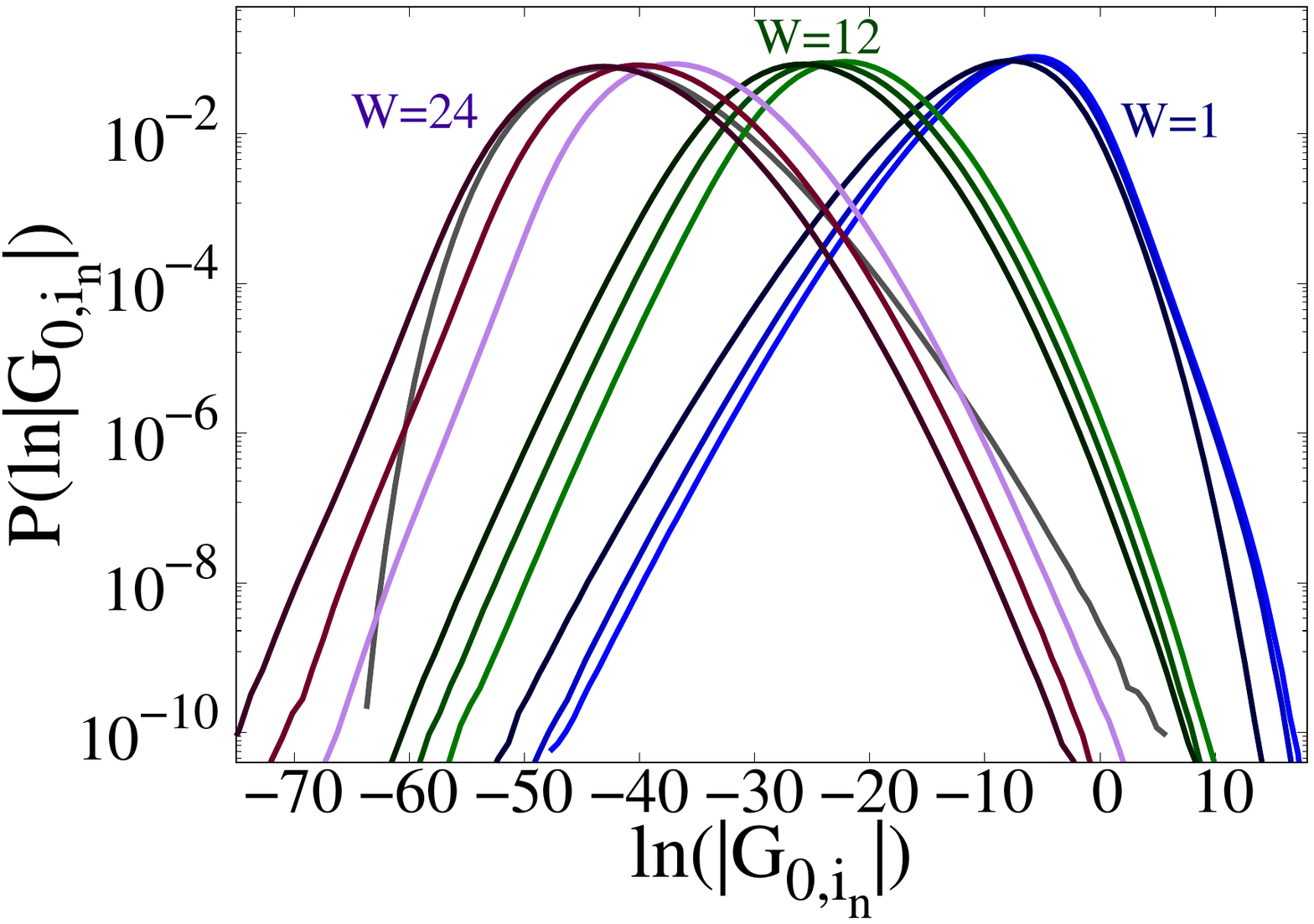} \hspace{0.5cm} \includegraphics[width=0.42\textwidth]{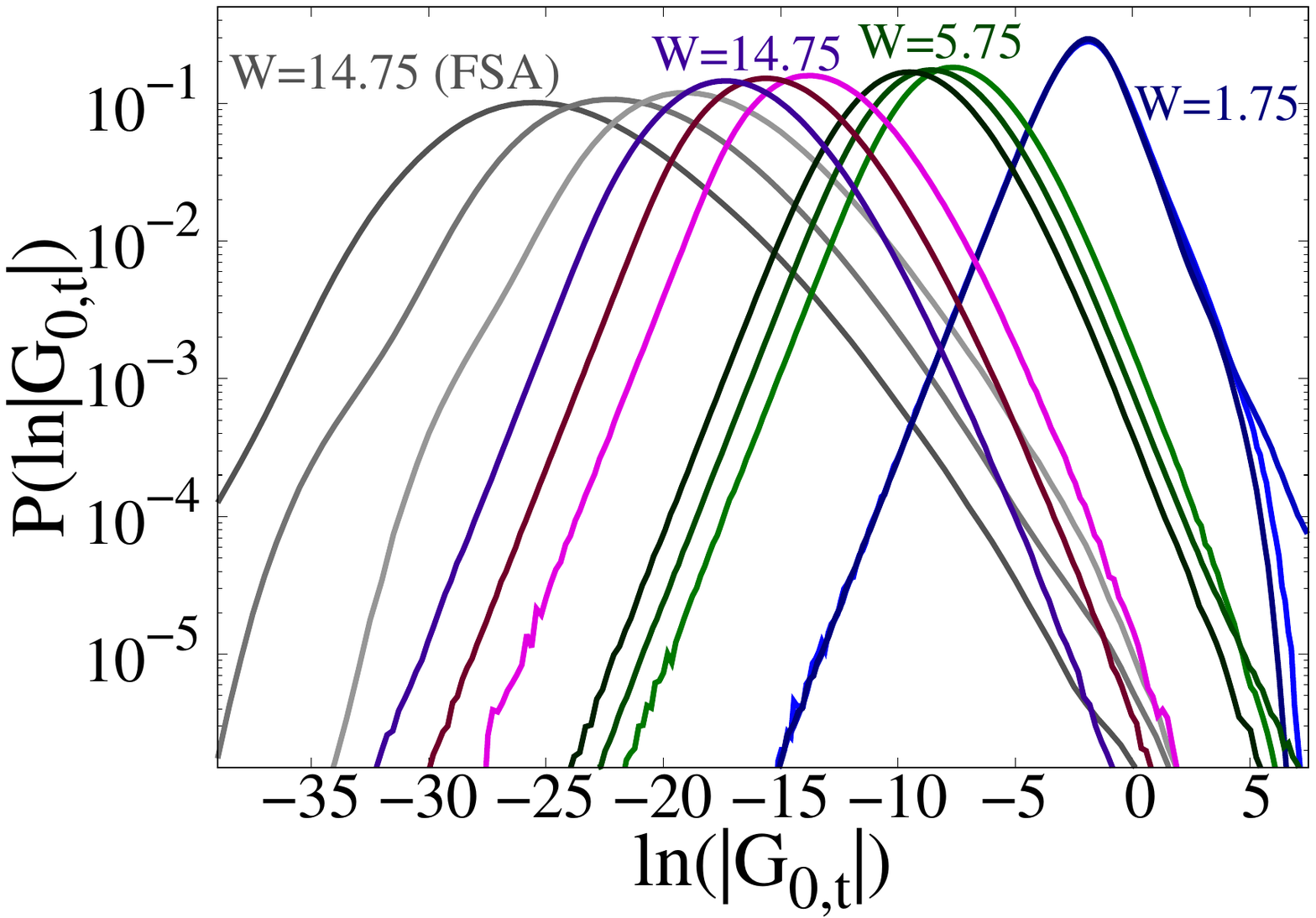} 
\caption{Left: Probability distributions of the transmission amplitudes $|G_{0,i_n}|$ for the Anderson model on the Cayley tree for $W=1$ (blue), $W=12$ (green), and $W=24$ (purple) and for $n=24$, $n=26$, and $n=28$ (from light to dark). The gray curve show the distribution obtained within the FSA for $W=24$ and $n=28$. Right: Probability distributions of the matrix elements $|G_{0,t}|$ for the interacting model~\eqref{eq:HMBL} for $W=1.75$ (blue), $W=5.75$ (green), and $W=14.75$ (purple) and for $n$ ranging from $12$ to $16$ from light to dark. The gray curves show the FSA results for $W=14.75$ and $n=18$, $n=22$, and $n=26$ (from light to dark).
\label{fig:pg0tCT}}
\end{figure*}

The probability distributions of the propagators between the root and the leaves for the Anderson model on the Cayley tree 
are plotted in the left panel  of Fig.~\ref{fig:pg0tCT} for three values of the disorder respectively in the weak disorder regime ($W=1$), in the ``glassy'' regime ($W=12$), and in the localized regime ($W=24$). These probability distributions are characterized by power-law tails $P(|G_{0,i_n}|) \propto |G_{0,i_n}|^{-\mu}$ with an exponent $\mu$ that decreases as the disorder is increased. 
In addition, at weak disorder the typical value of $|G_{0,i_n}|$ is independent of $n$, while at strong disorder it shifts to smaller and smaller values as $n$ is increased. In the figure we also plot the distribution of the transmission amplitudes obtained within the FSA for the largest value of the disorder ($W=24$) and the largest available size ($n=28$). The FSA accounts reasonably well for the exact distribution, although it overestimates the weight of large matrix elements in the tails of the distribution. 

Analogously, we have inspected the probability distributions of the transmission amplitudes $|G_{0,t}|$ for the interacting problem. The probability distributions are plotted in the right panel of Fig.~\ref{fig:pg0tCT} 
for $W=1.75$ (blue), $W=5.75$ (green), and $W=14.75$ (purple) and for $n$ ranging from $12$ to $16$ spins (from light to dark). These probability distributions are qualitatively similar similar to the ones found on the Cayley tree: They are  characterized by power-law tails $P(|G_{0,t}|) \propto |G_{0,t}|^{-\mu}$ with an exponent $\mu$ that seems to depend only very weakly on $W$ and varies from $\mu \approx 2$ at weak disorder to $\mu \approx 1.9$ deep into the MBL phase. However at weak disorder ($W=1.75$) and for $n$ large enough the tails of the distributions exhibit a sharp cut-off at large arguments above which $P(|G_{0,t}|)$ decays fast to zero. The existence of such cut-off is due to the real part of the self-energies that regularize the poles in the perturbative expansion of the Green's function in the metallic regime (see Eq.~\eqref{eq:SAW}), and signal the onset of the ergodic regime. The cut-off occurs at smaller values of $|G_{0,t}|$ when the system size is increased. The other important difference between weak and strong disorder is that in the former case the typical value of $|G_{0,t}|$ seems to be independent of $n$, while in the latter case it shifts to smaller and smaller values as $n$ is increased. 
In the figure we also show the probability distributions obtained within the FSA for the largest value of the disorder ($W=14.75$) and for $n=18$, $n=22$, and $n=26$ (from light gray to dark gray). Similarly to the non-interacting case, the FSA reproduces reasonably well the exact distributions, although it overestimates the number of large transmission amplitudes producing a heavier tail.




\bibliography{references}

\begin{thebibliography}{136}%
\makeatletter
\providecommand \@ifxundefined [1]{%
 \@ifx{#1\undefined}
}%
\providecommand \@ifnum [1]{%
 \ifnum #1\expandafter \@firstoftwo
 \else \expandafter \@secondoftwo
 \fi
}%
\providecommand \@ifx [1]{%
 \ifx #1\expandafter \@firstoftwo
 \else \expandafter \@secondoftwo
 \fi
}%
\providecommand \natexlab [1]{#1}%
\providecommand \enquote  [1]{``#1''}%
\providecommand \bibnamefont  [1]{#1}%
\providecommand \bibfnamefont [1]{#1}%
\providecommand \citenamefont [1]{#1}%
\providecommand \href@noop [0]{\@secondoftwo}%
\providecommand \href [0]{\begingroup \@sanitize@url \@href}%
\providecommand \@href[1]{\@@startlink{#1}\@@href}%
\providecommand \@@href[1]{\endgroup#1\@@endlink}%
\providecommand \@sanitize@url [0]{\catcode `\\12\catcode `\$12\catcode
  `\&12\catcode `\#12\catcode `\^12\catcode `\_12\catcode `\%12\relax}%
\providecommand \@@startlink[1]{}%
\providecommand \@@endlink[0]{}%
\providecommand \url  [0]{\begingroup\@sanitize@url \@url }%
\providecommand \@url [1]{\endgroup\@href {#1}{\urlprefix }}%
\providecommand \urlprefix  [0]{URL }%
\providecommand \Eprint [0]{\href }%
\providecommand \doibase [0]{http://dx.doi.org/}%
\providecommand \selectlanguage [0]{\@gobble}%
\providecommand \bibinfo  [0]{\@secondoftwo}%
\providecommand \bibfield  [0]{\@secondoftwo}%
\providecommand \translation [1]{[#1]}%
\providecommand \BibitemOpen [0]{}%
\providecommand \bibitemStop [0]{}%
\providecommand \bibitemNoStop [0]{.\EOS\space}%
\providecommand \EOS [0]{\spacefactor3000\relax}%
\providecommand \BibitemShut  [1]{\csname bibitem#1\endcsname}%
\let\auto@bib@innerbib\@empty
\bibitem [{\citenamefont {Anderson}(1958)}]{anderson1958absence}%
  \BibitemOpen
  \bibfield  {author} {\bibinfo {author} {\bibfnamefont {P.~W.}\ \bibnamefont
  {Anderson}},\ }\href {\doibase 10.1103/PhysRev.109.1492} {\bibfield
  {journal} {\bibinfo  {journal} {Phys. Rev.}\ }\textbf {\bibinfo {volume}
  {109}},\ \bibinfo {pages} {1492} (\bibinfo {year} {1958})}\BibitemShut
  {NoStop}%
\bibitem [{\citenamefont {Abrahams}\ \emph {et~al.}(1979)\citenamefont
  {Abrahams}, \citenamefont {Anderson}, \citenamefont {Licciardello},\ and\
  \citenamefont {Ramakrishnan}}]{abrahams1979scaling}%
  \BibitemOpen
  \bibfield  {author} {\bibinfo {author} {\bibfnamefont {E.}~\bibnamefont
  {Abrahams}}, \bibinfo {author} {\bibfnamefont {P.}~\bibnamefont {Anderson}},
  \bibinfo {author} {\bibfnamefont {D.}~\bibnamefont {Licciardello}}, \ and\
  \bibinfo {author} {\bibfnamefont {T.}~\bibnamefont {Ramakrishnan}},\
  }\href@noop {} {\bibfield  {journal} {\bibinfo  {journal} {Physical Review
  Letters}\ }\textbf {\bibinfo {volume} {42}},\ \bibinfo {pages} {673}
  (\bibinfo {year} {1979})}\BibitemShut {NoStop}%
\bibitem [{\citenamefont {Altman}\ and\ \citenamefont
  {Vosk}(2015)}]{altman2015universal}%
  \BibitemOpen
  \bibfield  {author} {\bibinfo {author} {\bibfnamefont {E.}~\bibnamefont
  {Altman}}\ and\ \bibinfo {author} {\bibfnamefont {R.}~\bibnamefont {Vosk}},\
  }\href@noop {} {\bibfield  {journal} {\bibinfo  {journal} {Annu. Rev.
  Condens. Matter Phys.}\ }\textbf {\bibinfo {volume} {6}},\ \bibinfo {pages}
  {383} (\bibinfo {year} {2015})}\BibitemShut {NoStop}%
\bibitem [{\citenamefont {Nandkishore}\ and\ \citenamefont
  {Huse}(2015)}]{nandkishore2015many}%
  \BibitemOpen
  \bibfield  {author} {\bibinfo {author} {\bibfnamefont {R.}~\bibnamefont
  {Nandkishore}}\ and\ \bibinfo {author} {\bibfnamefont {D.~A.}\ \bibnamefont
  {Huse}},\ }\href@noop {} {\bibfield  {journal} {\bibinfo  {journal} {Annu.
  Rev. Condens. Matter Phys.}\ }\textbf {\bibinfo {volume} {6}},\ \bibinfo
  {pages} {15} (\bibinfo {year} {2015})}\BibitemShut {NoStop}%
\bibitem [{\citenamefont {Abanin}\ and\ \citenamefont
  {Papi{\'c}}(2017)}]{abanin2017recent}%
  \BibitemOpen
  \bibfield  {author} {\bibinfo {author} {\bibfnamefont {D.~A.}\ \bibnamefont
  {Abanin}}\ and\ \bibinfo {author} {\bibfnamefont {Z.}~\bibnamefont
  {Papi{\'c}}},\ }\href@noop {} {\bibfield  {journal} {\bibinfo  {journal}
  {Annalen der Physik}\ }\textbf {\bibinfo {volume} {529}},\ \bibinfo {pages}
  {1700169} (\bibinfo {year} {2017})}\BibitemShut {NoStop}%
\bibitem [{\citenamefont {Alet}\ and\ \citenamefont
  {Laflorencie}(2018)}]{alet2018many}%
  \BibitemOpen
  \bibfield  {author} {\bibinfo {author} {\bibfnamefont {F.}~\bibnamefont
  {Alet}}\ and\ \bibinfo {author} {\bibfnamefont {N.}~\bibnamefont
  {Laflorencie}},\ }\href@noop {} {\bibfield  {journal} {\bibinfo  {journal}
  {Comptes Rendus Physique}\ }\textbf {\bibinfo {volume} {19}},\ \bibinfo
  {pages} {498} (\bibinfo {year} {2018})}\BibitemShut {NoStop}%
\bibitem [{\citenamefont {Abanin}\ \emph {et~al.}(2019)\citenamefont {Abanin},
  \citenamefont {Altman}, \citenamefont {Bloch},\ and\ \citenamefont
  {Serbyn}}]{abanin2019colloquium}%
  \BibitemOpen
  \bibfield  {author} {\bibinfo {author} {\bibfnamefont {D.~A.}\ \bibnamefont
  {Abanin}}, \bibinfo {author} {\bibfnamefont {E.}~\bibnamefont {Altman}},
  \bibinfo {author} {\bibfnamefont {I.}~\bibnamefont {Bloch}}, \ and\ \bibinfo
  {author} {\bibfnamefont {M.}~\bibnamefont {Serbyn}},\ }\href@noop {}
  {\bibfield  {journal} {\bibinfo  {journal} {Reviews of Modern Physics}\
  }\textbf {\bibinfo {volume} {91}},\ \bibinfo {pages} {021001} (\bibinfo
  {year} {2019})}\BibitemShut {NoStop}%
\bibitem [{\citenamefont {Basko}\ \emph {et~al.}(2006)\citenamefont {Basko},
  \citenamefont {Aleiner},\ and\ \citenamefont {Altshuler}}]{basko2006metal}%
  \BibitemOpen
  \bibfield  {author} {\bibinfo {author} {\bibfnamefont {D.~M.}\ \bibnamefont
  {Basko}}, \bibinfo {author} {\bibfnamefont {I.~L.}\ \bibnamefont {Aleiner}},
  \ and\ \bibinfo {author} {\bibfnamefont {B.~L.}\ \bibnamefont {Altshuler}},\
  }\href {\doibase 10.1016/j.aop.2005.11.014} {\bibfield  {journal} {\bibinfo
  {journal} {Ann. Phys.}\ }\textbf {\bibinfo {volume} {321}},\ \bibinfo {pages}
  {1126} (\bibinfo {year} {2006})}\BibitemShut {NoStop}%
\bibitem [{\citenamefont {Gornyi}\ \emph {et~al.}(2005)\citenamefont {Gornyi},
  \citenamefont {Mirlin},\ and\ \citenamefont
  {Polyakov}}]{gornyi2005interacting}%
  \BibitemOpen
  \bibfield  {author} {\bibinfo {author} {\bibfnamefont {I.~V.}\ \bibnamefont
  {Gornyi}}, \bibinfo {author} {\bibfnamefont {A.~D.}\ \bibnamefont {Mirlin}},
  \ and\ \bibinfo {author} {\bibfnamefont {D.~G.}\ \bibnamefont {Polyakov}},\
  }\href@noop {} {\bibfield  {journal} {\bibinfo  {journal} {Physical review
  letters}\ }\textbf {\bibinfo {volume} {95}},\ \bibinfo {pages} {206603}
  (\bibinfo {year} {2005})}\BibitemShut {NoStop}%
\bibitem [{\citenamefont
  {Imbrie}(2016{\natexlab{a}})}]{imbrie2016diagonalization}%
  \BibitemOpen
  \bibfield  {author} {\bibinfo {author} {\bibfnamefont {J.~Z.}\ \bibnamefont
  {Imbrie}},\ }\href@noop {} {\bibfield  {journal} {\bibinfo  {journal}
  {Physical review letters}\ }\textbf {\bibinfo {volume} {117}},\ \bibinfo
  {pages} {027201} (\bibinfo {year} {2016}{\natexlab{a}})}\BibitemShut
  {NoStop}%
\bibitem [{\citenamefont {Imbrie}\ \emph {et~al.}(2017)\citenamefont {Imbrie},
  \citenamefont {Ros},\ and\ \citenamefont {Scardicchio}}]{imbrie2017local}%
  \BibitemOpen
  \bibfield  {author} {\bibinfo {author} {\bibfnamefont {J.~Z.}\ \bibnamefont
  {Imbrie}}, \bibinfo {author} {\bibfnamefont {V.}~\bibnamefont {Ros}}, \ and\
  \bibinfo {author} {\bibfnamefont {A.}~\bibnamefont {Scardicchio}},\
  }\href@noop {} {\bibfield  {journal} {\bibinfo  {journal} {Annalen der
  Physik}\ }\textbf {\bibinfo {volume} {529}},\ \bibinfo {pages} {1600278}
  (\bibinfo {year} {2017})}\BibitemShut {NoStop}%
\bibitem [{\citenamefont {Serbyn}\ \emph {et~al.}(2014)\citenamefont {Serbyn},
  \citenamefont {Papi{\'c}},\ and\ \citenamefont {Abanin}}]{serbyn2014quantum}%
  \BibitemOpen
  \bibfield  {author} {\bibinfo {author} {\bibfnamefont {M.}~\bibnamefont
  {Serbyn}}, \bibinfo {author} {\bibfnamefont {Z.}~\bibnamefont {Papi{\'c}}}, \
  and\ \bibinfo {author} {\bibfnamefont {D.~A.}\ \bibnamefont {Abanin}},\
  }\href@noop {} {\bibfield  {journal} {\bibinfo  {journal} {Physical Review
  B}\ }\textbf {\bibinfo {volume} {90}},\ \bibinfo {pages} {174302} (\bibinfo
  {year} {2014})}\BibitemShut {NoStop}%
\bibitem [{\citenamefont {Serbyn}\ \emph {et~al.}(2013)\citenamefont {Serbyn},
  \citenamefont {Papi{\'c}},\ and\ \citenamefont {Abanin}}]{serbyn2013local}%
  \BibitemOpen
  \bibfield  {author} {\bibinfo {author} {\bibfnamefont {M.}~\bibnamefont
  {Serbyn}}, \bibinfo {author} {\bibfnamefont {Z.}~\bibnamefont {Papi{\'c}}}, \
  and\ \bibinfo {author} {\bibfnamefont {D.~A.}\ \bibnamefont {Abanin}},\
  }\href@noop {} {\bibfield  {journal} {\bibinfo  {journal} {Physical review
  letters}\ }\textbf {\bibinfo {volume} {111}},\ \bibinfo {pages} {127201}
  (\bibinfo {year} {2013})}\BibitemShut {NoStop}%
\bibitem [{\citenamefont {Ros}\ \emph {et~al.}(2015)\citenamefont {Ros},
  \citenamefont {M{\"u}ller},\ and\ \citenamefont
  {Scardicchio}}]{ros2015integrals}%
  \BibitemOpen
  \bibfield  {author} {\bibinfo {author} {\bibfnamefont {V.}~\bibnamefont
  {Ros}}, \bibinfo {author} {\bibfnamefont {M.}~\bibnamefont {M{\"u}ller}}, \
  and\ \bibinfo {author} {\bibfnamefont {A.}~\bibnamefont {Scardicchio}},\
  }\href@noop {} {\bibfield  {journal} {\bibinfo  {journal} {Nuclear Physics
  B}\ }\textbf {\bibinfo {volume} {891}},\ \bibinfo {pages} {420} (\bibinfo
  {year} {2015})}\BibitemShut {NoStop}%
\bibitem [{\citenamefont {Berthier}\ and\ \citenamefont
  {Biroli}(2011)}]{berthier2011theoretical}%
  \BibitemOpen
  \bibfield  {author} {\bibinfo {author} {\bibfnamefont {L.}~\bibnamefont
  {Berthier}}\ and\ \bibinfo {author} {\bibfnamefont {G.}~\bibnamefont
  {Biroli}},\ }\href@noop {} {\bibfield  {journal} {\bibinfo  {journal}
  {Reviews of modern physics}\ }\textbf {\bibinfo {volume} {83}},\ \bibinfo
  {pages} {587} (\bibinfo {year} {2011})}\BibitemShut {NoStop}%
\bibitem [{\citenamefont {Gornyi}\ \emph {et~al.}(2017)\citenamefont {Gornyi},
  \citenamefont {Mirlin}, \citenamefont {Polyakov},\ and\ \citenamefont
  {Burin}}]{gornyi2017spectral}%
  \BibitemOpen
  \bibfield  {author} {\bibinfo {author} {\bibfnamefont {I.}~\bibnamefont
  {Gornyi}}, \bibinfo {author} {\bibfnamefont {A.}~\bibnamefont {Mirlin}},
  \bibinfo {author} {\bibfnamefont {D.}~\bibnamefont {Polyakov}}, \ and\
  \bibinfo {author} {\bibfnamefont {A.}~\bibnamefont {Burin}},\ }\href
  {\doibase 10.1002/andp.201600360} {\bibfield  {journal} {\bibinfo  {journal}
  {Ann. Phys.}\ }\textbf {\bibinfo {volume} {529}},\ \bibinfo {pages} {1600360}
  (\bibinfo {year} {2017})}\BibitemShut {NoStop}%
\bibitem [{\citenamefont {Vosk}\ \emph {et~al.}(2015)\citenamefont {Vosk},
  \citenamefont {Huse},\ and\ \citenamefont {Altman}}]{vosk2015theory}%
  \BibitemOpen
  \bibfield  {author} {\bibinfo {author} {\bibfnamefont {R.}~\bibnamefont
  {Vosk}}, \bibinfo {author} {\bibfnamefont {D.~A.}\ \bibnamefont {Huse}}, \
  and\ \bibinfo {author} {\bibfnamefont {E.}~\bibnamefont {Altman}},\
  }\href@noop {} {\bibfield  {journal} {\bibinfo  {journal} {Physical Review
  X}\ }\textbf {\bibinfo {volume} {5}},\ \bibinfo {pages} {031032} (\bibinfo
  {year} {2015})}\BibitemShut {NoStop}%
\bibitem [{\citenamefont {Potter}\ \emph {et~al.}(2015)\citenamefont {Potter},
  \citenamefont {Vasseur},\ and\ \citenamefont
  {Parameswaran}}]{potter2015universal}%
  \BibitemOpen
  \bibfield  {author} {\bibinfo {author} {\bibfnamefont {A.~C.}\ \bibnamefont
  {Potter}}, \bibinfo {author} {\bibfnamefont {R.}~\bibnamefont {Vasseur}}, \
  and\ \bibinfo {author} {\bibfnamefont {S.}~\bibnamefont {Parameswaran}},\
  }\href@noop {} {\bibfield  {journal} {\bibinfo  {journal} {Physical Review
  X}\ }\textbf {\bibinfo {volume} {5}},\ \bibinfo {pages} {031033} (\bibinfo
  {year} {2015})}\BibitemShut {NoStop}%
\bibitem [{\citenamefont {Dumitrescu}\ \emph {et~al.}(2017)\citenamefont
  {Dumitrescu}, \citenamefont {Vasseur},\ and\ \citenamefont
  {Potter}}]{dumitrescu2017scaling}%
  \BibitemOpen
  \bibfield  {author} {\bibinfo {author} {\bibfnamefont {P.~T.}\ \bibnamefont
  {Dumitrescu}}, \bibinfo {author} {\bibfnamefont {R.}~\bibnamefont {Vasseur}},
  \ and\ \bibinfo {author} {\bibfnamefont {A.~C.}\ \bibnamefont {Potter}},\
  }\href@noop {} {\bibfield  {journal} {\bibinfo  {journal} {Physical review
  letters}\ }\textbf {\bibinfo {volume} {119}},\ \bibinfo {pages} {110604}
  (\bibinfo {year} {2017})}\BibitemShut {NoStop}%
\bibitem [{\citenamefont {Thiery}\ \emph {et~al.}(2017)\citenamefont {Thiery},
  \citenamefont {M{\"u}ller},\ and\ \citenamefont
  {De~Roeck}}]{thiery2017microscopically}%
  \BibitemOpen
  \bibfield  {author} {\bibinfo {author} {\bibfnamefont {T.}~\bibnamefont
  {Thiery}}, \bibinfo {author} {\bibfnamefont {M.}~\bibnamefont {M{\"u}ller}},
  \ and\ \bibinfo {author} {\bibfnamefont {W.}~\bibnamefont {De~Roeck}},\
  }\href@noop {} {\bibfield  {journal} {\bibinfo  {journal} {arXiv preprint
  arXiv:1711.09880}\ } (\bibinfo {year} {2017})}\BibitemShut {NoStop}%
\bibitem [{\citenamefont {Goremykina}\ \emph {et~al.}(2019)\citenamefont
  {Goremykina}, \citenamefont {Vasseur},\ and\ \citenamefont
  {Serbyn}}]{goremykina2019analytically}%
  \BibitemOpen
  \bibfield  {author} {\bibinfo {author} {\bibfnamefont {A.}~\bibnamefont
  {Goremykina}}, \bibinfo {author} {\bibfnamefont {R.}~\bibnamefont {Vasseur}},
  \ and\ \bibinfo {author} {\bibfnamefont {M.}~\bibnamefont {Serbyn}},\
  }\href@noop {} {\bibfield  {journal} {\bibinfo  {journal} {Physical review
  letters}\ }\textbf {\bibinfo {volume} {122}},\ \bibinfo {pages} {040601}
  (\bibinfo {year} {2019})}\BibitemShut {NoStop}%
\bibitem [{\citenamefont {Dumitrescu}\ \emph {et~al.}(2019)\citenamefont
  {Dumitrescu}, \citenamefont {Goremykina}, \citenamefont {Parameswaran},
  \citenamefont {Serbyn},\ and\ \citenamefont
  {Vasseur}}]{dumitrescu2019kosterlitz}%
  \BibitemOpen
  \bibfield  {author} {\bibinfo {author} {\bibfnamefont {P.~T.}\ \bibnamefont
  {Dumitrescu}}, \bibinfo {author} {\bibfnamefont {A.}~\bibnamefont
  {Goremykina}}, \bibinfo {author} {\bibfnamefont {S.~A.}\ \bibnamefont
  {Parameswaran}}, \bibinfo {author} {\bibfnamefont {M.}~\bibnamefont
  {Serbyn}}, \ and\ \bibinfo {author} {\bibfnamefont {R.}~\bibnamefont
  {Vasseur}},\ }\href@noop {} {\bibfield  {journal} {\bibinfo  {journal}
  {Physical Review B}\ }\textbf {\bibinfo {volume} {99}},\ \bibinfo {pages}
  {094205} (\bibinfo {year} {2019})}\BibitemShut {NoStop}%
\bibitem [{\citenamefont {Morningstar}\ and\ \citenamefont
  {Huse}(2019)}]{morningstar2019renormalization}%
  \BibitemOpen
  \bibfield  {author} {\bibinfo {author} {\bibfnamefont {A.}~\bibnamefont
  {Morningstar}}\ and\ \bibinfo {author} {\bibfnamefont {D.~A.}\ \bibnamefont
  {Huse}},\ }\href@noop {} {\bibfield  {journal} {\bibinfo  {journal} {Physical
  Review B}\ }\textbf {\bibinfo {volume} {99}},\ \bibinfo {pages} {224205}
  (\bibinfo {year} {2019})}\BibitemShut {NoStop}%
\bibitem [{\citenamefont {Morningstar}\ \emph {et~al.}(2020)\citenamefont
  {Morningstar}, \citenamefont {Huse},\ and\ \citenamefont
  {Imbrie}}]{morningstar2020many}%
  \BibitemOpen
  \bibfield  {author} {\bibinfo {author} {\bibfnamefont {A.}~\bibnamefont
  {Morningstar}}, \bibinfo {author} {\bibfnamefont {D.~A.}\ \bibnamefont
  {Huse}}, \ and\ \bibinfo {author} {\bibfnamefont {J.~Z.}\ \bibnamefont
  {Imbrie}},\ }\href@noop {} {\bibfield  {journal} {\bibinfo  {journal}
  {Physical Review B}\ }\textbf {\bibinfo {volume} {102}},\ \bibinfo {pages}
  {125134} (\bibinfo {year} {2020})}\BibitemShut {NoStop}%
\bibitem [{\citenamefont {{\v{Z}}nidari{\v{c}}}\ \emph
  {et~al.}(2008)\citenamefont {{\v{Z}}nidari{\v{c}}}, \citenamefont {Prosen},\
  and\ \citenamefont {Prelov{\v{s}}ek}}]{vznidarivc2008many}%
  \BibitemOpen
  \bibfield  {author} {\bibinfo {author} {\bibfnamefont {M.}~\bibnamefont
  {{\v{Z}}nidari{\v{c}}}}, \bibinfo {author} {\bibfnamefont {T.}~\bibnamefont
  {Prosen}}, \ and\ \bibinfo {author} {\bibfnamefont {P.}~\bibnamefont
  {Prelov{\v{s}}ek}},\ }\href@noop {} {\bibfield  {journal} {\bibinfo
  {journal} {Physical Review B}\ }\textbf {\bibinfo {volume} {77}},\ \bibinfo
  {pages} {064426} (\bibinfo {year} {2008})}\BibitemShut {NoStop}%
\bibitem [{\citenamefont {Pal}\ and\ \citenamefont {Huse}(2010)}]{pal2010many}%
  \BibitemOpen
  \bibfield  {author} {\bibinfo {author} {\bibfnamefont {A.}~\bibnamefont
  {Pal}}\ and\ \bibinfo {author} {\bibfnamefont {D.~A.}\ \bibnamefont {Huse}},\
  }\href@noop {} {\bibfield  {journal} {\bibinfo  {journal} {Physical review
  B}\ }\textbf {\bibinfo {volume} {82}},\ \bibinfo {pages} {174411} (\bibinfo
  {year} {2010})}\BibitemShut {NoStop}%
\bibitem [{\citenamefont {Kj{\"a}ll}\ \emph {et~al.}(2014)\citenamefont
  {Kj{\"a}ll}, \citenamefont {Bardarson},\ and\ \citenamefont
  {Pollmann}}]{kjall2014many}%
  \BibitemOpen
  \bibfield  {author} {\bibinfo {author} {\bibfnamefont {J.~A.}\ \bibnamefont
  {Kj{\"a}ll}}, \bibinfo {author} {\bibfnamefont {J.~H.}\ \bibnamefont
  {Bardarson}}, \ and\ \bibinfo {author} {\bibfnamefont {F.}~\bibnamefont
  {Pollmann}},\ }\href@noop {} {\bibfield  {journal} {\bibinfo  {journal}
  {Physical review letters}\ }\textbf {\bibinfo {volume} {113}},\ \bibinfo
  {pages} {107204} (\bibinfo {year} {2014})}\BibitemShut {NoStop}%
\bibitem [{\citenamefont {Luitz}\ \emph {et~al.}(2015)\citenamefont {Luitz},
  \citenamefont {Laflorencie},\ and\ \citenamefont {Alet}}]{luitz2015many}%
  \BibitemOpen
  \bibfield  {author} {\bibinfo {author} {\bibfnamefont {D.~J.}\ \bibnamefont
  {Luitz}}, \bibinfo {author} {\bibfnamefont {N.}~\bibnamefont {Laflorencie}},
  \ and\ \bibinfo {author} {\bibfnamefont {F.}~\bibnamefont {Alet}},\
  }\href@noop {} {\bibfield  {journal} {\bibinfo  {journal} {Physical Review
  B}\ }\textbf {\bibinfo {volume} {91}},\ \bibinfo {pages} {081103} (\bibinfo
  {year} {2015})}\BibitemShut {NoStop}%
\bibitem [{\citenamefont {{\v{S}}untajs}\ \emph
  {et~al.}(2020{\natexlab{a}})\citenamefont {{\v{S}}untajs}, \citenamefont
  {Bon{\v{c}}a}, \citenamefont {Prosen},\ and\ \citenamefont
  {Vidmar}}]{vsuntajs2020quantum}%
  \BibitemOpen
  \bibfield  {author} {\bibinfo {author} {\bibfnamefont {J.}~\bibnamefont
  {{\v{S}}untajs}}, \bibinfo {author} {\bibfnamefont {J.}~\bibnamefont
  {Bon{\v{c}}a}}, \bibinfo {author} {\bibfnamefont {T.}~\bibnamefont {Prosen}},
  \ and\ \bibinfo {author} {\bibfnamefont {L.}~\bibnamefont {Vidmar}},\
  }\href@noop {} {\bibfield  {journal} {\bibinfo  {journal} {Physical Review
  E}\ }\textbf {\bibinfo {volume} {102}},\ \bibinfo {pages} {062144} (\bibinfo
  {year} {2020}{\natexlab{a}})}\BibitemShut {NoStop}%
\bibitem [{\citenamefont {{\v{S}}untajs}\ \emph
  {et~al.}(2020{\natexlab{b}})\citenamefont {{\v{S}}untajs}, \citenamefont
  {Bon{\v{c}}a}, \citenamefont {Prosen},\ and\ \citenamefont
  {Vidmar}}]{vsuntajs2020ergodicity}%
  \BibitemOpen
  \bibfield  {author} {\bibinfo {author} {\bibfnamefont {J.}~\bibnamefont
  {{\v{S}}untajs}}, \bibinfo {author} {\bibfnamefont {J.}~\bibnamefont
  {Bon{\v{c}}a}}, \bibinfo {author} {\bibfnamefont {T.}~\bibnamefont {Prosen}},
  \ and\ \bibinfo {author} {\bibfnamefont {L.}~\bibnamefont {Vidmar}},\
  }\href@noop {} {\bibfield  {journal} {\bibinfo  {journal} {Physical Review
  B}\ }\textbf {\bibinfo {volume} {102}},\ \bibinfo {pages} {064207} (\bibinfo
  {year} {2020}{\natexlab{b}})}\BibitemShut {NoStop}%
\bibitem [{\citenamefont {Sierant}\ and\ \citenamefont
  {Zakrzewski}(2022)}]{sierant2022challenges}%
  \BibitemOpen
  \bibfield  {author} {\bibinfo {author} {\bibfnamefont {P.}~\bibnamefont
  {Sierant}}\ and\ \bibinfo {author} {\bibfnamefont {J.}~\bibnamefont
  {Zakrzewski}},\ }\href@noop {} {\bibfield  {journal} {\bibinfo  {journal}
  {Physical Review B}\ }\textbf {\bibinfo {volume} {105}},\ \bibinfo {pages}
  {224203} (\bibinfo {year} {2022})}\BibitemShut {NoStop}%
\bibitem [{\citenamefont {Sierant}\ \emph
  {et~al.}(2020{\natexlab{a}})\citenamefont {Sierant}, \citenamefont
  {Lewenstein},\ and\ \citenamefont {Zakrzewski}}]{sierant2020polynomially}%
  \BibitemOpen
  \bibfield  {author} {\bibinfo {author} {\bibfnamefont {P.}~\bibnamefont
  {Sierant}}, \bibinfo {author} {\bibfnamefont {M.}~\bibnamefont {Lewenstein}},
  \ and\ \bibinfo {author} {\bibfnamefont {J.}~\bibnamefont {Zakrzewski}},\
  }\href@noop {} {\bibfield  {journal} {\bibinfo  {journal} {Physical Review
  Letters}\ }\textbf {\bibinfo {volume} {125}},\ \bibinfo {pages} {156601}
  (\bibinfo {year} {2020}{\natexlab{a}})}\BibitemShut {NoStop}%
\bibitem [{\citenamefont {Sierant}\ \emph {et~al.}(2022)\citenamefont
  {Sierant}, \citenamefont {Chanda}, \citenamefont {Lewenstein},\ and\
  \citenamefont {Zakrzewski}}]{sierant2022slow}%
  \BibitemOpen
  \bibfield  {author} {\bibinfo {author} {\bibfnamefont {P.}~\bibnamefont
  {Sierant}}, \bibinfo {author} {\bibfnamefont {T.}~\bibnamefont {Chanda}},
  \bibinfo {author} {\bibfnamefont {M.}~\bibnamefont {Lewenstein}}, \ and\
  \bibinfo {author} {\bibfnamefont {J.}~\bibnamefont {Zakrzewski}},\
  }\href@noop {} {\bibfield  {journal} {\bibinfo  {journal} {arXiv preprint
  arXiv:2212.07107}\ } (\bibinfo {year} {2022})}\BibitemShut {NoStop}%
\bibitem [{\citenamefont {Sels}\ and\ \citenamefont
  {Polkovnikov}(2023)}]{sels2023thermalization}%
  \BibitemOpen
  \bibfield  {author} {\bibinfo {author} {\bibfnamefont {D.}~\bibnamefont
  {Sels}}\ and\ \bibinfo {author} {\bibfnamefont {A.}~\bibnamefont
  {Polkovnikov}},\ }\href@noop {} {\bibfield  {journal} {\bibinfo  {journal}
  {Physical Review X}\ }\textbf {\bibinfo {volume} {13}},\ \bibinfo {pages}
  {011041} (\bibinfo {year} {2023})}\BibitemShut {NoStop}%
\bibitem [{\citenamefont {Sels}(2021)}]{sels2021markovian}%
  \BibitemOpen
  \bibfield  {author} {\bibinfo {author} {\bibfnamefont {D.}~\bibnamefont
  {Sels}},\ }\href@noop {} {\bibfield  {journal} {\bibinfo  {journal} {arXiv
  preprint arXiv:2108.10796}\ } (\bibinfo {year} {2021})}\BibitemShut {NoStop}%
\bibitem [{\citenamefont {Sierant}\ \emph
  {et~al.}(2020{\natexlab{b}})\citenamefont {Sierant}, \citenamefont
  {Delande},\ and\ \citenamefont {Zakrzewski}}]{sierant2020thouless}%
  \BibitemOpen
  \bibfield  {author} {\bibinfo {author} {\bibfnamefont {P.}~\bibnamefont
  {Sierant}}, \bibinfo {author} {\bibfnamefont {D.}~\bibnamefont {Delande}}, \
  and\ \bibinfo {author} {\bibfnamefont {J.}~\bibnamefont {Zakrzewski}},\
  }\href@noop {} {\bibfield  {journal} {\bibinfo  {journal} {Physical Review
  Letters}\ }\textbf {\bibinfo {volume} {124}},\ \bibinfo {pages} {186601}
  (\bibinfo {year} {2020}{\natexlab{b}})}\BibitemShut {NoStop}%
\bibitem [{\citenamefont {Sels}\ and\ \citenamefont
  {Polkovnikov}(2021)}]{sels2021dynamical}%
  \BibitemOpen
  \bibfield  {author} {\bibinfo {author} {\bibfnamefont {D.}~\bibnamefont
  {Sels}}\ and\ \bibinfo {author} {\bibfnamefont {A.}~\bibnamefont
  {Polkovnikov}},\ }\href@noop {} {\bibfield  {journal} {\bibinfo  {journal}
  {Physical Review E}\ }\textbf {\bibinfo {volume} {104}},\ \bibinfo {pages}
  {054105} (\bibinfo {year} {2021})}\BibitemShut {NoStop}%
\bibitem [{\citenamefont {Sels}(2022)}]{sels2022bath}%
  \BibitemOpen
  \bibfield  {author} {\bibinfo {author} {\bibfnamefont {D.}~\bibnamefont
  {Sels}},\ }\href@noop {} {\bibfield  {journal} {\bibinfo  {journal} {Physical
  Review B}\ }\textbf {\bibinfo {volume} {106}},\ \bibinfo {pages} {L020202}
  (\bibinfo {year} {2022})}\BibitemShut {NoStop}%
\bibitem [{\citenamefont {Kiefer-Emmanouilidis}\ \emph
  {et~al.}(2021)\citenamefont {Kiefer-Emmanouilidis}, \citenamefont {Unanyan},
  \citenamefont {Fleischhauer},\ and\ \citenamefont {Sirker}}]{kiefer2021slow}%
  \BibitemOpen
  \bibfield  {author} {\bibinfo {author} {\bibfnamefont {M.}~\bibnamefont
  {Kiefer-Emmanouilidis}}, \bibinfo {author} {\bibfnamefont {R.}~\bibnamefont
  {Unanyan}}, \bibinfo {author} {\bibfnamefont {M.}~\bibnamefont
  {Fleischhauer}}, \ and\ \bibinfo {author} {\bibfnamefont {J.}~\bibnamefont
  {Sirker}},\ }\href@noop {} {\bibfield  {journal} {\bibinfo  {journal}
  {Physical Review B}\ }\textbf {\bibinfo {volume} {103}},\ \bibinfo {pages}
  {024203} (\bibinfo {year} {2021})}\BibitemShut {NoStop}%
\bibitem [{\citenamefont {Kiefer-Emmanouilidis}\ \emph
  {et~al.}(2020)\citenamefont {Kiefer-Emmanouilidis}, \citenamefont {Unanyan},
  \citenamefont {Fleischhauer},\ and\ \citenamefont
  {Sirker}}]{kiefer2020evidence}%
  \BibitemOpen
  \bibfield  {author} {\bibinfo {author} {\bibfnamefont {M.}~\bibnamefont
  {Kiefer-Emmanouilidis}}, \bibinfo {author} {\bibfnamefont {R.}~\bibnamefont
  {Unanyan}}, \bibinfo {author} {\bibfnamefont {M.}~\bibnamefont
  {Fleischhauer}}, \ and\ \bibinfo {author} {\bibfnamefont {J.}~\bibnamefont
  {Sirker}},\ }\href@noop {} {\bibfield  {journal} {\bibinfo  {journal}
  {Physical Review Letters}\ }\textbf {\bibinfo {volume} {124}},\ \bibinfo
  {pages} {243601} (\bibinfo {year} {2020})}\BibitemShut {NoStop}%
\bibitem [{\citenamefont {De~Roeck}\ and\ \citenamefont
  {Huveneers}(2017)}]{de2017stability}%
  \BibitemOpen
  \bibfield  {author} {\bibinfo {author} {\bibfnamefont {W.}~\bibnamefont
  {De~Roeck}}\ and\ \bibinfo {author} {\bibfnamefont {F.}~\bibnamefont
  {Huveneers}},\ }\href@noop {} {\bibfield  {journal} {\bibinfo  {journal}
  {Physical Review B}\ }\textbf {\bibinfo {volume} {95}},\ \bibinfo {pages}
  {155129} (\bibinfo {year} {2017})}\BibitemShut {NoStop}%
\bibitem [{\citenamefont {Thiery}\ \emph {et~al.}(2018)\citenamefont {Thiery},
  \citenamefont {Huveneers}, \citenamefont {M{\"u}ller},\ and\ \citenamefont
  {De~Roeck}}]{thiery2018many}%
  \BibitemOpen
  \bibfield  {author} {\bibinfo {author} {\bibfnamefont {T.}~\bibnamefont
  {Thiery}}, \bibinfo {author} {\bibfnamefont {F.}~\bibnamefont {Huveneers}},
  \bibinfo {author} {\bibfnamefont {M.}~\bibnamefont {M{\"u}ller}}, \ and\
  \bibinfo {author} {\bibfnamefont {W.}~\bibnamefont {De~Roeck}},\ }\href@noop
  {} {\bibfield  {journal} {\bibinfo  {journal} {Physical review letters}\
  }\textbf {\bibinfo {volume} {121}},\ \bibinfo {pages} {140601} (\bibinfo
  {year} {2018})}\BibitemShut {NoStop}%
\bibitem [{\citenamefont {Luitz}\ \emph {et~al.}(2017)\citenamefont {Luitz},
  \citenamefont {Huveneers},\ and\ \citenamefont {De~Roeck}}]{luitz2017small}%
  \BibitemOpen
  \bibfield  {author} {\bibinfo {author} {\bibfnamefont {D.~J.}\ \bibnamefont
  {Luitz}}, \bibinfo {author} {\bibfnamefont {F.}~\bibnamefont {Huveneers}}, \
  and\ \bibinfo {author} {\bibfnamefont {W.}~\bibnamefont {De~Roeck}},\
  }\href@noop {} {\bibfield  {journal} {\bibinfo  {journal} {Physical review
  letters}\ }\textbf {\bibinfo {volume} {119}},\ \bibinfo {pages} {150602}
  (\bibinfo {year} {2017})}\BibitemShut {NoStop}%
\bibitem [{\citenamefont {Goihl}\ \emph {et~al.}(2019)\citenamefont {Goihl},
  \citenamefont {Eisert},\ and\ \citenamefont
  {Krumnow}}]{goihl2019exploration}%
  \BibitemOpen
  \bibfield  {author} {\bibinfo {author} {\bibfnamefont {M.}~\bibnamefont
  {Goihl}}, \bibinfo {author} {\bibfnamefont {J.}~\bibnamefont {Eisert}}, \
  and\ \bibinfo {author} {\bibfnamefont {C.}~\bibnamefont {Krumnow}},\
  }\href@noop {} {\bibfield  {journal} {\bibinfo  {journal} {Physical Review
  B}\ }\textbf {\bibinfo {volume} {99}},\ \bibinfo {pages} {195145} (\bibinfo
  {year} {2019})}\BibitemShut {NoStop}%
\bibitem [{\citenamefont {Crowley}\ and\ \citenamefont
  {Chandran}(2020)}]{crowley2020avalanche}%
  \BibitemOpen
  \bibfield  {author} {\bibinfo {author} {\bibfnamefont {P.~J.}\ \bibnamefont
  {Crowley}}\ and\ \bibinfo {author} {\bibfnamefont {A.}~\bibnamefont
  {Chandran}},\ }\href@noop {} {\bibfield  {journal} {\bibinfo  {journal}
  {Physical Review Research}\ }\textbf {\bibinfo {volume} {2}},\ \bibinfo
  {pages} {033262} (\bibinfo {year} {2020})}\BibitemShut {NoStop}%
\bibitem [{\citenamefont {L{\'e}onard}\ \emph {et~al.}(2023)\citenamefont
  {L{\'e}onard}, \citenamefont {Kim}, \citenamefont {Rispoli}, \citenamefont
  {Lukin}, \citenamefont {Schittko}, \citenamefont {Kwan}, \citenamefont
  {Demler}, \citenamefont {Sels},\ and\ \citenamefont
  {Greiner}}]{leonard2023probing}%
  \BibitemOpen
  \bibfield  {author} {\bibinfo {author} {\bibfnamefont {J.}~\bibnamefont
  {L{\'e}onard}}, \bibinfo {author} {\bibfnamefont {S.}~\bibnamefont {Kim}},
  \bibinfo {author} {\bibfnamefont {M.}~\bibnamefont {Rispoli}}, \bibinfo
  {author} {\bibfnamefont {A.}~\bibnamefont {Lukin}}, \bibinfo {author}
  {\bibfnamefont {R.}~\bibnamefont {Schittko}}, \bibinfo {author}
  {\bibfnamefont {J.}~\bibnamefont {Kwan}}, \bibinfo {author} {\bibfnamefont
  {E.}~\bibnamefont {Demler}}, \bibinfo {author} {\bibfnamefont
  {D.}~\bibnamefont {Sels}}, \ and\ \bibinfo {author} {\bibfnamefont
  {M.}~\bibnamefont {Greiner}},\ }\href@noop {} {\bibfield  {journal} {\bibinfo
   {journal} {Nature Physics}\ ,\ \bibinfo {pages} {1}} (\bibinfo {year}
  {2023})}\BibitemShut {NoStop}%
\bibitem [{\citenamefont {Peacock}\ and\ \citenamefont
  {Sels}(2023)}]{peacock2023many}%
  \BibitemOpen
  \bibfield  {author} {\bibinfo {author} {\bibfnamefont {J.~C.}\ \bibnamefont
  {Peacock}}\ and\ \bibinfo {author} {\bibfnamefont {D.}~\bibnamefont {Sels}},\
  }\href@noop {} {\bibfield  {journal} {\bibinfo  {journal} {Physical Review
  B}\ }\textbf {\bibinfo {volume} {108}},\ \bibinfo {pages} {L020201} (\bibinfo
  {year} {2023})}\BibitemShut {NoStop}%
\bibitem [{\citenamefont {Morningstar}\ \emph {et~al.}(2022)\citenamefont
  {Morningstar}, \citenamefont {Colmenarez}, \citenamefont {Khemani},
  \citenamefont {Luitz},\ and\ \citenamefont
  {Huse}}]{morningstar2022avalanches}%
  \BibitemOpen
  \bibfield  {author} {\bibinfo {author} {\bibfnamefont {A.}~\bibnamefont
  {Morningstar}}, \bibinfo {author} {\bibfnamefont {L.}~\bibnamefont
  {Colmenarez}}, \bibinfo {author} {\bibfnamefont {V.}~\bibnamefont {Khemani}},
  \bibinfo {author} {\bibfnamefont {D.~J.}\ \bibnamefont {Luitz}}, \ and\
  \bibinfo {author} {\bibfnamefont {D.~A.}\ \bibnamefont {Huse}},\ }\href@noop
  {} {\bibfield  {journal} {\bibinfo  {journal} {Physical Review B}\ }\textbf
  {\bibinfo {volume} {105}},\ \bibinfo {pages} {174205} (\bibinfo {year}
  {2022})}\BibitemShut {NoStop}%
\bibitem [{\citenamefont {Ha}\ \emph {et~al.}(2023)\citenamefont {Ha},
  \citenamefont {Morningstar},\ and\ \citenamefont {Huse}}]{Ha2023many}%
  \BibitemOpen
  \bibfield  {author} {\bibinfo {author} {\bibfnamefont {H.}~\bibnamefont
  {Ha}}, \bibinfo {author} {\bibfnamefont {A.}~\bibnamefont {Morningstar}}, \
  and\ \bibinfo {author} {\bibfnamefont {D.~A.}\ \bibnamefont {Huse}},\
  }\href@noop {} {\bibfield  {journal} {\bibinfo  {journal} {arXiv preprint
  arXiv:2301.04658}\ } (\bibinfo {year} {2023})}\BibitemShut {NoStop}%
\bibitem [{\citenamefont {Long}\ \emph {et~al.}(2022)\citenamefont {Long},
  \citenamefont {Crowley}, \citenamefont {Khemani},\ and\ \citenamefont
  {Chandran}}]{long2022phenomenology}%
  \BibitemOpen
  \bibfield  {author} {\bibinfo {author} {\bibfnamefont {D.~M.}\ \bibnamefont
  {Long}}, \bibinfo {author} {\bibfnamefont {P.~J.}\ \bibnamefont {Crowley}},
  \bibinfo {author} {\bibfnamefont {V.}~\bibnamefont {Khemani}}, \ and\
  \bibinfo {author} {\bibfnamefont {A.}~\bibnamefont {Chandran}},\ }\href@noop
  {} {\bibfield  {journal} {\bibinfo  {journal} {arXiv preprint
  arXiv:2207.05761}\ } (\bibinfo {year} {2022})}\BibitemShut {NoStop}%
\bibitem [{\citenamefont {Doggen}\ \emph {et~al.}(2021)\citenamefont {Doggen},
  \citenamefont {Gornyi}, \citenamefont {Mirlin},\ and\ \citenamefont
  {Polyakov}}]{doggen2021many}%
  \BibitemOpen
  \bibfield  {author} {\bibinfo {author} {\bibfnamefont {E.~V.}\ \bibnamefont
  {Doggen}}, \bibinfo {author} {\bibfnamefont {I.~V.}\ \bibnamefont {Gornyi}},
  \bibinfo {author} {\bibfnamefont {A.~D.}\ \bibnamefont {Mirlin}}, \ and\
  \bibinfo {author} {\bibfnamefont {D.~G.}\ \bibnamefont {Polyakov}},\
  }\href@noop {} {\bibfield  {journal} {\bibinfo  {journal} {Annals of
  Physics}\ }\textbf {\bibinfo {volume} {435}},\ \bibinfo {pages} {168437}
  (\bibinfo {year} {2021})}\BibitemShut {NoStop}%
\bibitem [{\citenamefont {Luitz}\ and\ \citenamefont
  {Lev}(2017)}]{luitz2017ergodic}%
  \BibitemOpen
  \bibfield  {author} {\bibinfo {author} {\bibfnamefont {D.~J.}\ \bibnamefont
  {Luitz}}\ and\ \bibinfo {author} {\bibfnamefont {Y.~B.}\ \bibnamefont
  {Lev}},\ }\href@noop {} {\bibfield  {journal} {\bibinfo  {journal} {Annalen
  der Physik}\ }\textbf {\bibinfo {volume} {529}},\ \bibinfo {pages} {1600350}
  (\bibinfo {year} {2017})}\BibitemShut {NoStop}%
\bibitem [{\citenamefont {Agarwal}\ \emph {et~al.}(2017)\citenamefont
  {Agarwal}, \citenamefont {Altman}, \citenamefont {Demler}, \citenamefont
  {Gopalakrishnan}, \citenamefont {Huse},\ and\ \citenamefont
  {Knap}}]{agarwal2017rare}%
  \BibitemOpen
  \bibfield  {author} {\bibinfo {author} {\bibfnamefont {K.}~\bibnamefont
  {Agarwal}}, \bibinfo {author} {\bibfnamefont {E.}~\bibnamefont {Altman}},
  \bibinfo {author} {\bibfnamefont {E.}~\bibnamefont {Demler}}, \bibinfo
  {author} {\bibfnamefont {S.}~\bibnamefont {Gopalakrishnan}}, \bibinfo
  {author} {\bibfnamefont {D.~A.}\ \bibnamefont {Huse}}, \ and\ \bibinfo
  {author} {\bibfnamefont {M.}~\bibnamefont {Knap}},\ }\href@noop {} {\bibfield
   {journal} {\bibinfo  {journal} {Annalen der Physik}\ }\textbf {\bibinfo
  {volume} {529}},\ \bibinfo {pages} {1600326} (\bibinfo {year}
  {2017})}\BibitemShut {NoStop}%
\bibitem [{\citenamefont {Agarwal}\ \emph {et~al.}(2015)\citenamefont
  {Agarwal}, \citenamefont {Gopalakrishnan}, \citenamefont {Knap},
  \citenamefont {M{\"u}ller},\ and\ \citenamefont
  {Demler}}]{agarwal2015anomalous}%
  \BibitemOpen
  \bibfield  {author} {\bibinfo {author} {\bibfnamefont {K.}~\bibnamefont
  {Agarwal}}, \bibinfo {author} {\bibfnamefont {S.}~\bibnamefont
  {Gopalakrishnan}}, \bibinfo {author} {\bibfnamefont {M.}~\bibnamefont
  {Knap}}, \bibinfo {author} {\bibfnamefont {M.}~\bibnamefont {M{\"u}ller}}, \
  and\ \bibinfo {author} {\bibfnamefont {E.}~\bibnamefont {Demler}},\
  }\href@noop {} {\bibfield  {journal} {\bibinfo  {journal} {Physical review
  letters}\ }\textbf {\bibinfo {volume} {114}},\ \bibinfo {pages} {160401}
  (\bibinfo {year} {2015})}\BibitemShut {NoStop}%
\bibitem [{\citenamefont {{\v{Z}}nidari{\v{c}}}\ \emph
  {et~al.}(2016)\citenamefont {{\v{Z}}nidari{\v{c}}}, \citenamefont
  {Scardicchio},\ and\ \citenamefont {Varma}}]{vznidarivc2016diffusive}%
  \BibitemOpen
  \bibfield  {author} {\bibinfo {author} {\bibfnamefont {M.}~\bibnamefont
  {{\v{Z}}nidari{\v{c}}}}, \bibinfo {author} {\bibfnamefont {A.}~\bibnamefont
  {Scardicchio}}, \ and\ \bibinfo {author} {\bibfnamefont {V.~K.}\ \bibnamefont
  {Varma}},\ }\href@noop {} {\bibfield  {journal} {\bibinfo  {journal}
  {Physical review letters}\ }\textbf {\bibinfo {volume} {117}},\ \bibinfo
  {pages} {040601} (\bibinfo {year} {2016})}\BibitemShut {NoStop}%
\bibitem [{\citenamefont {Lev}\ and\ \citenamefont
  {Reichman}(2014)}]{lev2014dynamics}%
  \BibitemOpen
  \bibfield  {author} {\bibinfo {author} {\bibfnamefont {Y.~B.}\ \bibnamefont
  {Lev}}\ and\ \bibinfo {author} {\bibfnamefont {D.~R.}\ \bibnamefont
  {Reichman}},\ }\href@noop {} {\bibfield  {journal} {\bibinfo  {journal}
  {Physical Review B}\ }\textbf {\bibinfo {volume} {89}},\ \bibinfo {pages}
  {220201} (\bibinfo {year} {2014})}\BibitemShut {NoStop}%
\bibitem [{\citenamefont {Luitz}\ and\ \citenamefont
  {Lev}(2016)}]{luitz2016anomalous}%
  \BibitemOpen
  \bibfield  {author} {\bibinfo {author} {\bibfnamefont {D.~J.}\ \bibnamefont
  {Luitz}}\ and\ \bibinfo {author} {\bibfnamefont {Y.~B.}\ \bibnamefont
  {Lev}},\ }\href@noop {} {\bibfield  {journal} {\bibinfo  {journal} {Physical
  review letters}\ }\textbf {\bibinfo {volume} {117}},\ \bibinfo {pages}
  {170404} (\bibinfo {year} {2016})}\BibitemShut {NoStop}%
\bibitem [{\citenamefont {Doggen}\ \emph {et~al.}(2018)\citenamefont {Doggen},
  \citenamefont {Schindler}, \citenamefont {Tikhonov}, \citenamefont {Mirlin},
  \citenamefont {Neupert}, \citenamefont {Polyakov},\ and\ \citenamefont
  {Gornyi}}]{doggen2018many}%
  \BibitemOpen
  \bibfield  {author} {\bibinfo {author} {\bibfnamefont {E.~V.}\ \bibnamefont
  {Doggen}}, \bibinfo {author} {\bibfnamefont {F.}~\bibnamefont {Schindler}},
  \bibinfo {author} {\bibfnamefont {K.~S.}\ \bibnamefont {Tikhonov}}, \bibinfo
  {author} {\bibfnamefont {A.~D.}\ \bibnamefont {Mirlin}}, \bibinfo {author}
  {\bibfnamefont {T.}~\bibnamefont {Neupert}}, \bibinfo {author} {\bibfnamefont
  {D.~G.}\ \bibnamefont {Polyakov}}, \ and\ \bibinfo {author} {\bibfnamefont
  {I.~V.}\ \bibnamefont {Gornyi}},\ }\href@noop {} {\bibfield  {journal}
  {\bibinfo  {journal} {Physical Review B}\ }\textbf {\bibinfo {volume} {98}},\
  \bibinfo {pages} {174202} (\bibinfo {year} {2018})}\BibitemShut {NoStop}%
\bibitem [{\citenamefont {Bera}\ \emph {et~al.}(2017)\citenamefont {Bera},
  \citenamefont {De~Tomasi}, \citenamefont {Weiner},\ and\ \citenamefont
  {Evers}}]{bera2017density}%
  \BibitemOpen
  \bibfield  {author} {\bibinfo {author} {\bibfnamefont {S.}~\bibnamefont
  {Bera}}, \bibinfo {author} {\bibfnamefont {G.}~\bibnamefont {De~Tomasi}},
  \bibinfo {author} {\bibfnamefont {F.}~\bibnamefont {Weiner}}, \ and\ \bibinfo
  {author} {\bibfnamefont {F.}~\bibnamefont {Evers}},\ }\href@noop {}
  {\bibfield  {journal} {\bibinfo  {journal} {Physical Review Letters}\
  }\textbf {\bibinfo {volume} {118}},\ \bibinfo {pages} {196801} (\bibinfo
  {year} {2017})}\BibitemShut {NoStop}%
\bibitem [{\citenamefont {Schreiber}\ \emph {et~al.}(2015)\citenamefont
  {Schreiber}, \citenamefont {Hodgman}, \citenamefont {Bordia}, \citenamefont
  {L{\"u}schen}, \citenamefont {Fischer}, \citenamefont {Vosk}, \citenamefont
  {Altman}, \citenamefont {Schneider},\ and\ \citenamefont
  {Bloch}}]{schreiber2015observation}%
  \BibitemOpen
  \bibfield  {author} {\bibinfo {author} {\bibfnamefont {M.}~\bibnamefont
  {Schreiber}}, \bibinfo {author} {\bibfnamefont {S.~S.}\ \bibnamefont
  {Hodgman}}, \bibinfo {author} {\bibfnamefont {P.}~\bibnamefont {Bordia}},
  \bibinfo {author} {\bibfnamefont {H.~P.}\ \bibnamefont {L{\"u}schen}},
  \bibinfo {author} {\bibfnamefont {M.~H.}\ \bibnamefont {Fischer}}, \bibinfo
  {author} {\bibfnamefont {R.}~\bibnamefont {Vosk}}, \bibinfo {author}
  {\bibfnamefont {E.}~\bibnamefont {Altman}}, \bibinfo {author} {\bibfnamefont
  {U.}~\bibnamefont {Schneider}}, \ and\ \bibinfo {author} {\bibfnamefont
  {I.}~\bibnamefont {Bloch}},\ }\href@noop {} {\bibfield  {journal} {\bibinfo
  {journal} {Science}\ }\textbf {\bibinfo {volume} {349}},\ \bibinfo {pages}
  {842} (\bibinfo {year} {2015})}\BibitemShut {NoStop}%
\bibitem [{\citenamefont {Bordia}\ \emph {et~al.}(2017)\citenamefont {Bordia},
  \citenamefont {L{\"u}schen}, \citenamefont {Scherg}, \citenamefont
  {Gopalakrishnan}, \citenamefont {Knap}, \citenamefont {Schneider},\ and\
  \citenamefont {Bloch}}]{bordia2017probing}%
  \BibitemOpen
  \bibfield  {author} {\bibinfo {author} {\bibfnamefont {P.}~\bibnamefont
  {Bordia}}, \bibinfo {author} {\bibfnamefont {H.}~\bibnamefont {L{\"u}schen}},
  \bibinfo {author} {\bibfnamefont {S.}~\bibnamefont {Scherg}}, \bibinfo
  {author} {\bibfnamefont {S.}~\bibnamefont {Gopalakrishnan}}, \bibinfo
  {author} {\bibfnamefont {M.}~\bibnamefont {Knap}}, \bibinfo {author}
  {\bibfnamefont {U.}~\bibnamefont {Schneider}}, \ and\ \bibinfo {author}
  {\bibfnamefont {I.}~\bibnamefont {Bloch}},\ }\href@noop {} {\bibfield
  {journal} {\bibinfo  {journal} {Physical Review X}\ }\textbf {\bibinfo
  {volume} {7}},\ \bibinfo {pages} {041047} (\bibinfo {year}
  {2017})}\BibitemShut {NoStop}%
\bibitem [{\citenamefont {L{\"u}schen}\ \emph {et~al.}(2017)\citenamefont
  {L{\"u}schen}, \citenamefont {Bordia}, \citenamefont {Scherg}, \citenamefont
  {Alet}, \citenamefont {Altman}, \citenamefont {Schneider},\ and\
  \citenamefont {Bloch}}]{luschen2017observation}%
  \BibitemOpen
  \bibfield  {author} {\bibinfo {author} {\bibfnamefont {H.~P.}\ \bibnamefont
  {L{\"u}schen}}, \bibinfo {author} {\bibfnamefont {P.}~\bibnamefont {Bordia}},
  \bibinfo {author} {\bibfnamefont {S.}~\bibnamefont {Scherg}}, \bibinfo
  {author} {\bibfnamefont {F.}~\bibnamefont {Alet}}, \bibinfo {author}
  {\bibfnamefont {E.}~\bibnamefont {Altman}}, \bibinfo {author} {\bibfnamefont
  {U.}~\bibnamefont {Schneider}}, \ and\ \bibinfo {author} {\bibfnamefont
  {I.}~\bibnamefont {Bloch}},\ }\href@noop {} {\bibfield  {journal} {\bibinfo
  {journal} {Physical review letters}\ }\textbf {\bibinfo {volume} {119}},\
  \bibinfo {pages} {260401} (\bibinfo {year} {2017})}\BibitemShut {NoStop}%
\bibitem [{\citenamefont {Smith}\ \emph {et~al.}(2016)\citenamefont {Smith},
  \citenamefont {Lee}, \citenamefont {Richerme}, \citenamefont {Neyenhuis},
  \citenamefont {Hess}, \citenamefont {Hauke}, \citenamefont {Heyl},
  \citenamefont {Huse},\ and\ \citenamefont {Monroe}}]{smith2016many}%
  \BibitemOpen
  \bibfield  {author} {\bibinfo {author} {\bibfnamefont {J.}~\bibnamefont
  {Smith}}, \bibinfo {author} {\bibfnamefont {A.}~\bibnamefont {Lee}}, \bibinfo
  {author} {\bibfnamefont {P.}~\bibnamefont {Richerme}}, \bibinfo {author}
  {\bibfnamefont {B.}~\bibnamefont {Neyenhuis}}, \bibinfo {author}
  {\bibfnamefont {P.~W.}\ \bibnamefont {Hess}}, \bibinfo {author}
  {\bibfnamefont {P.}~\bibnamefont {Hauke}}, \bibinfo {author} {\bibfnamefont
  {M.}~\bibnamefont {Heyl}}, \bibinfo {author} {\bibfnamefont {D.~A.}\
  \bibnamefont {Huse}}, \ and\ \bibinfo {author} {\bibfnamefont
  {C.}~\bibnamefont {Monroe}},\ }\href@noop {} {\bibfield  {journal} {\bibinfo
  {journal} {Nature Physics}\ }\textbf {\bibinfo {volume} {12}},\ \bibinfo
  {pages} {907} (\bibinfo {year} {2016})}\BibitemShut {NoStop}%
\bibitem [{\citenamefont {Xu}\ \emph {et~al.}(2018)\citenamefont {Xu},
  \citenamefont {Chen}, \citenamefont {Zeng}, \citenamefont {Zhang},
  \citenamefont {Song}, \citenamefont {Liu}, \citenamefont {Guo}, \citenamefont
  {Zhang}, \citenamefont {Xu}, \citenamefont {Deng} \emph
  {et~al.}}]{xu2018emulating}%
  \BibitemOpen
  \bibfield  {author} {\bibinfo {author} {\bibfnamefont {K.}~\bibnamefont
  {Xu}}, \bibinfo {author} {\bibfnamefont {J.-J.}\ \bibnamefont {Chen}},
  \bibinfo {author} {\bibfnamefont {Y.}~\bibnamefont {Zeng}}, \bibinfo {author}
  {\bibfnamefont {Y.-R.}\ \bibnamefont {Zhang}}, \bibinfo {author}
  {\bibfnamefont {C.}~\bibnamefont {Song}}, \bibinfo {author} {\bibfnamefont
  {W.}~\bibnamefont {Liu}}, \bibinfo {author} {\bibfnamefont {Q.}~\bibnamefont
  {Guo}}, \bibinfo {author} {\bibfnamefont {P.}~\bibnamefont {Zhang}}, \bibinfo
  {author} {\bibfnamefont {D.}~\bibnamefont {Xu}}, \bibinfo {author}
  {\bibfnamefont {H.}~\bibnamefont {Deng}},  \emph {et~al.},\ }\href@noop {}
  {\bibfield  {journal} {\bibinfo  {journal} {Physical review letters}\
  }\textbf {\bibinfo {volume} {120}},\ \bibinfo {pages} {050507} (\bibinfo
  {year} {2018})}\BibitemShut {NoStop}%
\bibitem [{\citenamefont {Schulz}\ \emph {et~al.}(2020)\citenamefont {Schulz},
  \citenamefont {Taylor}, \citenamefont {Scardicchio},\ and\ \citenamefont
  {{\v{Z}}nidari{\v{c}}}}]{schulz2020phenomenology}%
  \BibitemOpen
  \bibfield  {author} {\bibinfo {author} {\bibfnamefont {M.}~\bibnamefont
  {Schulz}}, \bibinfo {author} {\bibfnamefont {S.~R.}\ \bibnamefont {Taylor}},
  \bibinfo {author} {\bibfnamefont {A.}~\bibnamefont {Scardicchio}}, \ and\
  \bibinfo {author} {\bibfnamefont {M.}~\bibnamefont {{\v{Z}}nidari{\v{c}}}},\
  }\href@noop {} {\bibfield  {journal} {\bibinfo  {journal} {Journal of
  Statistical Mechanics: Theory and Experiment}\ }\textbf {\bibinfo {volume}
  {2020}},\ \bibinfo {pages} {023107} (\bibinfo {year} {2020})}\BibitemShut
  {NoStop}%
\bibitem [{\citenamefont {Schir{\'o}}\ and\ \citenamefont
  {Tarzia}(2020)}]{schiro2020toy}%
  \BibitemOpen
  \bibfield  {author} {\bibinfo {author} {\bibfnamefont {M.}~\bibnamefont
  {Schir{\'o}}}\ and\ \bibinfo {author} {\bibfnamefont {M.}~\bibnamefont
  {Tarzia}},\ }\href@noop {} {\bibfield  {journal} {\bibinfo  {journal}
  {Physical Review B}\ }\textbf {\bibinfo {volume} {101}},\ \bibinfo {pages}
  {014203} (\bibinfo {year} {2020})}\BibitemShut {NoStop}%
\bibitem [{\citenamefont {Garratt}\ \emph {et~al.}(2021)\citenamefont
  {Garratt}, \citenamefont {Roy},\ and\ \citenamefont
  {Chalker}}]{garratt2021local}%
  \BibitemOpen
  \bibfield  {author} {\bibinfo {author} {\bibfnamefont {S.}~\bibnamefont
  {Garratt}}, \bibinfo {author} {\bibfnamefont {S.}~\bibnamefont {Roy}}, \ and\
  \bibinfo {author} {\bibfnamefont {J.}~\bibnamefont {Chalker}},\ }\href@noop
  {} {\bibfield  {journal} {\bibinfo  {journal} {Physical Review B}\ }\textbf
  {\bibinfo {volume} {104}},\ \bibinfo {pages} {184203} (\bibinfo {year}
  {2021})}\BibitemShut {NoStop}%
\bibitem [{\citenamefont {Villalonga}\ and\ \citenamefont
  {Clark}(2020)}]{villalonga2020eigenstates}%
  \BibitemOpen
  \bibfield  {author} {\bibinfo {author} {\bibfnamefont {B.}~\bibnamefont
  {Villalonga}}\ and\ \bibinfo {author} {\bibfnamefont {B.~K.}\ \bibnamefont
  {Clark}},\ }\href@noop {} {\bibfield  {journal} {\bibinfo  {journal} {arXiv
  preprint arXiv:2005.13558}\ } (\bibinfo {year} {2020})}\BibitemShut {NoStop}%
\bibitem [{\citenamefont {Khemani}\ \emph {et~al.}(2017)\citenamefont
  {Khemani}, \citenamefont {Lim}, \citenamefont {Sheng},\ and\ \citenamefont
  {Huse}}]{khemani2017critical}%
  \BibitemOpen
  \bibfield  {author} {\bibinfo {author} {\bibfnamefont {V.}~\bibnamefont
  {Khemani}}, \bibinfo {author} {\bibfnamefont {S.-P.}\ \bibnamefont {Lim}},
  \bibinfo {author} {\bibfnamefont {D.}~\bibnamefont {Sheng}}, \ and\ \bibinfo
  {author} {\bibfnamefont {D.~A.}\ \bibnamefont {Huse}},\ }\href@noop {}
  {\bibfield  {journal} {\bibinfo  {journal} {Physical Review X}\ }\textbf
  {\bibinfo {volume} {7}},\ \bibinfo {pages} {021013} (\bibinfo {year}
  {2017})}\BibitemShut {NoStop}%
\bibitem [{\citenamefont {De~Tomasi}\ \emph {et~al.}(2021)\citenamefont
  {De~Tomasi}, \citenamefont {Khaymovich}, \citenamefont {Pollmann},\ and\
  \citenamefont {Warzel}}]{de2021rare}%
  \BibitemOpen
  \bibfield  {author} {\bibinfo {author} {\bibfnamefont {G.}~\bibnamefont
  {De~Tomasi}}, \bibinfo {author} {\bibfnamefont {I.~M.}\ \bibnamefont
  {Khaymovich}}, \bibinfo {author} {\bibfnamefont {F.}~\bibnamefont
  {Pollmann}}, \ and\ \bibinfo {author} {\bibfnamefont {S.}~\bibnamefont
  {Warzel}},\ }\href {\doibase 10.1103/PhysRevB.104.024202} {\bibfield
  {journal} {\bibinfo  {journal} {Phys. Rev. B}\ }\textbf {\bibinfo {volume}
  {104}},\ \bibinfo {pages} {024202} (\bibinfo {year} {2021})}\BibitemShut
  {NoStop}%
\bibitem [{\citenamefont {Derrida}\ and\ \citenamefont
  {Spohn}(1988)}]{derrida1988polymers}%
  \BibitemOpen
  \bibfield  {author} {\bibinfo {author} {\bibfnamefont {B.}~\bibnamefont
  {Derrida}}\ and\ \bibinfo {author} {\bibfnamefont {H.}~\bibnamefont
  {Spohn}},\ }\href@noop {} {\bibfield  {journal} {\bibinfo  {journal} {Journal
  of Statistical Physics}\ }\textbf {\bibinfo {volume} {51}},\ \bibinfo {pages}
  {817} (\bibinfo {year} {1988})}\BibitemShut {NoStop}%
\bibitem [{\citenamefont {Monthus}\ and\ \citenamefont
  {Garel}(2008)}]{monthus2008anderson}%
  \BibitemOpen
  \bibfield  {author} {\bibinfo {author} {\bibfnamefont {C.}~\bibnamefont
  {Monthus}}\ and\ \bibinfo {author} {\bibfnamefont {T.}~\bibnamefont
  {Garel}},\ }\href@noop {} {\bibfield  {journal} {\bibinfo  {journal} {Journal
  of Physics A: Mathematical and Theoretical}\ }\textbf {\bibinfo {volume}
  {42}},\ \bibinfo {pages} {075002} (\bibinfo {year} {2008})}\BibitemShut
  {NoStop}%
\bibitem [{\citenamefont {Monthus}\ and\ \citenamefont
  {Garel}(2011)}]{monthus2011anderson}%
  \BibitemOpen
  \bibfield  {author} {\bibinfo {author} {\bibfnamefont {C.}~\bibnamefont
  {Monthus}}\ and\ \bibinfo {author} {\bibfnamefont {T.}~\bibnamefont
  {Garel}},\ }\href@noop {} {\bibfield  {journal} {\bibinfo  {journal} {Journal
  of Physics A: Mathematical and Theoretical}\ }\textbf {\bibinfo {volume}
  {44}},\ \bibinfo {pages} {145001} (\bibinfo {year} {2011})}\BibitemShut
  {NoStop}%
\bibitem [{\citenamefont {Biroli}\ and\ \citenamefont
  {Tarzia}(2020)}]{biroli2020anomalous}%
  \BibitemOpen
  \bibfield  {author} {\bibinfo {author} {\bibfnamefont {G.}~\bibnamefont
  {Biroli}}\ and\ \bibinfo {author} {\bibfnamefont {M.}~\bibnamefont
  {Tarzia}},\ }\href@noop {} {\bibfield  {journal} {\bibinfo  {journal}
  {Physical Review B}\ }\textbf {\bibinfo {volume} {102}},\ \bibinfo {pages}
  {064211} (\bibinfo {year} {2020})}\BibitemShut {NoStop}%
\bibitem [{\citenamefont {Kravtsov}\ \emph {et~al.}(2018)\citenamefont
  {Kravtsov}, \citenamefont {Altshuler},\ and\ \citenamefont
  {Ioffe}}]{kravtsov2018non}%
  \BibitemOpen
  \bibfield  {author} {\bibinfo {author} {\bibfnamefont {V.}~\bibnamefont
  {Kravtsov}}, \bibinfo {author} {\bibfnamefont {B.}~\bibnamefont {Altshuler}},
  \ and\ \bibinfo {author} {\bibfnamefont {L.}~\bibnamefont {Ioffe}},\
  }\href@noop {} {\bibfield  {journal} {\bibinfo  {journal} {Annals of
  Physics}\ }\textbf {\bibinfo {volume} {389}},\ \bibinfo {pages} {148}
  (\bibinfo {year} {2018})}\BibitemShut {NoStop}%
\bibitem [{\citenamefont {Lemari{\'e}}(2019)}]{lemarie2019glassy}%
  \BibitemOpen
  \bibfield  {author} {\bibinfo {author} {\bibfnamefont {G.}~\bibnamefont
  {Lemari{\'e}}},\ }\href@noop {} {\bibfield  {journal} {\bibinfo  {journal}
  {Physical review letters}\ }\textbf {\bibinfo {volume} {122}},\ \bibinfo
  {pages} {030401} (\bibinfo {year} {2019})}\BibitemShut {NoStop}%
\bibitem [{\citenamefont {Aizenman}\ and\ \citenamefont
  {Molchanov}(1993)}]{aizenman1993localization}%
  \BibitemOpen
  \bibfield  {author} {\bibinfo {author} {\bibfnamefont {M.}~\bibnamefont
  {Aizenman}}\ and\ \bibinfo {author} {\bibfnamefont {S.}~\bibnamefont
  {Molchanov}},\ }\href@noop {} {\bibfield  {journal} {\bibinfo  {journal}
  {Communications in Mathematical Physics}\ }\textbf {\bibinfo {volume}
  {157}},\ \bibinfo {pages} {245} (\bibinfo {year} {1993})}\BibitemShut
  {NoStop}%
\bibitem [{\citenamefont {Warzel}\ and\ \citenamefont
  {Aizenman}(2013)}]{warzel2013resonant}%
  \BibitemOpen
  \bibfield  {author} {\bibinfo {author} {\bibfnamefont {S.}~\bibnamefont
  {Warzel}}\ and\ \bibinfo {author} {\bibfnamefont {M.}~\bibnamefont
  {Aizenman}},\ }\href@noop {} {\bibfield  {journal} {\bibinfo  {journal}
  {Journal of the European Mathematical Society}\ }\textbf {\bibinfo {volume}
  {15}},\ \bibinfo {pages} {1167} (\bibinfo {year} {2013})}\BibitemShut
  {NoStop}%
\bibitem [{\citenamefont {Tikhonov}\ and\ \citenamefont
  {Mirlin}(2016)}]{tikhonov2016fractality}%
  \BibitemOpen
  \bibfield  {author} {\bibinfo {author} {\bibfnamefont {K.~S.}\ \bibnamefont
  {Tikhonov}}\ and\ \bibinfo {author} {\bibfnamefont {A.~D.}\ \bibnamefont
  {Mirlin}},\ }\href@noop {} {\bibfield  {journal} {\bibinfo  {journal}
  {Physical Review B}\ }\textbf {\bibinfo {volume} {94}},\ \bibinfo {pages}
  {184203} (\bibinfo {year} {2016})}\BibitemShut {NoStop}%
\bibitem [{\citenamefont {Sonner}\ \emph {et~al.}(2017)\citenamefont {Sonner},
  \citenamefont {Tikhonov},\ and\ \citenamefont
  {Mirlin}}]{sonner2017multifractality}%
  \BibitemOpen
  \bibfield  {author} {\bibinfo {author} {\bibfnamefont {M.}~\bibnamefont
  {Sonner}}, \bibinfo {author} {\bibfnamefont {K.}~\bibnamefont {Tikhonov}}, \
  and\ \bibinfo {author} {\bibfnamefont {A.}~\bibnamefont {Mirlin}},\
  }\href@noop {} {\bibfield  {journal} {\bibinfo  {journal} {Physical Review
  B}\ }\textbf {\bibinfo {volume} {96}},\ \bibinfo {pages} {214204} (\bibinfo
  {year} {2017})}\BibitemShut {NoStop}%
\bibitem [{\citenamefont {Kravtsov}\ \emph {et~al.}(2015)\citenamefont
  {Kravtsov}, \citenamefont {Khaymovich}, \citenamefont {Cuevas},\ and\
  \citenamefont {Amini}}]{Kravtsov_2015}%
  \BibitemOpen
  \bibfield  {author} {\bibinfo {author} {\bibfnamefont {V.~E.}\ \bibnamefont
  {Kravtsov}}, \bibinfo {author} {\bibfnamefont {I.~M.}\ \bibnamefont
  {Khaymovich}}, \bibinfo {author} {\bibfnamefont {E.}~\bibnamefont {Cuevas}},
  \ and\ \bibinfo {author} {\bibfnamefont {M.}~\bibnamefont {Amini}},\ }\href
  {\doibase 10.1088/1367-2630/17/12/122002} {\bibfield  {journal} {\bibinfo
  {journal} {New J. Phys.}\ }\textbf {\bibinfo {volume} {17}},\ \bibinfo
  {pages} {122002} (\bibinfo {year} {2015})}\BibitemShut {NoStop}%
\bibitem [{\citenamefont {von Soosten}\ and\ \citenamefont
  {Warzel}(2019)}]{vonSoosten_2019}%
  \BibitemOpen
  \bibfield  {author} {\bibinfo {author} {\bibfnamefont {P.}~\bibnamefont {von
  Soosten}}\ and\ \bibinfo {author} {\bibfnamefont {S.}~\bibnamefont
  {Warzel}},\ }\href {\doibase 10.1007/s11005-018-1131-7} {\bibfield  {journal}
  {\bibinfo  {journal} {Lett. Math. Phys.}\ }\textbf {\bibinfo {volume}
  {109}},\ \bibinfo {pages} {905} (\bibinfo {year} {2019})}\BibitemShut
  {NoStop}%
\bibitem [{\citenamefont {Facoetti}\ \emph {et~al.}(2016)\citenamefont
  {Facoetti}, \citenamefont {Vivo},\ and\ \citenamefont
  {Biroli}}]{Facoetti_2016}%
  \BibitemOpen
  \bibfield  {author} {\bibinfo {author} {\bibfnamefont {D.}~\bibnamefont
  {Facoetti}}, \bibinfo {author} {\bibfnamefont {P.}~\bibnamefont {Vivo}}, \
  and\ \bibinfo {author} {\bibfnamefont {G.}~\bibnamefont {Biroli}},\ }\href
  {\doibase 10.1209/0295-5075/115/47003} {\bibfield  {journal} {\bibinfo
  {journal} {Europhys. Lett.}\ }\textbf {\bibinfo {volume} {115}},\ \bibinfo
  {pages} {47003} (\bibinfo {year} {2016})}\BibitemShut {NoStop}%
\bibitem [{\citenamefont {Truong}\ and\ \citenamefont
  {Ossipov}(2016)}]{Truong_2016}%
  \BibitemOpen
  \bibfield  {author} {\bibinfo {author} {\bibfnamefont {K.}~\bibnamefont
  {Truong}}\ and\ \bibinfo {author} {\bibfnamefont {A.}~\bibnamefont
  {Ossipov}},\ }\href {\doibase 10.1209/0295-5075/116/37002} {\bibfield
  {journal} {\bibinfo  {journal} {Europhys. Lett.}\ }\textbf {\bibinfo {volume}
  {116}},\ \bibinfo {pages} {37002} (\bibinfo {year} {2016})}\BibitemShut
  {NoStop}%
\bibitem [{\citenamefont {Bogomolny}\ and\ \citenamefont
  {Sieber}(2018)}]{Bogomolny_2018}%
  \BibitemOpen
  \bibfield  {author} {\bibinfo {author} {\bibfnamefont {E.}~\bibnamefont
  {Bogomolny}}\ and\ \bibinfo {author} {\bibfnamefont {M.}~\bibnamefont
  {Sieber}},\ }\href {\doibase 10.1103/PhysRevE.98.032139} {\bibfield
  {journal} {\bibinfo  {journal} {Phys. Rev. E}\ }\textbf {\bibinfo {volume}
  {98}},\ \bibinfo {pages} {032139} (\bibinfo {year} {2018})}\BibitemShut
  {NoStop}%
\bibitem [{\citenamefont {Tomasi}\ \emph {et~al.}(2019)\citenamefont {Tomasi},
  \citenamefont {Amini}, \citenamefont {Bera}, \citenamefont {Khaymovich},\
  and\ \citenamefont {Kravtsov}}]{DeTomasi_2019}%
  \BibitemOpen
  \bibfield  {author} {\bibinfo {author} {\bibfnamefont {G.~D.}\ \bibnamefont
  {Tomasi}}, \bibinfo {author} {\bibfnamefont {M.}~\bibnamefont {Amini}},
  \bibinfo {author} {\bibfnamefont {S.}~\bibnamefont {Bera}}, \bibinfo {author}
  {\bibfnamefont {I.~M.}\ \bibnamefont {Khaymovich}}, \ and\ \bibinfo {author}
  {\bibfnamefont {V.~E.}\ \bibnamefont {Kravtsov}},\ }\href {\doibase
  10.21468/SciPostPhys.6.1.014} {\bibfield  {journal} {\bibinfo  {journal}
  {SciPost Phys.}\ }\textbf {\bibinfo {volume} {6}},\ \bibinfo {pages} {14}
  (\bibinfo {year} {2019})}\BibitemShut {NoStop}%
\bibitem [{\citenamefont {Amini}(2017)}]{amini2017spread}%
  \BibitemOpen
  \bibfield  {author} {\bibinfo {author} {\bibfnamefont {M.}~\bibnamefont
  {Amini}},\ }\href {\doibase 10.1209/0295-5075/117/30003} {\bibfield
  {journal} {\bibinfo  {journal} {Europhy. Lett.}\ }\textbf {\bibinfo {volume}
  {117}},\ \bibinfo {pages} {30003} (\bibinfo {year} {2017})}\BibitemShut
  {NoStop}%
\bibitem [{\citenamefont {Pino}\ \emph {et~al.}(2019)\citenamefont {Pino},
  \citenamefont {Tabanera},\ and\ \citenamefont {Serna}}]{pino2019ergodic}%
  \BibitemOpen
  \bibfield  {author} {\bibinfo {author} {\bibfnamefont {M.}~\bibnamefont
  {Pino}}, \bibinfo {author} {\bibfnamefont {J.}~\bibnamefont {Tabanera}}, \
  and\ \bibinfo {author} {\bibfnamefont {P.}~\bibnamefont {Serna}},\ }\href
  {\doibase 10.1088/1751-8121/ab4b76} {\bibfield  {journal} {\bibinfo
  {journal} {J. Phys. A - Mat. Theor.}\ }\textbf {\bibinfo {volume} {52}},\
  \bibinfo {pages} {475101} (\bibinfo {year} {2019})}\BibitemShut {NoStop}%
\bibitem [{\citenamefont {Berkovits}(2020)}]{berkovits2020super}%
  \BibitemOpen
  \bibfield  {author} {\bibinfo {author} {\bibfnamefont {R.}~\bibnamefont
  {Berkovits}},\ }\href {\doibase 10.1103/PhysRevB.102.165140} {\bibfield
  {journal} {\bibinfo  {journal} {Phys. Rev. B}\ }\textbf {\bibinfo {volume}
  {102}},\ \bibinfo {pages} {165140} (\bibinfo {year} {2020})}\BibitemShut
  {NoStop}%
\bibitem [{\citenamefont {Venturelli}\ \emph {et~al.}(2022)\citenamefont
  {Venturelli}, \citenamefont {Cugliandolo}, \citenamefont {Schehr},\ and\
  \citenamefont {Tarzia}}]{venturelli2022replica}%
  \BibitemOpen
  \bibfield  {author} {\bibinfo {author} {\bibfnamefont {D.}~\bibnamefont
  {Venturelli}}, \bibinfo {author} {\bibfnamefont {L.~F.}\ \bibnamefont
  {Cugliandolo}}, \bibinfo {author} {\bibfnamefont {G.}~\bibnamefont {Schehr}},
  \ and\ \bibinfo {author} {\bibfnamefont {M.}~\bibnamefont {Tarzia}},\
  }\href@noop {} {\bibfield  {journal} {\bibinfo  {journal} {arXiv preprint
  arXiv:2209.11732}\ } (\bibinfo {year} {2022})}\BibitemShut {NoStop}%
\bibitem [{\citenamefont {Altshuler}\ \emph {et~al.}(1997)\citenamefont
  {Altshuler}, \citenamefont {Gefen}, \citenamefont {Kamenev},\ and\
  \citenamefont {Levitov}}]{altshuler1997quasiparticle}%
  \BibitemOpen
  \bibfield  {author} {\bibinfo {author} {\bibfnamefont {B.~L.}\ \bibnamefont
  {Altshuler}}, \bibinfo {author} {\bibfnamefont {Y.}~\bibnamefont {Gefen}},
  \bibinfo {author} {\bibfnamefont {A.}~\bibnamefont {Kamenev}}, \ and\
  \bibinfo {author} {\bibfnamefont {L.~S.}\ \bibnamefont {Levitov}},\ }\href
  {\doibase 10.1103/PhysRevLett.78.2803} {\bibfield  {journal} {\bibinfo
  {journal} {Phys. Rev. Lett.}\ }\textbf {\bibinfo {volume} {78}},\ \bibinfo
  {pages} {2803} (\bibinfo {year} {1997})}\BibitemShut {NoStop}%
\bibitem [{\citenamefont {Tikhonov}\ and\ \citenamefont
  {Mirlin}(2021)}]{tikhonov2021anderson}%
  \BibitemOpen
  \bibfield  {author} {\bibinfo {author} {\bibfnamefont {K.~S.}\ \bibnamefont
  {Tikhonov}}\ and\ \bibinfo {author} {\bibfnamefont {A.~D.}\ \bibnamefont
  {Mirlin}},\ }\href@noop {} {\bibfield  {journal} {\bibinfo  {journal} {Annals
  of Physics}\ }\textbf {\bibinfo {volume} {435}},\ \bibinfo {pages} {168525}
  (\bibinfo {year} {2021})}\BibitemShut {NoStop}%
\bibitem [{\citenamefont {De~Luca}\ and\ \citenamefont
  {Scardicchio}(2013)}]{de2013ergodicity}%
  \BibitemOpen
  \bibfield  {author} {\bibinfo {author} {\bibfnamefont {A.}~\bibnamefont
  {De~Luca}}\ and\ \bibinfo {author} {\bibfnamefont {A.}~\bibnamefont
  {Scardicchio}},\ }\href@noop {} {\bibfield  {journal} {\bibinfo  {journal}
  {EPL (Europhysics Letters)}\ }\textbf {\bibinfo {volume} {101}},\ \bibinfo
  {pages} {37003} (\bibinfo {year} {2013})}\BibitemShut {NoStop}%
\bibitem [{\citenamefont {Biroli}\ and\ \citenamefont
  {Tarzia}(2017)}]{biroli2017delocalized}%
  \BibitemOpen
  \bibfield  {author} {\bibinfo {author} {\bibfnamefont {G.}~\bibnamefont
  {Biroli}}\ and\ \bibinfo {author} {\bibfnamefont {M.}~\bibnamefont
  {Tarzia}},\ }\href@noop {} {\bibfield  {journal} {\bibinfo  {journal}
  {Physical Review B}\ }\textbf {\bibinfo {volume} {96}},\ \bibinfo {pages}
  {201114} (\bibinfo {year} {2017})}\BibitemShut {NoStop}%
\bibitem [{\citenamefont {Logan}\ and\ \citenamefont
  {Welsh}(2019)}]{logan2019many}%
  \BibitemOpen
  \bibfield  {author} {\bibinfo {author} {\bibfnamefont {D.~E.}\ \bibnamefont
  {Logan}}\ and\ \bibinfo {author} {\bibfnamefont {S.}~\bibnamefont {Welsh}},\
  }\href@noop {} {\bibfield  {journal} {\bibinfo  {journal} {Physical Review
  B}\ }\textbf {\bibinfo {volume} {99}},\ \bibinfo {pages} {045131} (\bibinfo
  {year} {2019})}\BibitemShut {NoStop}%
\bibitem [{\citenamefont {Biroli}\ \emph {et~al.}(2021)\citenamefont {Biroli},
  \citenamefont {Facoetti}, \citenamefont {Schir{\'o}}, \citenamefont
  {Tarzia},\ and\ \citenamefont {Vivo}}]{biroli2021out}%
  \BibitemOpen
  \bibfield  {author} {\bibinfo {author} {\bibfnamefont {G.}~\bibnamefont
  {Biroli}}, \bibinfo {author} {\bibfnamefont {D.}~\bibnamefont {Facoetti}},
  \bibinfo {author} {\bibfnamefont {M.}~\bibnamefont {Schir{\'o}}}, \bibinfo
  {author} {\bibfnamefont {M.}~\bibnamefont {Tarzia}}, \ and\ \bibinfo {author}
  {\bibfnamefont {P.}~\bibnamefont {Vivo}},\ }\href@noop {} {\bibfield
  {journal} {\bibinfo  {journal} {Physical Review B}\ }\textbf {\bibinfo
  {volume} {103}},\ \bibinfo {pages} {014204} (\bibinfo {year}
  {2021})}\BibitemShut {NoStop}%
\bibitem [{\citenamefont {Tarzia}(2020)}]{tarzia2020many}%
  \BibitemOpen
  \bibfield  {author} {\bibinfo {author} {\bibfnamefont {M.}~\bibnamefont
  {Tarzia}},\ }\href@noop {} {\bibfield  {journal} {\bibinfo  {journal}
  {Physical Review B}\ }\textbf {\bibinfo {volume} {102}},\ \bibinfo {pages}
  {014208} (\bibinfo {year} {2020})}\BibitemShut {NoStop}%
\bibitem [{\citenamefont {Faoro}\ \emph {et~al.}(2019)\citenamefont {Faoro},
  \citenamefont {Feigel’man},\ and\ \citenamefont {Ioffe}}]{faoro2019non}%
  \BibitemOpen
  \bibfield  {author} {\bibinfo {author} {\bibfnamefont {L.}~\bibnamefont
  {Faoro}}, \bibinfo {author} {\bibfnamefont {M.~V.}\ \bibnamefont
  {Feigel’man}}, \ and\ \bibinfo {author} {\bibfnamefont {L.}~\bibnamefont
  {Ioffe}},\ }\href@noop {} {\bibfield  {journal} {\bibinfo  {journal} {Annals
  of Physics}\ }\textbf {\bibinfo {volume} {409}},\ \bibinfo {pages} {167916}
  (\bibinfo {year} {2019})}\BibitemShut {NoStop}%
\bibitem [{\citenamefont {Pietracaprina}\ \emph {et~al.}(2016)\citenamefont
  {Pietracaprina}, \citenamefont {Ros},\ and\ \citenamefont
  {Scardicchio}}]{pietracaprina2016forward}%
  \BibitemOpen
  \bibfield  {author} {\bibinfo {author} {\bibfnamefont {F.}~\bibnamefont
  {Pietracaprina}}, \bibinfo {author} {\bibfnamefont {V.}~\bibnamefont {Ros}},
  \ and\ \bibinfo {author} {\bibfnamefont {A.}~\bibnamefont {Scardicchio}},\
  }\href@noop {} {\bibfield  {journal} {\bibinfo  {journal} {Physical Review
  B}\ }\textbf {\bibinfo {volume} {93}},\ \bibinfo {pages} {054201} (\bibinfo
  {year} {2016})}\BibitemShut {NoStop}%
\bibitem [{\citenamefont {Abou-Chacra}\ \emph {et~al.}(1973)\citenamefont
  {Abou-Chacra}, \citenamefont {Thouless},\ and\ \citenamefont
  {Anderson}}]{abou1973selfconsistent}%
  \BibitemOpen
  \bibfield  {author} {\bibinfo {author} {\bibfnamefont {R.}~\bibnamefont
  {Abou-Chacra}}, \bibinfo {author} {\bibfnamefont {D.}~\bibnamefont
  {Thouless}}, \ and\ \bibinfo {author} {\bibfnamefont {P.}~\bibnamefont
  {Anderson}},\ }\href@noop {} {\bibfield  {journal} {\bibinfo  {journal}
  {Journal of Physics C: Solid State Physics}\ }\textbf {\bibinfo {volume}
  {6}},\ \bibinfo {pages} {1734} (\bibinfo {year} {1973})}\BibitemShut
  {NoStop}%
\bibitem [{\citenamefont {Imbrie}(2016{\natexlab{b}})}]{imbrie2016many}%
  \BibitemOpen
  \bibfield  {author} {\bibinfo {author} {\bibfnamefont {J.~Z.}\ \bibnamefont
  {Imbrie}},\ }\href@noop {} {\bibfield  {journal} {\bibinfo  {journal}
  {Journal of Statistical Physics}\ }\textbf {\bibinfo {volume} {163}},\
  \bibinfo {pages} {998} (\bibinfo {year} {2016}{\natexlab{b}})}\BibitemShut
  {NoStop}%
\bibitem [{\citenamefont {Srednicki}(1994)}]{srednicki1994chaos}%
  \BibitemOpen
  \bibfield  {author} {\bibinfo {author} {\bibfnamefont {M.}~\bibnamefont
  {Srednicki}},\ }\href {\doibase 10.1103/PhysRevE.50.888} {\bibfield
  {journal} {\bibinfo  {journal} {Phys. Rev. E}\ }\textbf {\bibinfo {volume}
  {50}},\ \bibinfo {pages} {888} (\bibinfo {year} {1994})}\BibitemShut
  {NoStop}%
\bibitem [{\citenamefont {Rigol}\ \emph {et~al.}(2008)\citenamefont {Rigol},
  \citenamefont {Dunjko},\ and\ \citenamefont
  {Olshanii}}]{rigol2008thermalization}%
  \BibitemOpen
  \bibfield  {author} {\bibinfo {author} {\bibfnamefont {M.}~\bibnamefont
  {Rigol}}, \bibinfo {author} {\bibfnamefont {V.}~\bibnamefont {Dunjko}}, \
  and\ \bibinfo {author} {\bibfnamefont {M.}~\bibnamefont {Olshanii}},\ }\href
  {\doibase 10.1038/nature06838} {\bibfield  {journal} {\bibinfo  {journal}
  {Nature}\ }\textbf {\bibinfo {volume} {452}},\ \bibinfo {pages} {854}
  (\bibinfo {year} {2008})}\BibitemShut {NoStop}%
\bibitem [{\citenamefont {Abanin}\ \emph {et~al.}(2021)\citenamefont {Abanin},
  \citenamefont {Bardarson}, \citenamefont {De~Tomasi}, \citenamefont
  {Gopalakrishnan}, \citenamefont {Khemani}, \citenamefont {Parameswaran},
  \citenamefont {Pollmann}, \citenamefont {Potter}, \citenamefont {Serbyn},\
  and\ \citenamefont {Vasseur}}]{abanin2021distinguishing}%
  \BibitemOpen
  \bibfield  {author} {\bibinfo {author} {\bibfnamefont {D.}~\bibnamefont
  {Abanin}}, \bibinfo {author} {\bibfnamefont {J.~H.}\ \bibnamefont
  {Bardarson}}, \bibinfo {author} {\bibfnamefont {G.}~\bibnamefont
  {De~Tomasi}}, \bibinfo {author} {\bibfnamefont {S.}~\bibnamefont
  {Gopalakrishnan}}, \bibinfo {author} {\bibfnamefont {V.}~\bibnamefont
  {Khemani}}, \bibinfo {author} {\bibfnamefont {S.}~\bibnamefont
  {Parameswaran}}, \bibinfo {author} {\bibfnamefont {F.}~\bibnamefont
  {Pollmann}}, \bibinfo {author} {\bibfnamefont {A.}~\bibnamefont {Potter}},
  \bibinfo {author} {\bibfnamefont {M.}~\bibnamefont {Serbyn}}, \ and\ \bibinfo
  {author} {\bibfnamefont {R.}~\bibnamefont {Vasseur}},\ }\href@noop {}
  {\bibfield  {journal} {\bibinfo  {journal} {Annals of Physics}\ }\textbf
  {\bibinfo {volume} {427}},\ \bibinfo {pages} {168415} (\bibinfo {year}
  {2021})}\BibitemShut {NoStop}%
\bibitem [{\citenamefont {Roy}\ and\ \citenamefont
  {Logan}(2021)}]{roy2021fock}%
  \BibitemOpen
  \bibfield  {author} {\bibinfo {author} {\bibfnamefont {S.}~\bibnamefont
  {Roy}}\ and\ \bibinfo {author} {\bibfnamefont {D.~E.}\ \bibnamefont
  {Logan}},\ }\href@noop {} {\bibfield  {journal} {\bibinfo  {journal}
  {Physical Review B}\ }\textbf {\bibinfo {volume} {104}},\ \bibinfo {pages}
  {174201} (\bibinfo {year} {2021})}\BibitemShut {NoStop}%
\bibitem [{\citenamefont {Creed}\ \emph {et~al.}(2022)\citenamefont {Creed},
  \citenamefont {Logan},\ and\ \citenamefont {Roy}}]{creed2022probability}%
  \BibitemOpen
  \bibfield  {author} {\bibinfo {author} {\bibfnamefont {I.}~\bibnamefont
  {Creed}}, \bibinfo {author} {\bibfnamefont {D.~E.}\ \bibnamefont {Logan}}, \
  and\ \bibinfo {author} {\bibfnamefont {S.}~\bibnamefont {Roy}},\ }\href@noop
  {} {\bibfield  {journal} {\bibinfo  {journal} {arXiv preprint
  arXiv:2212.14333}\ } (\bibinfo {year} {2022})}\BibitemShut {NoStop}%
\bibitem [{\citenamefont {Herre}\ \emph {et~al.}(2023)\citenamefont {Herre},
  \citenamefont {Karcher}, \citenamefont {Tikhonov},\ and\ \citenamefont
  {Mirlin}}]{herre2023ergodicity}%
  \BibitemOpen
  \bibfield  {author} {\bibinfo {author} {\bibfnamefont {J.-N.}\ \bibnamefont
  {Herre}}, \bibinfo {author} {\bibfnamefont {J.~F.}\ \bibnamefont {Karcher}},
  \bibinfo {author} {\bibfnamefont {K.~S.}\ \bibnamefont {Tikhonov}}, \ and\
  \bibinfo {author} {\bibfnamefont {A.~D.}\ \bibnamefont {Mirlin}},\
  }\href@noop {} {\bibfield  {journal} {\bibinfo  {journal} {arXiv preprint
  arXiv:2302.06581}\ } (\bibinfo {year} {2023})}\BibitemShut {NoStop}%
\bibitem [{\citenamefont {Mac{\'e}}\ \emph {et~al.}(2019)\citenamefont
  {Mac{\'e}}, \citenamefont {Alet},\ and\ \citenamefont
  {Laflorencie}}]{mace2019multifractal}%
  \BibitemOpen
  \bibfield  {author} {\bibinfo {author} {\bibfnamefont {N.}~\bibnamefont
  {Mac{\'e}}}, \bibinfo {author} {\bibfnamefont {F.}~\bibnamefont {Alet}}, \
  and\ \bibinfo {author} {\bibfnamefont {N.}~\bibnamefont {Laflorencie}},\
  }\href@noop {} {\bibfield  {journal} {\bibinfo  {journal} {Physical review
  letters}\ }\textbf {\bibinfo {volume} {123}},\ \bibinfo {pages} {180601}
  (\bibinfo {year} {2019})}\BibitemShut {NoStop}%
\bibitem [{\citenamefont {Biroli}\ \emph {et~al.}(2012)\citenamefont {Biroli},
  \citenamefont {Ribeiro-Teixeira},\ and\ \citenamefont
  {Tarzia}}]{biroli2012difference}%
  \BibitemOpen
  \bibfield  {author} {\bibinfo {author} {\bibfnamefont {G.}~\bibnamefont
  {Biroli}}, \bibinfo {author} {\bibfnamefont {A.}~\bibnamefont
  {Ribeiro-Teixeira}}, \ and\ \bibinfo {author} {\bibfnamefont
  {M.}~\bibnamefont {Tarzia}},\ }\href@noop {} {\bibfield  {journal} {\bibinfo
  {journal} {arXiv preprint arXiv:1211.7334}\ } (\bibinfo {year}
  {2012})}\BibitemShut {NoStop}%
\bibitem [{\citenamefont {Altshuler}\ \emph {et~al.}(2016)\citenamefont
  {Altshuler}, \citenamefont {Cuevas}, \citenamefont {Ioffe},\ and\
  \citenamefont {Kravtsov}}]{altshuler2016nonergodic}%
  \BibitemOpen
  \bibfield  {author} {\bibinfo {author} {\bibfnamefont {B.}~\bibnamefont
  {Altshuler}}, \bibinfo {author} {\bibfnamefont {E.}~\bibnamefont {Cuevas}},
  \bibinfo {author} {\bibfnamefont {L.}~\bibnamefont {Ioffe}}, \ and\ \bibinfo
  {author} {\bibfnamefont {V.}~\bibnamefont {Kravtsov}},\ }\href@noop {}
  {\bibfield  {journal} {\bibinfo  {journal} {Physical review letters}\
  }\textbf {\bibinfo {volume} {117}},\ \bibinfo {pages} {156601} (\bibinfo
  {year} {2016})}\BibitemShut {NoStop}%
\bibitem [{\citenamefont {Tikhonov}\ \emph {et~al.}(2016)\citenamefont
  {Tikhonov}, \citenamefont {Mirlin},\ and\ \citenamefont
  {Skvortsov}}]{tikhonov2016anderson}%
  \BibitemOpen
  \bibfield  {author} {\bibinfo {author} {\bibfnamefont {K.~S.}\ \bibnamefont
  {Tikhonov}}, \bibinfo {author} {\bibfnamefont {A.~D.}\ \bibnamefont
  {Mirlin}}, \ and\ \bibinfo {author} {\bibfnamefont {M.~A.}\ \bibnamefont
  {Skvortsov}},\ }\href@noop {} {\bibfield  {journal} {\bibinfo  {journal}
  {Physical Review B}\ }\textbf {\bibinfo {volume} {94}},\ \bibinfo {pages}
  {220203} (\bibinfo {year} {2016})}\BibitemShut {NoStop}%
\bibitem [{\citenamefont {Tikhonov}\ and\ \citenamefont
  {Mirlin}(2019{\natexlab{a}})}]{Tikhonov_2019}%
  \BibitemOpen
  \bibfield  {author} {\bibinfo {author} {\bibfnamefont {K.~S.}\ \bibnamefont
  {Tikhonov}}\ and\ \bibinfo {author} {\bibfnamefont {A.~D.}\ \bibnamefont
  {Mirlin}},\ }\href {\doibase 10.1103/PhysRevB.99.024202} {\bibfield
  {journal} {\bibinfo  {journal} {Phys. Rev. B}\ }\textbf {\bibinfo {volume}
  {99}},\ \bibinfo {pages} {024202} (\bibinfo {year}
  {2019}{\natexlab{a}})}\BibitemShut {NoStop}%
\bibitem [{gar(2017)}]{garcia2017scaling}%
  \BibitemOpen
  \href@noop {} {\bibfield  {journal} {\bibinfo  {journal} {Physical review
  letters}\ }\textbf {\bibinfo {volume} {118}},\ \bibinfo {pages} {166801}
  (\bibinfo {year} {2017})}\BibitemShut {NoStop}%
\bibitem [{\citenamefont {Biroli}\ and\ \citenamefont
  {Tarzia}(2018)}]{biroli2018delocalization}%
  \BibitemOpen
  \bibfield  {author} {\bibinfo {author} {\bibfnamefont {G.}~\bibnamefont
  {Biroli}}\ and\ \bibinfo {author} {\bibfnamefont {M.}~\bibnamefont
  {Tarzia}},\ }\href@noop {} {\bibfield  {journal} {\bibinfo  {journal} {arXiv
  preprint arXiv:1810.07545}\ } (\bibinfo {year} {2018})}\BibitemShut {NoStop}%
\bibitem [{gar(2022)}]{garcia2022critical}%
  \BibitemOpen
  \href@noop {} {\bibfield  {journal} {\bibinfo  {journal} {Physical Review B}\
  }\textbf {\bibinfo {volume} {106}},\ \bibinfo {pages} {214202} (\bibinfo
  {year} {2022})}\BibitemShut {NoStop}%
\bibitem [{\citenamefont {Vanoni}\ \emph {et~al.}(2023)\citenamefont {Vanoni},
  \citenamefont {Altshuler}, \citenamefont {Kravtsov},\ and\ \citenamefont
  {Scardicchio}}]{vanoni2023renormalization}%
  \BibitemOpen
  \bibfield  {author} {\bibinfo {author} {\bibfnamefont {C.}~\bibnamefont
  {Vanoni}}, \bibinfo {author} {\bibfnamefont {B.~L.}\ \bibnamefont
  {Altshuler}}, \bibinfo {author} {\bibfnamefont {V.~E.}\ \bibnamefont
  {Kravtsov}}, \ and\ \bibinfo {author} {\bibfnamefont {A.}~\bibnamefont
  {Scardicchio}},\ }\href@noop {} {\bibfield  {journal} {\bibinfo  {journal}
  {arXiv preprint arXiv:2306.14965}\ } (\bibinfo {year} {2023})}\BibitemShut
  {NoStop}%
\bibitem [{\citenamefont {Derrida}(1980)}]{derrida1980random}%
  \BibitemOpen
  \bibfield  {author} {\bibinfo {author} {\bibfnamefont {B.}~\bibnamefont
  {Derrida}},\ }\href@noop {} {\bibfield  {journal} {\bibinfo  {journal}
  {Physical Review Letters}\ }\textbf {\bibinfo {volume} {45}},\ \bibinfo
  {pages} {79} (\bibinfo {year} {1980})}\BibitemShut {NoStop}%
\bibitem [{\citenamefont {M{\'e}zard}\ \emph {et~al.}(1987)\citenamefont
  {M{\'e}zard}, \citenamefont {Parisi},\ and\ \citenamefont
  {Virasoro}}]{mezard1987spin}%
  \BibitemOpen
  \bibfield  {author} {\bibinfo {author} {\bibfnamefont {M.}~\bibnamefont
  {M{\'e}zard}}, \bibinfo {author} {\bibfnamefont {G.}~\bibnamefont {Parisi}},
  \ and\ \bibinfo {author} {\bibfnamefont {M.~A.}\ \bibnamefont {Virasoro}},\
  }\href@noop {} {\emph {\bibinfo {title} {Spin glass theory and beyond: An
  Introduction to the Replica Method and Its Applications}}},\ Vol.~\bibinfo
  {volume} {9}\ (\bibinfo  {publisher} {World Scientific Publishing Company},\
  \bibinfo {year} {1987})\BibitemShut {NoStop}%
\bibitem [{\citenamefont {Kolmogorov}\ \emph {et~al.}(1937)\citenamefont
  {Kolmogorov}, \citenamefont {Petrovsky},\ and\ \citenamefont
  {Piskounov}}]{kpp1937}%
  \BibitemOpen
  \bibfield  {author} {\bibinfo {author} {\bibfnamefont {A.}~\bibnamefont
  {Kolmogorov}}, \bibinfo {author} {\bibfnamefont {I.}~\bibnamefont
  {Petrovsky}}, \ and\ \bibinfo {author} {\bibfnamefont {B.}~\bibnamefont
  {Piskounov}},\ }\href@noop {} {\bibfield  {journal} {\bibinfo  {journal}
  {Moscow Univ. Bull. Math.}\ }\textbf {\bibinfo {volume} {1}},\ \bibinfo
  {pages} {1} (\bibinfo {year} {1937})}\BibitemShut {NoStop}%
\bibitem [{\citenamefont {Parisi}\ \emph {et~al.}(2019)\citenamefont {Parisi},
  \citenamefont {Pascazio}, \citenamefont {Pietracaprina}, \citenamefont
  {Ros},\ and\ \citenamefont {Scardicchio}}]{parisi2019anderson}%
  \BibitemOpen
  \bibfield  {author} {\bibinfo {author} {\bibfnamefont {G.}~\bibnamefont
  {Parisi}}, \bibinfo {author} {\bibfnamefont {S.}~\bibnamefont {Pascazio}},
  \bibinfo {author} {\bibfnamefont {F.}~\bibnamefont {Pietracaprina}}, \bibinfo
  {author} {\bibfnamefont {V.}~\bibnamefont {Ros}}, \ and\ \bibinfo {author}
  {\bibfnamefont {A.}~\bibnamefont {Scardicchio}},\ }\href@noop {} {\bibfield
  {journal} {\bibinfo  {journal} {Journal of Physics A: Mathematical and
  Theoretical}\ }\textbf {\bibinfo {volume} {53}},\ \bibinfo {pages} {014003}
  (\bibinfo {year} {2019})}\BibitemShut {NoStop}%
\bibitem [{\citenamefont {Fisher}\ and\ \citenamefont
  {Lee}(1981)}]{fisher1981relation}%
  \BibitemOpen
  \bibfield  {author} {\bibinfo {author} {\bibfnamefont {D.~S.}\ \bibnamefont
  {Fisher}}\ and\ \bibinfo {author} {\bibfnamefont {P.~A.}\ \bibnamefont
  {Lee}},\ }\href@noop {} {\bibfield  {journal} {\bibinfo  {journal} {Physical
  Review B}\ }\textbf {\bibinfo {volume} {23}},\ \bibinfo {pages} {6851}
  (\bibinfo {year} {1981})}\BibitemShut {NoStop}%
\bibitem [{\citenamefont {Miller}\ and\ \citenamefont
  {Derridda}(1994)}]{miller1994weak}%
  \BibitemOpen
  \bibfield  {author} {\bibinfo {author} {\bibfnamefont {J.~D.}\ \bibnamefont
  {Miller}}\ and\ \bibinfo {author} {\bibfnamefont {B.}~\bibnamefont
  {Derridda}},\ }\href@noop {} {\bibfield  {journal} {\bibinfo  {journal}
  {Journal of statistical physics}\ }\textbf {\bibinfo {volume} {75}},\
  \bibinfo {pages} {357} (\bibinfo {year} {1994})}\BibitemShut {NoStop}%
\bibitem [{\citenamefont {Feigel’man}\ \emph {et~al.}(2010)\citenamefont
  {Feigel’man}, \citenamefont {Ioffe},\ and\ \citenamefont
  {M{\'e}zard}}]{feigel2010superconductor}%
  \BibitemOpen
  \bibfield  {author} {\bibinfo {author} {\bibfnamefont {M.}~\bibnamefont
  {Feigel’man}}, \bibinfo {author} {\bibfnamefont {L.}~\bibnamefont {Ioffe}},
  \ and\ \bibinfo {author} {\bibfnamefont {M.}~\bibnamefont {M{\'e}zard}},\
  }\href@noop {} {\bibfield  {journal} {\bibinfo  {journal} {Physical Review
  B}\ }\textbf {\bibinfo {volume} {82}},\ \bibinfo {pages} {184534} (\bibinfo
  {year} {2010})}\BibitemShut {NoStop}%
\bibitem [{\citenamefont {Ioffe}\ and\ \citenamefont
  {M{\'e}zard}(2010)}]{ioffe2010disorder}%
  \BibitemOpen
  \bibfield  {author} {\bibinfo {author} {\bibfnamefont {L.}~\bibnamefont
  {Ioffe}}\ and\ \bibinfo {author} {\bibfnamefont {M.}~\bibnamefont
  {M{\'e}zard}},\ }\href@noop {} {\bibfield  {journal} {\bibinfo  {journal}
  {Physical review letters}\ }\textbf {\bibinfo {volume} {105}},\ \bibinfo
  {pages} {037001} (\bibinfo {year} {2010})}\BibitemShut {NoStop}%
\bibitem [{\citenamefont {Chakrabarti}\ \emph {et~al.}(2022)\citenamefont
  {Chakrabarti}, \citenamefont {Martins}, \citenamefont {Laflorencie},
  \citenamefont {Georgeot}, \citenamefont {Brunet},\ and\ \citenamefont
  {Lemari{\'e}}}]{chakrabarti2022traveling}%
  \BibitemOpen
  \bibfield  {author} {\bibinfo {author} {\bibfnamefont {A.}~\bibnamefont
  {Chakrabarti}}, \bibinfo {author} {\bibfnamefont {C.}~\bibnamefont
  {Martins}}, \bibinfo {author} {\bibfnamefont {N.}~\bibnamefont
  {Laflorencie}}, \bibinfo {author} {\bibfnamefont {B.}~\bibnamefont
  {Georgeot}}, \bibinfo {author} {\bibfnamefont {{\'E}.}~\bibnamefont
  {Brunet}}, \ and\ \bibinfo {author} {\bibfnamefont {G.}~\bibnamefont
  {Lemari{\'e}}},\ }\href@noop {} {\bibfield  {journal} {\bibinfo  {journal}
  {arXiv preprint arXiv:2212.13593}\ } (\bibinfo {year} {2022})}\BibitemShut
  {NoStop}%
\bibitem [{\citenamefont {Ros}\ and\ \citenamefont
  {M{\"u}ller}(2021)}]{ros2021fluctuation}%
  \BibitemOpen
  \bibfield  {author} {\bibinfo {author} {\bibfnamefont {V.}~\bibnamefont
  {Ros}}\ and\ \bibinfo {author} {\bibfnamefont {M.}~\bibnamefont
  {M{\"u}ller}},\ }\href@noop {} {\bibfield  {journal} {\bibinfo  {journal}
  {Physical Review B}\ }\textbf {\bibinfo {volume} {104}},\ \bibinfo {pages}
  {094205} (\bibinfo {year} {2021})}\BibitemShut {NoStop}%
\bibitem [{\citenamefont {Tikhonov}\ and\ \citenamefont
  {Mirlin}(2019{\natexlab{b}})}]{tikhonov2019critical}%
  \BibitemOpen
  \bibfield  {author} {\bibinfo {author} {\bibfnamefont {K.}~\bibnamefont
  {Tikhonov}}\ and\ \bibinfo {author} {\bibfnamefont {A.}~\bibnamefont
  {Mirlin}},\ }\href@noop {} {\bibfield  {journal} {\bibinfo  {journal}
  {Physical Review B}\ }\textbf {\bibinfo {volume} {99}},\ \bibinfo {pages}
  {214202} (\bibinfo {year} {2019}{\natexlab{b}})}\BibitemShut {NoStop}%
\bibitem [{\citenamefont {Ostilli}\ \emph {et~al.}(2022)\citenamefont
  {Ostilli}, \citenamefont {Bezerra},\ and\ \citenamefont
  {Viswanathan}}]{ostilli2022spectrum}%
  \BibitemOpen
  \bibfield  {author} {\bibinfo {author} {\bibfnamefont {M.}~\bibnamefont
  {Ostilli}}, \bibinfo {author} {\bibfnamefont {C.~G.}\ \bibnamefont
  {Bezerra}}, \ and\ \bibinfo {author} {\bibfnamefont {G.}~\bibnamefont
  {Viswanathan}},\ }\href@noop {} {\bibfield  {journal} {\bibinfo  {journal}
  {Physical Review E}\ }\textbf {\bibinfo {volume} {105}},\ \bibinfo {pages}
  {034123} (\bibinfo {year} {2022})}\BibitemShut {NoStop}%
\bibitem [{\citenamefont {Sade}\ and\ \citenamefont
  {Berkovits}(2003)}]{sade2003localization}%
  \BibitemOpen
  \bibfield  {author} {\bibinfo {author} {\bibfnamefont {M.}~\bibnamefont
  {Sade}}\ and\ \bibinfo {author} {\bibfnamefont {R.}~\bibnamefont
  {Berkovits}},\ }\href@noop {} {\bibfield  {journal} {\bibinfo  {journal}
  {Physical Review B}\ }\textbf {\bibinfo {volume} {68}},\ \bibinfo {pages}
  {193102} (\bibinfo {year} {2003})}\BibitemShut {NoStop}%
\bibitem [{\citenamefont {Derrida}\ \emph {et~al.}(1993)\citenamefont
  {Derrida}, \citenamefont {Evans},\ and\ \citenamefont
  {Speer}}]{derrida1993mean}%
  \BibitemOpen
  \bibfield  {author} {\bibinfo {author} {\bibfnamefont {B.}~\bibnamefont
  {Derrida}}, \bibinfo {author} {\bibfnamefont {M.~R.}\ \bibnamefont {Evans}},
  \ and\ \bibinfo {author} {\bibfnamefont {E.~R.}\ \bibnamefont {Speer}},\
  }\href@noop {} {\bibfield  {journal} {\bibinfo  {journal} {Communications in
  mathematical physics}\ }\textbf {\bibinfo {volume} {156}},\ \bibinfo {pages}
  {221} (\bibinfo {year} {1993})}\BibitemShut {NoStop}%
\bibitem [{\citenamefont {De~Tomasi}\ and\ \citenamefont
  {Khaymovich}(2020)}]{de2020multifractality}%
  \BibitemOpen
  \bibfield  {author} {\bibinfo {author} {\bibfnamefont {G.}~\bibnamefont
  {De~Tomasi}}\ and\ \bibinfo {author} {\bibfnamefont {I.~M.}\ \bibnamefont
  {Khaymovich}},\ }\href@noop {} {\bibfield  {journal} {\bibinfo  {journal}
  {Physical Review Letters}\ }\textbf {\bibinfo {volume} {124}},\ \bibinfo
  {pages} {200602} (\bibinfo {year} {2020})}\BibitemShut {NoStop}%
\bibitem [{\citenamefont {Oganesyan}\ and\ \citenamefont
  {Huse}(2007)}]{oganesyan2007localization}%
  \BibitemOpen
  \bibfield  {author} {\bibinfo {author} {\bibfnamefont {V.}~\bibnamefont
  {Oganesyan}}\ and\ \bibinfo {author} {\bibfnamefont {D.~A.}\ \bibnamefont
  {Huse}},\ }\href@noop {} {\bibfield  {journal} {\bibinfo  {journal} {Physical
  review b}\ }\textbf {\bibinfo {volume} {75}},\ \bibinfo {pages} {155111}
  (\bibinfo {year} {2007})}\BibitemShut {NoStop}%
\bibitem [{\citenamefont {Baldwin}\ and\ \citenamefont
  {Laumann}(2018)}]{baldwin2018quantum}%
  \BibitemOpen
  \bibfield  {author} {\bibinfo {author} {\bibfnamefont {C.}~\bibnamefont
  {Baldwin}}\ and\ \bibinfo {author} {\bibfnamefont {C.}~\bibnamefont
  {Laumann}},\ }\href@noop {} {\bibfield  {journal} {\bibinfo  {journal}
  {Physical Review B}\ }\textbf {\bibinfo {volume} {97}},\ \bibinfo {pages}
  {224201} (\bibinfo {year} {2018})}\BibitemShut {NoStop}%
\bibitem [{\citenamefont {Smelyanskiy}\ \emph {et~al.}(2019)\citenamefont
  {Smelyanskiy}, \citenamefont {Kechedzhi}, \citenamefont {Boixo},
  \citenamefont {Neven},\ and\ \citenamefont
  {Altshuler}}]{smelyanskiy2019intermittency}%
  \BibitemOpen
  \bibfield  {author} {\bibinfo {author} {\bibfnamefont {V.~N.}\ \bibnamefont
  {Smelyanskiy}}, \bibinfo {author} {\bibfnamefont {K.}~\bibnamefont
  {Kechedzhi}}, \bibinfo {author} {\bibfnamefont {S.}~\bibnamefont {Boixo}},
  \bibinfo {author} {\bibfnamefont {H.}~\bibnamefont {Neven}}, \ and\ \bibinfo
  {author} {\bibfnamefont {B.}~\bibnamefont {Altshuler}},\ }\href@noop {}
  {\bibfield  {journal} {\bibinfo  {journal} {arXiv preprint arXiv:1907.01609}\
  } (\bibinfo {year} {2019})}\BibitemShut {NoStop}%
\bibitem [{\citenamefont {Parolini}\ and\ \citenamefont
  {Mossi}(2020)}]{parolini2020multifractal}%
  \BibitemOpen
  \bibfield  {author} {\bibinfo {author} {\bibfnamefont {T.}~\bibnamefont
  {Parolini}}\ and\ \bibinfo {author} {\bibfnamefont {G.}~\bibnamefont
  {Mossi}},\ }\href@noop {} {\bibfield  {journal} {\bibinfo  {journal} {arXiv
  preprint arXiv:2007.00315}\ } (\bibinfo {year} {2020})}\BibitemShut {NoStop}%
\bibitem [{\citenamefont {M{\'e}zard}\ and\ \citenamefont
  {Parisi}(2001)}]{mezard2001bethe}%
  \BibitemOpen
  \bibfield  {author} {\bibinfo {author} {\bibfnamefont {M.}~\bibnamefont
  {M{\'e}zard}}\ and\ \bibinfo {author} {\bibfnamefont {G.}~\bibnamefont
  {Parisi}},\ }\href@noop {} {\bibfield  {journal} {\bibinfo  {journal} {The
  European Physical Journal B-Condensed Matter and Complex Systems}\ }\textbf
  {\bibinfo {volume} {20}},\ \bibinfo {pages} {217} (\bibinfo {year}
  {2001})}\BibitemShut {NoStop}%
\end{thebibliography}%

\end{document}